\documentstyle[epsfig,12pt]{article}
\newlength{\pubnumber} 

\catcode`\@=11
\@addtoreset{equation}{section}

\def\section{\@startsection{section}{1}{\z@}{3.5ex plus 1ex minus .2ex}
 {2.3ex plus .2ex}{\large\bf}}
\def\subsection{\@startsection{subsection}{2}{\z@}{2.3ex plus .2ex}
 {2.3ex plus .2ex}{\bf}}

\newcommand{\IR}[1]{\rotatebox{90}{\ #1}}
\def\beq{\begin{equation}}
\def\eeq{\end{equation}}
\def\beqn{\begin{eqnarray}}

\def\eeqn{\end{eqnarray}}

\def\IZ{\relax{\bf Z}}\def\IC{\relax{\bf C}}
\def\IR{\relax{\rm I\kern-.18em R}}

\def\abstract#1{
\vskip .5in\vfil\centerline
{\bf Abstract}\penalty1000
{{\smallskip\ifx\answ\bigans\leftskip 2pc \rightskip 2pc
\else\leftskip 5pc \rightskip 5pc\fi
\noindent\abstractfont \baselineskip=12pt
{#1} \smallskip}}
\penalty-1000}
\def\us#1{\underline{#1}}
\def\hth/#1#2#3#4#5#6#7{{\tt hep-th/#1#2#3#4#5#6#7}}
\def\nup#1({Nucl.\ Phys.\ $\us {B#1}$\ (}
\def\plt#1({Phys.\ Lett.\ $\us  {B#1}$\ (}
\def\cmp#1({Commun.\ Math.\ Phys.\ $\us  {#1}$\ (}
\def\prp#1({Phys.\ Rep.\ $\us  {#1}$\ (}
\def\prl#1({Phys.\ Rev.\ Lett.\ $\us  {#1}$\ (}
\def\prv#1({Phys.\ Rev.\ $\us  {#1}$\ (}
\def\mpl#1({Mod.\ Phys.\ Let.\ $\us  {A#1}$\ (}
\def\ijmp#1({Int.\ J.\ Mod.\ Phys.\ $\us{A#1}$\ (}

\def\IZ{\relax{\bf Z}}\def\IC{\relax{\bf C}}
\def\IR{\relax{\rm I\kern-.18em R}}

\def\abstract#1{
\vskip .5in\vfil\centerline
{\bf Abstract}\penalty1000
{{\smallskip\ifx\answ\bigans\leftskip 2pc \rightskip 2pc
\else\leftskip 5pc \rightskip 5pc\fi
\noindent\abstractfont \baselineskip=12pt
{#1} \smallskip}}
\penalty-1000}
\def\us#1{\underline{#1}}
\def\hth/#1#2#3#4#5#6#7{{\tt hep-th/#1#2#3#4#5#6#7}}
\def\nup#1({Nucl.\ Phys.\ $\us {B#1}$\ (}
\def\plt#1({Phys.\ Lett.\ $\us  {B#1}$\ (}
\def\cmp#1({Commun.\ Math.\ Phys.\ $\us  {#1}$\ (}
\def\prp#1({Phys.\ Rep.\ $\us  {#1}$\ (}
\def\prl#1({Phys.\ Rev.\ Lett.\ $\us  {#1}$\ (}
\def\prv#1({Phys.\ Rev.\ $\us  {#1}$\ (}
\def\mpl#1({Mod.\ Phys.\ Let.\ $\us  {A#1}$\ (}
\def\ijmp#1({Int.\ J.\ Mod.\ Phys.\ $\us{A#1}$\ (}

\def\inbar{\,\vrule height1.5ex width.4pt depth0pt}

\def\IC{\relax\hbox{$\inbar\kern-.3em{\rm C}$}}
\def\IQ{\relax\hbox{$\inbar\kern-.3em{\rm Q}$}}
\def\IR{\relax{\rm I\kern-.18em R}}
 \font\cmss=cmss10 \font\cmsss=cmss10 at 7pt
\def\IZ{\relax\ifmmode\mathchoice
 {\hbox{\cmss Z\kern-.4em Z}}{\hbox{\cmss Z\kern-.4em Z}}
 {\lower.9pt\hbox{\cmsss Z\kern-.4em Z}}
 {\lower1.2pt\hbox{\cmsss Z\kern-.4em Z}}\else{\cmss Z\kern-.4em Z}\fi}

\begin{document}
\begin{titlepage}

\samepage{
\setcounter{page}{1}
\rightline{CERN--TH--2000/049}
\rightline{ACT--3/2K}
\rightline{CPT--TAMU--05/2K}
\rightline{\tt hep-th/0002102}
\rightline{February 2000}
\vfill

\begin{center}
 {\Large \bf Towards an Algebraic Classification }\\
\vspace{.05in}
{\Large \bf of 
Calabi-Yau Manifolds }\\
\vspace{.05in}
{\Large \bf I: Study of K3 Spaces}\\ 
\vspace{.15in}

\vfill
 \vspace{.15in}
 {\large F. Anselmo$^{1}$, J. Ellis$^{2}$, D.V. Nanopoulos$^{3}$
   $\,$and$\,$ G. Volkov$^{4}$\\}
 \vspace{.25in}
 {\it $^{1}$ INFN-Bologna, Bologna, Italy\\}
 \vspace{.05in}
 {\it  $^{2}$ Theory Division, CERN, CH-1211 Geneva, Switzerland \\}
 \vspace{.05in}
 {\it  $^{3}$ Dept. of Physics,
 Texas A \& M University, College Station, TX~77843-4242, USA,  \\
 HARC, The Mitchell Campus, Woodlands, TX~77381, USA, and \\
 Academy of Athens, 28~Panepistimiou Avenue,
 Athens 10679, Greece\\}
 \vspace{.05in}
 {\it  $^{4}$Theory Division, CERN, CH-1211 Geneva, Switzerland, and \\
 Theory Division, Institute for High-Energy Physics, Protvino, Russia\\}

\vspace{.25in}

{\bf Abstract}


\end{center}


 We present an inductive algebraic approach to the systematic construction
and classification of generalized Calabi-Yau (CY)  manifolds in different
numbers of complex dimensions, based on Batyrev's formulation of CY
manifolds as toric varieties in weighted complex projective spaces
associated with reflexive polyhedra. We show how the allowed weight
vectors in lower dimensions may be extended to higher dimensions,
emphasizing the roles of projection and intersection in their dual
description, and the natural appearance of Cartan-Lie algebra structures. 
The 50 allowed extended four-dimensional vectors may be combined in pairs
(triples) to form 22 (4) chains containing 90 (91) $K3$ spaces, of which
94 are distinct, and one further $K3$ space is found using duality. In the
case of $CY_3$ spaces, pairs (triples) of the 10~270 allowed extended
vectors yield 4242 (259) chains with $K3$ (elliptic) fibers containing 730
additional $K3$ polyhedra. A more complete study of $CY_3$ spaces is left
for later work. 



\vfill
\smallskip }

\end{titlepage}
\tableofcontents
%
%
%

\section{Introduction}

One of the outstanding issues in both string theory and phenomenology
is the choice of vacuum. Recent dramatic advances in the 
non-perturbative understanding of strings
have demonstrated that all string
theories, thought previously to be distinct, are in fact 
related by various dualities, and can be regarded as different phases of
a single underlying theory, called variously $M$ and/or $F$
theory~\cite{MFreview}.
This deeper non-perturbative understanding does not alter the
fact that
many classical string vacua appear equally consistent at the
perturbative level. However, the new non-perturbative methods may provide
us with new tools to understand transitions between 
these classical vacua, and perhaps eventually provide a
dynamical criterion for deciding which vacuum is preferred 
physically \cite{rev1,rev3}.

Consistent string vacua are constrained by
the principles of quantum mechanics applied 
to extended objects. At the classical level, these are
expressed in the conformal symmetry of the supersymmetric
world-sheet field theory.
Consistent 
quantization of the string must confront a possible anomaly in
conformal symmetry, as manifested in a net non-zero central charge of the
Virasoro algebra. 
Early studies of the quantum 
mechanics of extended objects indicated that strings could not survive in
the familiar dimension $D = 3+1$ of our space-time. 
The way initially used to cancel the conformal
anomaly was to choose appropriately
the dimension of the ambient 
space-time,
for example, 
${D=25+1}$ for bosonic strings and ${D=9+1}$ for
the supersymmetric and heterotic strings.

This suggested that the surplus
${n=6}$ real dimensions should be compactified.
The simplest possibility is on a
Calabi-Yau manifold~\cite{CY}, which is defined by the following
conditions:
\begin{itemize}
\item{It has a complex structure, with $N = 3$ complex dimensions
required for the $D = 9 + 1 \rightarrow 3 + 1$ case of most
direct interest, though all the cases $N = 1,2,3,4,..$ have some
interest.}
\item{It is compact.}
\item{It has a K\"ahler structure.}
\item{It has holonomy group $SU(n)$ or $Sp(n)$, e.g., $SU(3)$ in the
$N = 3$ case.}
\end{itemize}
It has subsequently been realized that one could compactify on
an orbifold~\cite{orbifold}, rather than a manifold, and also that
generalized
heterotic strings could be formulated directly in $D = 3 + 1$
dimensions, with extra world-sheet degrees of freedom
replacing the surplus space coordinates. More recently, the
non-perturbative formulation of the theory in eleven or twelve
dimensions, as $M$ or $F$ theory, has opened up new
possibilities~\cite{othermodels}.
However, Calabi-Yau compactifications continue to play a key role
in the search for realistic four-dimensional string models,
motivating us to revisit their classification.

One of the most important tools in the investigation of such
complex manifolds is the feature that
their singularities are connected with the structure of Lie algebras.
Kaluza was the first to attempt to understand this circumstance, 
and used this idea to embark on the unification of all the
gauge interactions known at that time, namely electromagnetism and 
gravitation. These ideas were subsequently extended to non-Abelian
gauge theories, and string theory can be regarded as the latest
stage in the evolution of this programme.

The three-complex-dimensional CY manifolds can be situated
in a sequence of complex spaces of increasing dimensions: two-real-
(one-complex-)dimensional tori $T_2$, the two-complex-dimensional
$K3$ spaces, the three-complex-dimensional $CY_3$ themselves,
four-complex-dimensional $CY_4$, etc., whose topological
structure and classification become progressively more complicated.
Their topologies may be described by the Betti-Hodge numbers 
which count the numbers of
distinct one-,  two-,
three-dimensional, ... cycles (holes,...). 
The topological data of the different CY manifolds
determine their physical properties, such as the different numbers
of generations $N_g$ (which are related to the Euler
characteristics of $CY_3$ spaces), etc.. This emphasizes the desirability
of
approaching systematically the problem of their classification
and the relations between, e.g., $CY_3$ manifolds with different values 
of the Euler characteristic and hence the number of generations $N_g$.
Since some non-perturbative tools now exist for studying transitions
between different CY manifolds, one could hope eventually
to find some dynamical criterion for determining $N_g$.

The topologies and classification of the
lower-dimensional spaces in this sequence are better known:
although our ultimate objective is deeper understanding of
$CY_3$ spaces, in this paper we study as a warm-up problem
the simpler case of the two-complex-dimensional ${K3}$ hypersurfaces.
These are of considerable interest in their own right, 
since, for example, they may
appear as fibrations of higher-dimensional $CY_n$ spaces.
It is well known  that any two $K3$ spaces
are diffeomorphic to each other. This can be seen, for example, by using
the polyhedron techniques of Batyrev~\cite{Batyrev} discussed in Sections
2 and 3, to
calculate
the Betti-Hodge 
invariants for all the {$K3$} hypersurfaces corresponding
to the ${\vec {k}_4}$ vectors we find. It is easy to check that Batyrev's
results yield the same Euler number 24 for all $K3$ manifolds \cite{K3}.

The quasi-homogeneous polynomial equations 
(hereafter called CY equations) whose zeroes define
the CY spaces as hypersurfaces in complex projective space
are defined (\ref{quasihom}, \ref{poly1}, \ref{poly2}, \ref{poly3}) by
projective vectors $\vec{k}$, whose components specify 
the exponents of the polynomials. The 
number of CY manifolds
is large but finite, as follows from the property of reflexivity
introduced in Section 2. The central problem in the understanding of 
classifying these manifolds may be expressed as
that of understanding the set of possible projective vectors 
${\vec {k}=(k_1, \ldots , k_{n+1})}$, the corresponding Lie algebras  
and their representations.  More precisely, the classification of all
CY manifolds contains the following problems: \\
$\bullet$
{To study the structure of the ${K3, CY_3, ...}$
projective vectors ${\vec{k_n}}$, in particular, to find the links with the
projective  vectors of lower dimensions: $D = n-1, n-2,...$.} \\
$\bullet$
{To establish the web of connections between all the 
projective vectors ${\vec {k_n}}$ of the same dimension.} \\
$\bullet$
{To find an algebraic description of the geometrical 
structure for all projective vectors, and 
calculate the corresponding Betti-Hodge invariants.} \\
$\bullet$
{To establish the connections between the projective vectors ${\vec
{k_n}}$, 
the singularities of the corresponding CY 
hypersurfaces, the gauge groups and their matter representations,
such as the number of generations, $N_g$.} \\
$\bullet$
{To study the duality symmetries and hypermodular
transformations of the projective vectors ${\vec{k_n}}$.}

In addition to the topological properties and
gauge symmetries already mentioned, it is now well known that string vacua
may be related by duality symmetries. This feature is familiar even from
simple compactifications on $S_1$ spaces of radius $R$, which revealed a 
symmetry:
$R \rightarrow 1/R$~\cite{duality}. In the case of compactifications on
tori,
there are known to be
${S}, {T}$, and ${U}$ dualities that interrelate
five string theories and play key roles in the formulations of
$M$ and $F$ theories~\cite{STUdualities}. Compactifications
on different types of CY manifolds have also been used extensively
in verifying these string dualities~\cite{STUdualities}. For example, in
proving the duality between type-IIA and type-IIB string theories,
essential use was made of the very important observation
that all CY manifolds have mirror partners 
\cite{Dix,Ler1,can5,Roan,Batyrev,Asp2}.
Thus, duality in string theory found its origins in 
a duality of complex geometry.
  
Further information about string/$M$/$F$ theory and its compactifications on
CY manifolds can be obtained using the methods of {\it toric  geometry}. 
The set of homogeneous polynomials of degree $d$ in the complex projective
space ${CP^n}$ defined 
by the vector $\vec{k_{n+1}}$ with ${d=k_1+...k_{n+1}}$ defines a 
convex {\it reflexive} polyhedron~\footnote{The
notion of a reflexive polyhedron is introduced and defined in
Section 2.}, whose intersection with the integer 
lattice corresponds to the polynomials of the CY equation.
Therefore, instead of studying the complex hypersurfaces 
directly, one can study the 
geometry of polyhedrons. This method was first used to look for the 
solutions of the algebraic equations of degree five or more in terms of 
radicals~\cite{radicals}. 
Thus, the problem of classifying CY hypersurfaces is also connected with the 
problem of solving high-degree polynomial 
equations in terms of radicals.
The solutions of quintic- and higher-degree algebraic equations 
in terms of radicals may be expressed using elliptic and 
hyperelliptic functions, respectively.
Specifically, it is known that CY manifolds may be represented using 
double-periodic elliptic or multi-periodic hyperelliptic
functions~\cite{elliptics}. 
 These functions have therefore been used 
to describe the behaviour of strings, and they should also be used to 
construct the ambient space-time in which strings  move.

\begin{figure}[th!]
   \begin{center}
   \mbox{
   \epsfig{figure=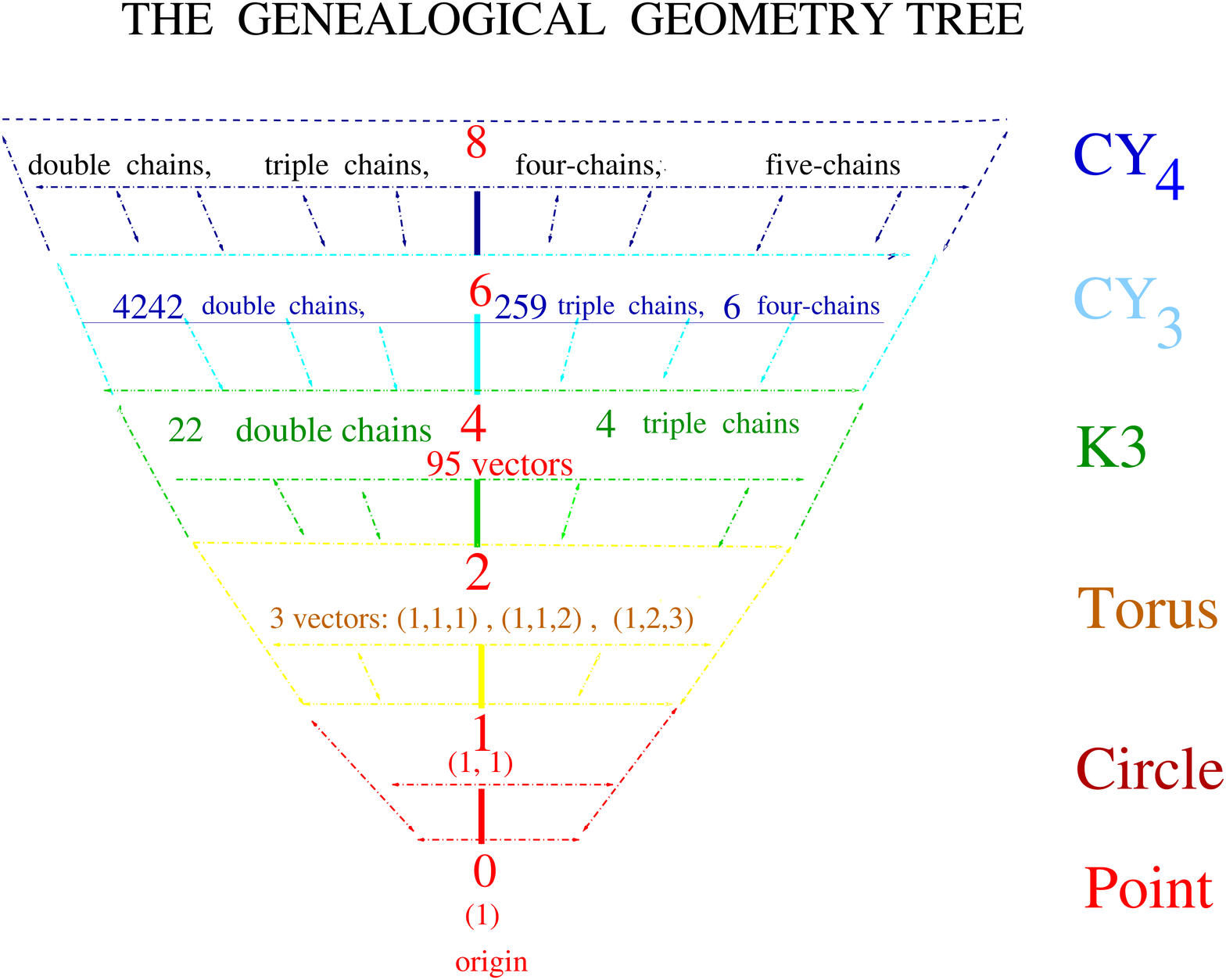,height=16cm,width=16cm}}
   \end{center}
   \caption{\it The genealogical tree of reflexive projective vectors
in different dimensions up to $d=4$.}
\label{gen1}
\end{figure}

We embark here on a systematic classification of
$K3$ manifolds, as a prelude to a subsequent classification of
$CY_3$ manifolds, based on their
construction in the framework of toric geometry. Within this approach, 
CY manifolds and their mirrors are toric varieties that can
be associated with polyhedra in spaces of various dimensions.
We propose here an inductive algebraic-geometric construction of the
projective vectors ${\vec k}$ that define these polyhedra and the related
$K3$ and CY spaces. This method has the potential to become exhaustive up
to any desired complex dimensionality $d =1,2,3,4,5,6,...$ 
(see Figure~\ref{gen1}), limited
essentially by the available computer power.
As a first step in this programme, we present 
in this article a construction of
$K3$ spaces, which is complete for those described by
simple polynomial zeroes, and
in principle for $K3$ spaces obtained as the complete intersections of
pairs or triples of such polynomial zero loci. 
In the construction of projective vectors
corresponding to hypersurfaces without an intersection with
one internal point, the duality between a
complex manifold and its mirror (which does contain
an intersection) plays an important role.
We discuss here also aspects of the $CY_3$ construction that are
relevant for the classification of $K3$ spaces. We also
indicate already how one may generate
${CY_3}$ manifolds with elliptic fibrations or
$K3$ fibers. More aspects of our $CY_3$ construction are left for
later work.

To get the flavour of our construction, which is based on the
formalism reviewed in Sections 2 and 3~\cite{Batyrev}, and is discussed in
more detail in Sections 4 {\it et seq.}, consider first $CP^1$ space.
Starting from the trivial unit `vector' ${\vec{k}_1 \equiv (1)}$,
we introduce two {\it singly-extended} basic vectors
\begin{equation}
\vec {k}^{ex'}_1= (0,1), \; \vec {k}^{ex''}_1 = (1,0),
\label{CP1basics}
\end{equation}
obtained by combining ${\vec{k}_1}$ with zero in the two
possible ways. The basic vectors (\ref{CP1basics})
correspond to the sets of polynomials
\begin{eqnarray}
x^n \cdot y \,\,&\Longrightarrow &\,\,
\{\,\vec{\mu}_1\,\}\,=\,(n,\,1):\,\,\,\,\,\,
\vec{\mu}_1 \cdot \vec{k}^{ex'}_{1}\,\,=\,d\,=\,1, \nonumber\\
x \cdot y^m \,\,&\Longrightarrow &\,\,
\{\,\vec{\mu}_2\,\}\,=\,(1,\,m):\,\,\,\,\,\,   
\vec{\mu}_2 \cdot \vec{k}^{ex''}_{1}\,=d\,=\,1,
\end{eqnarray}
respectively.
The only polynomial common to these two sequences is $xy$, which
may be considered as corresponding to the trivial `vector'
${\vec k}_1 = (1)$.
Consider now the composite vector ${\vec{k}_2 = (1,1)}$, 
which can be constructed out of the basic vectors
(\ref{CP1basics}), and is easily seen to
correspond the following three monomials of two complex
arguments $(x,y)$:
\begin{eqnarray}
\{\,x^2, x \cdot y,\, y^2\,\}
\,\Longrightarrow \,
\vec{\mu}|_{i=1,2,3}\,&=&\,\{\,(2,0), (1,1),(0,2)\,\}
\,\,\Longrightarrow \nonumber\\
\vec{\mu}^{'}|_{i=1,2,3} \, \equiv \,
\vec{\mu}|_{i=1,2,3}-\vec{1}\,&=&\,\{\, (1,-1), (0,0),(-1,+1)\,\},
\label{CP1vectors}
\end{eqnarray}
where we have used the condition:
 $\vec{\mu} \cdot \vec{k}_2=\mu_1 \cdot 1 + \mu_2 \cdot 1=
d=2$, corresponding to $\vec{\mu}~{'} \cdot \vec{k}_2= 0$,
and we denote by $d$ the dimensionality of the
projective vectors. It is convenient to parametrize (\ref{CP1vectors})
in terms of the new basis vector $\vec{e}=(-1,1)$:
\begin{equation}
\vec{\mu}^{'}|_{i=1,2,3}
\,\,\Longrightarrow \vec (e)|_{i=1,2,3}\,=\,\{\, (-1), (0),(+1)\,\}
\times \vec{e}
\end{equation}
The three points $(2,0), (1,1),(0,2)$ (or $-1, 0, +1$)
corresponding to the composite vector 
$k_2=(1,1)$ may be considered as composing a degenerate linear
polyhedron with two integer vertices $\{ (2,0), (0,2) \}$ $( \pm 1)$
and one central interior point $(1,1)$ $(0)$. As we see in more
detail later, this polyhedron is self-dual, or reflexive as
defined in Section 2.

To describe $CY_1$ spaces in ${CP^2}$ projective space, 
via the analogous projective vectors
${\vec {k}_3 = (1,1,1), (1,1,2), (1,2,3)}$ that are associated with
the corresponding polynomial zero loci,
one may introduce the
two following types of extended vectors:
the {\it doubly-extended basic} vectors
\begin{equation}
\vec {k}_1^{ex}= (0,0,1), \; (0,1,0), \; (1,0,0)
\label{doubasex}
\end{equation}
obtained by adding zero to the two-dimensional basic vectors
(\ref{CP1basics}) in all possible ways,
and the three simple
extensions of the composite vector ${\vec{k}_2 = (1,1)}$:
\begin{equation}
\vec {k}_2^{ex}= (0,1,1), \; (1,0,1), \; (1,1,0).
\label{sincomex}
\end{equation}
Then, out of all the extended vectors (\ref{doubasex})
and (\ref{sincomex}) and the corresponding sets of
monomials, one should 
consider only those pairs (triples)  whose common monomials
correspond to
the composite vector $\vec{k}_2=(1,1)$
(to the unit vector) which produces the  reflexive  linear polyhedron 
with three integer points (a single point). The condition  of reflexivity
restricted to the extended vector pairs (triples), ... will also be
very important for constructing the closed  sets of
higher-dimensional projective vectors (again reflexive).

For example, consider one such `good' pair, 
\begin{eqnarray}
\vec{k}_2^{ex}\,=\,(0,1,1) \,\,\Longleftrightarrow\,\,
\vec{k}_1^{ex}\,=\,(1,0,0),
\end{eqnarray}
with the corresponding set of monomials, 
\begin{eqnarray}
\{x^m \cdot y^2\}\,\,       &\Longrightarrow &\,\, 
\vec{\mu}= (m,2,0),\,\,\nonumber\\
\{x^n \cdot y \cdot z\}\,\,&\Longrightarrow &\,\,
\vec{\mu}=(n,1,1),\,\,\,\nonumber\\
\{x^p \cdot z^2\}\,\,      &\Longrightarrow &\,\,
\vec{\mu}=(p,0,2),\,\, \nonumber\\
\,\,\,\vec{\mu}_i \cdot \vec{k}_2^{ex}\,&=&\,2, 
\end{eqnarray}
and
\begin{eqnarray}
\{x \cdot y^k \cdot z^l\}
\,\,&\Longrightarrow &\,\,\vec{\mu}=(1,k,l),\nonumber\\
\,\,\,\vec{\mu} \cdot \vec{k}_1^{ex}&=&1.  
\end{eqnarray}
 The common action of
these two extended vectors, (0,1,1) and (1,0,0),
gives as results only the following three monomials:
\begin{eqnarray}
&&\{x \cdot y^2,\, x \cdot y \cdot z,\,  x \cdot z^2 \}
\,\,\Longrightarrow       \nonumber\\ 
\vec{\mu}|_{i=1,2,3}\,&=&\,\{(1,2,0),\,(1,1,1),\,(1,0,2)\}
\,\,\Longrightarrow       \nonumber\\
\vec{\mu}|_{i=1,2,3}-\vec{1}\,&=&\,\{(0,1,-1),\,(0,0,0),\,(0,-1,1)\}
\,\,\Longrightarrow       \nonumber\\
\vec{e}|_{i=1,2,3}\,&=&\,\{(-1),\,(0),\,(1)\}
\end{eqnarray} 
which correspond to the $CP^1$ case.
Such pairs or triples may be termed `reflexive' pairs or triples, because
the vertices $\vec{e}|_{i=1,2,3}$ above generate a
(degenerate) reflexive polyhedron.

Such pairs, triples and higher-order sets of
projective vectors $\vec k_1$ may be used to define {\it chains}
of integer-linear combinations, as explained in more detail in
subsection 4.1:
\begin{equation}
m_1 {\vec k}_1 + m_2 {\vec k}_2 + \dots
\end{equation}
We use the term {\it eldest vector} for the leading entry
in any such chain, with minimal values of $m_1, m_2, \dots$.
In the above case, there are just two
distinct types of `reflexive' pairs:
$\{(0,0,1), (1,1,0)\}$ and $\{(0,1,1), (1,0,1)\}$, which give rise
to two such chains: $\{(1,1,1),(1,1,2)\}$ and
$\{(1,1,2)$, $(1,2,3)\}$. There is only one useful `reflexive'
triple: $\{(0,0,1),(0,1,0),(1,0,0)\}$ defining a non-trivial three-vector
chain. Together, these chains can be used to
construct all three projective $\vec{k}_2$ vectors. The second
possible `reflexive' triple  $\{(0,1,1),(1,0,1),$ $(1,1,0)\}$ 
produces a chain that consists of only one projective $\vec{k}_3$
vector: $(1,1,1)$.

In addition to the zero loci of single polynomials, CY spaces may be
found by higher-level contructions as the intersections
of the zero loci of two or more polynomial loci. The higher-level
$CY_1$ spaces found in this way are given in the last Section of
this paper.

In the case of the $K3$ hypersurfaces in ${CP^3}$ projective space,
our construction starts from the five possible
types of extended vectors, with all their possible 
{Galois} groups of permutations. These types are
the triply-extended basic vectors with the cyclic ${C_4}$
group of permutations,
\begin{equation}
\vec {k}_1^{ex}= (0,0,0,1):\,\,\,\,\,|C_4|=4,
\end{equation}
the doubly-extended composite vectors with the $D_3$ dihedral group of
permutations,
\begin{equation} 
\vec {k}_2^{ex}= (0,0,1,1):\,\,\,\,\,|D_3|=6,
\end{equation}
and the following singly-extended composite vectors  with the cyclic
${C_4}$, 
alternating ${A_4}$ and symmetric ${S_4}$
groups of permutations, respectively:
\begin{eqnarray}
\vec {k}_3^{ex}\,&=&\, (0,1,1,1):\,\,\,\,\, |C_4|=4, \\
\vec {k}_3^{ex}\,&=&\, (0,1,1,2):\,\,\,\,\, |A_4|=12, \\ 
\vec {k}_3^{ex}\,&=&\, (0,1,2,3):\,\,\,\,\, |S_4|=24.
\end{eqnarray}  
The ${A_4}$ and ${S_4}$ groups of permutations can be identified 
with the tetrahedral ${T}$ and octahedral ${O}$ rotation groups, 
respectively.
Combining these 50 extended vectors in pairs, we find 22
pairs whose common actions correspond to reflexive polyhedra in the
plane. These give rise to {22} chains (lattices parametrized by 
two positive integers), which together yield
90 vectors ${\vec {k}_4}$ based on such extended
structures, that are discussed in more detail in Section 5.
In addition, there exist just four triples
constructed from the 10
extended 
vectors $(0,0,0,1)~+$ permutations and $(0,0,1,1)~+$ permutations whose
common actions give a unique reflexive polyhedron on the line:
$(-1),(0),(+1)$. The corresponding
four triple chains
(lattices parametrized by three positive integers) yield
{91} ${\vec {k}_4}$ vectors, as discussed in Section 6.
As also discussed there, it turns out that most of the vectors ${\vec
{k}_4}$
obtained from the triple combinations are already included among
those found in the double chains, so that the combined number of
distinct vectors is just 94. The total number of vectors is,
however, 95 (see Table \ref{list95}), because there exists, in addition 
to the above enumeration, a
single vector ${\vec {k}_4=(7,8,9,12)}$ which has only a 
trivial intersection consisting just of the zero point.
This can be found within our approach
using the non-trivial projection structure of its dual,
which is an example of the importance of duality
in our classification, as discussed in Section 7.

To find all CY manifolds, and thereby to close their algebra with respect
the duality between {\it intersection} and {\it projection} that is
described in
more detail in Sections 3 and 4, one must consider how to 
classify the projective structures of CY manifolds. Some 
of the {22} chains are dual with respect to the `intersection-projection'
structure, but more analysis is required to close the CY algebra. 
As discussed in Section 7, it is useful for this purpose
to look for
so-called {\it invariant} directions. 
To find all such {\it invariant}
directions in the case of $K3$ spaces, one should consider all
triples selected from the following five extended
vectors: $(0,0,0,1), (0,0,1,1), (0,1,1,1), (0,1,1,2), (0,1,2,3)$,
and their possible permutations, whose intersections give the 
following five types of {\it invariant}
directions defined by two monomials: 
\begin{eqnarray}
\vec {\pi}_1^{\alpha}\,&=&\,\{(1,1,1,1) \,\rightarrow \,  
(0,1,1,3)\},\,\,\,{\alpha}\,=\,1,\,2,\nonumber\\
\vec {\pi}_2^{\alpha}\,&=&\,\{(1,1,1,1) \,\rightarrow \,  
(0,0,0,3)\},\,\,\,{\alpha}\,=\,1,\,2,\,3,\,4, \nonumber\\
\vec {\pi}_3^{\alpha}\,&=&\,\{(1,1,1,1) \,\rightarrow \,  
(0,0,1,3)\},\,\,\,{\alpha}\,=\,1,\,2,\,3,\,4, \nonumber\\
\vec {\pi}_4^{\alpha}\,&=&\,\{(1,1,1,1) \,\rightarrow \,  
(0,0,0,4)\},\,\,\,{\alpha}\,=\,1,\,2,\,3,\,4, \nonumber\\
\vec {\pi}_5^{\alpha}\,&=&\,\{(1,1,1,1) \,\rightarrow \,  
(0,0,1,4)\},\,\,\,{\alpha}\,=\,1, \nonumber\\
\end{eqnarray}
and the following three types of {\it invariant}
directions defined by three monomials:
\begin{eqnarray}
\vec {\pi}_6^{\alpha}\,&=&\,\{(0,2,1,1)  \,\rightarrow \,
(1,1,1,1) \,\rightarrow \,(2,0,1,1) \},
\,\,\,{\alpha}\,=\,1,\,2,\nonumber\\
\vec {\pi}_7^{\alpha}\,&=&\,\{(0,0,1,2) 
\,\rightarrow \,
(1,1,1,1) \,\rightarrow \,  (2,2,1,0)\}, \,\,\,{\alpha}\,=\,1,\,2,\,3,\,4,
\nonumber\\
\vec {\pi}_8^{\alpha}\,&=&\,\{(0,0,0,2)\,\rightarrow \,
(1,1,1,1) \,\rightarrow \,  (2,2,2,0)\},
\,\,\,{\alpha}\,=\,1,\,2,\,3,\,4,\nonumber\\
\end{eqnarray}
respectively.
Each double intersection of a pair of extended vectors
from one of these triples
gives the same `good' planar polyhedron whose intersection with the plane
integer lattice $Z_2$ has just one interior point.

By this method, one can 
classify the projective vectors by projections, finding
78 projective vectors which can be characterized by their invariant 
directions.
Taking into account the projective vectors 
with intersection-projection duality that have already been found by the 
double-intersection method, one can recover all 95 $K3$ projective
vectors, including the exceptional vector $(7,8,9,12)$ that was not
found previously among the double and triple chains. 

\scriptsize
\begin{table}[!ht]
\centering
\caption{\it  The algebraic structure of the 95 projective vectors
characterizing $K3$ spaces. The numbers of points/vertices in the
corresponding
polyhedra (their duals) are denoted by $N/V$ ($N^*/V^*$), and 
their Picard numbers are denoted by $Pic$ ($Pic^*$). In each case,
we also list the double, triple chains and projective chains
where the corresponding $K3$ vector may be found.}
\vspace{.05in}
\label{list95}
\begin{tabular}{|c|c||c|c||c|c||c|c||c|c|c|}
\hline
${ \aleph}  $&$ {\vec k}_4   $&$N $&$ N^*       
$&$V $&$  V^*          $&$ Pic $&$ Pic^*
$&$ Double~chains    $&$ Triple~chains
$&$ Projective~chains  $\\ \hline\hline
 $ 1$&$( 1, 1, 1, 1)$&$ 35$&$ 5$&$ 4$&$ 4$&$ 1$&$ 19$&$ I, VII, X, XII
 $&$ I$&$ \vec{\pi}_3, \vec{\pi}_4, \vec{\pi}_6$\\
 \hline
 $ 2$&$( 1, 1, 1, 2)$&$ 34$&$ 6$&$ 6$&$ 5$&$ 2$&$ 18$&$ I, IV, XI, XIV
 $&$ I, II$&$ \vec{\pi}_1,\vec{\pi}_3, \vec{\pi}_5, \vec{\pi}_6, \vec{\pi}_7$\\
 \hline
 $ 3$&$( 1, 1, 1, 3)$&$ 39$&$ 6$&$ 4$&$ 4$&$ 1$&$ 19$&$ I, V, XX$&$
 I, III$&$ \vec{\pi}_8$\\
 \hline
 $ 4$&$( 1, 1, 2, 2)$&$ 30$&$ 6$&$ 4$&$ 4$&$ 4$&$ 18$&$ II, IV, X, XXI
 XXII$&$ I, II, IV$&$ \vec{\pi}_2, \vec{\pi}_6, \vec{\pi}_7$\\
 \hline
 $ 5$&$( 1, 1, 2, 3)$&$ 31$&$ 8$&$ 7$&$ 6$&$ 4$&$ 16$&$ IV, XI, XIII, XV
 $&$ I, II$&$ \vec{\pi}_1, \vec{\pi}_3, \vec{\pi}_6, \vec{\pi}_7$\\
 \hline
 $ 6$&$( 1, 1, 2, 4)$&$ 35$&$ 7$&$ 4$&$ 4$&$ 3$&$ 18$&$ IV, V, VI, XVI
 $&$ I, II, III$&$ \vec{\pi}_8$\\
 \hline
 $ 7$&$( 1, 1, 3, 4)$&$ 33$&$ 9$&$ 5$&$ 5$&$ 4$&$ 16$&$ XI, XVII$&$ I,
 II$&$ \vec{\pi}_2, \vec{\pi}_6$\\
 \hline
 $ 8$&$( 1, 1, 3, 5)$&$ 36$&$ 9$&$ 5$&$ 5$&$ 3$&$ 17$&$ V, XVIII$&$ I,
 III$&$ \vec{\pi}_8$\\
 \hline
 $ 9$&$( 1, 1, 4, 6)$&$ 39$&$ 9$&$ 4$&$ 4$&$ 2$&$ 18$&$ V, XIX$&$ I,
 III$&$$\\
 \hline
 $ 10$&$( 1, 2, 2, 3)$&$ 24$&$ 8$&$ 6$&$ 5$&$ 7$&$ 16$&$ VII, VIII, XI,
 XV, XXII$&$ I, II, IV$&$ \vec{\pi}_1, \vec{\pi}_3, \vec{\pi}_4, \vec{\pi}_7$\\
 \hline
 $ 11$&$( 1, 2, 2, 5)$&$ 28$&$ 8$&$ 4$&$ 4$&$ 6$&$ 18$&$ V, IX, XVI$&$
 I, III$&$ \vec{\pi}_8$\\
 \hline
 $ 12$&$( 1, 2, 3, 3)$&$ 23$&$ 8$&$ 6$&$ 5$&$ 8$&$ 16$&$ II, III, XIV,
 XV$&$ I, II$&$ \vec{\pi}_1, \vec{\pi}_2, \vec{\pi}_7$\\
 \hline
 $ 13$&$( 1, 2, 3, 4)$&$ 23$&$ 11$&$ 7$&$ 6$&$ 8$&$ 13$&$ XII, XIII, XV,
 XXII$&$ II, IV$&$ \vec{\pi}_1, \vec{\pi}_3, \vec{\pi}_7$\\
 \hline
 $ 14$&$( 1, 2, 3, 5)$&$ 24$&$ 13$&$ 8$&$ 7$&$ 8$&$ 12$&$ XIII, XIV, XV$&
 $ II$&$ \vec{\pi}_1, \vec{\pi}_3, \vec{\pi}_5, \vec{\pi}_7$\\
 \hline
 $ 15$&$( 1, 2, 3, 6)$&$ 27$&$ 9$&$ 4$&$ 4$&$ 7$&$ 16$&$ VI, XV, XVI,
 XX$&$ II, III$&$ \vec{\pi}_8$\\
 \hline
 $ 16$&$( 1, 2, 4, 5)$&$ 24$&$ 12$&$ 5$&$ 5$&$ 8$&$ 14$&$ XVII, XXI, XXII$&
 $ II, IV$&$ \vec{\pi}_1, \vec{\pi}_2$\\
 \hline
 $ 17$&$( 1, 2, 4, 7)$&$ 27$&$ 12$&$ 5$&$ 5$&$ 7$&$ 15$&$ XVI, XVIII$&$
 III$&$ \vec{\pi}_8$\\
 \hline
 $ 18$&$( 1, 2, 5, 7)$&$ 26$&$ 17$&$ 6$&$ 6$&$ 8$&$ 12$&$ XVII$&$ II$&$
 \vec{\pi}_1, \vec{\pi}_2$\\
 \hline
 $ 19$&$( 1, 2, 5, 8)$&$ 28$&$ 14$&$ 5$&$ 5$&$ 7$&$ 14$&$ XVI, XVIII$&$
 III$&$ \vec{\pi}_8$\\
 \hline
 $ 20$&$( 1, 2, 6, 9)$&$ 30$&$ 12$&$ 4$&$ 4$&$ 6$&$ 16$&$ XVI, XIX$&$
 III$&$$\\
 \hline
 $ 21$&$( 1, 3, 4, 4)$&$ 21$&$ 9$&$ 4$&$ 4$&$ 10$&$ 16$&$ II, VIII$&$
 I, II$&$ \vec{\pi}_2, \vec{\pi}_7$\\
 \hline
 $ 22$&$( 1, 3, 4, 5)$&$ 20$&$ 15$&$ 7$&$ 7$&$ 10$&$ 10$&$ XIII, XIV$&$
 II$&$ \vec{\pi}_7$\\
 \hline
 $ 23$&$( 1, 3, 4, 7)$&$ 22$&$ 17$&$ 6$&$ 6$&$ 10$&$ 10$&$ XIII$&$ II$&
 $ \vec{\pi}_3, \vec{\pi}_7$\\
 \hline
 $ 24$&$( 1, 3, 4, 8)$&$ 24$&$ 12$&$ 5$&$ 5$&$ 9$&$ 14$&$ VI, IX$&$
 II, III$&$ \vec{\pi}_8$\\
 \hline
 $ 25$&$( 1, 3, 5, 6)$&$ 21$&$ 15$&$ 5$&$ 5$&$ 10$&$ 12$&$ III, XVII$&$
 II$&$$\\
 \hline
 $ 26$&$( 1, 3, 5, 9)$&$ 24$&$ 15$&$ 5$&$ 5$&$ 9$&$ 13$&$ XVIII, XX$&$
 III$&$ \vec{\pi}_8$\\
 \hline
 $ 27$&$( 1, 3, 7, 10)$&$ 24$&$ 24$&$ 4$&$ 4$&$ 10$&$ 10$&$ XVII$&$ II
 $&$ \vec{\pi}_2$\\
 \hline
 $ 28$&$( 1, 3, 7, 11)$&$ 25$&$ 20$&$ 5$&$ 5$&$ 9$&$ 11$&$ XVIII$&$ III$&
 $ \vec{\pi}_8$\\
 \hline
 $ 29$&$( 1, 3, 8, 12)$&$ 27$&$ 15$&$ 4$&$ 4$&$ 8$&$ 14$&$ XIX$&$ III$&
 $$\\
 \hline
 $ 30$&$( 1, 4, 5, 6)$&$ 19$&$ 17$&$ 6$&$ 6$&$ 11$&$ 9$&$ VIII, XIII$&$
 II$&$ \vec{\pi}_7$\\
 \hline
 $ 31$&$( 1, 4, 5, 10)$&$ 23$&$ 13$&$ 4$&$ 4$&$ 10$&$ 14$&$ VI$&$ II,
 III$&$ \vec{\pi}_8$\\
 \hline
 $ 32$&$( 1, 4, 6, 7)$&$ 19$&$ 20$&$ 6$&$ 6$&$ 11$&$ 9$&$ XVII$&$ II$&$
 $\\
 \hline
 $ 33$&$( 1, 4, 6, 11)$&$ 22$&$ 20$&$ 6$&$ 6$&$ 10$&$ 10$&$ IX, XVIII$&$
 III$&$$\\
 \hline
 $ 34$&$( 1, 4, 9, 14)$&$ 24$&$ 24$&$ 4$&$ 4$&$ 10$&$ 10$&$ XVIII$&$ III
 $&$ \vec{\pi}_8$\\
 \hline
 $ 35$&$( 1, 4, 10, 15)$&$ 25$&$ 20$&$ 5$&$ 5$&$ 9$&$ 11$&$ XIX$&$ III
 $&$$\\
 \hline
 $ 36$&$( 1, 5, 7, 8)$&$ 18$&$ 24$&$ 5$&$ 5$&$ 12$&$ 8$&$ XVII$&$ II$&$
 $\\
 \hline
 $ 37$&$( 1, 5, 7, 13)$&$ 21$&$ 24$&$ 5$&$ 5$&$ 11$&$ 9$&$ XVIII$&$ III$&
 $$\\
 \hline
 $ 38$&$( 1, 5, 12, 18)$&$ 24$&$ 24$&$ 4$&$ 4$&$ 10$&$ 10$&$ XIX$&$ III
 $&$$\\
 \hline
 $ 39$&$( 1, 6, 8, 9)$&$ 18$&$ 24$&$ 5$&$ 5$&$ 12$&$ 8$&$ XVII$&$ II$&$
 $\\
 \hline
 $ 40$&$( 1, 6, 8, 15)$&$ 21$&$ 24$&$ 5$&$ 5$&$ 11$&$ 9$&$ XVIII$&$ III$&
 $$\\
 \hline
 $ 41$&$( 1, 6, 14, 21)$&$ 24$&$ 24$&$ 4$&$ 4$&$ 10$&$ 10$&$ XIX$&$ III
 $&$$\\
 \hline
 $ 42$&$( 2, 2, 3, 5)$&$ 17$&$ 11$&$ 5$&$ 5$&$ 11$&$ 14$&$ VIII, XI$&$
 I, II$&$ \vec{\pi}_4, \vec{\pi}_6, \vec{\pi}_7$\\
 \hline
 $ 43$&$( 2, 2, 3, 7)$&$ 19$&$ 11$&$ 5$&$ 5$&$ 10$&$ 16$&$ V, IX$&$
 I, III$&$ \vec{\pi}_8$\\
 \hline
 $ 44$&$( 2, 3, 3, 4)$&$ 15$&$ 9$&$ 4$&$ 4$&$ 12$&$ 16$&$ III, VII, XXI$&
 $ I, IV$&$ \vec{\pi}_1, \vec{\pi}_2, \vec{\pi}_3, \vec{\pi}_4, \vec{\pi}_6$\\
 \hline
 $ 45$&$( 2, 3, 4, 5)$&$ 13$&$ 16$&$ 7$&$ 7$&$ 13$&$ 9$&$ XII, XIV, XXII$&
 $ II, IV$&$ \vec{\pi}_1, \vec{\pi}_3, \vec{\pi}_5, \vec{\pi}_7$\\
 \hline
 \end{tabular}
 \end{table}
 \begin{table}[!ht]
 \caption{\it  Continuation of Table~1.}
\vspace{.05in}
 \begin{tabular}{|c|c||c|c||c|c||c|c||c|c|c|}
 \hline
 $ 46$&$( 2, 3, 4, 7)$&$ 14$&$ 18$&$ 6$&$ 6$&$ 13$&$ 10$&$ VIII, XIV$&$
 II$&$ \vec{\pi}_1, \vec{\pi}_3, \vec{\pi}_4, \vec{\pi}_5, \vec{\pi}_7$\\
 \hline
 $ 47$&$( 2, 3, 4, 9)$&$ 16$&$ 14$&$ 5$&$ 5$&$ 12$&$ 13$&$ IX, XVI, XX
 $&$ III$&$ \vec{\pi}_8$\\
 \hline
 $ 48$&$( 2, 3, 5, 5)$&$ 14$&$ 11$&$ 6$&$ 5$&$ 14$&$ 14$&$ II$&$ I,
 II$&$ \vec{\pi}_2, \vec{\pi}_7$\\
 \hline
 $ 49$&$( 2, 3, 5, 7)$&$ 13$&$ 20$&$ 8$&$ 8$&$ 14$&$ 6$&$ XIII$&$ II$&$
 \vec{\pi}_3, \vec{\pi}_5, \vec{\pi}_7$\\
 \hline
 $ 50$&$( 2, 3, 5, 8)$&$ 14$&$ 20$&$ 6$&$ 6$&$ 14$&$ 7$&$ XIII$&$ II$&$
 \vec{\pi}_3, \vec{\pi}_7$\\
 \hline
 $ 51$&$( 2, 3, 5, 10)$&$ 16$&$ 14$&$ 5$&$ 5$&$ 13$&$ 12$&$ VI$&$ II,
 III$&$ \vec{\pi}_8$\\
 \hline
 $ 52$&$( 2, 3, 7, 9)$&$ 14$&$ 23$&$ 6$&$ 6$&$ 14$&$ 8$&$ XVII$&$ II$&$
 \vec{\pi}_2$\\
 \hline
 $ 53$&$( 2, 3, 7, 12)$&$ 16$&$ 20$&$ 5$&$ 5$&$ 13$&$ 10$&$ XVIII$&$ III
 $&$ \vec{\pi}_8$\\
 \hline
 $ 54$&$( 2, 3, 8, 11)$&$ 15$&$ 27$&$ 4$&$ 4$&$ 14$&$ 8$&$ XVII$&$ II$&
 $ \vec{\pi}_2$\\
 \hline
 $ 55$&$( 2, 3, 8, 13)$&$ 16$&$ 23$&$ 5$&$ 5$&$ 13$&$ 9$&$ XVIII$&$ III$&
 $ \vec{\pi}_8$\\
 \hline
 $ 56$&$( 2, 3, 10, 15)$&$ 18$&$ 18$&$ 4$&$ 4$&$ 12$&$ 12$&$ XIX$&$ III
 $&$$\\
 \hline
 $ 57$&$( 2, 4, 5, 9)$&$ 13$&$ 23$&$ 4$&$ 4$&$ 14$&$ 10$&$ VIII$&$ II$&
 $ \vec{\pi}_1, \vec{\pi}_4, \vec{\pi}_7$\\
 \hline
 $ 58$&$( 2, 4, 5, 11)$&$ 14$&$ 19$&$ 5$&$ 5$&$ 13$&$ 11$&$ IX, XVI$&$
 III$&$ \vec{\pi}_8$\\
 \hline
 $ 59$&$( 2, 5, 6, 7)$&$ 11$&$ 23$&$ 5$&$ 5$&$ 15$&$ 7$&$ VIII$&$ II$&$
 \vec{\pi}_3, \vec{\pi}_4, \vec{\pi}_7$\\
 \hline
 $ 60$&$( 2, 5, 6, 13)$&$ 13$&$ 23$&$ 5$&$ 5$&$ 14$&$ 9$&$ IX$&$ III$&
 $ \vec{\pi}_8$\\
 \hline
 $ 61$&$( 2, 5, 9, 11)$&$ 11$&$ 32$&$ 6$&$ 6$&$ 16$&$ 4$&$ XVII$&$ II$&
 $ \vec{\pi}_2$\\
 \hline
 $ 62$&$( 2, 5, 9, 16)$&$ 13$&$ 29$&$ 5$&$ 5$&$ 15$&$ 6$&$ XVIII$&$ III$&
 $ \vec{\pi}_8$\\
 \hline
 $ 63$&$( 2, 5, 14, 21)$&$ 15$&$ 27$&$ 4$&$ 4$&$ 14$&$ 8$&$ XIX$&$ III
 $&$$\\
 \hline
 $ 64$&$( 2, 6, 7, 15)$&$ 13$&$ 23$&$ 4$&$ 4$&$ 14$&$ 10$&$ IX$&$ III
 $&$ \vec{\pi}_8$\\
 \hline
 $ 65$&$( 3, 3, 4, 5)$&$ 12$&$ 12$&$ 5$&$ 5$&$ 14$&$ 14$&$ III$&$ I$&
 $  \vec{\pi}_2, \vec{\pi}_3, \vec{\pi}_5, \vec{\pi}_6$\\
 \hline
 $ 66$&$( 3, 4, 5, 6)$&$ 10$&$ 17$&$ 6$&$ 6$&$ 15$&$ 9$&$ III, XII, XXI$&
 $ IV$&$ \vec{\pi}_1, \vec{\pi}_2, \vec{\pi}_3$\\
 \hline
 $ 67$&$( 3, 4, 5, 7)$&$ 9$&$ 24$&$ 7$&$ 8$&$ 16$&$ 4$&$ XIV$&$ II$&$
 \vec{\pi}_3, \vec{\pi}_5, \vec{\pi}_7$\\
 \hline
 $ 68$&$( 3, 4, 5, 8)$&$ 10$&$ 22$&$ 6$&$ 6$&$ 16$&$ 7$&$ VIII$&$ II$&$
 \vec{\pi}_1, \vec{\pi}_3, \vec{\pi}_4, \vec{\pi}_7$\\
 \hline
 $ 69$&$( 3, 4, 5, 12)$&$ 12$&$ 18$&$ 5$&$ 5$&$ 15$&$ 10$&$ IX, XX$&$
 III$&$ \vec{\pi}_8$\\
 \hline
 $ 70$&$( 3, 4, 7, 10)$&$ 10$&$ 26$&$ 5$&$ 6$&$ 17$&$ 3$&$ XIII$&$ II$&
 $ \vec{\pi}_3, \vec{\pi}_7$\\
 \hline
 $ 71$&$( 3, 4, 7, 14)$&$ 12$&$ 18$&$ 5$&$ 5$&$ 16$&$ 10$&$ VI$&$ II,
 III$&$ \vec{\pi}_8$\\
 \hline
 $ 72$&$( 3, 4, 10, 13)$&$ 10$&$ 35$&$ 5$&$ 5$&$ 17$&$ 3$&$ XVII$&$ II
 $&$ \vec{\pi}_2$\\
 \hline
 $ 73$&$( 3, 4, 10, 17)$&$ 11$&$ 31$&$ 6$&$ 6$&$ 16$&$ 4$&$ XVIII$&$ III
 $&$ \vec{\pi}_8$\\
 \hline
 $ 74$&$( 3, 4, 11, 18)$&$ 12$&$ 30$&$ 4$&$ 4$&$ 16$&$ 6$&$ XVIII$&$ III
 $&$ \vec{\pi}_8$\\
 \hline
 $ 75$&$( 3, 4, 14, 21)$&$ 13$&$ 26$&$ 5$&$ 5$&$ 15$&$ 7$&$ XIX$&$ III
 $&$$\\
 \hline
 $ 76$&$( 3, 5, 6, 7)$&$ 9$&$ 21$&$ 5$&$ 5$&$ 16$&$ 8$&$ III$&$$&$
 \vec{\pi}_1, \vec{\pi}_2 $\\
 \hline
 $ 77$&$( 3, 5, 11, 14)$&$ 9$&$ 39$&$ 4$&$ 4$&$ 18$&$ 2$&$ XVII$&$ II$&
 $ \vec{\pi}_2$\\
 \hline
 $ 78$&$( 3, 5, 11, 19)$&$ 10$&$ 35$&$ 5$&$ 5$&$ 17$&$ 3$&$ XVIII$&$ III
 $&$ \vec{\pi}_8$\\
 \hline
 $ 79$&$( 3, 5, 16, 24)$&$ 12$&$ 30$&$ 4$&$ 4$&$ 16$&$ 6$&$ XIX$&$ III
 $&$$\\
 \hline
 $ 80$&$( 3, 6, 7, 8)$&$ 9$&$ 21$&$ 4$&$ 4$&$ 16$&$ 10$&$ III$&$$&$ 
\vec{\pi}_1, \vec{\pi}_2, \vec{\pi}_3, \vec{\pi}_4$\\
 \hline
 $ 81$&$( 4, 5, 6, 9)$&$ 8$&$ 26$&$ 5$&$ 6$&$ 17$&$ 4$&$ XIV$&$ II$&$
 \vec{\pi}_3, \vec{\pi}_4, \vec{\pi}_5, \vec{\pi}_7$\\
 \hline
 $ 82$&$( 4, 5, 6, 15)$&$ 10$&$ 20$&$ 5$&$ 5$&$ 16$&$ 9$&$ XX$&$ III$&
 $ \vec{\pi}_8$\\
 \hline
 $ 83$&$( 4, 5, 7, 9)$&$ 7$&$ 32$&$ 5$&$ 6$&$ 18$&$ 2$&$$&$ II$&$ 
 \vec{\pi}_3, \vec{\pi}_7$\\
 \hline
 $ 84$&$( 4, 5, 7, 16)$&$ 9$&$ 27$&$ 5$&$ 5$&$ 17$&$ 6$&$ IX$&$ III$&$
 \vec{\pi}_8$\\
 \hline
 $ 85$&$( 4, 5, 13, 22)$&$ 9$&$ 39$&$ 4$&$ 4$&$ 18$&$ 2$&$ XVIII$&$ III$&
 $ \vec{\pi}_8$\\
 \hline
 $ 86$&$( 4, 5, 18, 27)$&$ 10$&$ 35$&$ 5$&$ 5$&$ 17$&$ 3$&$ XIX$&$ III
 $&$$\\
 \hline
 $ 87$&$( 4, 6, 7, 11)$&$ 7$&$ 35$&$ 4$&$ 4$&$ 18$&$ 3$&$ VIII$&$ II$&$
 \vec{\pi}_4, \vec{\pi}_5, \vec{\pi}_7$\\
 \hline
 $ 88$&$( 4, 6, 7, 17)$&$ 8$&$ 31$&$ 5$&$ 5$&$ 17$&$ 4$&$ IX$&$ III$&$
 \vec{\pi}_8$\\
 \hline
 $ 89$&$( 5, 6, 7, 9)$&$ 6$&$ 30$&$ 5$&$ 6$&$ 18$&$ 2$&$ III$&$$&$
 \vec{\pi}_2, \vec{\pi}_3, \vec{\pi}_5$\\
 \hline
 $ 90$&$( 5, 6, 8, 11)$&$ 6$&$ 39$&$ 4$&$ 4$&$ 19$&$ 1$&$$&$ II$&$
 \vec{\pi}_3, \vec{\pi}_7$\\
 \hline
 $ 91$&$( 5, 6, 8, 19)$&$ 7$&$ 35$&$ 5$&$ 5$&$ 18$&$ 2$&$ IX$&$ III$&$
 \vec{\pi}_8$\\
 \hline
 $ 92$&$( 5, 6, 22, 33)$&$ 9$&$ 39$&$ 4$&$ 4$&$ 18$&$ 2$&$ XIX$&$ III$&
 $$\\
 \hline
 $ 93$&$( 5, 7, 8, 20)$&$ 8$&$ 28$&$ 4$&$ 4$&$ 18$&$ 6$&$$&$ III$&$
 \vec{\pi}_8$\\
 \hline
 $ 94$&$( 7, 8, 10, 25)$&$ 6$&$ 39$&$ 4$&$ 4$&$ 19$&$ 1$&$$&$ III$&$
 \vec{\pi}_8$\\
 \hline
 $ 95$&$( 7, 8, 9, 12)$&$ 5$&$ 35$&$ 4$&$ 4$&$ 19$&$ 1$&$$&$$&$ \vec{\pi}_2,
 \vec{\pi}_3, \vec{\pi}_4, \vec{\pi}_5$\\
 \hline
 \end{tabular}
 \end{table}

\normalsize

Section 8 of this paper contains a systematic description how
various gauge groups emerge associated with singularities in our
construction of $K3$ spaces \cite{block1}. 
These are interesting because of their
possible role in studies of $F$ theory. Since this
may be regarded as a decompactification of type-$IIA$
string, understanding of duality between the heterotic string
and type $IIA$  string in $D=6$ dimensions can be used to
help understand the duality between the
heterotic string on $T^2$ and $F$ theory on an elliptically-fibered $K3$
hypersurface \cite{block2}. 
The gauge group is directly defined by the ${ADE}$ classification
of the quotient singularities of hypersurfaces. The Cartan matrix of the Lie
group in this case  coincides up to a sign with the intersection 
matrix of the blown-down divisors.
There are two different mechanisms leading to enhanced gauge groups
on the $F$-theory side and on the heterotic side. On 
the $F$-theory side, the singularities of the CY hypersurface give rise 
to the gauge groups, but on the heterotic side the singularities
can give an enhancement of the gauge group if `small' instantons of the
gauge bundle lie on these singularities \cite{block3}.
This question has been studied in terms of the numbers of
instantons  placed on a singularity of type $G$, where $G$ is a
simply-laced group. Studies of groups associated with singularities
of $K3$ spaces are also interesting because
elliptic $CY_n$ ($n=3,4$) manifolds with $K3$ fibers can be
considered to study $F$-theory dual compactifications of the 
$E_8 \times E_8$ or $SO(32)$ string theory. To do this in toric
geometry,
it is possible to consider the $K3$ polyhedron fiber as a
subpolyhedron 
of the  $CY_n$ polyhedron, and the Dynkin diagrams of the gauge groups of
the type-$IIA$ string ($F$-theory) compactifications on the corresponding
threefold (fourfold) can then be seen precisely in the polyhedron of
this $K3$ hypersurface. By extension, one could consider the case of
an elliptic $CY_4$ with $CY_3$ fiber, where the last is a
CY hypersurface with $K3$ fiber. We give in Section 8 several
detailed examples of group structures associated with chains of
$K3$ spaces, which our algebraic approach equips us to study
systematically.

Finally, Section 9 provides a
brief discussion of $CY_3$ manifolds and describes how
additional CY spaces can be constructed at higher levels as the
intersections of multiple polynomial loci. This discussion is
illustrated by the examples of higher-level $CY_1$ and $K3$
spaces obtained via our construction of lower-level
$K3$ and $CY_3$ spaces. We find, for example, 7 new 
polyhedra describing $CY_1$ spaces given by `level-one' intersections of
pairs of polynomial loci, and three new `level-two' polyhedra given
by triple intersections of polynomial loci. In looking for higher-level
$K3$ spaces, we start from 100 types of extended vectors in five
dimensions, corresponding to 10~270 distinct vectors when permutations
are taken into account. We find that these give rise to 4242
two-vector chains of $CY_3$ spaces, 259 triple-vector chains and 6
quadruple-vector chains. Analyzing their internal structures, we
we find 730 new $K3$ polyhedra at level one, of which 146 can be obtained 
as intersections
of polynomials corresponding to simple polyhedra (points, line segments,
triangles and tetrahedra). A complete characterization of
higher-level $K3$ spaces given by multiple intersections of
polynomial loci lies beyond our present computing scope, and we leave
their further study to later work.

\section{Calabi-Yau Spaces as Toric Varieties}

We recall that an $n$-dimensional complex manifold is a 
${2\cdot n}$-dimensional {Riemannian} space with 
a Hermitean metric
\begin{equation}
d s^2\,=\, g_{i \bar j}\cdot d z^i \cdot d {\bar z}^{\bar j}: 
{g_{i j}=g_{{\bar i}{\bar j}}=0},
{g_{i \bar j}= {\bar g}_{\bar j i}}.
\label{metric}
\end{equation}
on its $n$ complex coordinates $z_i$. Such a complex manifold is {K\"ahler} if the 
$(1,1)$ differential two-form
\begin{equation}
\Omega\,=\, \frac{1}{2} \cdot i \cdot g_{i \bar j} \cdot
d z^i \Lambda d {\bar z}^{\bar j},
\end{equation}   
is closed, i.e., ${d \Omega = 0}$. In the case of a 
{K\"ahler} manifold, the metric 
(\ref{metric}) is defined by a {K{\"a}hler } potential:
\begin{equation}
g_{i \bar j}\,=\, \frac {{\partial}^2 K(z^i,{\bar z}^{\bar j})}
{\partial z^i \partial {\bar z}^{\bar j}}.
\end{equation}
The {K{\"a}hler} property yields  the following 
constraints on components of the {Cristoffel} symbols: 
\begin{eqnarray}
{\Gamma}^i_{\bar j k} \,&=&\, {\Gamma}^{\bar i}_{j \bar k}\,=\,
{\Gamma}^i_{\bar j k}\,=\,0,                         \nonumber\\
{\Gamma}^{\bar i}_{\bar j \bar k} \,&=&\,
 \bar {{\Gamma}^i_{j k}}\,=\, g^{{\bar i} s} \cdot 
\frac{ \partial g_{\bar k s}}{\partial {\bar z}^{\bar j}}.
\end{eqnarray}
yielding in turn the following form
\begin{equation}
R_{\bar i j}\,=\, - \frac{\partial {\Gamma}_{\bar i \bar k}^{\bar k}}
{\partial z^j}.
\end{equation}
for the Ricci tensor.

Since the only compact submanifold of $C^n$ is a point~\cite{Sch},
in order to find non-trivial compact submanifolds,
one considers weighted complex projective spaces,
${CP^n(k_1,k_2,...,k_{n+1})}$,
 which are characterized by
{$(n+1)$} quasihomogeneous coordinates ${z_1,...,z_{n+1}}$,
with the identification:
\begin{equation}
(z_1, \ldots ,z_{n+1})\,\sim\, (\lambda ^{k_1} \cdot z_1, \ldots ,
\lambda ^{k_{n+1}} \cdot z_{n+1}).
\label{quasihom}
\end{equation}
The loci of zeroes of quasihomogeneous polynomial equations in such
weighted projective spaces
yield compact submanifolds, as we explain in more detail in
the rest of Section 2, where we introduce and review several of the
geometric and algebraic techniques used in our
subsequent classification.
Other compact submanifolds may be obtained as the complete
intersections of such polynomial zero constraints,
as we discuss in more detail in Section 9.

\subsection{The Topology of Calabi-Yau Manifolds in the 
Polyhedron Method}

A CY variety $X$ in a weighted projective space 
$CP^n(\vec {k}) = CP^n (k_1,...,k_{n+1})$ is given by the
locus of zeroes of a transversal quasihomogeneous polynomial 
${\wp}$ of degree $deg ({\wp})=d$, with $d = \sum _{j=1}^{n+1} k_j$ 
\cite{Batyrev,Sch,Sha1,Sha2,can5,Roan,Dim,Bar,Mir,Asp2,Bgr1,Lam,Cox}:
\begin{eqnarray}
X \equiv X_{d}({k}) \equiv
\{[x_1,...,x_{n+1}]\in CP^n({k})|{\wp}(x_1,...,x_{n+1}) = 0\}.
\label{poly1}
\end{eqnarray}
The general polynomial of degree $d$ is a linear combination
\begin{equation}
{\wp} = \sum_{\vec {\mu}} c_{\vec{\mu}} 
x^{\vec{\mu}}
\label{poly2}
\end{equation}
of monomials $x^{\vec{\mu}} = x_1^{{\mu}_1} 
x_2^{{\mu}_2}...x_{r+1}^{{\mu}_{r+1}}$ with the condition:
\begin{equation}
\vec {\mu} \cdot \vec {k}\,=\,d.
\label{poly3}
\end{equation} 
We recall that the existence of a {\it mirror symmetry},
according to which
each Calabi-Yau manifold should have a dual partner, was first observed
pragmatically in the literature \cite{Ler1,Bgr1,Roan,can5,Dix}.
Subsequently, Batyrev~\cite{Batyrev} found a very elegant way of
describing any Calabi-Yau
hypersurface in terms of the corresponding {\it Newton polyhedron},
associated with degree-${d}$ monomials in the CY equation,
which is the convex hull of all the
vectors $\vec{\mu}$ of degree ${d}$. The Batyrev description provides
a systematic approach to duality and mirror symmetry.

To each monomial associated with a
vector $\vec {\mu}$ of degree $d$,
i.e., $ \vec {\mu} \cdot \vec {k} = d $, one can associate a vector 
$\vec {\mu}^{'} \equiv \vec {\mu} - {\vec e}_0: {\vec e}_0 \equiv 
(1,1,...,1)$, 
so that $\vec {\mu}^{'} \cdot \vec {k} = 0$.
Using the new vector $\vec {\mu}^{'}$,
hereafter denoted without the prime $(')$, it is useful to define
the lattice $\Lambda $: 
\begin{equation}
\Lambda = \{ \overrightarrow {\mu} \in Z^{r+1} : 
\overrightarrow {\mu} \cdot \overrightarrow {k} = 0 \}
\end{equation}
with basis vectors $e_i$,
and the dual lattice ${\Lambda }^{*}$ with basis $e^{*}_j$, where
$e^{*}_j \cdot e_i= {\delta }_{ij} $.
Consider the polyhedron $\triangle$, defined to be
the convex hull of $\{ \vec {\mu} \in \Lambda:
{\mu}_i \geq - 1, \forall i\}$.
Batyrev~\cite{Batyrev} showed that to describe a Calabi-Yau 
hypersurface~\footnote{I.e., with 
trivial canonical bundle and at worst 
Gorenstein canonical singularities only.},
such a polyhedron should satisfy the following conditions:
\begin{itemize}
\item{the vertices of the polyhedron should correspond to the vectors
$\vec{\mu}$ with integer components;}
\item{there should be only one interior integer point, called the center;}
\item{the distance of any face of this polyhedron 
from the center should be equal to unity.} 
\end{itemize}
Such an integral polyhedron $\triangle$ is called {\it reflexive},
and the only interior point
of $\triangle (k_1 + ... +k_{r+1} =d)$ 
may be taken as the origin $(0,...,0)$.
Batyrev~\cite{Batyrev} showed that the mirror polyhedron
\begin{eqnarray}
{\triangle}^{*} \equiv \{\vec {\nu} \in {\Lambda}^{*} :
  \vec {\nu} \cdot \vec {\mu}
\geq -1, \forall \vec {\mu} \in \triangle \}
\end{eqnarray}
of any reflexive integer polyhedron is also reflexive, i.e.,
is also integral and contains one interior point only.
Thus Batyrev proved the existence of dual pairs of hypersurfaces $M$ and
$M^{\prime}$ with
dual Newton polyhedra, ${\triangle}$ and ${\triangle}^{*}$.

Following Batyrev~\cite{Batyrev}, to obtain all the topological invariants
of the 
${K3}, {CY}_3$, etc.,
manifolds, one should study the reflexive regular polyhedra in three,
four, etc., dimensions. For this purpose, it is useful to
recall the types of polyhedra 
and their duality properties. In  three dimensions,
the Descartes-Euler polyhedron formula
relates the numbers of vertices, ${N_0}$, the number of edges,
${N_1}$ and numbers of faces, ${N_2}$:
\begin{eqnarray}
1\,-\,N_0\,+\,N_1\,-\,N_2\,+1\,=\,0.
\end{eqnarray}
This formula yields:
\begin{eqnarray}
1\,-\,4\,+\,6\,-\,4\,+1\,=\,0\,\,
 &\Rightarrow &\,\, \{3,3\}: {\rm Tetrahedron} \nonumber\\
1\,-\,8\,+\,12\,-\,6\,+1\,=\,0\,\,
& \Rightarrow &\,\,\{3,4\}: {\rm Cube} \nonumber\\
1\,-\,6\,+\,12\,-\,8\,+1\,=\,0\,\,
& \Rightarrow &\,\,\{4,3\}: {\rm Octahedron} \nonumber\\
1\,-\,20\,+\,30\,-\,12\,+1\,=\,0\,\,
& \Rightarrow &\,\,\{5,3\}: {\rm Dodecahedron} \nonumber\\
1\,-\,12\,+\,30\,-\,20\,+1\,=\,0\,\,
& \Rightarrow &\,\,\{3,5\}: {\rm Icosahedron} 
\end{eqnarray}
in the particular cases of the five Platonic solids, with the duality
relations $T \leftrightarrow T, C \leftrightarrow O, D \leftrightarrow
I$.

As we shall see later when we consider the ${K3}$
classification, it is interesting to recall the link
between the
classification of the five ADE simply-laced 
Cartan-Lie algebras and the finite rotation groups 
in three dimensions, namely the cyclic and dihedral groups and the 
groups of the tetrahedron, octahedron (cube) and icosahedron
(dodecahedron): $G_M \equiv C_n, D_n, T, O, I$, corresponding to
the $A_n, D_n$ series and the exceptional groups $E_{6,7,8}$,
respectively~\cite{CLAlink}. Any cyclic group  ${C_n}$  of order ${n}$
may be represented as the rotations in a plane around an axis  ${0x}$
through angles
${(2 \cdot m  \cdot \pi) / n}$ for ${m=0,1,2,..., n-1}$. 
This symmetry is realized
by the group of symmetries of an oriented regular ${n}$-gon. 
The dihedral group ${D_n}$
consists of the transformations in ${C_n}$
and in addition ${n}$ rotations through angles ${\pi}$
around axes lying in planes orthogonal to ${0x}$, crossing ${0x}$
and making angles with
one another that are multiples of ${(2 \cdot  \pi)/n}$. This group
has order 
${2 \cdot n}$. In the case of three-dimensional space, there are three
exceptional examples ${T, O, I}$ of finite groups,
related to the corresponding regular polyhedra.
The order of the corresponding
${G_M}$ is equal to the product of the number of the
vertexes of the regular polyhedra with
the number of edges leaving the vertex:
\begin{eqnarray}
|T|\,&=&\,|A_4|\,=\,12; \nonumber\\
|O|\,&=&\,|S_4|\,=\,24;\nonumber\\
|I|\,&=&\,|A_5|\,=\,60.\nonumber\\
\end{eqnarray}  
The dual polyhedron, whose vertexes are the midpoints of the faces of the 
corresponding polyhedron,
has the same group of symmetry, ${G_M}$. 
The finite groups of orthogonal transformations in
three-dimensional space do not consist only of rotations. It is
remarkable to note 
that every finite group of rotations of three-space that preserves the
sphere centred at the origin
can be interpreted as a fractional-linear transformation of the Riemann 
sphere of a complex variable. 

Finally, we recall that all ${K3}$ hypersurfaces 
have the following common values of the topological
invariants: the Hodge number $h_{1,1}$  is 20, the Betti number 
${b_2=22}$, and we have
\begin{eqnarray}
{ Pic}\, = \, { h}_{1,1}\, -\, ( l(\Delta)\,-4\,
-\sum_{\theta \in \Delta} l^{\prime} (\theta))\,
\leq\, 20 .
\end{eqnarray}
for the Picard number, where $l(\Delta)$ is the number of integer points
in the polyhedron and $l^{\prime}(\theta)$ is the number of integer
interior points on the facets.

In the case of the ${CY_3}$ classification, a corresponding important 
role will be played by the structure and 
the duality properties of the four regular polyhedra known in
four-dimensional Euclidean space~\cite{Schlafli}. 
The Descartes-Euler formulae for these cases become:
\begin{eqnarray}
1\,-\,5\,+\,10\,-\,10\,+\,5\,-\,1\,=\,0\,\,
 &\Rightarrow &\,\,\{3,3,3\}: {\rm Pentahedroid} \nonumber\\
1\,-\,16\,+\,32\,-\,24\,+\,8\,-\,1\,=\,0\,\,
& \Rightarrow &\,\,\{3,3,4\}: {\rm Hypercube} \nonumber\\
1\,-\,8\,+\,24\,-\,32\,+\,16\,-\,1\,=\,0\,\,
& \Rightarrow &\,\,\{4,3,3\}: {\rm 16-hedroid} \nonumber\\
1\,-\,24\,+\,96\,-\,96\,+\,24\,-\,1\,=\,0\,\,
& \Rightarrow &\,\,\{3,4,3\}: {\rm 24-hedroid} \nonumber\\
1\,-\,600\,+\,1200\,-\,720\,+\,120\,-\,1\,=\,0\,\,
& \Rightarrow &\,\,\{3,3,5\}: {\rm 120-hedroid}\nonumber\\
1\,-\,120\,+\,720\,-\,1200\,+\,600\,-\,1\,=\,0\,\,
& \Rightarrow &\,\,\{5,3,3\}: {\rm 600-hedroid}.\nonumber\\
\end{eqnarray}
with the duality relations 
$P \leftrightarrow P$, 
$H \leftrightarrow {\rm 16-hedroid}$, 
${\rm 24-hedroid}$ $\leftrightarrow$ ${\rm 24-hedroid}$, \\
${\rm 120-hedroid}$ $\leftrightarrow$ ${\rm 600-hedroid}$.

We do not discuss these relations further in this paper, 
but do recall that
each mirror  pair of CY spaces, $M_{CY}$ and $M_{CY}^*$, has Hodge
numbers that
satisfying the mirror symmetry relation~\cite{Batyrev,Asp2}:
\begin{eqnarray}
h_{1,1}(M) \,&=&\, h_{d-1,1}(M^*),\nonumber\\
h_{d-1,1}(M)\,&=&\, h_{1,1}(M^*) 
\end{eqnarray}
This means that the {Hodge diamond} of $M_{CY}^*$ is a mirror
reflection through a diagonal axis of the Hodge diamond of $M_{CY}$.
The existence of mirror symmetry is a consequence of the dual
properties of {CY} manifolds. 
A pair of reflexive polyhedra $(\triangle ,\triangle ^*)$
gives a pair of mirror CY manifolds and the following identities for the 
Hodge numbers for $n \geq 4$:

\begin{eqnarray}
h_{1,1}(\triangle) &=& h_{d-1,1} (\triangle ^{*})= \nonumber\\
&=& l(\triangle ^{*}) - ( d + 2 ) \, - \, 
\sum_{codim {\Theta}^{*}= 1} l^{'}(\Theta^{*}) \nonumber\\
&+&\, \sum_{codim  {\Theta^{*}} =2} l'(\Theta^{*}) l^{'}{\Theta},
\label{twist}
\end{eqnarray}

\begin{eqnarray}
h_{1,1}(\triangle ^{*}) &=& h_{d-1,1} (\triangle)\, = \nonumber\\
&=& l(\triangle) - ( d + 2) \,-\,
\sum_{codim \Theta =1} l'(\Theta) \nonumber\\
&+& \, \sum _{codim \Theta =2} l'(\Theta) l'(\Theta ^*),
\label{twistagain}
\end{eqnarray}

\begin{eqnarray}
h_{p,1} = \sum_{{codim \Theta ^*}=p+1} l^{'}(\Theta) \cdot l^{'}(\Theta ^*),
\,\,\, 1 < p < d - 1.
\end{eqnarray}
Here, the quantities $l(\Theta )$ and $l'(\Theta)$ are the numbers of
integer points on a 
face $\Theta $ of $\triangle$ and in its interior, and
similarly for $\Theta ^*$ and $\triangle ^*$.  
An $l-$dimensional face $\Theta$ can be defined by its vertices $(v_{i_1}=
...=v_{i_k})$, and the dual face defined by  $\Theta ^* =\{
m\in \triangle^* : ( m,v_{i_1})=,...,=(m,v_{i_k}) = -1\}$
is an $(n - l - 1 )$-dimensional face of $\triangle ^*$. Thus, we have a 
duality between the $l$-dimensional faces of $\triangle$ and the $(n - l
- 1)$-dimensional faces of $\triangle ^{*}$. 
The last terms in (\ref{twist}, \ref{twistagain})
correspond to the `twisted' contributions, and
the last term corresponds to
$d=4$. In this case, if the manifold has $SU(4)$ group holonomy, then
$h_{2,0} = h_{1,0} =0 $, and the remaining non-trivial Hodge number
$h_{2,2}$ is determined by:
\begin{equation}
h_{2,2} = 2 [22 + 2 h_{1,1} + h_{3,1} - h_{2,1}].
\end{equation}
Some further comments about $CY_3$ spaces are made in Section 9.


\subsection{The Web of CY Manifolds in the Holomorphic-Quotient 
Approach to Toric Geometry}

It is well known that weighted projective spaces are 
examples of {\it toric varieties}~\cite{Fulton}. The complex weighted
projective space 
${CP^n}$ can be defined as
\begin{equation}
CP^n\, \equiv \,\frac{C^{n+1}\,-\,\vec {0}}{C^*},
\end{equation}
with the action ${C^*}$: 
\begin{equation}
(x_1,...,x_{n+1})\,\Rightarrow\, (\lambda^{k_1} \cdot x_1,
...., \lambda^{k_{n+1}} \cdot x_{n+1}),\,\,\,\,\,\,
\lambda \,\in\, C \backslash {0}.
\end{equation}
The generalization of the projective space ${CP^n}$ to a toric variety
can be expressed in the following form:
\begin{eqnarray}
\mho\, \equiv \,\frac{C^n\,-\,Z_{\Sigma}}{(C^*)^p},
\end{eqnarray}
where, instead of removing the origin, as in the case of 
a simple projective space, here
one removes a point set ${Z_{\Sigma}}$, and one takes the 
quotient by a suitable set of ${C^*}$ actions. Thus, to understand the 
structure of certain geometrical spaces in the framework of toric
geometry, 
one must specify the combinatorical properties 
of the ${Z_{\Sigma}}$ and the actions $C^*$.

In the toric-geometry approach, algebraic varieties are described by a
dual pair of
lattices ${M}$ and ${N}$, each isomorphic to $Z^n$, and a {\it fan}
$\Sigma^*$~\cite{Fulton}
defined on ${N_R}$, the real extension of the lattice ${N}$.
In the toric-variety description, the equivalence relations of projective
vectors can be  considered as diagrams in the lattice $N$, in which some
vectors
${\vec v}_i$ satisfy linear relations (see later some
examples
in $P^2(1,1,1)$, $P^2(1,1,2)$, $P^2(1,2,3)$ projective spaces).
The complex dimension of the variety 
coincides with the 
dimension of the lattice ${N}$. To determine the structure of 
a toric variety in higher dimensions $d > 2$, it is useful to
introduce the notion of a {\it fan}~\cite{Greene1,Fulton}.
A fan $\Sigma^*$ is defined as a collection of 
$r$-dimensional ($0 \leq r \leq d $) convex polyhedral cones 
with apex in { 0}, with the properties that with every cone it
contains also a face,
and that the intersection of any two cones is a face of each one.

In the holomorphic-quotient
approach of Batyrev~\cite{Batyrev} and Cox~\cite{Cox},
a single homogeneous coordinate is assigned to the
system $\mho_{\Sigma}$ of varieties,
in a way similar to the usual construction of $P^n$.
This holomorphic-quotient construction gives immediately the usual
description in terms of projective spaces, and turns out to be more
natural in the descriptions of the elliptic, $K3$ and other fibrations 
of higher-dimensional CY spaces.

One can assign a coordinate $z_k: k = 1,...,{N}$ to each
one-dimensional cone in $\Sigma$.
The integer points of ${\Delta}^{*} \cap {N} $ define these 
one-dimensional cones
\begin{equation}
({v}_1,...,{v}_{N}) = {{\Sigma}_1}^{*}
\end{equation}
of the fan $\Sigma^{*}$.
The  one-dimensional cones span the vector space $N_R$ and satisfy
$({N} -n)$ linear relations with non-negative integer coefficients:
\begin{equation}
\sum_l k_j^l {v_l} = 0, \,\,\, k_j^l \geq 0.
\end{equation}
These linear relations can be used to determine
equivalence relations on the space $C^{N} \backslash Z_{\Sigma^{*}}$.
A variety ${\mho}_{\Sigma^{*}} $ is the space 
$C^N \backslash Z_{{\Sigma}^{*}}$ modulo the action of a group 
which is the product 
of a finite Abelian group and the torus $(C^{*})^{(N-n)}$ :
\begin{equation}
(z_1,...,z_{\it N} )  \sim ({\lambda} ^{k^1_j} z_1,...., 
{\lambda }^{k^N_j} z_N),\,\,\, j = 1,...,{\it N}-n.
\end{equation}
The set $Z_{\Sigma^{*}}$ is defined by the fan in the following way:
\begin{equation}
Z_{\Sigma^{*}} \, \equiv \, \bigcup_ I((z_1,...,z_{\it N}) | z_i =0 ,
\forall i \in  I)
\end{equation}
where the union is taken over all index sets 
$I = ( i_1,...,i_k)$ such that $({v}_{i_1},..., {v}_{i_k})$
do not belong to the same maximal 
cone in ${\Sigma}^{*}$, or several ${z_i}$ can vanish simultaneously
only if the corresponding one-dimensional cones 
${v_i}$ are from the same cone. 

It is clear from the above definitions that toric varieties 
can  have often singularities, which will be very important for 
understanding the link between the topological properties of 
Calabi-Yau hypersurfaces and Cartan-Lie algebras:
see the more systematic discussion in Section 8. 
The method of blowing up (blowing down) these singularities
was developed in algebraic geometry: it consists of
replacing the singular point or curve by a higher-dimensional 
(lower-dimensional) variety. The structure of the fan $\Sigma^*$
determines what kind of singularities will appear in 
Calabi-Yau hypersurfaces. For example, if the fan $\Sigma^*$ is 
simplicial, one can get only orbifold singularities in the 
corresponding variety \cite{Greene1}.

The elements of $ {\Sigma}_1^{*}$
are in one-to-one correspondence with divisors 
\begin{equation}
D_{v_i}\,=\, \mho_{\Sigma_{1i}^{*}},
\end{equation}
which are subvarieties  
given simply by ${z_i=0}$. This circumstance was used~\cite{Can1} to give
a simple graphic explanation of Cartan-Lie algebra
(CLA) diagrams, whose Coxeter number could be identified with the
intersections of the divisors $D_{v_i}$. 

Two divisors, $D_{v_i}$ and $D_{v_j}$, 
can intersect only when the  corresponding 
one-dimensional cones $v_i$ and $v_j$ lie in a single
higher-dimensional cone of the fan $\Sigma^*$.
The divisors ${D_{v_i}}$ form a free Abelian group 
${Div (\mho_{\Sigma^*})}$. In general, a  divisor
${D \in  Div (\mho_{\Sigma^*})}$
is a linear combination of some irreducible hypersurfaces with integer
coefficients:
\begin{equation}
q{D = \sum a_i \cdot D_{v_i}}.
\end{equation}
If ${a_i\,\geq \,0}$ for every ${i}$, one can say that 
${D\,>\,0}$.
For a meromorphic function $f$ on a toric variety, one can define a 
principal divisor
\begin{equation}
(f)\, \equiv \,\sum_i ord_{D_i}(f) \cdot D_i,
\end{equation}
where ${ord_{D_i}(f)}$ is the order of the meromorphic function ${f}$
at ${D_i}$.  
One can further define the zero 
divisor ${(f)_0}$ and the polar divisor ${(f)_{\inf}}$
of the meromorphic function ${f}$, such that 
\begin{equation}
{(f)\,=\,(f)_0\,-\,(f)_{\inf}}.
\end{equation}  
Any two divisors ${D_1,\,D_2}$ are linearly equivalent:
${D_1 \sim D_2}$, if their difference is a principal divisor,
${D - D'=(f)}$ for some appropriate  ${f}$. The quotient of all 
divisors ${Div (\mho_{\Sigma^*})}$
by the principal divisors forms the {Picard} group. 


The points of 
${\Delta \cap M}$ are in one-to-one correspondence with the monomials
in the homogeneous coordinates $z_i$. A general polynomial
is given by
\begin{equation}
\wp\,=\, \sum_{{m} \in {\Delta \cap M}} c_{m} 
\prod_{l=1}^{N}  z_l^{\langle {v}_l, {m} \rangle +1}.
\end{equation}   
The equation ${\wp=0}$ is well defined and ${\wp}$
is holomorphic if the condition
\begin{equation}
\langle {v}_l, {m} \rangle \, \geq \,-1 \,\,\, for \,\,all\,\,l
\end{equation}
is satisfied.
The $c_{m}$ parametrize a family ${M_{\Delta}}$ of 
CY surfaces defined by the zero locus of $\wp$.

\subsection{Three Examples of $CY_1$ Spaces}
 
\begin{figure}[th!]
   \begin{center}
   \mbox{
   \epsfig{figure=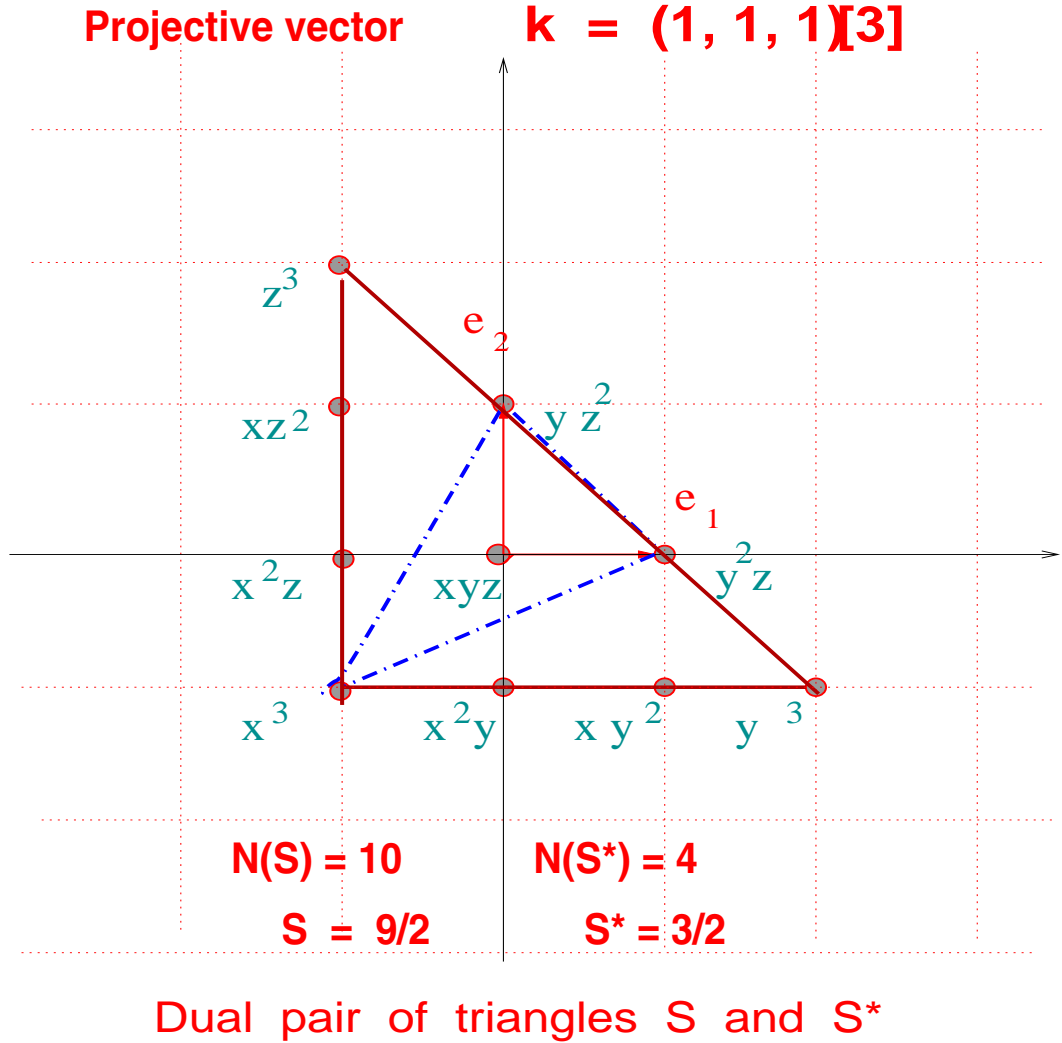,height=14 cm,width=14cm}}
   \end{center}
   \caption{\it The dual pair of reflexive plane polyhedra
 defined by the projective
vector $(1,1,1)$ with $N(S)=10$ and $N(S^*)=4$ integer
points, respectively.
$SL(2,Z)$ transformations produce an infinite
number of dual-pair triangles, conserving the 
areas $S=9/2$ and $S^*=3/2$, respectively.}
\label{vec1}
\end{figure}

As discussed in Section 1, three $CY_1$ spaces may be
obtained as simple loci of polynomial zeroes associated with
refelxive polyhedra.
For a better understanding of the preceding formalism,
we consider as warm-up examples the three elliptic reflexive 
polyhedron  pairs ${\Delta}_i$ and ${\Delta}_i^{*}$, which define the 
$CY_1$ surfaces $P^2(1,1,1)[3]$, $P^2(1,1,2)[4]$, and
$P^2(1,2,3)[6]$~\footnote{Here and subsequently, 
we use the conventional notation for such surfaces in n-dimensional
projective space:
$P^n (k_1, k_2, ...)[k_1 +k_2 + ...]$.}.
The first polyhedron
${\Delta}_I \equiv {\Delta}(P^2(1,1,1)[3])$ consists of the following
ten integer points:

\begin{eqnarray}
z^3    \,&\Longrightarrow &\,{ \mu}_1^{(I)}   \,=\,(-1,2),   \nonumber\\
x z^2  \,&\Longrightarrow &\,{ \mu}_2^{(I)}   \,=\,(-1,1),   \nonumber\\
x^2z   \,&\Longrightarrow &\,{ \mu}_3^{(I)}   \,=\,(-1,0),   \nonumber\\
x^3    \,&\Longrightarrow &\,{ \mu}_4^{(I)}   \,=\,(-1,-1),  \nonumber\\
yz^2   \,&\Longrightarrow &\,{ \mu}_5^{(I)}   \,=\,(0,1),    \nonumber\\
xyz    \,&\Longrightarrow &\,{ \mu}_6^{(I)}   \,=\,(0,0),    \nonumber\\
x^2y   \,&\Longrightarrow &\,{ \mu}_7^{(I)}   \,=\,(0,-1),   \nonumber\\
y^2z   \,&\Longrightarrow &\,{ \mu}_8^{(I)}   \,=\,(1,0),    \nonumber\\
x y^2  \,&\Longrightarrow &\,{ \mu}_9^{(I)}   \,=\,(1,-1),   \nonumber\\
y^3    \,&\Longrightarrow &\,{ \mu}_{10}^{(I)}\,=\,(2,-1).
\end{eqnarray}
and the mirror polyhedron 
${\Delta}_I^{*} \equiv {\Delta^*}(P^2(1,1,1)[3])$ consists of one
interior point and three one-dimensional cones:
\begin{eqnarray}
{ v}_1^{(I)}\,&=&\, (0,1),\nonumber\\
{ v}_2^{(I)}\,&=&\, (1,0),\nonumber\\
{ v}_3^{(I)}\,&=&\, (-1,-1).
\end{eqnarray}
We use as a basis the exponents of the following monomials:
\begin{eqnarray}
y^2 z\,&\Longrightarrow &\,\vec{e}_1\,=\,(-1,1,0), \nonumber\\
yz^2 \,&\Longrightarrow &\, \vec{e}_2\,=\,(-1,0,1).
\end{eqnarray}
where the determinant of this lattice
coincides with the dimension of
the projective vector $\vec{k}=(1,1,1)$ (see Figure~\ref{vec1}): 

\begin{eqnarray}
det \{\vec{e}_1,\vec{e}_2, \vec{e}_0\}\,=\,dim(\vec{k})\,=\,3,
\label{area3}
\end{eqnarray}
where $\vec{e}_0$ is the unit vector (1,1,1).

For this projective vector there exist 27 possibilities of
choising two monomials for constructing the basis. Of course,
all these bases are equivalent, i.e., they are connected by the $ SL(2,Z)$
modular transformations:
$$ L_{i,j}\,=\,\pmatrix{
a&b\cr
c&d\cr}$$
where $a,b,c,d \in Z$ and $ ad-bc=1$.
For the mirror polyhedron obtained from this vector, the basis should
correspond to
a lattice with determinant three times greater than (\ref{area3}),
namely 9, for example:

\begin{eqnarray}
\vec{e}_1\,=\,(-1,2,-1), \nonumber\\
\vec{e}_2\,=\,(-1,-1,2).
\end{eqnarray}
with
\begin{eqnarray}
det \{\vec{e}_1,\vec{e}_2, \vec{e}_0\}\,=\,dim(\vec{k})\,=\,9.
\end{eqnarray}
where $\vec{e}_0$ is again the unit vector (1,1,1).

To describe this toric curve, one should embed it in the toric variety 
\begin{equation}
P^2\,=\, (C^3\backslash {0})/(C\backslash {0}),
\end{equation}
where the  equivalence relation
\begin{equation}
(x_1, x_2, x_3)\,\sim \, (\lambda x_1, \lambda x_2, 
\lambda x_3)\,\,\,\, for \,\, \lambda \in C\backslash {0}
\end{equation}
is a consequence of the equation:
\begin{eqnarray}
q_1 \cdot {v}_1^{(I)}\,+\,q_2 \cdot {v}_2^{(I)}\,
+\,q_3 \cdot {v}_3^{(I)}\,=\,0,
\end{eqnarray}
where the
${q_i=1,\, i=1,2,3}$ are the exponents of ${\lambda }$.
The corresponding general polynomial describing a CY surface is
(setting $z_l \equiv x_l$):
\begin{eqnarray}
{\wp}_I\,& =&\, x_1^3 \,+\,x_2^3 \,+ \, x_3^3 
\,+\, x_1x_2x_3 \,\nonumber\\
&+&\,x_1^2x_2\,+\,x_1^2x_3\,+\,x_2^2x_1\,+
\,x_2^2x_3\,+\,x_3^2x_1\,+\,x_3^2x_2.
\end{eqnarray}
and the Weierstrass  equation can be written in the 
following form:
\begin{equation}\label{eq11}
 y^2 \cdot z\,=\, x^3 \,+\,a \cdot x \cdot y^3 \,+ \,b \cdot z^3 
\end{equation}
where we have set $x_1=x,x_2=y,x_3=z$.

\begin{figure}[th!]
   \begin{center}
   \mbox{
   \epsfig{figure=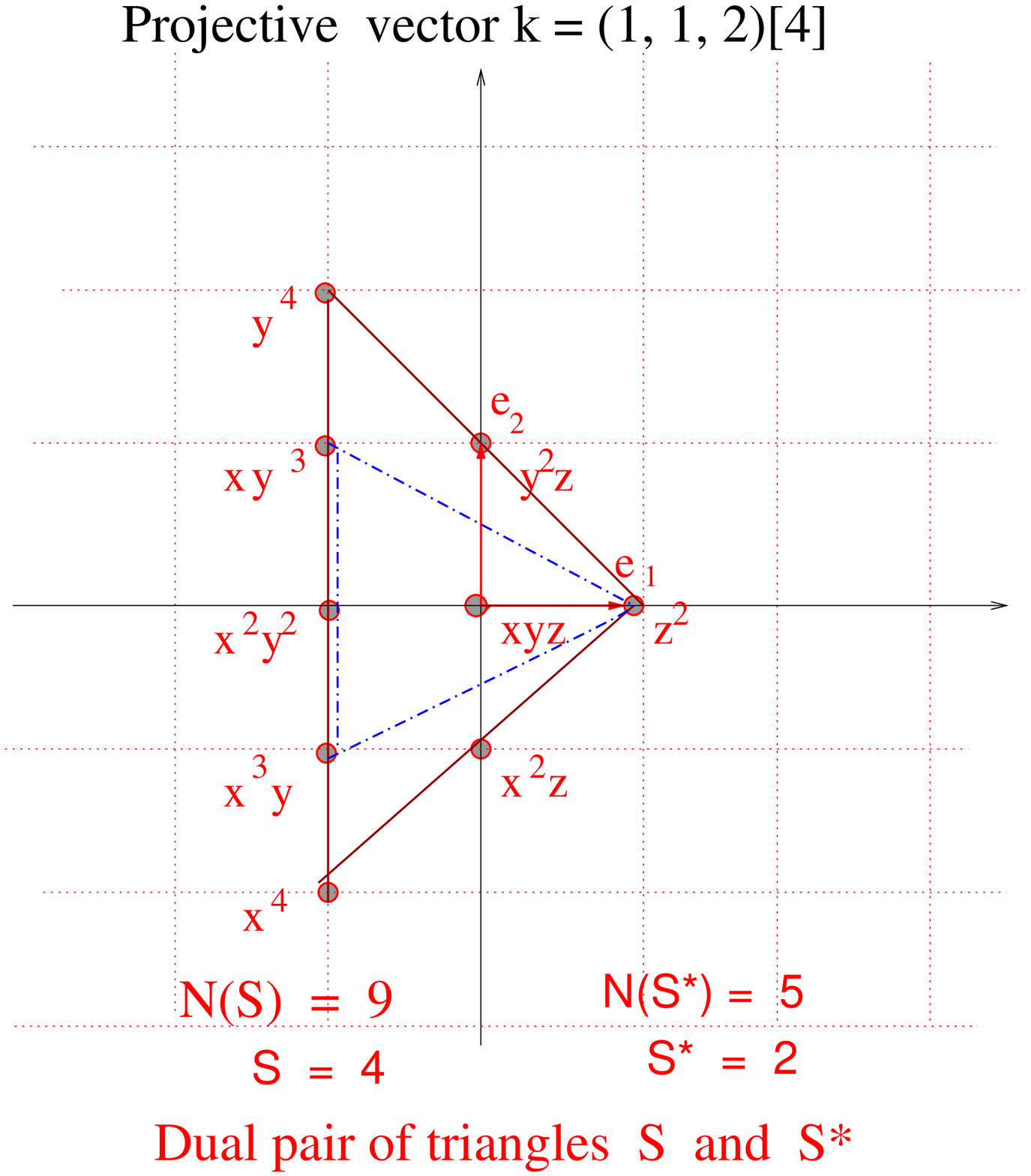,height=14 cm,width=14cm}}
   \end{center}
   \caption{\it The dual pair of reflexive plane polyhedra
 defined by the projective
vector $(1,1,2)$ with $N(S)=9$ and $N(S^*)=5$ integer points,
respectively.  
$SL(2,Z)$ transformations produce an infinite
number of dual-pair triangles, conserving the areas
$S=4$ and $S^*=2$, respectively.}
\label{vec2}
\end{figure}

The second dual pair of triangle polyhedra  
${\Delta}_{II} \equiv {\Delta}(P^2(1,1,2)[4])$ 
and its mirror  ${\Delta}_{II}^{*} \equiv {\Delta^*}(P^2(1,1,2)[4])$
have nine points
\begin{eqnarray}
y^4    \,&\Longrightarrow &\,{ \mu}_1^{(II)}\, =\,(-1,2),    \nonumber\\
xy^3   \,&\Longrightarrow &\,{ \mu}_2^{(II)}\, =\,(-1,1),    \nonumber\\
x^2y^2 \,&\Longrightarrow &\,{ \mu}_3^{(II)}\, =\,(-1,0),    \nonumber\\
x^3y   \,&\Longrightarrow &\,{ \mu}_4^{(II)}\, =\,(-1,-1),   \nonumber\\
x^4    \,&\Longrightarrow &\,{ \mu}_5^{(II)}\, =\,(-1,-2),   \nonumber\\ 
y^2z   \,&\Longrightarrow &\,{ \mu}_6^{(II)}\, =\,(0,1),     \nonumber\\
xyz    \,&\Longrightarrow &\,{ \mu}_7^{(II)}\, =\,(0,0),     \nonumber\\
x^2z   \,&\Longrightarrow &\,{ \mu}_8^{(II)}\, =\,(0,-1),    \nonumber\\
z^2      \,&\Longrightarrow &\,{ \mu}_9^{(II)}\, =\,(1,0).
\end{eqnarray}
and five points, respectively (see Figure~\ref{vec2}). 

We use as a basis the exponents of the following monomials:
\begin{eqnarray}
z^2  \,&\Longrightarrow &\,\vec{e}_1\,=\,(-1,-1,1), \nonumber\\
y^2z \,&\Longrightarrow &\, \vec{e}_2\,=\,(-1,1,0).
\end{eqnarray}
where   the determinant of this lattice
coincides with the dimension of
the projective vector $\vec{k}=(1,1,2)$: 

\begin{eqnarray}
det \{\vec{e}_1,\vec{e}_2, \vec{e}_0\}\,=\,dim(\vec{k})\,=\,4,
\label{area4}
\end{eqnarray}
where $\vec{e}_0$ is again the unit vector (1,1,1).

To get the mirror polyhedron with 5 integer points, 4 on the edges
and one interior point, one should find a basis with lattice 
determinant twice (\ref{area4}), namely 8, for example:
 \begin{eqnarray}
\vec{e}_1\,&=&\,(-1,-1,1), \nonumber\\
\vec{e}_2\,&=&\,(-2,2,0).
\end{eqnarray}
The  following four points  define four 
one-dimensional cones in ${\Sigma}_1({\Delta}_{II}^{*})$:
\begin{eqnarray}
{v}_1^{(3)} &=& (1,0),  \nonumber\\
{v}_2^{(3)} &=& (-1,0), \nonumber\\
{v}_3^{(3)} &=& (-1,-1),\nonumber\\
{v}_4^{(3)} &=& (-1,1).
\end{eqnarray}
Using the linear relations between the four one-dimensional cones,
the corresponding
$(C^{*})^2$ is seen to be given by ($z_l \equiv \chi_l$):
\begin{eqnarray}
(\chi_1,\chi_2,\chi_3,\chi_4) \Longrightarrow 
(\lambda  {\mu}^2 \chi_1, \lambda \chi_2,
\mu \chi_3, \mu \chi_4).
\end{eqnarray}
and the general polynomial has the following nine terms:
\begin{eqnarray}
{\wp}_{II} &=& {\chi}_2^2 {\chi}_3^4\, +\,{\chi}_2^2{\chi}_3^3{\chi}_4\,
+\,{\chi}_2^2{\chi}_3^2{\chi}_4^2\,+\,{\chi}_2^2{\chi}_3{\chi}_4^3 
+\,{\chi}_2^2{\chi}_4^4 \nonumber\\
&+&\, {\chi}_1{\chi}_2{\chi}_3^2\,+\,
{\chi}_1{\chi}_2{\chi}_3{\chi}_4\,+\,{\chi}_1{\chi}_2{\chi}_4^2\,+
\,{\chi}_1^2
\end{eqnarray}
in this case.

The vectors $\vec{k}=(1,1,1)$ and $\vec{k}=(1,1,2)$ have three common
monomials  and a related reflexive segment-polyhedron, corresponding
to the projective vector  $\vec{k}_2=(1,1)$ of  $CP^1$. This circumstance 
can be used further in the construction of the projective algebra
in which these two vectors appear in the same chain. 

\begin{figure}[th!]
   \begin{center}
   \mbox{
   \epsfig{figure=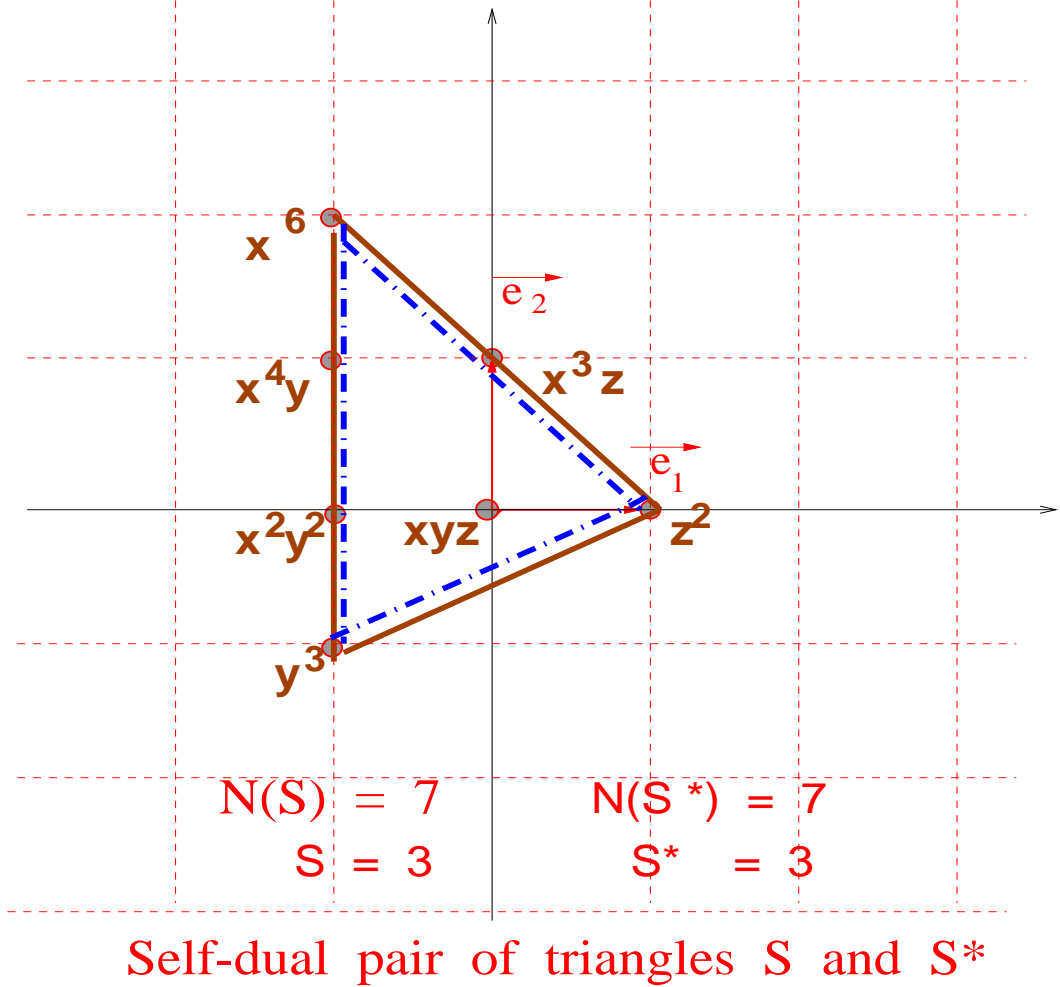,height=14 cm,width=14cm}}
   \end{center}
   \caption{\it The self-dual pair of reflexive plane polyhedra
 defined by the projective
vector (1,2,3) with $N(S)=7$ and $N(S^*)=7$ integer points.  
$SL(2,Z)$ transformations produce an infinite
number of the dual-pair triangles, conserving the areas
$S=3$ and $S^*=3$, respectively.}
\label{vec3}
\end{figure}

The last $CY_1$ example involves the plane of  the 
projective vector ${\vec {k}\,=\,(1,2,3)}$,
whose polyhedron ${\Delta}_{III} \equiv {\Delta}(P^2(1,2,3)[6])$  
and its mirror partner  
${\Delta}_{III}^{*} \equiv {\Delta^*}(P^2(1,2,3)[6]) $  
both have seven self-dual points,
and one can check the existence of the 
following six one-dimensional cones (see Figure~\ref{vec3}):
\begin{eqnarray}
z^2      \,&\Longrightarrow &\,{v}_1^{(III)}\,=\,(1,0),   \nonumber\\
x^2 y^2  \,&\Longrightarrow &\,{v}_2^{(III)}\,=\,(-1,0),  \nonumber\\
x^3z     \,&\Longrightarrow &\,{v}_3^{(III)}\,=\,(0,1),   \nonumber\\
y^3      \,&\Longrightarrow &\,{v}_4^{(III)}\,=\,(-1,-1), \nonumber\\
x^4 y    \,&\Longrightarrow &\,{v}_5^{(III)}\,=\,(-1,1),  \nonumber\\
x^6      \,&\Longrightarrow &\,{v}_6^{(III)}\,=\,(-1,2).
\end{eqnarray}
We use as a basis the exponents of the following monomials:
\begin{eqnarray}
z^2   \,&\Longrightarrow &\, \vec{e}_1\,=\,(-1,-1,1), \nonumber\\
x^3 z \,&\Longrightarrow &\, \vec{e}_2\,=\,(2,-1,0).
\end{eqnarray}
where   the determinant of this lattice
coincides with the dimension of
the projective vector $\vec{k}=(1,2,3)$: 

\begin{eqnarray}
det \{\vec{e}_1,\vec{e}_2, \vec{e}_0\}\,=\,dim(\vec{k})\,=\,6.
\end{eqnarray}
As in the case of the two projective vectors $\vec{k}=(1,1,1)$ and
$\vec{k}=(1,1,2)$, the vectors $\vec{k}=(1,1,2)$ and $\vec{k}=(1,2,3)$
also have three common monomials, corresponding to the 
reflexive segment polyhedron described by the vector $\vec{k}_2=(1,1)$
in $CP^1$ projective space. Hence these vectors will appear in the second
chain of the plane projective algebra.  

Thus one can see that,
with these three plane projective vectors,
$\vec{k}=(1,1,1)$, $\vec{k}=(1,1,2)$, $\vec{k}=(1,2,3)$, one
finds only triangle reflexive polyhedra intersecting the integer 
planar lattice in $10+4^*$, $9+5^*$, $7+7^*$ points. Of course, on the
plane
one can find other reflexive polyhedra, whose intersection with 
the integer plane lattice will give new $CP^1$ surfaces
corresponding to different polygons with more then three
vertices,
such as a reflexive pair of
square and rhombus. These new figures can be obtained using the techniques
of extended vectors. 

In the following, we will go on to study
reflexive polyhedron pairs in three-dimensional space.
The corresponding general polynomial can be expressed in terms of six
variables, and contains seven monomials:
\begin{eqnarray}
{\wp}_{III} &=& z_1^2 z_3 + z_2^2 z_3 z_4^2 z_5^2 z_6^2 + 
z_1 z_2 z_3^2 z_5^2 z_6^3 +
z_2^2 z_4^3 z_5 + \nonumber \\
&+&z_2^2 z_3^2 z_4 z_5^3 z_6^4 + z_2^2 z_3^3 z_5^4 z_6^6+
z_1 z_2 z_3 z_4 z_5 z_6.
\end{eqnarray}
The ${C^*}^4$ action is determined by the following linear relations:
\begin{eqnarray}
{v}_1^{(III)} \,&+&\, {v}_2^{(III)} \,=\,{0},\nonumber\\
 2 {v}_1^{(III)}\,&+& \,{v}_4^{(III)}\, +\, 
{v}_5^{(III)}\, =\,{0},\nonumber \\
{v}_1^{(III)}\, &+&\, {v}_3^{(III)} \,+\, {v}_4^{(III)}\, =
\, {0},\nonumber\\
3 {v}_1^{(III)}\, &+&\, 2 {v}_4^{(III)}\, +\, {v}_6^{(III)}\, =
\, {0} 
\end{eqnarray}
between the elements of ${\Sigma}_1({\Delta}_{III}^{*})$, and is given
by
\begin{eqnarray}
(z_1,z_2,z_3,z_4,z_5,z_6)
\longrightarrow (\lambda {\mu}^2 {\nu} {\rho}^3 z_1,
\lambda z_2, \nu z_3, \mu \nu {\rho}^2 z_4, \mu z_5, \rho z_6).
\end{eqnarray}
One can introduce the following  birational map between
$P^2(1,2,3)[6]$ and $ {\mho}_{{\Sigma}^{*}}$: 
\begin{eqnarray}
            z_1^2z_3 &=& y_3^2\\
       z_2^2z_4^3z_5 &=& y_2^3\\
z_2^2z_3^3z_5^4z_6^6 &=& y_1^6
\end{eqnarray}
Then, a dimensionally-reduced example of a CY manifold embedded in a toric
variety 
is described by the  weight vector ${k}=(1,2,3)$ and the zero 
locus of the Weierstrass polynomial
\begin{eqnarray}
{\wp}_{III} = y_1^6 \,+\, y_2^3\,+\, y_3^2 \,+\, y_1y_2y_3 
+\,y_1^4y_2 \,+
\,y_1^2y_2^2 \,+\, y_1^3y_3.
\end{eqnarray}  
The elliptic Weierstrass equation can be written
in the weighted projective space ${P^2(1,2,3)[6]}$ as
\begin{equation}
y^2\,=\,x^3\,+\,a \cdot x \cdot z^4 \, +\, b \cdot z^6
\label{Weier1}
\end{equation}
with the following equivalence relation
\begin{eqnarray}\label{eq13}
(x, y, z) \, \sim \, ( \lambda^2 x,\lambda^3 y, \lambda z),\,
\,\,\, \lambda \in C\backslash {0}
\end{eqnarray}  
in this case.

\begin{figure}[ht!]
   \begin{center}
   \mbox{
   \epsfig{figure=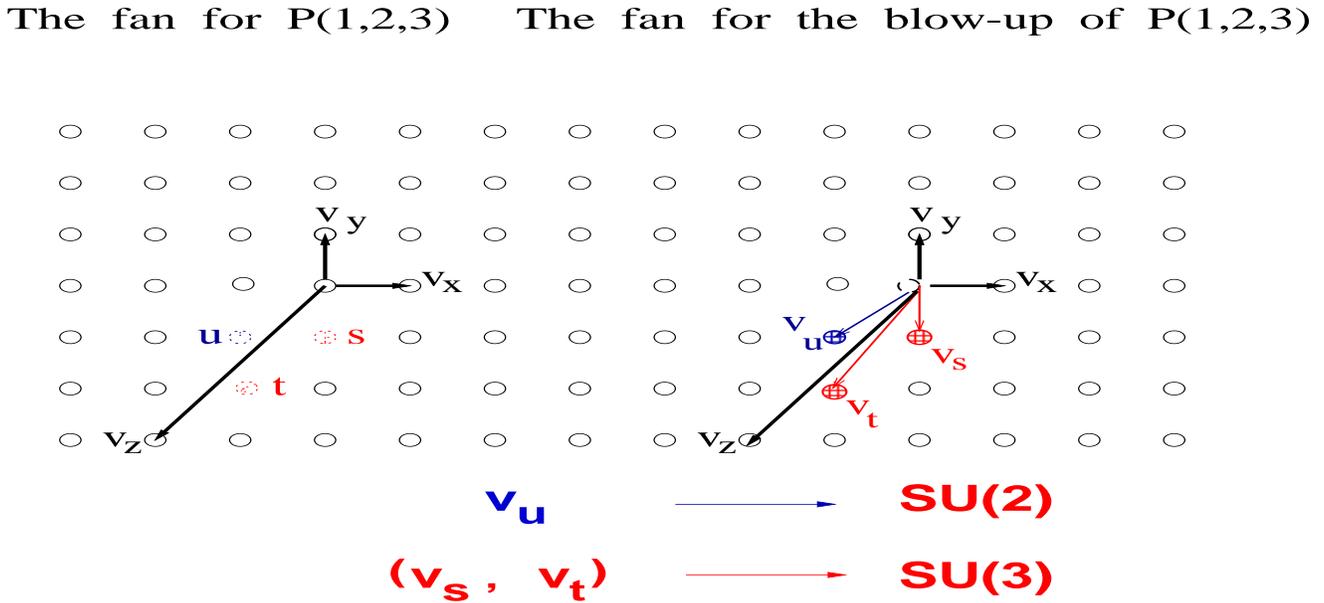,height=8cm,width=18cm}}
   \end{center}
   \caption{\it The toric variety $P(1,2,3)$ with two orbifold
singularities at the points $y=z=0$ and $x=z=0$ can be blown up by extra
divisors $D(v_u)$ and $D(v_s)$, $D(v_t)$, respectively.}
\label{div}
\end{figure}

These examples illustrate how toric varieties can be defined by the 
quotient of $C^k \backslash Z_{\Sigma}$, and not only by a group 
$(C \backslash {0})^{k-n}$. One should divide 
$C^k \backslash Z_{\Sigma}$ also by a finite Abelian 
group $G(v_1,...,v_k)$, which is determined by the relations between 
the $D_{v_i}$ divisors. In this case,
the toric varieties can often have orbifold singularities, 
$C^k \backslash G$. For example, the toric variety defined by 
(\ref{Weier1}) looks near the points ${y=z=0}$ and ${x=z=0}$
locally like ${C^2\backslash Z_2}$ (related to the $SU(2)$ algebra) and  
${C^2\backslash Z_3}$ (related to the
$SU(3)$ algebra), respectively, as seen in  Figure~\ref{div}.

%
%
%
%

\section{Gauge Group Identifications from Toric Geometry}

\subsection{ Calabi-Yau Spaces as Toric Fibrations}

As discussed in section 2, any Calabi-Yau manifold
can be considered as a hypersurface in a toric variety,
with a corresponding reflexive polyhedron
${\Delta}$ with a
positive-integer lattice ${\Lambda}$, associated with a dual 
polyhedron ${\Delta^*}$ in the dual lattice ${\Lambda^*}$.
The toric variety is determined by a fan  $\Sigma^*$, 
consisting of the cones which are given by a triangulation of $\Delta^*$.
A large subset of reflexive polyhedra and their corresponding
Calabi-Yau manifolds can be classified in terms of their
fibration structures. In this way, it is possible, as we
discuss later,
to connect the structures of all the projective vectors of the one
dimensionality
with the projective vectors of other dimensionalities, and
thereby construct a new algebra in the set of all
`reflexive' projective vectors that gives the full set of ${CY_d}$
hypersurfaces in all dimensions: $d = 1, 2, 3, ...$.

In order to embark on this programme, it is useful first to review two
key operations, {\it intersection} and {\it projection},
which can give possible fibration structures for reflexive 
polyhedra~\cite{Can1}:
\begin{itemize}
\item{There may exist a projection operation 
$\pi:\Lambda \rightarrow \Lambda _{n-k}$, where  $\Lambda_{n-k}$ is an 
$(n-k)$-dimensional sublattice, and $\pi (\Delta)$ is also a reflexive
polyhedron, and}
\item{there may exist an intersection projection $J$ through the
origin of a reflexive polyhedron, such that $J(\Delta)$ is again an
$(n-l)$-dimensional reflexive polyhedron, and}
\item{these operations may exhibit the following duality properties:}
\end{itemize}
\begin{eqnarray}
 \Pi(\Delta)\,&  \Leftrightarrow &\,  J(\Delta^*)  \nonumber\\
   J(\Delta)\,&  \Leftrightarrow &\,\Pi(\Delta^*).
\end{eqnarray}
For a reflexive polyhedron $\Delta$ with fan $\Sigma$ over a triangulation 
of the facets of ${\Delta}^*$, the CY hypersurface in variety 
${\mho}_{\Sigma}$ is given by the zero locus of the polynomial:
\begin{equation}
\wp \,=\, \sum_{\mu \in {\Delta \cap M}}\, c_{\vec {\mu}}\,\cdot
\prod_{i=1}^{N}  z_i^{\langle \vec{v}_i \cdot \vec{\mu} \rangle +1}. 
\end{equation}
One can consider the variety 
${\mho}_{\Sigma}$ as a fibration over the base ${\mho}_{{\Sigma}_{base}}$
with generic fiber ${\mho}_{{\Sigma}_{fiber}}$. 
This fibration structure can be written in terms of homogeneous 
coordinates. The fiber as an algebraic subvariety is determined by the 
polyhedron ${\Delta}^*_{fiber} \subset {\Delta}^*_{CY}$, whereas the base
can be seen as a projection of the fibration along the fiber. The set of 
one-dimensional cones in ${\Sigma}_{base}$ (the primitive generator of a 
cone is zero or ${\tilde {v}}_i$) is the set of images of 
one-dimensional cones in ${\Sigma}_{CY}$ (with primitive generator
${v}_j$) that do not lie in $N_{fiber}$. The image ${\Sigma}_{base}$
of ${\Sigma}_{CY}$ under $\Pi : N_{CY} \rightarrow N_{base}$
gives us the following relation:
\begin{eqnarray}
\Pi {v}_i\, =\, r_i^j \cdot {\tilde {v}}_j,
\end{eqnarray}
if $\Pi {v}_i$ is in the set of one-dimensional cones determined by 
${\tilde{v}}_j$ $r^j_i \in N$, otherwise $r_i^j=0$.

Similarly, the base space is the weighted projective space with the torus
transformation:
\begin{eqnarray}
({\tilde x}_1,..., {\tilde x}_{\tilde N})\,\, \, \sim
({\lambda}^{{\tilde k}^1_j} \cdot {\tilde x}_1,...,
{\lambda}^{{\tilde k}^{\tilde N}_j} \cdot {\tilde x}_{\tilde N}),
\,\,\, j=1,...,\tilde N - \tilde n,
\end{eqnarray} 
where the ${\tilde k}^i_j$ are integers such that
$\sum_j {{\tilde k}_i^j} {\tilde {v}}_j = 0$.
The projection map from the variety ${\mho}_{\Sigma}$ to the base can be 
written as
\begin{eqnarray}
{\tilde x}_i\,=\, \prod_j x_j^{r_j^i},
\end{eqnarray}
corresponding to the following redefinitions of the torus transformation
for ${\tilde x}_i$: 
\begin{eqnarray}
\Pi:
{\tilde x}_i \rightarrow {\lambda}^{{k^j_l}\cdot {r^i_j}} \cdot 
{\tilde x}_i ,\,\,\,\, \sum{ k_l^j} \cdot {r_j^i} \cdot {\tilde {v}}_i \,=\,0.
\end{eqnarray}

\noindent
In the toric description of $K3$ surfaces with elliptic fibers, denoted by
${\Delta^*}_{fiber}$,  one can
consider the following divisors: $D_{fiber}$,  $D_{section}$, 
$D_{v_a}$ and $D_{v_b}$. The last pair of divisors correspond to
lattice points of
$\Delta^*$ that are `above' or `below' the fiber, respectively. Let us
consider 
the case when all divisors $D_{v_a}$ (or $D_{v_b}$) shrink to zero size.
In this case, there appears a $K3$ hypersurface with two point 
singularities, which belong to the {ADE} classification. 
The process of blowing up these singularities gives the primordial
$K3$ manifold, and its intersection structure is given by the structure of
the edges. The Cartan-Lie algebra (CLA) diagrams of the gauge groups that
appear when the exceptional
fibers are blown down to points are nothing but the edge diagrams of the 
upper and lower parts of $\Delta^*$ without vertices, respectively.
A simple well-known example with elliptic fiber and with base ${P^1}$
is given by the following Weierstrass equation for the  fiber:
\begin{equation}
{y^2\,=\, x^3\,+\,f(z_1,z_2)\cdot x \cdot z^4\,+\,g(z_1,z_2)\cdot z^6,}
\end{equation}
where the coefficients ${f(z_1,z_2), g(z_1,z_2)}$ are functions on the
base.

In the following parts of this Section, we discuss some examples of
$K3$ spaces from our general classification, and explain the
identification of their corresponding gauge groups.

\subsection{Examples of $K3$ Toric Fibrations with $J=\Pi$ Weierstrass
structure}

As a first example, we
consider the case of the elliptic $K3$ hypersurface
 with elliptic fiber $P^2(1,2,3)[6]$ defined by the integer positive lattice 
with basis (we explain this lattice basis later in terms of
our algebraic description):
$$\pmatrix{
\vec{e}_1\cr
\vec{e}_2\cr
\vec{e}_3\cr}\,=\,\pmatrix{
-m&n&0&0\cr
-2&-2&1&0\cr
-1&-1&-1&1\cr},
$$
where we  consider the following 12 pairs of integer numbers 
$(m,n)$ which are 
taken from the numbers: $1,2,3,4,5,6$,
 $$\{\,(1,1),\,(1,2),\, 
(1,3),\,(1,4),\,(1,5),\,(1,6),\,(2,3),\,(2,5)\,(3,4),\,(4,5),\,
(5,6)\}.$$
With this choice of the pairs, the basis above determines 
a self-dual set of 12 projective 
$\vec{k}_4$-vectors:
\begin{eqnarray}
m\,=1,\,n\,=1\, &\Longrightarrow &\,
\vec{k}_4\,=\, (1,1,4,\,\,6\,\,)[12],
\,\,\,\,\,\,\, \Longleftrightarrow\,\,\,\,\,\,\,(5,6,22,33) \nonumber\\
m\,=1,\,n\,=2\, &\Longrightarrow &\,
\vec{k}_4\,=\, (1,2,6,\,\,9\,\,)[18],
\,\,\,\,\,\,\,\,\Longleftrightarrow\,\,\,\,\,\,\,(3,5,16,24) \nonumber\\
m\,=1,\,n\,=3\, &\Longrightarrow &\,
\vec{k}_4\,=\, (1,3,8,\,\,12)   [24],
\,\,\,\,\,\,\, \Longleftrightarrow\,\,\,\,\,\,\,(2,5,14,21) \nonumber\\
m\,=1,\,n\,=4\, &\Longrightarrow &\,
\vec{k}_4\,=\, (1,4,10,15)      [30],
\,\,\,\,\,\,\,\Longleftrightarrow\,\,\,\,\,\,\, DI^{'} \nonumber\\
m\,=1,\,n\,=5\, &\Longrightarrow &\,
\vec{k}_4\,=\, (1,5,12,16)      [36],
\,\,\,\,\,\,\,\Longleftrightarrow\,\,\,\,\,\,\, self-dual \nonumber\\
m\,=1,\,n\,=6\, &\Longrightarrow &\,
\vec{k}_4\,=\, (1,6,14,21)      [42], 
\,\,\,\,\,\,\,\Longleftrightarrow\,\,\,\,\,\,\, self-dual \nonumber\\
m\,=2,\,n\,=3\, &\Longrightarrow &\,
\vec{k}_4\,=\, (2,3,10,15)      [30],
\,\,\,\,\,\,\,\Longleftrightarrow\,\,\,\,\,\,\, self-dual \nonumber\\
m\,=2,\,n\,=5\, &\Longrightarrow &\,
\vec{k}_4\,=\, (2,5,14,21)      [42], 
\,\,\,\,\,\,\,\,\Longleftrightarrow\,\,\,\,\,\,\,(1,3,8,12)  \nonumber\\
m\,=3,\,n\,=4\, &\Longrightarrow &
\vec{k}_4\,=\, (3,4,14,21)      [42],
\,\,\,\,\,\,\,\Longleftrightarrow\,\,\,\,\,\,\, DI^{''} \nonumber\\
m\,=3,\,n\,=5\, &\Longrightarrow &\,
\vec{k}_4\,=\, (3,5,16,24)      [48],
\,\,\,\,\,\,\,\,\Longleftrightarrow\,\,\,\,\,\,\,(1,2,6,9) \nonumber\\
m\,=4,\,n\,=5\, &\Longrightarrow &\,
\vec{k}_4\,=\, (4,5,18,27)      [54], 
\,\,\,\,\,\,\,\Longleftrightarrow\,\,\,\,\,\,\, DI^{'''} \nonumber\\
m\,=5,\,n\,=6\, &\Longrightarrow &\,
\vec{k}_4\,=\, (5,6,22,33)      [66], 
\,\,\,\,\,\,\,\,\Longleftrightarrow\,\,\,\,\,\,\,(1,1,4,6)  \nonumber\\
\label{selfdset}
\end{eqnarray}
Later this set will emerge as the intersection-projection
symmetric $XIX$ chain ($J=\Pi$) of our 
algebraic classification.
In this example, one can see that the projective vectors corresponding to
the tetrahedra produce a self-dual set. We also show in
(\ref{selfdset}) the duality
relations between 6 other vectors and some of the vectors in Table~1.

However, three of the projective vectors in (\ref{selfdset}),
$\vec{k}_4\,=\,(1,4,10,15)[30]$, 
$ (3,4,14,21)[42]$ and $(4,5,18,27)[54]$,  correspond
to polyhedra with 5 vertices, and their duals can be found among
higher-level $K3$ spaces. They are found by 
double intersections (DI) among the 
five-dimensional extensions of the $K3$ vectors
shown in Table 1:
\begin{eqnarray}
\vec{k}_4        =(1,4,10,15)  [30]      
&&\stackrel{DI^{'}}{ \Longleftrightarrow }
\{ \vec{k}^{ex}_5=(0,1,6,8,15) [30]\}      \bigcap
\{ \vec{k}^{ex}_5=(6,1,0,14,21)[42]\}      \nonumber\\ 
\vec{k}_4        =(3,4,14,21)  [42]      
&&\stackrel{DI^{''}}{\Longleftrightarrow} 
\{ \vec{k}^{ex}_5=(2,1,0,6,9)  [18]\}      \bigcap
\{ \vec{k}^{ex}_5=(0,1,2,4,7)  [14]\}      \nonumber\\ 
\vec{k}_4        =(4,5,18,27)  [54]      
&&\stackrel{DI^{'''}}{\Longleftrightarrow}  
\{ \vec{k}^{ex}_5=(1,0,1,4,6)  [12]\}      \bigcap
\{ \vec{k}^{ex}_5=(0,1,1,3,5)  [10]\}      \nonumber\\ 
\end{eqnarray}
as discussed in more detail in Section 6.

The ascending Picard numbers  for polyhedra
in this chain include:
\begin{eqnarray}
({\Delta}(P^3(1,6,14,21)[42]):\,\aleph &=& 24(24^*),
\,\,Pic=10(10^*)\,
\nonumber\\
\approx 
({\Delta}(P^3(1,5,12,18)[36]):\,\aleph &=& 24(24^*),
\,\,Pic=10(10^*)\,
\nonumber\\
\subset 
({\Delta}(P^3(1,4,10,15)[30]):\,\aleph &=& 25(20^*),
\,\,Pic=9(11^*) \,
\nonumber\\
\subset 
({\Delta}(P^3(1,3,8,12) [24]):\,\aleph &=& 27(15^*),
\,\,Pic=8(14^*) \,
\nonumber\\
\subset
({\Delta}(P^3(1,2,6,9)  [18]):\,\aleph &=& 30(12^*),
\,\,Pic=6(16^*) \,
\nonumber\\
\subset 
({\Delta}(P^3(1,1,4,6)  [12]):\,\aleph &=& 39(9),
\,\,Pic=2(18^*) \,\subset......... 
\end{eqnarray}
In the case of the  mirror polyhedron chain, there is the inverse
property:
${\Delta}^*(P^3(1,6,14,21)[42])$ corresponds to the maximal 
member of the set of mirror polyhedra. These Picard numbers 
are tabulated  in Table~1, together with those of the other $K3$ spaces.

\begin{figure}[th!]
   \begin{center}
   \mbox{
   \epsfig{figure=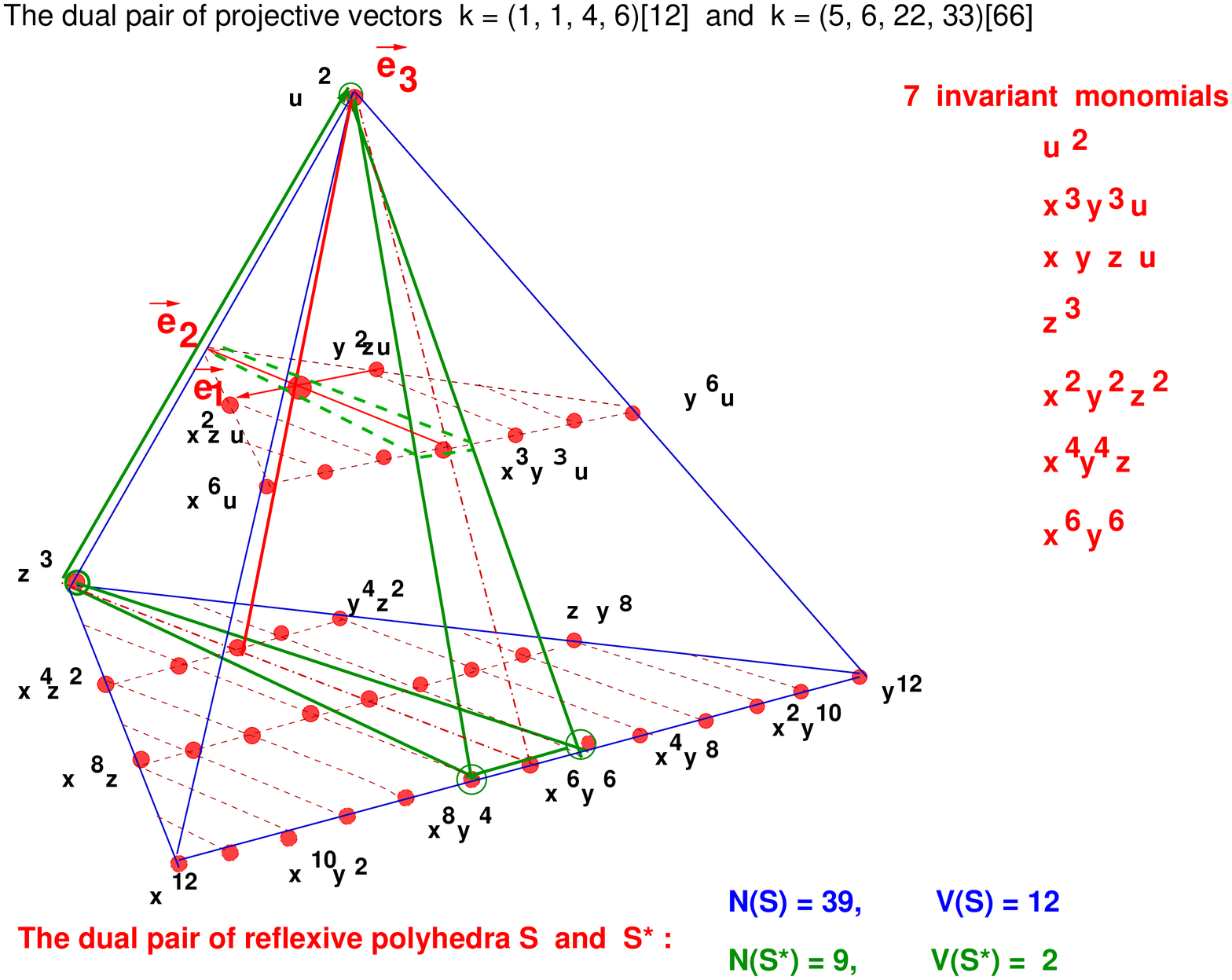,height=14 cm,width=14cm}}
   \end{center}
   \caption{\it The dual pair of reflexive planar
polyhedra defined by the eldest projective
vector (1,1,4,6) with $N(S)=39$ and the youngest projective
vector (5,6,22,33) with $N(S^*)=9$ 
integer points (marked by circles), respectively.  
$SL(3,Z)$ transformations produce an infinite
number of dual pairs of tetrahedra, conserving the volumes
$Vol(S) = 12$, $Vol(S^*) = 6$, respectively.}
\label{vec1146}
\end{figure}

\begin{figure}[th!]
   \begin{center}
   \mbox{
   \epsfig{figure=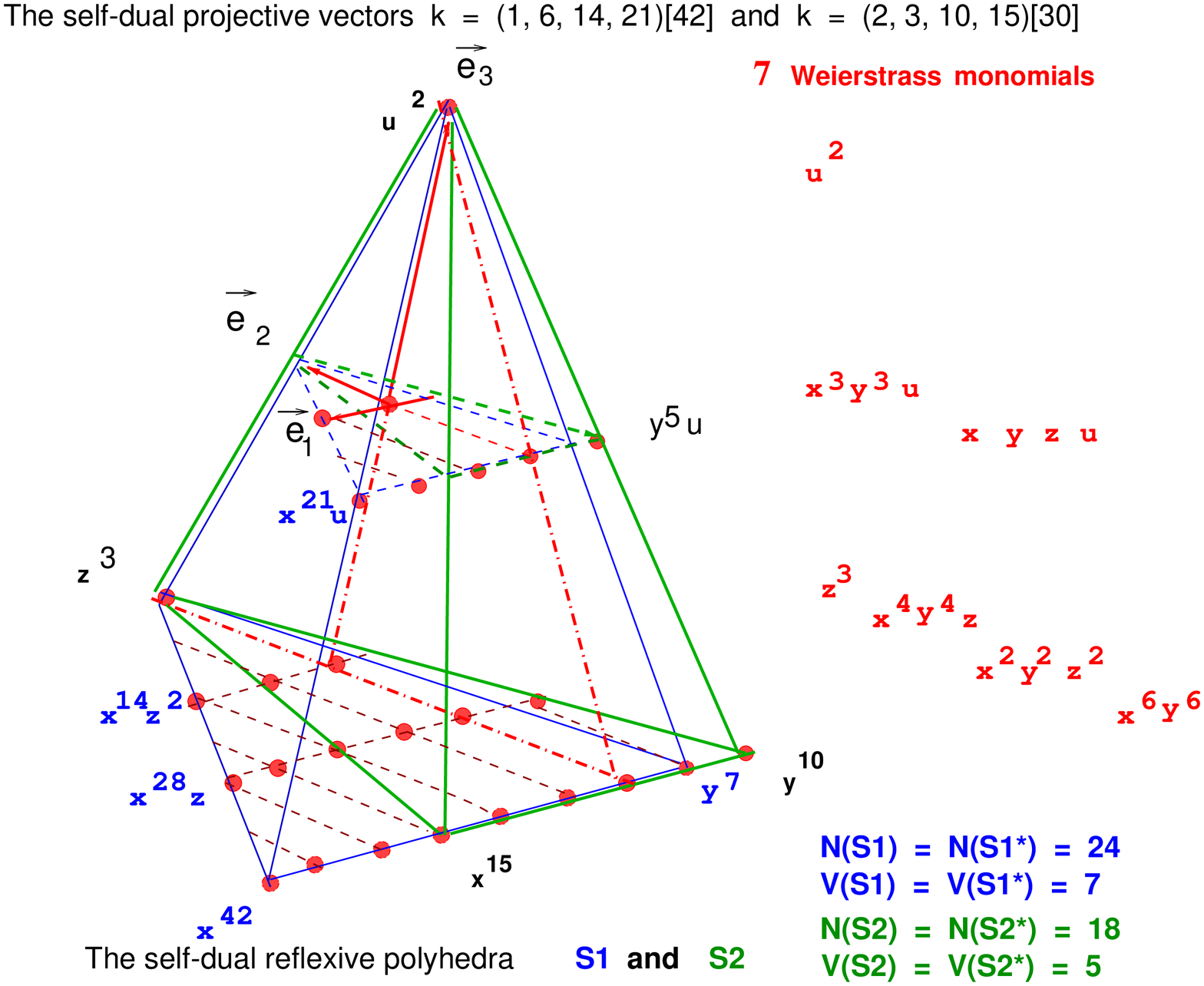,height=14 cm,width=14cm}}
   \end{center}
   \caption{\it The self-dual polyhedra in the chain XIX
determined by projective
vector (1,6,14,21) with $N(S1)=24$ and 
vector (2,3,10,15) with $N(S2)=18$ 
integer points, respectively.  
$SL(3,Z)$ transformations produce an infinite
number of dual pairs of tetrahedra, conserving the volumes
$Vol(S1)=7$  and $Vol(S2)=5$, respectively.}
\label{vec161421}
\end{figure}

In the chain (\ref{selfdset}), the mirror polyhedra,  ${\Delta }^{*}$,
have an
intersection plane
$H^*_{fiber}$ through the interior point which defines an 
elliptic-fiber triangle
with seven integer points, $P^2(1,2,3)[6]$ 
(see Figures \ref{vec1146},\ref{vec161421}):
\begin{equation}
\Delta^*_{fiber}\,=\,\Delta^* \,\bigcap \,  H^*_{fiber}.
\end{equation}
 By mirror symmetry in the 
polyhedron ${\Delta}$, a projection operator $\pi$ can be defined:
$\pi: M \rightarrow M_{n-1}$, where $M_{n-1}$ is an $(n-1)$-dimensional
sublattice, such that ${\pi}({\Delta})$ is a reflexive polyhedron in 
$M_{n-1}$. This reflexive polyhedron also consists of seven points,
so it is self-dual. Also, one can find a planar intersection $H$
through ${\Delta}$ and through the interior point, which also
produces the reflexive polyhedron with seven points, 
namely the fiber $P^2(1,2,3)[6]$
(see Figures \ref{vec1146},\ref{vec161421}):
\begin{equation}
\Delta_{fiber}\,=\,\Delta \,\bigcap\,  H_{fiber}.
\end{equation}
The dual pair of tetrahedra ${\Delta}(P^3(1,1,4,6)[12]$
and ${\Delta}(P^3(5,6,22,33)[66]$ consist of $39$ 
and $9$ points, respectively, as seen in Figure~\ref{vec1146}.
They are the biggest and smallest polyhedra in the chain (\ref{selfdset}),
and all other tetrahedra in this chain can be found in this Figure.
This contains, in particular,
the two  self-dual polyhedra ${\Delta}(P^3(1,6,14,21)[42]$
and ${\Delta}(P^3(2,3,10,15)[30]$ consist of $24+24^*$ 
and $18+18^*$ points,
respectively, as seen in Figure~\ref{vec161421}:
\begin{eqnarray}
&&\, (0,0,1),\, (0,1.-1), \,(-1,-2,-1), \, (6,-2,-1); \nonumber\\
&&\, (0,0,1),\, (0,1.-1), \,(-2,-2,-1), \, (3,-2,-1).\nonumber\\
\end{eqnarray}

\noindent
We now consider the intersection of the three-dimensional polyhedron 
${\Delta}(P^3(1,6,14,21)[42])$ with the two-dimensional
plane $H$ through the interior point. The intersection of this plane with the 
polyhedron, $H\bigcap {\Delta}$, forms a reflexive polyhedron fiber
$P^2(1,2,3)$ with seven points. The equation of this plane in canonical
coordinates $\mu_1,\mu_2,\mu_3$ is: $m_1\,=\,0.$
The fiber consists of the following polyhedron points:
\begin{eqnarray}
{ v}_0\,&=&\,( 0, \underline{  0, 0})   \nonumber\\
{ v}_1\,&=&\,( 0, \underline{ -1, 0})   \nonumber\\
{ v}_2\,&=&\,( 0, \underline{  0, 1})   \nonumber\\
{ v}_3\,&=&\,( 0, \underline{  1,-1})   \nonumber\\
{ v}_4\,&=&\,( 0, \underline{  0,-1})   \nonumber\\
{ v}_5\,&=&\,( 0, \underline{ -1,-1})   \nonumber\\
{ v}_6\,&=&\,( 0, \underline{ -2,-1}).   
\end{eqnarray}  
Here and subsequently, the components of the vector
corresponding to the fiber are underlined.

With respect to this fiber, the base is one-dimensional: $P^1$, 
and its fan $F_2$ consists of the divisors corresponding to the
interior point and two divisors corresponding 
to two rays, $R_1=+\vec {e}_1$ and 
$R_2=-\vec {e}_1$, with directions from the point $(0,-2,-1)$
to the point $(6,-2,-1)$ and from the point $(0,-2,-1)$ to $(-1,-2,-,1)$, 
respectively. The points of 
${\pi}_B^{-1}(R_i)$ ( i.e., the points projected
onto $R_i$ by $\pi_B$)
for the rays $R_i,\,\ (i=1=+,i=2=-)$ are of the form $(\pm..., b,c)$,
where $(0,b,c)$ is the point of the fiber. 

The 16 points of $\pi_B^{-1}(R_1)$ are listed in the
Table \ref{Coxeter}: they correspond to the divisors $D_{v_i}$, which produce
the  $E_8$ algebra~\cite{Can1}.
Also, from this Table one can easily read the 
Coxeter numbers/weights.
There is only one point in ${\pi}_B^{-1}(R_2)$, namely
\begin{equation}
{\tilde {v}_1}^{1}\,=\,({-1},\underline{ -2,-1}) 
\end{equation}
which therefore does not correspond to any non-trivial group.

\begin{table}[!ht]
\caption{\it The points of $\pi_B^{-1}(R_1)$.}   
\label{Coxeter}
\centering
\scriptsize
\vspace{.05in}
\begin{tabular}{|c|c|c|c|c|c|}
\hline
$ Coxeter~\# $&$ {v}_6^{(i)} $&$ {v}_5^{(i)} $&$ {v}_1^{(i)} $&$
{v}_4^{(i)} $&${v}_0^{(i)} $ \\
\hline\hline
$1 $&$({1},\underline{-2,-1}) $&$({1},\underline {-1,-1}) $&$
({1},\underline{-1,0}) $&$({1},\underline{0,-1}) $&$
({1},\underline{0,0})   $ \\
$2 $&$({2},\underline{-2,-1}) $&$ ({2},\underline {-1,-1}) $&$ 
({2},\underline{-1,0}) $&$ ({2},\underline{0,-1}) $&$ --- $ \\
$3 $&$({3} \underline{-2,-1}) $&$ ({3},\underline {-1,-1}) $&$ 
({3},\underline{-1,0})  $&$--- $&$--- $ \\
$4 $&$({4},\underline{-2,-1}) $&$ ({4},\underline {-1,-1}) $&$ 
 --- $&$--$&$----  $ \\
$5 $&$({5},\underline{-2,-1}) $&$ ---$&$--- $&$---$&$--- $ \\
$6 $&$({6},\underline{-2,-1}) $&$ ---$&$--- $&$---$&$--- $ \\
\hline
\end{tabular}
\end{table}
\normalsize

\subsection{Example of Gauge-Group Identification}

Consider again the toric variety determined by the dual pair of polyhedra
${\Delta}(P^3(1,1,4,6)[12])$ and its   dual ${\Delta^*}$ shown in
Figure~\ref{vec1146}.
The mirror polyhedron contains the intersection $H^*$ through the
interior point, the elliptic fiber $P^2(1,2,3)$. For all integer points
of ${\Delta}^*$ (apart from the interior point), one can define
in a convenient basis the corresponding complex variables:
\begin{eqnarray}
{v}_1\,&=&\,(0,\underline{-2,-3}) \rightarrow \, z_1 \nonumber \\
{v}_2\,&=&\,(0,\underline{-1,-2}) \rightarrow \, z_2 \nonumber \\
{v}_3\,&=&\,(0,\underline{-1,-1}) \rightarrow \, z_3 \nonumber \\
{v}_4\,&=&\,(0,\underline{0,-1})  \rightarrow \, z_4 \nonumber \\
{v}_0\,&=&\,(0,\underline{0,0})                      \nonumber \\
{v}_6\,&=&\,(0,\underline{1,0})   \rightarrow \, z_6 \nonumber \\
{v}_7\,&=&\,(0,\underline{0,1})   \rightarrow \, z_7
\end {eqnarray}
and
\begin{eqnarray}
{v}_8\,&=&\,(-1,\underline{-4,-6}) \rightarrow \, z_8 \nonumber \\
{v}_9\,&=&\,(1,\underline{0,0}) \rightarrow \, z_9.
\end{eqnarray}
There are some linear relations between integer points inside the fiber:
\begin{eqnarray}
{v}_1\,+\,2\cdot {v}_6\,+\,3 \cdot {v}_7\,&=&\,0,\nonumber\\
{v}_2\,+\,       {v}_6\,+\,2 \cdot {v}_7\,&=&\,0,\nonumber\\
{v}_3\,+\,       {v}_6\,+\, {v}_7       \,&=&\,0,\nonumber\\
{v}_4\,+\,       {v}_6\,+\,{v}_7        \,&=&\,0
\end{eqnarray}
and also the following relation between points in ${\Delta}^*$:
\begin{eqnarray}
{v}_8\,+\,{v}_9\,+\,4\cdot {v}_6\,+\,6 \cdot {v}_7\,&=&\,0
\end{eqnarray} 
The polyhedron ${\Delta}(P^3(1,1,4,6))$ contains 39 points, which can be 
subdivided as follows. There are seven points in the fiber $P^2(1,2,3)$,
determined by the intersection of the plane 
$m_1\,+\,2 \cdot  m_2\, +\,3 \cdot  m_3\,=\,0$
and the positive integer lattice. 
This plane separates the remaining 32 points in $16$ `left'
and $16$ `right' points.

These `left' and `right' points define singularities of the $E_{8_L}$ and
$E_{8_R}$ types, respectively,
which may be illustrated as follows.
The plane $H({\Delta})=m_1+2m_2+3m_3$ contains the following
seven points:
\begin{eqnarray}
t_1\,&=&\, ({5},\underline{-1,-1})\rightarrow ({z_8^6 z_9^6})\cdot
(\underline {z_1^6 z_2^4 z_3^3z_4^2}),\nonumber\\
t_2\,&=&\, ({3},\underline{0,-1})\rightarrow ({z_8^4 z_9^4})\cdot
(\underline {z_1^4 z_2^3 z_2^3 z_4^2 z_6}),\nonumber\\
t_3\,&=&\, ({2},\underline{-1,0})\rightarrow ({z_8^3 z_9^3})\cdot
(\underline {z_1^3 z_2^2 z_3^2 z_4 z_7}),\nonumber\\
t_4\,&=&\, ({1},\underline{1,-1})\rightarrow ({z_8^2 z_9^2})\cdot
(\underline {z_1^2 z_2^2 z_3 z_4^2 z_6^2}),\nonumber\\
t_5\,&=&\, ({0},\underline{0,0})\rightarrow ({z_8 z_9})\cdot
(\underline {z_1^6 z_2^4 z_3^3z_4^2}),\nonumber\\
t_6\,&=&\, ({-1},\underline{2,-1})\rightarrow 
(\underline {z_2 z_4^2 z_6^3}),\nonumber\\
t_7\,&=&\, ({-1},\underline{-1,1})\rightarrow 
(\underline {z_3 z_7^2}).\nonumber\\
\end{eqnarray}
The Weierstrass equation for the $E_{8_L}$ group
based on the polyhedron ${\Delta}(P^3(1,1,4,6))$
can be written in the form:
\begin{eqnarray}
&&{\underline {z}}_6^3\, +\, \,{\underline {z}}_6^2\cdot ( a_2^{(1)} 
{z_8 z_9^3} \,+\, a_2^{(2)} {z_9^4})\,+\nonumber\\
&& {\underline {z}}_1^4 \cdot {\underline {z}}_6 
\cdot( a_4^{(1)}{z_8^3z_9^5}\,+
\,a_4^{(2)} {z_8^2z_9^6}\,+\,
a_4^{(3)} {z_8 z_9^7}\,+\,a_4^{(4)}{z_9^8})\,+\nonumber\\
&&{\underline {z}}_1^6\cdot ( a_6^{(1)} {z_8^5z_9^7}\,
+\,a_6^{(2)} {z_8^4z_9^8}\,
+\, a_6^{(3)}{z_8^3z_9^9}\,+\, a_6^{(4)} {z_8^2z_9^{(10)}}\,+
a_6^{(5)}{z_8z_9^{(11)}}\,+\,a_6^{(6)}{z_9^{(12)}})\nonumber\\
&&=\, {\underline{z}}_7^2\,+\,a_1\cdot {\underline{z}}_6{\underline{z}}_7
\cdot {z_9^2}\,+  \,
{\underline{z}}_7\cdot (a_3^{(1)} {z_8^2z_9^4}\,
+\,a_3^{(2)}{z_8z_9^{(5)}}).
\end{eqnarray}
The second Weierstrass equation for the $E_{8_R}$ group
can be obtained from this equation by interchanging the variables
desrcibing the base:
$z_8 \leftrightarrow z_9$~\footnote{The coefficients $a_i$ correspond to
the notations of~\cite{Kodaira}.}.
The Weierstrass triangle equation 
can be presented in the following general form,
where we denote ${\underline {z}}_6 = x$, ${\underline{z}}_7 = y$:
\begin{eqnarray}\label{W2}
y^2 \,+\, a_1 \cdot x \cdot y\,+\,a_3 \cdot y \,=\,
x^3 \,+ \, a_2 \cdot x^2\, +\, a_4 \cdot x \,+\, a_6,
\end{eqnarray}
where the $a_i$ are polynomial functions on the base.
The Weierstrass equation can be written in more simplified form as:
\begin{equation}\label{W1}
y^2\, =\, x^3\,+\, x \cdot f \,+\, g,
\end{equation}
with discriminant 
\begin{equation}
\Delta \, =\, 4 f^3 \, +\, 27 g^2.
\end{equation}
In the zero locus of the discriminant, some  divisors $D_i$ define
the degeneration of the torus fiber.

In addition to the method~\cite{Can1} described above, there is a somewhat
different way to find the singularity type~\cite{Kodaira}.
As we saw in the above example, the polynomials ${f}$ and ${g}$ can  be
homogeneous of orders 8 and 12, respectively, with a fibration that is
degenerate over {24} points of the base. For this form of Weierstrass 
equation, there exists the {ADE} classification of degenerations of 
elliptic fibers. In this approach, the type of
degeneration
of the fiber is determined by the orders of vanishing of the functions 
${f}$, ${g}$ and ${\delta}$. 
In the case of the general Weierstrass equation,
a general algorithm for the {ADE}
classification of elliptic singularities was considered by Tate~\cite{TATE}.
For convenience, we repeat in the Table~\ref{Tate} some 
results of Tate's
algorithm, from which one can recover the
$ E_8 \times E_8$ type of Lie-algebra singularity for the
(1,1,4,6) polyhedron.  

\begin{table}[!ht]
\centering
\caption{\it Lie algebras obtained from Tate's algorithm~\cite{TATE}:
$a_{j;k}=a_j/{\sigma^k}$.}
\label{Tate}
\scriptsize
\vspace{.05in}
\begin{tabular}{|c|c|c|c|c|c|c|c|}
\hline\hline
$Type $&$Group$&$k(a_1)$&$k(a_2)$&$k(a_3)$&$k(a_4)$&$k(a_6)$&$k(\Delta)$\\
\hline\hline
$I_0            $&$---      $&$ 0 $&$ 0  $&$ 0   $&$ 0  $&$ 0   $&$ 0   $\\
$I_1            $&$---      $&$ 0 $&$ 0  $&$ 1   $&$ 1  $&$ 1   $&$ 1    $\\
$I_2            $&$SU(2)    $&$ 0 $&$ 0  $&$ 1   $&$ 1  $&$ 1   $&$ 1  $\\
$I^{ns}_{2k}    $&$Sp(2k)   $&$ 0 $&$ 0  $&$ k   $&$ k  $&$ 2k  $&$ 2k $\\
$I^s_{2k}       $&$SU(2k)   $&$ 0 $&$ 1  $&$ k   $&$ k  $&$ 2k  $&$ 2k $\\
$I^s_{2k+1}     $&$SU(2k+1) $&$ 0 $&$ 1  $&$ k   $&$ k+1$&$ 2k+1$&$ 2k+1 $\\
$III            $&$SU(2)    $&$ 1 $&$ 1  $&$ 1   $&$ 1  $&$ 2   $&$ 3 $\\
$IV_s           $&$SU(3)    $&$ 1 $&$ 1  $&$ 1   $&$ 2  $&$ 3   $&$ 4 $\\
$I^{*ns}_0      $&$G_2    $&$ 1 $&$ 1  $&$ 2   $&$ 2  $&$ 3   $&$ 6  $\\
$I^{*s}_1       $&$SO(10)   $&$ 1 $&$ 1  $&$ 2   $&$ 3  $&$ 5   $&$ 7 $\\
$I^{*ns}_{2k-3} $&$SO(4k+1) $&$ 1 $&$ 1  $&$ k   $&$ k+1$&$ 2k  $&$ 2k+3 $\\
$I^{*s}_{2k-3}  $&$SO(4k+2) $&$ 1 $&$ 1  $&$ k   $&$ k+1$&$ 2k+1$&$ 2k+3 $\\
$I^{*ns}_{2k-2} $&$SO(4k+3) $&$ 1 $&$ 1  $&$ k+1 $&$ k+1$&$ 2k+1$&$ 2k+4 $\\
$I^{*s}_{2k-2}  $&$SO(4k+4) $&$ 1 $&$ 1  $&$ k+1 $&$ k+1$&$ 2k+1$&$ 2k+4 $\\
$IV^{*ns}       $&$F_4      $&$ 1 $&$ 2  $&$ 2   $&$ 3  $&$ 4   $&$ 8    $\\
$IV^{*s}        $&$E_6     $&$ 1 $&$ 2  $&$ 2   $&$ 3  $&$ 5   $&$ 8     $\\
$III^*          $&$E_7     $&$ 1 $&$ 2  $&$ 3   $&$ 3  $&$ 5   $&$ 9 $\\
$II^*           $&$E_8     $&$ 1 $&$ 2  $&$ 3   $&$ 4  $&$ 5   $&$ 10 $\\
\hline\hline
\end{tabular}
\normalsize
\end{table}

\section{The Composite Structure of Projective Vectors}

We now embark in more detail on our construction of the projective
vectors ${ \vec k}$ which determine CY hypersurfaces, as
previewed briefly in the Introduction and based on the
polyhedron technique and the concept of duality~\cite{Batyrev}
reviewed in Section 2.
We develop this construction inductively, studying the structure of 
these vectors initially in
low dimensions and then proceeding to higher ones.

\subsection{Initiation to the Dual Algebra of CY Projective Vectors}

Our starting point is the trivial zero-dimensional  `vector', 
\begin{equation}
{ \vec {k}_1\,=\,(1)}.
\end{equation} 
which defines the trivial self-dual polyhedron comprising a single
point, with the simplest possible associated monomial:
\begin{equation}
{ x}\,\Rightarrow \,\mu_1\,=1\, \Rightarrow \, \mu'_1=(0).
\nonumber\\
\end{equation}  
The next step is to consider the only polyhedron on the line
$R^1$ which is also self-dual, and 
whose intersection with the integer lattice on the line 
contains three integer points:
\begin{equation}
  \mu'_1=(-1),\,\mu'_1=(0),\,\mu'_1=(+1).
\end{equation}
The projective vector corresponding to this 
linear polyhedron is
\begin{equation}
{ \vec {k}_2\,=\, (1,1)},
\end{equation}
which can be constructed from the ${ \vec {k}_1}$ vector, by the 
following procedure.

We extend the vector ${ \vec {k}_1}$ to a two-dimensional vector
in $CP_1$, by inserting  a zero component in  all possible ways:
\begin{eqnarray}
  { \vec {k}_1^{ex'}}\,&=& \,{ (0,1)}  \nonumber\\ 
   { \vec {k}_1^{ex''}}\,&=&\,{  (1,0)}.
\end{eqnarray}
The following monomials correspond to these `extended' vectors:
\begin{eqnarray}
  \mu^{'}\, =\,(\nu,1)\,& \Rightarrow & 
\,  x^{\nu} \cdot y                       \nonumber\\
  \mu^{''}\,=\,(1,\xi)\,& \Rightarrow & 
\, x\cdot y^{\xi}
\end{eqnarray} 
with the arbitrary integer numbers ${ \nu, \xi}$. 
From all the possible ${ \vec {k}}$ pairs:
\begin{eqnarray} 
( \vec {k}^{ex'} \, \Leftrightarrow  \, \vec {k}^{ex'} ),\,\,\,
( \vec {k}^{ex''}\, \Leftrightarrow \, \vec {k}^{ex''}) \,\,\,
( \vec {k}^{ex'} \, \Leftrightarrow \, \vec {k}^{ex''}), 
\end{eqnarray}
we select only those whose intersections give a reflexive polyhedron.
In  this simple two-dimensional case, only a single pair is so selected,
namely ${ \vec {k}_1^{ex'}}$ and  ${ \vec {k}_1^{ex''}}$:
\begin{eqnarray}
{ \vec {k}_1^{ex'} \bigcap \vec {k}_1^{ex''}\,=\,1}.
\end{eqnarray}
and the reflexive polyhedron comprises just a single point.
The corresponding monomial is ${ x \cdot y}$, whose
degree is unity for both variables: ${ deg_x=1}$ and ${ deg_y=1}$.

We now introduce a second operation on these `extended'
vectors ${ \vec{k}^{'...}}$, which is
`dual' to the intersection, namely the `sum' operation:
\begin{equation}
 \vec {k}_1^{ex'}\,\bigcup\,\vec {k}_1^{ex''}\,\, = \,\,
\vec {k}_2\,=\,(0,1)\,+\,(1,0)\,=\,(1,1).  
\end{equation}
In this simple case, one has three quadratic monomials:
\begin{eqnarray}
 { x^2}\,       & \Rightarrow &  \mu_1\, =\,(2,0)\,
\Rightarrow  \mu'_1=(-1);                    \nonumber\\
{ x \cdot y}\, & \Rightarrow &  \mu_2\, =\,(1,1)\, 
\Rightarrow  \mu'_2=(0);                     \nonumber\\
{ y^2} \,       & \Rightarrow &  \mu_3\, =\,(0,2)\,  
\Rightarrow  \mu'_3=(+1).                    \nonumber\\ 
\end{eqnarray}
If a projective vector is multiplied by a positive integer 
number $ m \in Z^+$, it still determines the same hypersurface.
Therefore, we should also consider sums of such vectors,
characterized by two positive integer numbers, ${ m,\,\, n}$:
\begin{equation}
 m \cdot \vec {k}_1^{ex'}\,+\,n\cdot \vec {k}_1^{ex''}.
\end{equation}
It turns out that, in order to get a reflexive polyhedron 
with only one interior point,
the numbers $m$ and $n$ have to be lower than 
certain maximal values: ${ m_{max}}$ and ${ n_{max}}$, 
respectively. In our first trivial example, we find that
\begin{equation}
 m_{max}\,=1,\, \,\,  n_{max}\,=\,1.
\end{equation} 
In general, the set of all pairs $(m,n)$ with
${ m \leq m_{max}}$ and ${ n \leq n_{max}}$ 
generate a `chain' of possible reflexive polyhedra,
which happens to be trivial in this simple case.
 

Following the previous procedure, to construct all possible vectors on the 
plane we should start from two  vectors, ${ \vec {k}_1}$ and 
${ \vec {k}_2}$, `extended' to dimension three in $CP_2$ space:
\begin{eqnarray}
 \vec {k}_1^{ex'}&=&(0,0,1),\,\vec {k}_1^{ex''}\,=\, (0,1,0),\,\,
\vec {k}_1^{ex'''}\,=\,(1,0,0); \nonumber\\
\vec {k}_2^{ex'} &=&(0,1,1),\,\vec {k}_2^{ex''}\,=\, (1,1,0),\,\,
\vec {k}_2^{ex'''}\,=\,(1,0,1).
\end{eqnarray}
The next step consists of finding all possible pairs of these 
three-dimensional vectors whose intersection gives the only 
reflexive polyhedron of dimension two, which
corresponds to the polyhedron projective vector 
${ \vec  {k}_2} = (1,1)$.
Only two pairs (plus cyclic permutations) satisfy this constraint:
\begin{eqnarray}
 [\vec {k}_1^{ex'}(0,0,1)] \,  \bigcap  \,
[\vec {k}_2^{ex''}(1,1,0)] \,  =  \, [ \vec {k}_2(1,1) ]_J
\end{eqnarray}
and
\begin{eqnarray}
  [ \vec {k}_2^{ex'}(0,1,1) ]  \, \bigcap  \,
[ \vec {k}_2^{ex''}(1,1,0)] \,&=&\,[ \vec {k}_2(1,1)]_J.
\end{eqnarray}
In these two cases, the corresponding monomials are:
\begin{eqnarray}
{ x^2 \cdot z} \,       & \Rightarrow &\, \mu_1=(2,0,1)\,
\Rightarrow \,   (-1);                \nonumber\\
{ x \cdot y \cdot z} \, & \Rightarrow &\, \mu_2=(1,1,1)\,
\Rightarrow \,   (0);                 \nonumber\\
{ y^2 \cdot z} \,       & \Rightarrow & \,\mu_3=(0,2,1)\,  
\Rightarrow \,   (+1);                   \nonumber\\  
\end{eqnarray}
and
\begin{eqnarray}
{ x^2 \cdot z^2\,}    & \Rightarrow &\, \mu_1=(2,0,2)\,
\Rightarrow \,  (-1);                \nonumber\\ 
{ x \cdot y \cdot z}\, & \Rightarrow &\, \mu_2=(1,1,1)\,
\Rightarrow \,  (0);                 \nonumber\\ 
{ y^2}  \,    & \Rightarrow & \,\mu_3=(0,2,0)\,  
\Rightarrow \,  (+1);                \nonumber\\  
\end{eqnarray}
respectively.           
These lead to the two following chains:
\begin{eqnarray}
 I.~~~~~~\, \vec {k}_3(1)\, &=& \,1 \cdot \vec {k}_1^{ex'}\,+
\,1 \cdot \vec {k}_2^{ex''}\,
=\,(1,1,1);\,\,m=1,\,n=1 \nonumber\\
 ~~~~~~~~\vec {k}_3(2)\, &=& \,2 \cdot \vec {k}_2^{ex'}\,+
\,1 \cdot \vec {k}_2^{ex''}\, =\,(1,1,2);\,\,m=2,\,n=1 \nonumber\\
 ~~~~~~~~ m_{max}\,&=&\,dim (\vec {k}_2^{ex''})\,=\,2,\,\,\,\,
n_{max}\,=\,dim (\vec {k}_2^{ex'})\,=\,1
\end{eqnarray}
and
\begin{eqnarray}
II.~~~~~~\, \vec {k}_3(2)\, &=& \,1 \cdot \vec {k}_2^{ex'}\,+
\,1 \cdot \vec {k}_2^{ex'''}\, =\,(1,1,2);\,\,m=1,\,n=1;\nonumber\\
~~~~~~~~~\vec {k}_3(3)\, &=& \,2 \cdot \vec {k}_2^{ex'}\,+
\,1 \cdot \vec {k}_2^{ex'''}\, =\,(1,2,3);\,\, m=2,\,n=1, \nonumber\\
 ~~~~~~~~ m_{max}\,&=&\,dim (\vec {k}_2^{ex''})\,=\,2,\,\,\,\,
n_{max}\,=\,dim (\vec {k}_2^{ex'})\,=\,2.
\end{eqnarray}
Where the eldest vectors are given on the first lines of
the two preceding equations, and we note that the
vector $(1,1,2)$ is common to both chains.

It turns out that, also in higher dimensions, some ${ \vec {k}}$ vectors 
are common to more than one chain. Thus it is possible to
make a transition from one chain to another by 
passing through the common vectors. The algebra of projective vectors 
with the two operations
$\bigcap$ and $\bigcup$ should be closed under 
{ duality symmetry}:
\begin{eqnarray}
J\,\,\,\Longleftrightarrow \,\,\,\Pi,
\end{eqnarray}
where the symbols ${ J}$ and ${ \Pi}$ denotes two dual conjugate 
operations: {\it intersection} and {\it projection}, respectively.
In this way, all vectors ${ \vec {k}_d}$ can be found.
This structure underpins the idea of a web
of transitions between all Calabi-Yau manifolds.    

\subsection{General Formulation of Calabi-Yau Algebra}

In the spirit of the simple constructions
of the previous subsection, we can also construct 
the corresponding  { closed} ${ \vec{k}_4}$ 
algebra in the case of $K3$ hypersurfaces. However,
before giving the results, we first briefly formulate a theorem
underlying the construction of a
$\vec {k}_{d+1}$ projective vector, determining an associated reflexive 
$d+1$-dimensional polyhedron and 
$CY_{d}$ hypersurface, starting from $\vec {k}_{d}$ projective
vectors, 
which determine a
$d$-dimensional reflexive polyhedron with one interior point and 
a corresponding $CY_{d-1}$ hypersurface. This theorem underlies
our systematic inductive algebraic construction of CY manifolds.

The theorem is based on two general points:

\begin{itemize}
\item{First: from the vector ${ \vec {k}_d}$, we construct 
the `extended' vectors ${ \vec {k}_{d+1}^{ex}}$ using the rule:
\begin{equation}
(*)\,\,\,\,\,{ \vec {k}_d}\,=\,{ (k_1,...,k_2)}\,\,
\stackrel{{\pi}^{-1}}{\Longrightarrow}
\,\, { \vec {k}_{d+1}^{ex(i)}}\,=\, 
{ (k_1,...,0^{i},...,k_d)}.
\end{equation}
}

\item{Second: we consider only those pairs of all possible 
`extended' vectors, ${ \vec {k}_{d+1}^{ex(i)}}$ 
and ${ \vec {k}_{d+1}^{ex(j)}}$ with 
${ 0 \,\leq \,i, j\, \leq d}$, whose intersection  
gives the reflexive polyhedron of dimension ${  d}$
with only one interior point. We denote this operation by:
\begin{equation}
(**)\,\,\,\,\,\,\,\,\,\,\,\, \vec k_{d+1}^{ex(i)} \, \bigcap \, 
\vec {k}_{d+1}^{ex(j)}\, =\,[\vec {k}_d]_J.
\end{equation}
}

\end{itemize}

The statement of the theorem is:

\begin{itemize}
\item {If by the rule (*) one can get, from
the projective ${ \vec {k}_d}$-vector, a set of `extended' vectors
${ \vec {k}_{d+1}^{ex(i)}}$, ${ 0 \leq i \leq d}$, and for any 
pair of such ``extended'' ${ \vec {k}_{d+1}^{ex(i)}}$-vectors the 
conditions (**) are fulfilled, 
then the sum of these two `extended' vectors will give  
an eldest projective vector ${ \vec {k}_{d+1}}$,
which determines a reflexive polyhedron with only one interior point.}

\item{Two finite positive integer numbers,
${ n_{max},m_{max} \in Z_+}$, exist such that any linear combination of  
two vectors ${ \vec {k}_{d+1}^{i,j}(n,m)}$,
with integer coefficients ${ m\leq m_{max};\, n \leq n_{max}}$ 
produce a CY hypersurface. We call `chain' the set of vectors 
generated by any such pair of `extended' vectors:
\begin{eqnarray}
p \cdot \vec {k}_{d+1}^{i,j}(n,m)\,&=&\, 
m \cdot \vec {k}_{d+1}^{ex(i)}\,+\,
n \cdot \vec {k}_{d+1}^{ex(j)}; \nonumber\\
\vec {k}_{d+1}^{i,j}(1,1)\,&=&\, 
\vec {k}_{d+1}^{i,j}(eld)
\end{eqnarray} 
}

\item{The intersection of the vector ${ \vec {k}_{d+1}^{i,j}(m,n)}$
with the vector ${ \vec {k}_{d+1}^{ex(i)}}$ is equal to the 
intersection of this vector with the vector 
${ \vec {k}_{d+1}^{ex(j)}}$:
\begin{eqnarray}
[\vec {k}_{d+1}^{i,j}(m,n)] \bigcap 
[\vec {k}_{d+1}^{ex(j)}]\,=\, 
[\vec {k}_{d+1}^{i,j}(m,n)] \bigcap 
[\vec {k}_{d+1}^{ex(i)}].
\end{eqnarray}
}
\end{itemize}

We can also formulate a converse theorem:

\begin{itemize}
\item {If one can decompose a reflexive projective vector 
${ \vec {k}_{d+1}}$ as the sum of two reflexive projective vectors
${ \vec {k}^{'}_{d+1}}$ and ${ \vec {k}^{``}_{d+1}}$, 
then there exists the intersection of the vector 
${ \vec {k}}_{d+1}$ with either of these two vectors, which defines a 
projective vector ${ \vec {k}_d}$ and a reflexive polyhedron with 
only one interior point.}
\end{itemize}

The above theorem provides a description of 
all  $CY_{d+1}$ hypersurfaces with ${d}$-dimensional fibers
in terms of two positive-integer parameters.
Similarly, one can also consider the intersections of   
three (or more) `doubly-extended' vectors ${ \vec {k}_{d+1}^{ex(')}}$, 
${ \vec {k}_{d+1}^{ex('')}}$, 
${\vec {k}_{d+1}^{ex(''')}}$ (by `doubly-extended' we mean
that they may be obtained by inserting two zero components 
in ${ \vec {k}_{d-1}}$ vectors).  
One should check that this  intersection gives a reflexive 
polyhedron in the $d-2$ space:
\begin{eqnarray}
 [\vec {k}_{d-1}^{ex(2^{'})}] \,\bigcap\, 
[\vec {k}_{d-1}^{ex(2^{''})}]\, \bigcap
\,[ \vec{k}_{d-1}^{ex(2^{'''})}]\,=\,[\vec {k}_{d-1}]_J.
\end{eqnarray} 
In this way, one may obtain a $3, 4, ..., \leq d$
positive-integer parameter description of the $(d+1)$-dimensional 
polyhedra with $(d-1), (d-2),...$-dimensional fiber sections:  
\begin{eqnarray}
p \cdot \vec  {k}_{d+1}\,=\, m \,\cdot \vec {k}_{d-1}^{ex(2^{'})}\,+
\, n \cdot \vec {k}_{d-1}^{ex(2^{''})}\,+
\, l \cdot \vec {k}_{d-1}^{ex(2^{'''})}. 
\end{eqnarray}
Finally, one can obtain additional lists of ${ \vec {k}_{d+1}}$
vectors by using three `extended' vectors,
${ \vec {k}_{d}^{ex^{r}}}$,
${ \vec {k}_{d}^{ex^{i}}}$
${ \vec {k}_{d}^{ex^{j}}}$
(and similarly using four ${ \vec {k}_{...}^{ex}}$, etc.),
and a special algebra of 
summing these vectors only if the following three conditions are 
fulfilled:.
\begin{eqnarray}
1. [\vec {k}_{d}^{ex^{r}}] \,\bigcap\, 
   [\vec {k}_{d}^{ex^{i}}] \,&=&\,[\vec {k}_{d-1}]_{J}^{'}; \nonumber\\
2. [\vec {k}_{d}^{ex^{i}}] \,\bigcap\, 
   [\vec {k}_{d}^{ex^{j}}] \,&=&\,[\vec {k}_{d-1}]_{J}^{''};\nonumber\\
3. [\vec {k}_{d}^{ex^{j}}] \,\bigcap\, 
   [\vec {k}_{d}^{ex^{r}}] \,&=&\,[\vec {k}_{d-1}]_{J}^{'''}.\nonumber\\
\end{eqnarray}
In this way, one may obtain a complete description of the positive-integer 
lattice which defines all reflexive ${ \vec {k}}$ vectors.  

\section{Two-Vector Chains of $K3$ Spaces}

We now embark on a parametrization of the ${ \vec
{k}_4}$
vectors defining $K3$ hypersurfaces with fiber sections.  
Our description of ${ K3}$ hypersurfaces is based on the above
understanding of the composite and dual structure of the projective 
${ \vec{k}_4}$ vectors. As already mentioned, we find a link between this
structure and  the finite subgroups
of the group of rotations of three-space, namely the cyclic and dihedral
groups
and the symmetry groups of the Platonic solids: the tetrahedron, the
octahedron-cube  and the icosahedron-dodecahedron:

\begin{itemize}
\item{$ { C_n}: n=1,2,3,...$, the cyclic group of finite rotations 
in the plane  around an axis `1' by the angles $\alpha=2\pi/n$;}
\item{$ { D_n}: n=2,3,4,...$, the dihedral group, comprising all
these rotations together
with the all reflections of a second axis `$n$' lying in this plane,
which is orthogonal  to the axis `l', and producing with 
respect to  each other the angle $\alpha/2$;}
\item{{ T} -The finite group of the transformations leaving invariant
the regular tetrahedron, with 12 parameters;}
\item{{ O}- The finite group of the transformations leaving invariant
the regular cube and octahedron, with 24 parameters;}
\item{{ I}- The finite group of the transformations leaving invariant
the regular icosahedron and dodecahedron, with 60 parameters.}
\end{itemize}

\noindent
We use the polyhedron technique introduced in the previous
Section,
taking into account all its duality, intersection and projection
properties to study the projective-vector classification of $K3$
spaces.

\subsection{Two-Dimensional Integer Chains of $K3$ Hypersurfaces}

In the $K3$ case, as already foreshadowed in the Introduction, the 
classification can start from 
a basis of  five types of `extended' vectors. We 
recall that the structure of the three `planar' projective vectors 
${ \vec{k}_3=(1,1,1), (1,1,2), (1,2,3)}$ can easily be understood
on the basis of the  doubly-extended vector
${ \vec{k}_1^{ext}=(0,0,1)}$ and the
singly-extended vector ${ \vec{k}_2^{ext}=(0,1,1)}$. The structure
of the underlying composite vector ${ \vec{k}_2=(1,1)}$ is also obvious.
The full list of ${ K3}$ projective vectors 
is obtainable from the algebra of the following five extended
vectors: the maximally-extended vector of the form 
\begin{equation}
{ \vec {k}_C^{ext}=(0,0,0,1)}
\end{equation}
with its 4 cyclic permutations, the doubly-extended dihedral vector of the
form
\begin{equation}
{ \vec {k}_D^{ext}=(0,0,1,1)}
\end{equation}
with its 6 dihedral permutations, the singly-extended tetrahedral
vector of the form
\begin{equation}
{ \vec {k}_T^{ext}=(0,1,1,1)}
\end{equation}
with its 4 cyclic permutations, the extended octahedral vector of the form
\begin{equation}
{ \vec {k}_O^{ext}=(0,1,1,2)}
\end{equation}
with its 12 permutations, and finally the extended icosahedral vector of
the
form
\begin{equation}
{ \vec {k}_I^{ext}=(0,1,2,3)}
\end{equation}
with its 24 permutations, for a total of 50 extended vectors.

Using the algebra of combining pairs of these 50 extended
${ \vec{k}^{i}}$ vectors, we obtain 90 distinct ${ \vec{k}_4}$ vectors 
in 22 double chains with a regular planar $k$-gon intersection:
${ k > 3}$ with only one interior point, as seen in Table 1.
Combining three extended ${ \vec{k}^{i}}$ vectors, we obtain
four triple chains with self-dual line-segment 
intersection-projections and  
one interior  point, which contain 91 distinct vectors, of
which only four ${ \vec{k}_4}$ vectors are different from
the { 90} vectors found previously, as also seen in Table 1. 
Of course, there are some vectors which have 
a regular planar $k$-gon  in their intersection
and no line-segment intersection.
Further, as we see later in Section 7, there is just one vector, ${ \vec
{k}_4=(7,8,9,12)}$,
which has only a single point intersection, i.e., the 
intersection consists of the zero point alone, and can be determined 
by the intersection-projection
\begin{eqnarray}
 J   (\Delta) & \leftrightarrow & \Pi (\Delta ^*)  \nonumber\\ 
 \Pi (\Delta) & \leftrightarrow &  J (\Delta ^*)      
\end{eqnarray}
duality, where the polyhedra ${ \Delta}$ and ${ \Delta ^*}$
determine a dual pair of ${K3}$ hypersurfaces. We recall that 
the sum of the integer points  in intersection, ${ J( \Delta)}$,
and in projection,  ${ \Pi({\Delta}^*)}$, is equal to 
${14 = 4+10, 5 + 9, 6 + 8, 7 + 7, 8 + 6, 9 + 5, 10 + 4}$ for the plane
intersection-projection and ${6=3+3}$ for the line-segment
intersection-projection.  
This duality plays a very important role in our description.  
Eleven of the 22 two-vector chains found previously
satisfy directly the following condition:
\begin{eqnarray}
J(\Delta ^n)   \,&=&\, \Pi (\Delta^n)   \,=\,{\Delta}^{n-1} ,\nonumber\\
\Pi({\Delta^n}^*)\,&=&\,  J ({\Delta^n}^*)\,=\,{{\Delta}^{n-1}}^*,
\end{eqnarray}
which means that the number of integer points in the intersection
of the polyhedron (mirror polyhedron) forming the reflexive polyhedron of
lower dimension is equal to the number of projective lines crossing these
integer points of the polyhedron (mirror).
The projections of these lines on a plane in the polyhedron 
and a plane in its mirror polyhedron reproduce,
of course, the reflexive polyhedra of lower dimension.
Only for self-dual polyhedra can one have
\begin{equation}
 J(\Delta ^n)  \,=\, \Pi (\Delta^n)   \,=\,
\Pi({\Delta^n}^*)\,=\,  J ({\Delta^n}^*)\,=\,
{\Delta}^{n-1}\,=\,{{\Delta}^{n-1}}^*,
\end{equation}
namely the most symmetrical form of these relations.

Following the recipe presented as our central Theorem in
Section 4, we 
present Table~\ref{tabl22}, 
which lists all the $\vec {k}_4$ projective
vectors derived from {\it pairs} of extended vectors of lower dimension,
which fall into the 22 chains listed. In each case, we list the maximum
integers ${m, n}$ in the chains, which
are determined by the dimensions of the extended ${ \vec {k}_i}$
vectors. This Table includes all the 90 projective 
${ \vec {k}_4}$ vectors found using our construction. All of
these ${ \vec {k}_4}$ vectors define $K3$ hypersurfaces which could
be obtained using the $Z^n$ symmetry coset action~\footnote{They
may also be used to construct higher-level $CY_1$ spaces as the
intersections of polynomial loci, as discussed in Section 9.}. 

\begin{table}[!ht]
\scriptsize
\centering
\caption{\it The 22 chains of $K3$ obtained using
pairs of ${\vec k}_4$ projective vectors. The number of $K3$ spaces in
each chain is denoted by $N$.}
\label{tabl22}
\vspace{.05in}
\begin{tabular}{|c|c|c|c|c|c|c|} \hline
${Chain} $&$ {\vec k}_i \bigcap \vec{k}_j $ &$ \Delta _{Int}, {\Delta
^*}_{Int}
$&$ { \vec k(K3)  =  m \cdot \vec{k}_i + n \cdot \vec{k}_j} 
$&$ \sum m + \sum n $&$ max (m, n)$&$
N $\\ \hline\hline
$I $&$ (0 0 0 1) \bigcap(1 1 1 0) $&$ (10,4) $&$ ({ n, n, n, m}) 
$&$ 1 \cdot m + 3\cdot n $&$ m=1,n=3$&$ 3$ \\
\hline
$IV $&$ (0 0 0 1) \bigcap(1 1 2 0) $&$ (9,5) $&$ ({ n, n, 2n, m}) 
$&$ 1 \cdot m + 4\cdot n $&$ m=1,n=4$&$ 4$ \\
\hline
$XV $&$ (0 0 0 1) \bigcap(1 2 3 0) $&$ (7,7) $&$ ({ n, 2n, 3n, m}) 
$&$ 1 \cdot m + 6\cdot n $&$ m=1,n=6$&$6$ \\
\hline\hline
$X $&$ (0 0 1 1) \bigcap(1 1 0 0) $&$ (9,5) $&$ ({ n, n, m, m}) 
$&$ 2 \cdot m + 2\cdot n $&$m=2, n=2 $&$4 $ \\
\hline
$XI $&$ (0 0 1 1) \bigcap(1 1 0 1) $&$ (9,5) $&$ ({ n, n, m, m+n}) 
$&$ 2\cdot  m + 3\cdot n $&$m=3, n=2 $&$6 $ \\
\hline
$V $&$ (0 0 1 1) \bigcap(1 1 0 2) $&$ (9,5) $&$ ({ n, n, m, m + 2n}) 
$&$2 \cdot  m + 4\cdot n $&$m=4, n=2 $&$6$ \\
$XXII $&$ (0 0 1 1) \bigcap(1 2 1 0) $&$ (7,7) $&$ ({ n, 2n, m+n, m}) 
$&$ 2 \cdot m + 4\cdot n $&$m=3, n=3 $&$6$ \\
\hline
$XVI $&$ (0 0 1 1) \bigcap(1 2 0 3 ) $&$ (7,7) $&$ ({ n, 2n, m, m+3n}) 
$&$ 2 \cdot m + 6\cdot n $&$m=6, n=2 $&$9$ \\
\hline\hline
$II $&$ (0 1 1 1) \bigcap(1 0 1 1) $&$ (10,4) $&$ ({ n, m, m+n, m+n}) 
$&$ 3\cdot m + 3\cdot n $&$m=3, n=3$&$4$ \\
\hline
$XIII $&$ (0 1 1 1) \bigcap(1 0 1 2) $&$ (8,6) $&$ ({ n, m, m+n, m+2n}) 
$&$ 3 \cdot m + 4\cdot n $&$m=4,n=3$&$9$ \\
\hline
$III $&$ (0 1 1 1) \bigcap(3 0 1 2) $&$ (4, 10) $&$ ({ 3n, m, m+n, m+2n}) 
$&$ 3 \cdot m + 6\cdot n $&$m=6,n=2$&$9 $ \\
$XVII $&$ (0 1 1 1) \bigcap(1 0 2 3) $&$ (7,7) $&$ ({ n, m, m+2n, m+3n}) 
$&$3 \cdot  m + 6\cdot n $&$m=6,n=3$&$12$ \\
\hline \hline
$VI $&$ (0 1 1 2) \bigcap(1 0 1 2) $&$ (9,5) $&$ ({ n, m, n+m,  2m+2n}) 
$&$ 4 \cdot m + 4\cdot n $&$m=4,n=4$&$6$ \\
$VIII $&$ (0 1 1 2) \bigcap(1 1 2 0) $&$ (5,9) $&$ ({ n, m+n, m+2n, 2m}) 
$&$ 4 \cdot m + 4\cdot n $&$m=3,n=4$&$9$ \\
$VII $&$ (0 1 1 2) \bigcap(2 1 1 0) $&$ (9,5) $&$ ({ 2n, m+n, m+n, 2m}) 
$&$ 4 \cdot m + 4\cdot n $&$m=3,n=1$&$3$ \\
\hline \hline
$XVIII $&$ (0 1 1 2) \bigcap(1 0 2 3) $&$ (7,7) $&$ ({ n, m, m+2n, 2m+3n}) 
$&$ 4\cdot m + 6\cdot n $&$m=6,n=4$&$16$ \\
$XIV $&$ (0 1 1 2) \bigcap(2 1 3 0) $&$ (6,8) $&$ ({ 2n, m+n, m+3n, 2m}) 
$&$ 4 \cdot m + 6\cdot n $&$m=5,n=3$&$8$ \\
$IX $&$ (0 1 1 2) \bigcap(2 1 0 3) $&$ (5,9) $&$ ({ 2n, m+n, m, 2m+3n}) 
$&$ 4\cdot m + 6\cdot n $&$m=6,n=3$&$12$ \\
\hline \hline
$XIX $&$ (0 1 2 3) \bigcap (1 0 2 3) $&$ (7,7) $&$ ({ n, m, 2m +2n, 3m +3n}
) 
$&$ 6 \cdot m + 6\cdot n $&$m=6,n=6$&$12$ \\
$XX $&$ (0 1 2 3) \bigcap (2 1 0 3) $&$ (7,7) $&$ ({ 2n, m+n, 2m, 3m+3n}) 
$&$ 6 \cdot m + 6\cdot n $&$m=3,n=3$&$12$ \\
$XXI $&$ (0 1 2 3) \bigcap (3 1 2 0) $&$ (7,7) $&$ ({ 3n, m+n, 2m + 2n, 3n})$&$ 6 \cdot m + 6\cdot n $&$m=2,n=2$&$4$ \\
$XII $&$ (0 1 2 3) \bigcap(3 2 1 0) $&$ (5,9) $&$ ({ 3n, m+2n, 2m+n, 3m}) 
$&$ 6 \cdot m + 6\cdot n $&$m=2,n=2$&$4$ \\
\hline
\end{tabular}
\end{table}
\normalsize

For illustration, we give in Table~\ref{Tabl22v} the 
eldest vectors in each chain, i.e., the first members of all
22 chains, which have
${ m=1, n=1}$. As one can see, some vectors 
are common to more than one chain. 
Using our understanding of
the origins of the intersections,
and duality, we can classify these 22 chains in five
classes, as indicated by the groupings in Table~\ref{Tabl22v},
which correspond to the intersections, as indicated. 

\begin{table}[!ht]
\scriptsize
\centering
\caption{\it The eldest vectors ${ \vec k_i}$ in the 22 $K3$ chains
which have ${ m=n=1}$.}
\label{Tabl22v}
\vspace{.05in}
\begin{tabular}{|c|c|c|c|c|c|} \hline
$ {Chain}$&$ { d} $&${  {\vec k}_i}$&$ {\vec k}_i $&$  
(\Delta,  {\Delta}^*) $&$ ({\Delta}_{Int},  {\Delta}^*_{Int})$\\
\hline\hline 
${ I}$&$ { d=1 \cdot m + 3 \cdot n} $&$( { n,n,n,m}) 
$&$({ 1,1,1,1})$&$ (35,5) $&$(10,4)$ \\
 \hline
${ II} $&$ { d=3 \cdot m + 3 \cdot n}$&$({ n,m,m+n,m+n}) 
$&$({ 1,1,2,2})$&$ (30,6) $&$(10,4)$ \\
\hline
${ III} $&$ { d=3 \cdot m + 6 \cdot n}$&$({ 3n,m,m+n,m+2n}) 
$&$({ 1,2,3,3})$&$ (23,8) $&$(4,10)$ \\ 
 \hline \hline
${ IV}$&$ { d=1 \cdot m + 4 \cdot n} $&$( { n,n,2n,m}) 
$&$({ 1,1,1,2})$&$ (34,6) $&$(9,5)$ \\ 
\hline 
${ V} $&$ { d=2 \cdot m + 2 \cdot n}$&$({ n,n,m,m}) 
$&$({ 1,1,1,1})$&$ (35,5) $&$(9,5)$ \\ 
\hline
${ VI} $&$ { d=2 \cdot m + 3 \cdot n}$&$({ n,n,m,m+n}) 
$&$({ 1,1,1,2})$&$ (34,6) $&$(9,5)$ \\ 
\hline
${ VII} $&$ { d=2 \cdot m + 4 \cdot n}$&$({ n,n,m,m+2n}) 
$&$({ 1,1,1,3})$&$ (39,6) $&$(9,5)$ \\ 
\hline 
${ VIII} $&$ { d=4 \cdot m + 4 \cdot n}$&$({ n,m,m+n,2m+2n}) 
$&$({ 1,1,2,4})$&$ (35,7) $&$(9,5)$ \\ 
\hline 
${ IX} $&$ { d=4 \cdot m + 4 \cdot n}$&$({ 2n,m+n,m+n,2m}) 
$&$({ 1,1,1,1})$&$ (35,5) $&$(9,5)$ \\ 
\hline 
${ X} $&$ { d=4 \cdot m + 4 \cdot n}$&$({ n,m+n,m+2n,2m}) 
$&$({ 1,2,2,3})$&$ (24,8) $&$(5,9)$ \\ 
\hline
 ${ XI} $&$ { d=4 \cdot m + 6 \cdot n}
$&$({ 2n,m+n,m,2m+3n}) 
$&$({ 1,2,2,5})$&$ (28,8) $&$(5,9)$ \\ 
\hline
 ${ XII} $&$ { d=6 \cdot m + 6 \cdot n}
$&$({ 3n,m+2n,2m+n,3m}) $&$({ 1,1,1,1})$&$ (35,5) $&$(5,9)$ \\ 
\hline\hline
${ XIII} $&$ { d=3 \cdot m + 4 \cdot n}$&$({ n,m,m+n,m+2n}) 
$&$({ 1,1,2,3})$&$ (31,8) $&$(8,6)$ \\
\hline 
 ${ XIV} $&$ { d=4 \cdot m + 6 \cdot n}$&$({ 2n,m+n,m+3n,2m}) 
$&$({ 1,1,1,2})$&$ (34,6) $&$(6,8)$ \\ 
\hline\hline
${ XV} $&$ { d=1 \cdot m + 6 \cdot n}$&$({ n,2n,3n,m}) 
$&$({ 1,1,2,3})$&$ (31,8) $&$(7,7)$ \\ 
\hline
${ XVI} $&$ { d=2 \cdot m + 6 \cdot n}$&$({ m,n,2n,3n+m}) 
$&$({ 1,1,2,4})$&$ (35,7) $&$(7,7)$ \\ 
\hline
${ XVII} $&$ { d=3 \cdot m + 6 \cdot n}$&$({ n,m,m+2n,,m+3n}) 
$&$({ 1,1,3,4})$&$ (33,9) $&$(7,7)$ \\ 
\hline
 ${ XVIII} $&$ { d=4 \cdot m + 6 \cdot n}
$&$({ n,m,m+2n,2m+3n}) 
$&$({ 1,1,3,5})$&$ (36,9) $&$(7,7)$ \\ 
\hline
 ${ XIX} $&$ { d=6 \cdot m + 6 \cdot n}
$&$({ n,m,2m+2n,3m+3n}) 
$&$({ 1,1,4,6})$&$ (39,9) $&$(7,7)$ \\ 
\hline
 ${ XX} $&$ { d=6 \cdot m + 6 \cdot n}
$&$({ 2n,m+n,2m,3m+3n}) 
$&$({ 1,1,1,3})$&$ (39,6) $&$(7,7)$ \\ 
\hline
 ${ XXI} $&$ { d=6 \cdot m + 6 \cdot n}
$&$({ 3n,m+n,2m+2n,3m}) 
$&$({ 2,3,3,4})$&$ (15,9) $&$(7,7)$ \\ 
\hline\hline
${ XXII} $&$ { d=2 \cdot m + 4 \cdot n}$&$({ n,2n,m+n,m}) 
$&$({ 1,1,2,2})$&$ (30,6) $&$(7,7)$ \\ 
\hline
\end{tabular}
\end{table}

It should be noted, however, that the above doubly-extended vector
structure does not exhaust the full list of possible ${K3}$ projective
vectors. The projective vectors
\begin{eqnarray}
{ (\vec {k}_4)_{91}}\, &=&\,  { (4,5,7,9)  },        \nonumber\\
{ (\vec {k}_4)_{92}}\, &=&\,  { (5,6,8,11) },        \nonumber\\
{ (\vec {k}_4)_{93}}\, &=&\,  { (5,7,8,20) },        \nonumber\\
{ (\vec {k}_4)_{94}}\, &=&\,  { (7,8,10,25)},        \nonumber\\ 
{ (\vec {k}_4)_{95}}\,&=&\,
{ (7,8,9,12)},                           
\label{morevecs}
\end{eqnarray}
have no planar reflexive polyhedron intersections, and therefore were not 
included in this list. 
To obtain most of the additional ${ \vec {k}_4}$- vectors
({\ref{morevecs}), one must consider chains constructed from 
three extended vectors of the type  ${ \vec{k}^{ex}=(0,0,1)}$ and  
${ \vec{k}^{ex}=(0,1,1)}$, with all possible permutations, having 
in the intersection the line-segment polyhedron consisting of
three integer points. All these chains will  
be  ${ J_1 - \Pi_1}$ self-dual: ${ J_1 = \Pi_1 = 3}$. It is easy
to see that only four different such
triple chains can be built, as discussed in Section 6. These chains are
much longer than the
previous two-vector chains, although their total number,
91, is also less than the full number of all ${K3}$ vectors.
The projective vectors
\begin{eqnarray}
{ (\vec {k}_4)_{12}}\,&=&\,{ (3,5,6,7) },             \nonumber\\
{ (\vec {k}_4)_{13}}\,&=&\,{ (3,6,7,8) },             \nonumber\\
{ (\vec {k}_4)_{14}}\,&=&\,{ (5,6,7,9) },             \nonumber\\
{ (\vec {k}_4)_{95}}\,&=&\,
{ (7,8,9,12)}
\end{eqnarray} 
are not involved in these chains. However, the union
of the doubly-extended
and triply-extended vector chains gives a total of 94
${ \vec{k}_4}$ projective vectors. Only the 
${ \vec{k}_4=(7,8,9,12)}$ vector has just a point-intersection structure,
and is not found by either the double- or triple-vector
constructions, as discussed in more detail in Section 7.

To preview how it arises, note that,
by ${ J -\Pi}$ duality, we know that to ${\vec{k}=(1,1,1,1)}$,
which has three intersection planes {(1,1,1)} with ten points, 
there must correspond
a ${\vec{k}}$ which has three different ${\pi}$ projections with four 
points.
Since it should have a non-trivial projection structure, namely
a four-point 
planar polyhedron with one interior point in three independent directions,
its external points should satisfy the following condition: 
\begin{eqnarray}
\frac{1}{4}\cdot 
\{ M_1\,+\,M_2\,+\,M_3\,+\,M_4\,\}=\,M_0\,=\,(1,1,1,1).
\end{eqnarray}
In three-space, these points can only be taken as:
\begin{eqnarray}
M_1\,=(4,1,0,0),\,\,\,M_2\,=\,(0,3,1,0),\,\,\,
M_3\,=(0,0,4,0),\,\,\,M_4\,=\,(0,0,0,3).
\label{fourpoints}
\end{eqnarray}
One can easily check that this polyhedron has three projections:
${ \pi_{x_1}}$, ${ \pi_{x_2}}$, ${ \pi_{x_3}}$,
with  four points giving the ${ (1,1,1)}$ planar polyhedron.
The four points ${ M_i}$ (\ref{fourpoints}) give the exceptional vector
 ${ \vec{k}} = (7,8,9,12)$.
By  projection, one can see that the five integer points
of this polyhedron produce the ${ (1,1,1)}$ planar polyhedron
with four points.


\subsection{Invariant Monomials and the ${ J-\Pi}$ Structure of Calabi-Yau
Equations}

The experience provided by
working with $K3$ hypersurfaces can aid in the
classification  of Calabi-Yau manifolds. Also for this more
complicated case, one should 
use the duality conditions: one must be prepared to study the
intersection structures of
polyhedra and their mirrors and/or to study the 
projection structures for polyhedra and
mirror polyhedra.

This `intersection-projection' structure of the ${ \vec {k}_4}$ 
vectors from doubly-, triply- and quadruply-extended vectors allows us
to introduce the concept of {\it invariant} monomials in the {CY}
equations. These invariant monomials
are homogeneous under the action of the extended vectors, i.e., if

\begin{eqnarray}
\vec {z'} \,=\, 
{\lambda }^{{\vec k}^{ex}_j}  \cdot \vec {z},\,\,\, j=1,2,3,...,
\end{eqnarray}
then

\begin{eqnarray}
{\vec {z'}}^{\vec {\mu}} \,=\, 
{\lambda}^{{\vec k}^{ex}_j \cdot \vec {\mu}}
\cdot {\vec z}^{\vec {\mu}}\,=\, 
{\lambda }^{d_j} \cdot {\vec z}^{\vec {\mu}},
\end{eqnarray}

\noindent
where ${ d_j=dim (\vec{k}^{ex}_j)}$ and $j=1,2,3,...$ is 
the number of extended ${ \vec{k}^{ex}_j}$ vectors. 
The invariant monomials,  ${ \wp_i}$,  correspond 
to the reflexive polyhedra produced by the invariant set  
${ \Psi_{inv}}$ which is the same for all the chains.

These extended vectors can be
formed from the following five familiar types of projective vectors of
lower dimensions:
\begin{eqnarray}
\vec {k}_1\,&=&\,(0,0,1), \nonumber\\
\vec {k}_1\,&=&\,(0,1,1), \nonumber\\ 
\vec {k}_3\,&=&\,(1,1,1),\,\, \vec {k}_3=(1,1,2),\,\, 
\vec {k}_3=(1,2,3).
\end{eqnarray}
A chain of ${ \vec{k}_4}$ projective vectors
can be generated from the linear sums of extended vectors, for
example, for $j=1,2$ one can get:
\begin{eqnarray}
\vec {k}_4(m,n)\,&=&\,m \cdot \vec {k}^{ex}_1\,+\,
n \cdot \vec {k}^{ex}_2\,\, \nonumber\\
{\rm if} \,\,\,\vec {k}^{ex}_1 \bigcap \vec {k}^{ex}_2\,&=&\,
\{\wp_i: \wp_i \in \Sigma_{inv} \}.
\end{eqnarray}
The invariant monomials are  universal for all the ${ \vec{k}_4}$
vectors in this chain. 

To construct the ${ \vec{k}_4}$ vectors determining $K3$ hypersurfaces,
i.e., determining the corresponding polyhedra with the property of
reflexivity,
one has to give a correct set of invariant monomials. We have constructed
the {22} sets of invariant monomials corresponding to the 
doubly-extended vector structures among the ${ \vec{k}_4}$ projective
vectors. In this case, these sets of the invariant monomials give in the 
intersection reflexive polyhedra of lower dimensions. The number of
invariant monomials for this doubly-extended vector structure is given by
\begin{equation}
31\,\,=\,\,{1}\,+\,{4} \times 2\,+\,
{{22}},
\end{equation}
where the last number corresponds to the Betti number for
${K3}$ hypersurfaces: ${ b_2=22}$.
The  structure of the ${ \vec{k}_4}$ projective vectors 
obtained from the
triply-extended vectors, namely ${ \vec{k}^{ex}=(0,0,0,1)}$
and ${ \vec{k}^{ex}=(0,0,1,1)}$, is given by the following
four types of invariant monomials:
\begin{eqnarray} 
 \Psi_{I_3}:(2,0,1,1)\,  , \,(0,2,1,1)\,,\,(1,1,1,1,) &\Rightarrow &   
 x^2 \cdot z    \cdot u   \,  , \,
 y^2 \cdot z    \cdot u   \,  , \,
 x \cdot y \cdot z \cdot u ;                                   \nonumber\\
 \Psi_{II_3}:(2,2,1,0)\,  , \,(0,0,1,2)\,,\,(1,1,1,1,) &\Rightarrow & 
 x^2 \cdot y^2  \cdot z   \,,\,
 z    \cdot u^2           \,, \,   
 x \cdot y \cdot z \cdot u ;                                   \nonumber\\
\Psi_{III_3}:(2,2,2,0)\,  , \,(0,0,0,2)\,,\,(1,1,1,1,) &\Rightarrow & 
x^2 \cdot y^2  \cdot z^2  \,,\,
u^2                       \,, \,                  
x \cdot y \cdot z \cdot u ;                                   \nonumber\\
\Psi_{IV_3}:(2,0,0,2)\,  , \,(0,2,2,0)\,,\,(1,1,1,1,)  &\Rightarrow & 
x^2 \cdot u^2             \,, \,
y^2 \cdot z^2             \,, \,
x \cdot y \cdot z \cdot u .
\end{eqnarray}  
The four chains corresponding to these sets of invariant monomials are
(see Tables~5,6,7 and 8):
\begin{eqnarray}
\vec {k}_4(\Psi_{I_3})  &=&M\cdot (1,1,0,0)\,+\,N\cdot (0,0,1,0)\,+\,
L\cdot (0,0,0,1) =  (M,M,N,L),                     \nonumber\\
\vec {k}_4(\Psi_{II_3}) &=&M\cdot (1,0,0,1)\,+\,N\cdot (0,1,0,1)\,+\,
L\cdot (0,0,1,0) =  (M,N,L,M+N),                   \nonumber\\ 
\vec {k}_4(\Psi_{III_3})&=&M\cdot (1,0,0,1)\,+\,N\cdot (0,1,0,1)\,+\,
L\cdot (0,0,1,1) =  (M,N,L,M+N+L),                 \nonumber\\ 
\vec {k}_4\Psi_{(IV_3}) &=&M\cdot (1,0,1,0)\,+\,N\cdot (0,1,0,1)\,+\,
L\cdot (0,0,1,1) =  (M,N,M+L,N+L).                  \nonumber\\
\end{eqnarray}
In these chains~\footnote{There is in fact another `good'
triple intersection, of the extended vectors
$(1,1,0,0), (0,0,1,1), (0,0,0,1)$, but the chain 
${I_3^{\prime}=(M,M,N,N+L)}$ it produces has the same three invariant
monomials,
$(0,2,1,1)+(2,0,1,1)+(1,1,1,1)$ as the ${I_3}$ chain, which includes
all its projective vectors.}, the sets of projective vectors are subject
to the following additional projective restrictions:
\begin{eqnarray}
\vec {k}_4(\Psi_{I_3}) \cdot \vec {e}_I\,&=&\,0,\,\,\,\,\,\,\,\,
\vec {e}_I\,=\,(-1,1,0,0)                          \nonumber\\
\vec {k}_4(\Psi_{II_3}) \cdot \vec {e}_{II}\,&=&\,0,\,\,\,\,\,\,\,\,
\vec {e}_{II}\,=\,(-1,-1,0,1)                      \nonumber\\
\vec {k}_4(\Psi_{III_3}) \cdot \vec {e}_{III}\,&=&\,0,\,\,\,\,\,\,\,\,
\vec {e}_{III}\,=\,(-1,-1,-1,1)                      \nonumber\\
\vec {k}_4(\Psi_{IV_3}) \cdot \vec {e}_{IV}\,&=&\,0,\,\,\,\,\,\,\,\,
\vec {e}_{IV}\,=\,(1,-1,-1,1)                      \nonumber\\
\end{eqnarray}
Corresponding to these chains, the following
triple intersections
\begin{eqnarray}
\vec {k}_M^{ex} \bigcap \vec {k}_N^{ex} \bigcap \vec {k}_L^{ex}
\,=\,\Psi_{I_3},\, \Psi_{II_3},\, \Psi_{III_3},\,\Psi_{IV_3}.  
\end{eqnarray}
have the above-mentioned invariant monomials.

The ${K3}$ algebra has the interesting consequence
that all the $\{1+ 4 + {22}\}$ invariant monomials
that give `good' planar reflexive polyhedra in the {22}
two-vector chains also can be found by triple constructions. 
Therefore it is interesting to list now the {22} types of 
invariant monomials
whose origin is also connected  with
the triple intersections of all types of projective vectors,
the triply-extended vectors ${ \vec{k}_1^{ex}=(0,0,0,1)}$, 
the doubly-extended vectors  ${ \vec{k}_2^{ex}=(0,0,1,1)}$,
and the singly-extended vectors, 
${ \vec{k}_3^{ex}=(0,1,1,1), (0,1,1,2),(0,1,2,3)}$.}

These monomials, ${ \vec {z}^{\vec {\mu}}}$, are invariant under 
action of the extended vectors 
\begin{eqnarray}
\vec {k}_i^{ex}\cdot \vec {\mu}\,&=&\,dim (\vec {k}_i^{ex}),\nonumber\\
\vec {k}_j^{ex}\cdot \vec {\mu}\,&=&\,dim (\vec {k}_j^{ex}),\nonumber\\
\vec {k}_l^{ex}\cdot \vec {\mu}\,&=&\,dim (\vec {k}_l^{ex}).
\end{eqnarray}

\noindent
The directions of the possible projections ${ \Pi}$ 
are determined~\footnote{Additional constraints on the
invariant monomials are given in Section 7, reducing their number to 9
= 1 + 3 + 5.} by the degenerate monomial 
${ (x \cdot y \cdot z \cdot u) \Rightarrow \vec {\mu}=(1,1,1,1)}$ 
and by the exponents of the following 22 invariant monomials, 
${ \mu =({\mu}_1,{\mu}_2,{\mu}_3,{\mu}_4)}$:

\begin{eqnarray} 
& &\underline{(3,0,0,0),\,(3,1,0,0),\,(3,1,1,0),}\,(3,2,0,0), \nonumber\\
& &(3,2,1,0),\,(3,3,0,0),\,(3,3,1,0),                         \nonumber\\
& &\underline {(4,0,0,0),\,(4,1,0,0),}\,(4,1,1,0),\,(4,2,0,0),\nonumber\\
& &(4,2,1,0),(4,3,0,0),\,(4,3,1,0),\,(4,4,0,0),\,
(4,4,1,0),                                                    \nonumber\\
&&(6,0,0,0),\,(6,1,0,0),\,(6,2,0,0),\,(6,3,0,0),              \nonumber\\
&&(6,4,0,0),\,(6,6,0,0).                                     
\end{eqnarray}
where the underlines pick out those triple intersections
where the intersections of pairs of vectors also specify
reflexive polyhedra, which will be important later.
The four other types of possible projections were already defined above.

The algebraic-geometry sense of  
${ (J,\Pi) (\Delta) \leftrightarrow (\Pi,J) (\Delta^*)}$ 
duality for $K3$ hypersurfaces can be 
interpreted through the invariant monomials:
the list of the invariant monomials for the two-extended-vector 
classification and the list of all of the 
three-extended-vector classification are the same, and the total
number of them is equal to ${ 31=1+4\times 2 + 22}$.
The ${ J (\Delta, \Delta ^*) \leftrightarrow 
\Pi (\Delta^*, \Delta)}$ duality
can be interpreted at a deeper level for ${ J =\Pi}$ chains: 
the invariant monomials are identical for corresponding {CY} 
equation and for its mirror equation.
The projection-projection structure gives additional
information about the form of the corresponding {CY} equation.
For example, this structure
determines the subset of monomials corresponding to the invariant
monomials. As result, the homogeneous CY equation 
can be written in according in terms of the 
intersection-projection structure of the projective 
${ \vec {k}}$ vectors:
\begin{eqnarray}
\wp(\vec {z})\,=\, \sum_{i}^{J} \vec {z}^{\vec {m}_0^i}
\{ \sum_p^{\Pi}  a_{\vec {m}_0^i}^p \, 
\vec {z}^{{n}_p \cdot \vec {e}^{\Pi}} \} \,=\,0. 
\end{eqnarray}
Here the $\vec {z}^{\vec {m}_0^i}$ are the invariant monomials
which are defined by intersection structure, the vector $\vec {e}^{\Pi}$
is the direction of the projection, and the
${n}_p$ are integer numbers.

\section{Three-Vector Chains of $K3$ Spaces}

As already mentioned, one can find additional 
chains of $K3$ projective vectors ${ \vec {k}_4}$
if one considers systems of three extended
vectors of the type 
${ \vec {k}^{ex}_1=(0,0,0,1)}$ and
${ \vec {k}^{ex}_2=(0,0,1,1)}$, which have in their intersections
only three integer points or only three invariant monomials.
As also already remarked, there are only four different chains,
corresponding 
to the four kinds of invariant monomial triples. 
We have also commented that these new chains yield
only four additional $K3$ vectors, whilst the
remaining vector, ${ \vec {k}_4=(7,8,9,12)}$, can be
constructed out of four extended vectors, as discussed
in the following Section. The relationship
between the two- and three-vector constructions, and their
substantial overlap, is the subject of this Section.

\subsection{The Three-Vector Chain $I_3$: $\vec{k}_4=(M,\,M,\,N,\,L)$}

\scriptsize    
\begin{table}[!ht]
\centering
\caption{\it The {18} $K3$ hypersurfaces
in the three-vector chain $I_3$: $\vec {k} = (M,M,N,L)= M \cdot (1,1,0,0)
+ N \cdot (0,0,1,0) + L \cdot (0,0,0,1)$. Here and subsequently,
the symbol $\aleph$ in the first column denotes the location of
the corresponding vector in Table 1. The numbers in the last columns
indicate their locations in the corresponding chains.}
\label{TabXLI}
\vspace{.05in}
\begin{tabular}{|c||c|c|c|c|c||c||c|c|c|c|c|c|c|c|} 
\hline
${ \aleph} 
$&$  M        $&$  M          $&$   N          $&$   L   
$&$ { [d]} $&$ (\Delta, \Delta ^*) 
$&$  I        $&$    II       $&$      IV      $&$   V  
$&$  VII      $&$    X        $&$      XI      $&$   III $ \\ \hline\hline
$ { J(\Delta)}     
$&$  -        $&$     -       $&$     -        $&$  -    
$&$ -         $&$     - 
$&${ 10}   $&$  { 10}   $&$  { 9 }     $&$  { 9 }     
$&${  9}   $&$  { 9 }   $&$  { 9 }     $&$   - $\\ \hline 
$ { \pi(\Delta^*)}     
$&$  -        $&$     -       $&$     -        $&$  -    
$&$ -         $&$     - 
$&${ 4 }   $&$  { 4}    $&$  { 5 }     $&$  { 5 }     
$&${  5}   $&$  { 5}    $&$  { 5 }     $&$   - $\\ \hline \hline
$ { \pi(\Delta^)}     
$&$  -        $&$     -       $&$     -        $&$    -  
$&$ -         $&$ -     
$&${ - }   $&$ { 10}    $&$  { - }     $&$  { - }      
$&${ 9'}   $&$ { - }    $&$  { - }     $&$   - $\\ \hline 
$ { J(\Delta^*)}     
$&$  -        $&$     -       $&$     -        $&$    -  
$&$ -         $&$ -     
$&${ - }   $&$ { 4}     $&$  { - }     $&$  { - }      
$&${ 5'}   $&$ { -}     $&$  { - }             $&$   -   $\\ \hline \hline
$1            
$&$ { 1} $&$ { 1}   $&$ {  1}    $&${ 1} 
$&${[ 4]}  $&$ (35, 5^*) 
$&$  { 1}  $&$    -        $&$        -        $&$   -   
$&$  { 1}  $&$  { 1}    $&$        -           $&$   -   $ \\
$2            
$&$ { 1} $&$ { 1}   $&$ {  1}    $&${ 2} 
$&${[ 5]}  $&$ (34, 6^*) 
$&$   2       $&$    -        $&$      { 1}    $&$   -   
$&$    -      $&$    -        $&$      { 1}    $&$   -   $ \\ 
$3            
$&$ { 1} $&$ { 1}   $&$ {  1}    $&${ 3} 
$&${[ 6]}  $&$ (39, 6^*) 
$&$   3      $&$    -        $&$        -      $&$ { 1} 
$&$    -      $&$    -        $&$      -       $&$   -   $ \\
$4            
$&$ { 1} $&$ { 1}   $&$ {  2}    $&${ 2} 
$&${[ 6]}  $&$ (30, 6^*) 
$&$    -      $&$   { 1}   $&$        2        $&$   - 
$&$    -      $&$    2       $&$        -      $&$   -   $ \\
$5           
$&$ { 1} $&$ { 1}   $&$ {  2}    $&${ 3} 
$&${[ 7]}  $&$ (31, 8^*) 
$&$    -      $&$     -       $&$        3     $&$   -  
$&$    -      $&$     -       $&$        2     $&$   -   $ \\ 
$6           
$&$ { 1} $&$ { 1}   $&$ {  2}    $&${ 4} 
$&${[ 8]}  $&$ (35, 7^*) 
$&$    -      $&$     -       $&$        4     $&$   2   
$&$    -      $&$     -       $&$        -     $&$   -   $ \\
$7           
$&$ { 1} $&$ { 1}   $&$ {  3}    $&${ 4} 
$&${[ 9]}  $&$ (33, 9^*) 
$&$    -      $&$     -       $&$        -     $&$   -   
$&$    -      $&$     -       $&$        4     $&$   -   $ \\ 
$8           
$&$ { 1} $&$ { 1}   $&$ {  3}    $&${ 5} 
$&${[ 10]} $&$ (36, 9^*) 
$&$    -      $&$     -       $&$        -     $&$   3
$&$    -      $&$     -       $&$        -     $&$   -   $ \\ 
$9           
$&$ { 1} $&$ { 1}   $&$ {  4}    $&${ 6} 
$&${[ 12]} $&$ (39, 9^*) 
$&$    -      $&$     -       $&$        -     $&$   5
$&$    -      $&$     -       $&$        -     $&$   -   $ \\
$10           
$&$ { 2} $&$ { 2}   $&$ {  1}    $&${ 3} 
$&${[ 8]}  $&$ (24, 8^*) 
$&$    -      $&$     -       $&$        -     $&$   -   
$&$    2      $&$     -       $&$        3     $&$   -   $ \\
$11           
$&$ { 2} $&$ { 2}   $&$ {  1}    $&${ 5} 
$&${[ 10]} $&$ (28, 8^*) 
$&$    -      $&$     -       $&$        -     $&$   4 
$&$    -      $&$     -       $&$        -     $&$   -   $ \\
$42           
$&$ { 2} $&$ { 2}   $&$ {  3}    $&${ 5} 
$&${[ 12]} $&$ (17,11^*) 
$&$    -      $&$     -       $&$        -     $&$   -  
$&$    -      $&$     -       $&$        5     $&$   -   $ \\ 
$43           
$&$ { 2} $&$ { 2}   $&$ {  3}    $&${ 7} 
$&${[ 14]} $&$ (19,11^*) 
$&$    -      $&$     -       $&$        -     $&$   6
$&$    -      $&$     -       $&$        -     $&$   -   $ \\
$12            
$&$ { 3} $&$ { 3}   $&$ {  1}    $&${ 2} 
$&${[ 9]}  $&$ (23, 8^*) 
$&$    -      $&$     2       $&$        -     $&$   -   
$&$    -      $&$     -       $&$        -     $&$   -   $ \\
$44            
$&$ { 3} $&$ { 3}   $&$ {  2}    $&${ 4} 
$&${[ 12]} $&$ (15, 9^*) 
$&$    -      $&$     -       $&$        -     $&$   -  
$&$    3    $&$     -       $&$        -       $&$   -   $ \\
$65           
$&$ { 3} $&$ { 3}   $&$ {  4}    $&${ 5} 
$&${[ 15]} $&$ (12,12^*) 
$&$   -        $&$     -       $&$        -    $&$   -       
$&$     -      $&$     -       $&$        -    $&$   1   $ \\         
$21           
$&$ { 4} $&$ { 4}   $&$ {  1}    $&${ 3} 
$&${[ 12]} $&$ (21, 9^*) 
$&$    -      $&$     3       $&$        -     $&$   - 
$&$    -      $&$     -       $&$        -     $&$   -   $ \\
$48            
$&$ { 5} $&$ { 5}   $&$ {  2}    $&${ 3} 
$&${[ 15]} $&$ (14,11^*) 
$&$    -      $&$     4      $&$        -      $&$   -  
$&$    -      $&$     -       $&$        -     $&$   -   $ \\ 
\hline
\end{tabular}
\end{table}

\normalsize

In this chain, the dimension
$({ d=2M+N+L})$ and the eldest vector is ${ \vec{k}_{eld}=(1,1,1,1)}$,
whose
invariant monomials are ${ (2,0,1,1) + (0,2,1,1)}$.  
The relations between this three-vector chain and 
the previously-discussed two-vector chains  
can easily be found. We consider the first three vectors
in Table~\ref{TabXLI}, which also form the two-vector chain $I$: 
\begin{eqnarray}
I:\,&&\,m\cdot (1,1,1,0)\,+\,n \cdot (0,0,0,1)\,
=\,(m,m,m,n)\,\rightarrow          \nonumber\\
&&M\,=\,N\,=\,m\,=\,[dim]\{(0,0,0,1)\}\,=\,1,\nonumber\\
&&L\,=\,n\,\leq\,[dim]\{(1,1,1,0)\}\,=\,3.
\end{eqnarray}
Similarly, one can consider four vectors
${ (2,2,1,1)}$, ${ (3,3,1,2)}$, 
${ (4,4,1,3)}$ and ${ (5,5,2,3)}$, which  form the 
two-vector chain $II$:
\begin{eqnarray}
II:\,&&\,m\cdot (1,1,1,0)\,+\,n \cdot (1,1,0,1)\,
=\,(m,n,m+n,m+n)\,\rightarrow                        \nonumber\\
&&N\,=\,m\,\leq\,[dim]\{(1,1,0,1)\}\,=\,3,           \nonumber\\
&&L\,=\,n\,\leq\,[dim]\{(1,1,1,0)\}\,=\,3,           \nonumber\\
&&M\,=\,m\,+\,n\,<\,6.                               \nonumber\\
\end{eqnarray}
The  four vectors ${ (1,1,2,1)}$, ${ (1,1,2,2)}$, 
 ${ (1,1,2,3)}$ and  ${ (1,1,2,4)}$ from the
two-vector chain $IV$ have the following relations with this triple chain: 
\begin{eqnarray}
IV:\,&&\,m\cdot (1,1,2,0)\,+\,n \cdot (0,0,0,1)\,
=\,(m,m,2m,n)\,\rightarrow                        \nonumber\\
&&M\,=\,m\,\leq\,[dim]\{(0,0,0,1)\}\,=\,1,        \nonumber\\
&&N\,=\,2m\,=\,2,                                 \nonumber\\
&&L\,=\,n\,\leq \,[dim]\{(1,1,2,0)\}\,=\,4.       \nonumber\\
\end{eqnarray}
The six vectors ${ (1,1,1,3)}$, ${ (1,1,2,4)}$, ${ (1,1,3,5)}$, 
${ (1,1,4,6)}$, ${ (2,2,1,5)}$  and ${ (2,2,3,7)}$
in Table~\ref{TabXLI} correspond to the two-vector chain $V$:
\begin{eqnarray}
V:\,&&\,m\cdot (1,1,0,2)\,+\,n \cdot (0,0,1,1)\,
=\,(m,m,n,2m+n)\,\rightarrow                      \nonumber\\
&&M\,=\,m\,\leq\,[dim]\{(0,0,1,1)\}\,=\,2,        \nonumber\\
&&N\,=\,n\,\leq\,[dim]\{(1,1,0,2)\}\,=\,4,        \nonumber\\
&&L\,=\,2m+n\,<\,=\,8.                            \nonumber\\
\end{eqnarray}
The next three vectors 
${ (1,1,1,1)}$,  ${ (3,3,2,4)}$ and ${ (2,2,1,3)}$ 
from the two-vector chain $VII$ have the following
connection to this triple chain:
\begin{eqnarray}
VII:\,&&\,m\cdot (1,1,2,0)\,+\,n \cdot (1,1,0,2)\,
=\,(m+n,m+n,2m,2n)\,\rightarrow                     \nonumber\\
&&M\,=\,m+n\,<\,4,                                  \nonumber\\
&&N\,=\,2m \,<\,4,                                  \nonumber\\
&&L\,=\,2n \,<\,4.                                  \nonumber\\
\end{eqnarray}
Two vectors ${ (1,1,1,1)}$ and ${ (1,1,2,2)}$ 
correspond to the two-vector chain $X$:
\begin{eqnarray}
X:\,&&\,m\cdot (1,1,0,0)\,+\,n \cdot (0,0,1,1)\,
=\,(m,m,n,n)\,\rightarrow                     \nonumber\\
&&M\,=\,m \,\leq \,2,                         \nonumber\\
&&N\,=\,n \,\leq \,2,                         \nonumber\\
&&L\,=\,n \,\leq \,2.                         \nonumber\\
\end{eqnarray}
Finally, the values of ${M,N,L}$ of the five projective vectors
${ (1,1,1,2)}$,    ${ (1,1,2,3)}$,    ${ (1,1,3,4)}$,    
${ (2,2,1,3)}$ and   ${ (2,2,3,5)}$  correspond to the
fact that they are also from the two-vector chain $XI$:  
\begin{eqnarray}
XI:\,&&\,m\cdot (1,1,0,1)\,+\,n \cdot (0,0,1,1)\,
=\,(m,m,n,m+n)\,\rightarrow                    \nonumber\\
&&M\,=\,m \,  \leq \,2,                         \nonumber\\
&&N\,=\,n \,  \leq \,3,                         \nonumber\\
&&L\,=\,m+n \,\leq \,5.                         \nonumber\\
\end{eqnarray}

\subsection{The Three-Vector Chain $ II_3$: $\vec{k}_4=(M,N,L,M+N)$}

\scriptsize
\begin{table}[!ht]
\centering
\caption{\it The {45} $K3$ hypersurfaces in the $II_3$ chain:
${\vec k} = (M,N,L,Q=N+M)=
M \cdot (1,0,0,1) + N \cdot (0,1,0,1) + L \cdot (0,0,1,0)$.}
\label{TabXLII}
\vspace{.05in}
\begin{tabular}{|c||c|c|c|c||c||c||c|c|c|c|c|c|c|c|c|c|} \hline
${ \aleph} 
$&$  M         $&$   N       $&$   L       $&$   Q      $&$ { [d]} 
$&$ (\Delta, \Delta ^*)
$&${ II}   $&${ IV}    $&${ VI}   $&${ VIII}$&${ XI} 
$&${ XIII} $&${ XIV}   $&${ XV}   $&${ XVII}$&${ XXII}
$\\ \hline\hline
$ -     
$&$  -        $&$     -      $&$   -       $&$     -    $&$     -     
$&${ J(\Delta)}           
$&${ 10}   $&${ 9}     $&${ 9}    $&${ 5}   $&$  { 9}  
$&${ 8 }   $&${ 6}     $&${ 9}    $&${ 9}   $&$  { 7}
$\\ \hline
$ -
$&$  -        $&$     -      $&$     -     $&$     -    $&$     -     
$&${ \pi(\Delta^*)}       
$&${ 4}    $&${ 5}     $&${ 5}    $&${ 9 }  $&$  { 5}  
$&${ 6}    $&${ 8}     $&${ 5}    $&${ 5}   $&$  { 7}
$\\ \hline \hline
$ -   
$&$  -        $&$     -      $&$    -      $&$     -    $&$     -     
$&${ \pi(\Delta)}         
$&${ 10}   $&${ -}     $&${ 9}    $&${ 5'}  $&$   - 
$&${ -}    $&${ -}     $&${ -}    $&${ - }  $&$   -
$\\  \hline 
$ -   
$&$  -        $&$     -      $&$     -     $&$     -    $&$     -     
$&${ J (\Delta^*)}           
$&${ 4 }   $&${ -}     $&${ 5}    $&${ 9'}  $&$  -  
$&${ -}    $&${ -}     $&${ -}    $&${ - }  $&$  -
$\\  \hline \hline
$2            
$&$  { 1}  $&$ { 1}    $&$ { 1}   $&${ 2}   $&$  [5]       
$&$ (34, 6^*) 
$&$           
$&${ 1}    $&$            $&$           $&$ { 1}
$&$           $&$  { 1}   $&$           $&$          $&$
$\\ \hline\hline
$4            
$&$   {1}  $&$ { 1}   $&$ { 2}   $&${ 2}    $&$ [6]       
$&$ (30, 6^*)  
$&$ { 1}   $&$   2        $&$           $&$          $&$    
$&$           $&$            $&$           $&$          $&${ 1}
$\\ \hline
$5           
$&$  { 1}  $&$ { 1}    $&$ { 3}   $&${ 2}   $&$[7]       
$&$ (31, 8^*) 
$&$           $&$   3        $&$           $&$          $&$ 2
$&$ { 1}   $&$            $&$   { 1} $&$          $&$         
$\\ \hline
$6           
$&$  { 1 } $&$ { 1}    $&$ { 4}   $&${ 2}   $&$[8]      
$&$ (35, 7^*)  
$&$           $&$   4       $&$  { 1}  $&$          $&$ 
$&$           $&$            $&$           $&$          $&$
$\\ \hline
$10           
$&$  { 1}  $&$ { 2}    $&$ { 2}   $&${ 3}   $&$[8]      
$&$ (24, 8^*) 
$&$           $&$            $&$           $&$ { 1}  $&$ 3
$&$           $&$            $&$     2     $&$          $&$ 2
$\\ \hline
$12            
$&$  { 1}   $&$ { 2}   $&$ { 3}   $&${ 3}   $&$ [ 9]      
$&$ (23, 8^*) 
$&$    2      $&$            $&$           $&$          $&$ 
$&$           $&$   2        $&$     3     $&$          $&$ 
$\\ \hline
$13           
$&$  { 1}  $&$ { 2}    $&$ { 4}   $&${ 3}   $&$ [10]      
$&$(23, 11^*)  
$&$           $&$            $&$            $&$         $&$ 
$&$  2        $&$            $&$   4        $&$         $&$ 3 
$\\ \hline
$14           
$&$  { 1}  $&$ { 2}    $&$   { 5}  $&${ 3}  $&$[11]      
$&$ (24, 13^*)
$&$           $&$            $&$            $&$         $&$ 
$&$  3        $&$   3        $&$    5       $&$         $&$    
$\\ \hline
$15           
$&$  { 1}  $&$ { 2}    $&$  { 6}   $&${ 3} $&$[12]
$&$ (27, 9^*) 
$&$           $&$            $&$     2      $&$        $&$   
$&$           $&$            $&$     6     $&$        $&$
$\\ \hline
$7           
$&$ { 1}   $&$ { 3}    $&$  { 1}   $&${ 4} $&$[12]      
$&$ (33, 9^*) 
$&$           $&$            $&$            $&$        $&$4 
$&$           $&$            $&$            $&${ 1} $&$
$\\ \hline
$21            
$&$ { 1}   $&$ { 3}   $&$  { 4}   $&${ 4} $&$[12]  
$&$ (21, 9^*)
$&$    3      $&$            $&$            $&$      3 $&$ 
$&$           $&$            $&$            $&$        $&$
$\\ \hline
$24           
$&$ {  1}  $&$ { 3}   $&$  { 8}    $&${ 4} $&$[16]
$&$ (24,12^*) 
$&$           $&$            $&$    3       $&$        $&$ 
$&$           $&$            $&$            $&$        $&$
$\\ \hline
$16           
$&$  { 1}  $&$ { 4}    $&$  { 2}   $&${ 5} $&$[12]
$&$ (24, 12^*)
$&$           $&$            $&$            $&$        $&$   
$&$           $&$            $&$            $&$    2   $&$ 4 
$\\ \hline
$22           
$&$  { 1}  $&$ { 4}    $&$  { 3}   $&${ 5} $&$[13]
$&$ (20,15^*) 
$&$           $&$            $&$            $&$        $&$ 
$&$    4      $&$   5        $&$            $&$        $&$
$\\ \hline
$31           
$&$  { 1}  $&$ { 4}    $&$  { 10}  $&${ 5} $&$[20]
$&$ (23,13^*)  
$&$           $&$            $&$    4       $&$        $&$ 
$&$           $&$            $&$            $&$        $&$
$\\ \hline
$25           
$&$  { 1}  $&$ { 5}    $&$ { 3}    $&${ 6} $&$[15]
$&$ (21,15^*)  
$&$           $&$            $&$            $&$        $&$ 
$&$           $&$            $&$            $&$    3   $&$
$\\ \hline
$30           
$&$   { 1} $&$ { 5}    $&$ { 4}    $&${ 6} $&$[16]
$&$ (19, 17^*)
$&$           $&$            $&$            $&$    4   $&$ 
$&$    5      $&$            $&$            $&$        $&$
$\\ \hline
$32           
$&$   { 1} $&$ {  6}   $&$ { 4}    $&${ 7} $&$[18]
$&$ (19,20^*) 
$&$           $&$            $&$            $&$        $&$ 
$&$           $&$            $&$            $&$    4   $&$
$\\ \hline
$36           
$&$   { 1} $&$ { 7}    $&$ { 5}    $&${ 8} $&$[21]
$&$ (18,24^*) 
$&$           $&$            $&$            $&$        $&$ 
$&$           $&$            $&$            $&$    5   $&$
$\\ \hline
$39           
$&$  {  1} $&$ { 8}    $&$ { 6}    $&${ 9} $&$[24]
$&$ (18,24^*) 
$&$           $&$            $&$            $&$        $&$ 
$&$           $&$            $&$            $&$    6   $&$
$\\ \hline 
$42           
$&$  { 2}  $&$ { 3}    $&$ { 2}   $&${ 5}  $&$[12]
$&$ (17, 11^*) 
$&$           $&$            $&$            $&$  2     $&$5
$&$           $&$            $&$            $&$        $&$   
$\\ \hline
$45           
$&$  { 2}  $&$ { 3}    $&$  { 4}   $&${ 5} $&$[14]
$&$ (13, 16^*)
$&$           $&$            $&$            $&$        $&$   
$&$           $&$   4        $&$            $&$        $&$   5
$\\ \hline
$48           
$&$ { 2}   $&$ { 3}    $&$  { 5}   $&${ 5} $&$[15]   
$&$ (14, 11^*) 
$&$    4     $&$            $&$            $&$        $&$ 
$&$           $&$            $&$            $&$        $&$
$\\ \hline
$51           
$&$ { 2}   $&$ { 3}    $&$  { 10}  $&${ 5} $&$[20]
$&$ (16,14^*) 
$&$           $&$            $&$    5       $&$        $&$ 
$&$           $&$            $&$            $&$        $&$
$\\ \hline
$18          
$&$ { 2}   $&$ { 5}    $&$ { 1}    $&${ 7} $&$[15]
$&$ (26,17^*) 
$&$           $&$            $&$            $&$        $&$ 
$&$           $&$            $&$            $&$    7   $&$
$\\ \hline
$49           
$&$ { 2}   $&$ { 5}    $&$   { 3}  $&${ 7} $&$[17]
$&$ (13,20^*) 
$&$           $&$            $&$            $&$        $&$ 
$&$   7       $&$            $&$            $&$        $&$
$\\ \hline
$59           
$&$  { 2}  $&$ { 5}    $&$  { 6}   $&${ 7} $&$[20]
$&$(11,23^*)  
$&$           $&$            $&$            $&$    7   $&$ 
$&$           $&$            $&$            $&$        $&$ 
$\\ \hline
$52           
$&$  { 2}  $&$ { 7}    $&$  { 3}   $&${ 9} $&$[21]
$&$ (14,23^*) 
$&$           $&$            $&$            $&$        $&$ 
$&$           $&$            $&$            $&$    9   $&$
$\\ \hline
$61           
$&$  { 2}  $&$ { 9}    $&$    { 5} $&${ 11}$&$[27]
$&$ (11,32^*) 
$&$           $&$            $&$            $&$        $&$ 
$&$           $&$            $&$            $&$    13 $&$
$\\ \hline
$23           
$&$  { 3}  $&$ { 4}    $&$ { 1}    $&${ 7} $&$[15]
$&$ (22,17^*) 
$&$           $&$            $&$            $&$        $&$ 
$&$   6       $&$            $&$            $&$        $&$
$\\ \hline
$46           
$&$ { 3}   $&$ { 4}    $&$   { 2}  $&${ 7} $&$[16]
$&$ (14,18^*)  
$&$           $&$            $&$            $&$   5    $&$ 
$&$           $&$   6        $&$            $&$        $&$ 
$\\ \hline
$67           
$&$ { 3}   $&$ { 4}    $&$   { 5}  $&${ 7} $&$[19]
$&$ (9,24^*)  
$&$           $&$            $&$            $&$        $&$ 
$&$           $&$   7        $&$            $&$        $&$
$\\ \hline
$71           
$&$ { 3}   $&$  { 4}   $&$   { 14} $&${ 7} $&$[28]
$&$ (12,18^*) 
$&$           $&$            $&$    6      $&$        $&$ 
$&$           $&$            $&$            $&$        $&$
$\\ \hline
$50           
$&$ { 3}   $&$ {  5}   $&$   { 2}  $&${ 8} $&$[18]
$&$ (14,20^*) 
$&$           $&$            $&$            $&$        $&$ 
$&$   8       $&$            $&$            $&$        $&$
$\\ \hline
$68           
$&$ { 3}   $&$ {  5}   $&$  {  4}  $&${ 8} $&$[20]
$&$ (10,22^*)   
$&$           $&$            $&$            $&$  8     $&$ 
$&$           $&$            $&$            $&$        $&$
$\\ \hline
$27          
$&$ {  3}  $&$  { 7}   $&$  { 1}   $&${ 10}$&$[21]
$&$ (24,24^*)  
$&$           $&$            $&$            $&$        $&$ 
$&$           $&$            $&$            $&$    8   $&$
$\\ \hline
$70           
$&$ { 3}   $&$ { 7}    $&$  { 4}   $&${ 10}$&$[24]
$&$ (10,26^*)
$&$           $&$            $&$            $&$        $&$ 
$&$   9      $&$            $&$            $&$        $&$
$\\ \hline
$54           
$&$ {  3}  $&$ { 8}    $&$  { 2}   $&${ 11}$&$[24]
$&$ (15,27^*) 
$&$           $&$            $&$            $&$        $&$ 
$&$           $&$            $&$            $&$    10  $&$
$\\ \hline
$72           
$&$ { 3}   $&$ { 10}   $&$  { 4}   $&${ 13}$&$[30]
$&$ (10,35^*)  
$&$           $&$            $&$            $&$        $&$ 
$&$           $&$            $&$            $&$    11  $&$
$\\ \hline
$77           
$&$  { 3}  $&$ { 11}   $&$  { 5}   $&${ 14}$&$[33]
$&$ (9,39^*)  
$&$           $&$            $&$            $&$        $&$ 
$&$           $&$            $&$            $&$    12  $&$
$\\ \hline
$57           
$&$ {  4}  $&$ { 5}    $&$  { 2}   $&${ 9 }$&$[20]
$&$ (13,23^*)
$&$           $&$            $&$            $&$   6    $&$ 
$&$           $&$            $&$            $&$        $&$ 
$\\ \hline
$81           
$&$ { 4}   $&$ { 5}    $&$  { 6}   $&${ 9 }$&$[24]
$&$ (8,26^*)   
$&$           $&$            $&$            $&$        $&$ 
$&$           $&$   8       $&$            $&$        $&$
$\\ \hline
$83           
$&$ { 4}   $&$ { 5}    $&$  { 7}   $&${ 9 } $&$[25]
$&$ (7,32^*)  
$&$           $&$            $&$            $&$         $&$ 
$&$           $&$            $&$            $&$         $&$
$\\ \hline
$87           
$&$ { 4}   $&$ { 7}    $&$  { 6}   $&${ 11} $&$[28]
$&$ (7,35^*)  
$&$           $&$            $&$            $&$  9     $&$ 
$&$           $&$            $&$            $&$         $&$
$\\ \hline
$90           
$&$ { 5}   $&$ { 6}    $&$  { 8}   $&${ 11} $&$[30]
$&$ (6,39^*)  
$&$           $&$            $&$            $&$         $&$ 
$&$           $&$            $&$            $&$         $&$
$\\ \hline
\end{tabular}
\end{table}

\normalsize

In this chain, shown in Table~\ref{TabXLII}, the dimension
$d=2M+2N+L$, there is a symmetry: $ M \leftrightarrow N $,
the eldest vector ${\vec k}_{eld}=(1,1,1,2)$, and
the invariant monomials are $(2,2,1,0)+(0,0,1,2)$.
Comparing this chain with the previous two-vector chains, one can see
clearly
the possible values of ${ M,N,L}$ for the projective vectors
${ (M,N,L,M+N)}$. For example, if one compares the four vectors
${ (1,1,2,2)}$, ${ (1,2,3,3)}$, 
${ (1,3,4,4)}$ and ${ (2,3,5,5)}$ in this triple chain
with their structure in the two-vector chain $II$, one finds
the following relations:

\begin{eqnarray}
II:\,&&\,m\cdot (1,0,1,1)\,+\,n \cdot (0,1,1,1)\,
=\,(m,n,m+n,m+n)\,\rightarrow                        \nonumber\\
&&M\,=\,m\,\leq\,[dim]\{(0,1,1,1)\}\,=\,3,           \nonumber\\
&&N\,=\,n\,\leq\,[dim]\{(1,0,1,1)\}\,=\,3,           \nonumber\\
&&L\,=\,m\,+\,n\,<\,6.                               \nonumber\\
\end{eqnarray}
Similarly, we find the following relations between the values of ${M,N,L}$
in the triple chain and the values of ${ m,n}$ for double
chains:

\begin{eqnarray}
IV:\,&&\,m\cdot (1,1,0,2)\,+\,n \cdot (0,0,1,0)\,
=\,(m,m,n,2m)\,\rightarrow                             \nonumber\\
&&M\,=\,N\,=\,m\,\leq\,[dim]\{(0,0,1,0)\}\,=\,1,       \nonumber\\
&&L\,=\,n\,\leq\,[dim]\{(1,1,0,2)\}\,=\,4.             \nonumber\\
\end{eqnarray}

\begin{eqnarray}
VI:\,&&\,m\cdot (1,0,2,1)\,+\,n \cdot (0,1,2,1)\,
=\,(m,n,2m+2n,m+n)\,\rightarrow                        \nonumber\\
&&M\,=\,m\,\leq\,[dim]\{(0,1,2,1)\}\,=\,4,             \nonumber\\
&&N\,=\,n\,\leq\,[dim]\{(1,0,2,1)\}\,=\,4,             \nonumber\\
&&L\,=\,2m+2n\,<\,8.                                   \nonumber\\
\end{eqnarray}

\begin{eqnarray}
VIII:\,&&\,m\cdot (1,0,2,1)\,+\,n \cdot (1,1,0,2)\,
=\,(m+n,n,2m,m+2n)\,\rightarrow                        \nonumber\\
&&M\,=\,m+n\,\leq \,8,                                 \nonumber\\
&&N\,=\,n\,\leq\,[dim]\{(1,0,2,1)\}\,=\,4,             \nonumber\\
&&L\,=\,2m\,\leq\,2[dim]\{(1,1,0,2)\}\,=\,8.           \nonumber\\
\end{eqnarray}

\begin{eqnarray}
XI:\,&&\,m\cdot (1,0,1,1)\,+\,n \cdot (0,1,0,1)\,
=\,(m,n,m,m+n)\,\rightarrow                              \nonumber\\
&&M\,=\,m\,\leq\, [dim]\{(0,1,0,1)\}\,=\,2,              \nonumber\\
&&N\,=\,n\,\leq\, [dim]\{(1,0,1,1)\}\,=\,3,              \nonumber\\
&&L\,=\,m.                                               \nonumber\\
\end{eqnarray}

\begin{eqnarray}
XIII:\,&&\, m \cdot (1,0,2,1)\,+\,n \cdot (0,1,1,1)\,
=\,(m,n,m,m+n)\,   \rightarrow                            \nonumber\\
&&M\,=\,m\,   \leq\, [dim]\{(0,1,1,1)\}\,=\,3,            \nonumber\\
&&N\,=\,n\,   \leq\, [dim]\{(1,0,2,1)\}\,=\,4,            \nonumber\\
&&L\,=\,2m+n.                                             \nonumber\\
\end{eqnarray}

\begin{eqnarray}
XIV:\,&&\,m\cdot (1,0,2,1)\,+\,n \cdot (1,2,0,3)\,
=\,(m+n,2n,2m,m+3n)\,   \rightarrow                       \nonumber\\
&&M\,=\,m+n,                                              \nonumber\\
&&N\,=\,2n\,\leq\, 2[dim]\{(1,0,2,1)\}\,=\,8,             \nonumber\\
&&L\,=\,2m\, \leq\, 2[dim]\{(1,2,0,3)\}\,=\,12.           \nonumber\\
\end{eqnarray}

\begin{eqnarray}
XV:\,&&\,m\cdot (1,2,0,3)\,+\,n \cdot (0,0,1,0)\,
=\,(m,2m,n,3m)\,        \rightarrow                       \nonumber\\
&&M\,=\,m,  \leq\, [dim]\{(0,0,1,0)\}\,=\,1,              \nonumber\\
&&N\,=\,2m\, \leq\, 2[dim]\{(0,0,1,0)\}\,=\,2,            \nonumber\\
&&L\,=\,n\, \leq\, [dim]\{(1,2,0,3)\}\,=\,6.              \nonumber\\
\end{eqnarray}

\begin{eqnarray}
XVII:\,&&\,m\cdot (1,2,0,3)\,+\,n \cdot (0,1,1,1)\,
=\,(m,2m+n,n,3m+n)\,        \rightarrow                       \nonumber\\
&&M\,=\,m,  \leq\, [dim]\{(0,1,1,1)\}\,=\,3,              \nonumber\\
&&N\,=\,2m+n                                            \nonumber\\
&&L\,=\,n\, \leq\, [dim]\{(1,2,0,3)\}\,=\,6.              \nonumber\\
\end{eqnarray}

\begin{eqnarray}
XXII:\,&&\,m\cdot (1,0,2,1)\,+\,n \cdot (0,1,0,1)\,
=\,(m,n,2m,m+n)\,        \rightarrow                       \nonumber\\
&&M\,=\,m\,  \leq\, [dim]\{(0,1,0,1)\}\,=\,2,              \nonumber\\
&&N\,=\,n\,  \leq\, [dim]\{(1,0,2,1)\}\,=\,6,              \nonumber\\
&&L\,=\,2m.                                                \nonumber\\
\end{eqnarray}

\normalsize
\subsection{The Three-Vector  Chain $ III_3$: $  \vec{k}_4=(M,N,L,M+N+L)$}

\scriptsize
\begin{table}[!ht]
\centering
\caption{\it The {48} $K3$ hypersurfaces in
the $III_3$ chain:  $\vec {k} =
(M,N,L,Q=N+M+L)=
M \cdot (1,0,0,1) + N \cdot (0,1,0,1) + L \cdot (0,0,1,1)$.}
\label{TabXLIII}
\scriptsize
\vspace{.05in}
\begin{tabular}{|c||c|c|c|c|c||c||c|c|c|c|c|c|c|} \hline
${ \aleph}      
$&${ M}            $&${ N}              $&${ L}        $&${ Q}
$&$ [d]        
$&$(\Delta,\Delta ^*)
$&$ V              $&$ VI               $&$ IX         $&$ XVI    
$&$ XVIII          $&$ XIX              $&$ XX           $\\ 
\hline
$   -
$&$  -             $&$     -            $&$     -      $&$  -      
$&$ -
$&$   { J(\Delta)}              
$&$  { 9 }         $&$ { 9}             $&$  { 5}      $&${ 7 } 
$&$  { 7 }         $&$ { 7}             $&$  { 7}        $\\
\hline 
$ -    
$&$  -             $&$     -            $&$     -      $&$  -      
$&$ -
$&$  { \pi(\Delta^*)}            
$&$  {  5}         $&$ { 5}             $&$  { 9}      $&${ 7 } 
$&$  { 7 }         $&$ { 7}             $&$  { 7}        $\\
\hline \hline
$ -   
$&$  -             $&$     -            $&$     -      $&$  -      
$&$ -
$&$   { \pi(\Delta)}              
$&$  { -}          $&$ { 9}             $&$  { -}      $&${ - } 
$&$  { 7}          $&$ { 7}             $&$  { 7}        $\\
\hline
$-     
$&$   -            $&$     -            $&$     -      $&$  -      
$&$   -
$&$   {  J(\Delta^*)}              
$&$  { -}          $&$ { 5}             $&$  { -}      $&${ - } 
$&$  { 7 }         $&$ { 7}             $&$  { 7}        $\\
\hline \hline
$3      
$&${ 1}            $&${ 1}              $&${ 1}        $&${ 3} 
$&$ [6]
$&$ (39, 6^*)      
$&$ 1              $&$                  $&$            $&$        
$&$                $&$                  $&$ 1            $\\
\hline \hline
$6      
$&${ 1}            $&${ 1}              $&${ 2}        $&${ 4} 
$&$ [8]
$&$ (35, 7^*)      
$&$ 2              $&$1                 $&$            $&$    1   
$&$                $&$                  $&$              $\\
\hline
$8      
$&${ 1}            $&${ 1}              $&${ 3}        $&${ 5} 
$&$ [10]
$&$ (36, 9^*)      
$&$ 3              $&$                  $&$            $&$        
$&$     1          $&$                  $&$              $\\
\hline
$9      
$&${ 1}            $&${ 1}           $&${ 4}   $&${ 6} 
$&$ [12]
$&$ (39, 9^*)      
$&$  4   $&$       $&$       $&$        
$&$          $&$ 1      $&$   $\\  
\hline
$11      
$&${ 1}            $&${ 2}           $&${ 2}   $&${ 5} 
$&$ [10]
$&$ (28, 8^*)         
$&$   5  $&$       $&$   1   $&$ 2      
$&$          $&$        $&$   $\\
\hline
$15      
$&${ 1}         $&${ 2}           $&${ 3}   $&${ 6} 
$&$ [12]
$&$ (27, 9^*)      
$&$      $&$2      $&$       $&$ 3      
$&$          $&$        $&$ 2 $\\
\hline
$17     
$&${ 1}         $&${ 2}           $&${ 4}   $&${ 7} 
$&$ [14]
$&$ (27, 12^*)     
$&$      $&$       $&$       $&$ 4      
$&$  2       $&$        $&$   $\\
\hline
$19     
$&${ 1}         $&${ 2}           $&${ 5}   $&${ 8} 
$&$ [16]
$&$ (28, 14^*)     
$&$      $&$       $&$       $&$  5     $
&$  3       $&$        $&$   $\\
\hline
$20      
$&${ 1}         $&${ 2}           $&${ 6}   $&${ 9} 
$&$ [18]
$&$ (30, 12^*)     
$&$      $&$       $&$       $&$  6     
$&$          $&$ 2      $&$   $\\
\hline
$24      
$&${ 1}         $&${ 3}           $&${ 4}   $&${ 8}
$&$ [16]
$&$ (24, 12^*)     
$&$      $&$ 3     $&$ 3     $&$        
$&$          $&$        $&$   $\\
\hline
$26      
$&${ 1}         $&${ 3}           $&${ 5}   $&${ 9} 
$&$ [18]
$&$ (24, 15^*)     
$&$      $&$       $&$       $&$        
$&$  4       $&$        $&$   $\\
\hline
$28     
$&${ 1}         $&${ 3}           $&${ 7}   $&${ 11}
$&$ [22]
$&$ (25,20^*)      
$&$      $&$       $&$       $&$        
$&$     8    $&$        $&$   $\\
\hline
$29      
$&${ 1}         $&${ 3}           $&${ 8}   $&${ 12}
$&$ [24]
$&$ (27,15^*)      
$&$      $&$       $&$       $&$        $&$          $&$ 3      $&$   $\\
\hline
$31      
$&${ 1}         $&${ 4}           $&${ 5}   $&${ 10}
$&$ [20]
$&$ (23, 13^*)        
$&$      $&$ 4     $&$       $&$        $&$          $&$        $&$   $\\
\hline
$33      
$&${ 1}         $&${ 4}           $&${ 6}   $&${ 11}
$&$ [22]
$&$ (22,20^*)           
$&$      $&$       $&$   5   $&$        $&$ 5        $&$        $&$   $\\
\hline
$34     
$&${ 1}         $&${ 4}           $&${ 9}   $&${ 14}
$&$ [28]
$&$ (24,24^*)      
$&$      $&$       $&$       $&$        $&$  9       $&$        $&$   $\\ 
\hline
$35     
$&${ 1}         $&${ 4}           $&${ 10}  $&${ 15}
$&$ [30]
$&$ (25,20^*)      
$&$      $&$       $&$       $&$        $&$          $&$ 4      $&$   $\\
\hline
$37     
$&${ 1}         $&${ 5}           $&${ 7}   $&${ 13}
$&$ [26]
$&$ (21, 24^*)      
$&$      $&$       $&$       $&$        $&$       6  $&$        $&$   $\\ 
\hline
$38     
$&${ 1}         $&${ 5}           $&${ 12}  $&${ 18}
$&$ [36]
$&$ (14,18^*)       
$&$      $&$       $&$       $&$        $&$          $&$ 5      $&$   $\\ 
\hline
$40      
$&${ 1}         $&${ 6}           $&${ 8}   $&${ 15}
$&$ [30]
$&$ (21,24^*)         
$&$      $&$       $&$       $&$        $&$ 7        $&$        $&$   $\\
\hline
$41     
$&${ 1}         $&${ 6}           $&${ 14}  $&${ 21}
$&$ [42]
$&$ (24,24^*)     
$&$      $&$       $&$       $&$        $&$          $&$ 6      $&$   $\\
\hline
$43      
$&${ 2}         $&${ 2}           $&${ 3}   $&${ 7} 
$&$ [14]
$&$ (19, 11^*)     
$&$6    $&$       $&$ 2     $&$        $&$          $&$        $&$   $\\
\hline
$47     
$&${ 2}         $&${ 3}           $&${ 4}   $&${ 9}
 $&$ [18]
$&$(16, 14^*)     
$&$      $&$       $&$ 4     $&$   7    $&$          $&$        $&$ 3 $\\ 
\hline
$51    
$&${ 2}         $&${ 3}           $&${ 5}   $&${ 10}
$&$ [20]
$&$ (16, 14^*)     
$&$      $&$ 5     $&$       $&$        $&$          $&$        $&$   $\\
\hline
$53      
$&${ 2}         $&${ 3}           $&${ 7}   $&${ 12}
$&$ [24]
$&$ (16,20^*)      
$&$      $&$       $&$       $&$        $&$ 11       $&$        $&$   $\\
\hline
$55      
$&${ 2}         $&${ 3}           $&${ 8}   $&${ 13}
$&$ [26]
$&$ (16,23^*)     
$&$      $&$       $&$       $&$        
$&$ 10       $&$        $&$   $\\
\hline
$56      
$&${ 2}         $&${ 3}           $&${ 10}  $&${ 15}
$&$ [30]
$&$ (18,18^*)     
$&$      $&$       $&$       $&$        
$&$          $&$ 7      $&$     $\\
\hline
$58      
$&${ 2}         $&${ 4}           $&${ 5}   $&${ 11}
$&$ [22]
$&$ (14,19^*)      
$&$      $&$       $&$   6   $&$        
$&$          $&$        $&$     $\\  
\hline
$60      
$&${ 2}         $&${ 5}           $&${ 6}   $&${ 13}
$&$ [26]
$&$ (13,23^*)     
$&$      $&$       $&$   7   $&$        
$&$          $&$        $&$  $\\  
\hline
$62     
$&${ 2}         $&${ 5}           $&${ 9}   $&${ 16}
$&$ [32]
$&$ (13,29^*)     
$&$      $&$       $&$       $&$        
$&$ 14       $&$        $&$  $\\
\hline
$63    
$&${ 2}         $&${ 5}           $&${ 14}  $&${ 21}
$&$ [42]
$&$ (15,27^*)     
$&$      $&$       $&$       $&$        
$&$          $&$ 9      $&$  $\\ 
\hline
$64     
$&${ 2}         $&${ 6}           $&${ 7}   $&${ 15}
$&$ [30]
$&$ (13,23^*)      
$&$      $&$       $&$  9    $&$        
$&$          $&$        $&$                                  $\\
\hline
$69      
$&${ 3}         $&${ 4}           $&${ 5}   $&${ 12}
$&$ [24]
$&$ (12,18^*)      
$&$             $&$               $&$  8                     $&$       
$&$             $&$               $&$                          $\\
\hline
$71     
$&${ 3}         $&${ 4}           $&${ 7}   $&${ 14}
$&$ [28]
$&$ (12,18^*)      
$&$             $&$  6            $&$                    $&$        
$&$             $&$               $&$                      $\\
\hline
$73      
$&${ 3}         $&${ 4}           $&${ 10}  $&${ 17}
$&$ [34]
$&$ (11,31^*)     
$&$      $&$       $&$       $&$        
$&$ 13       $&$        $&$   $\\
\hline
$74      
$&${ 3}         $&${ 4}           $&${ 11}  $&${ 18}
$&$ [36]
$&$ (12,30^*)     
$&$      $&$       $&$       $&$        
$&$ 12       $&$        $&$   $\\
\hline
$75      
$&${ 3}         $&${ 4}           $&${ 14}  $&${ 21}
$&$ [42]
$&$ (13,26^*)      
$&$               $&$              $&$            $&$        
$&$               $&$  8           $&$             $\\
\hline
$78     
$&${ 3}           $&${ 5}          $&${ 11}       $&${ 19}
$&$ [38]
$&$ (10,35^*)      
$&$               $&$              $&$             $&$        
$&$  15           $&$              $&$               $\\
\hline
$79     
$&${ 3}            $&${ 5}         $&${ 16}          $&${ 24}
$&$ [48]
$&$ (12,30^*)      
$&$                $&$             $&$               $&$        
$&$                $&$10           $&$                 $\\
\hline
$82     
$&${ 4}            $&${ 5}         $&${ 6}           $&${ 15}
$&$ [30]
$&$ (10,20^*)     
$&$                $&$             $&$                $&$        
$&$                $&$             $&$ 4                $\\
\hline
$84      
$&${ 4}         $&${ 5}           $&${ 7}   $&${ 16}
$&$ [32]
$&$(9,27^*)        
$&$      $&$       $&$ 10    $&$        
$&$          $&$        $&$   $\\
\hline
$85      
$&${ 4}         $&${ 5}           $&${ 13}  $&${ 22}
$&$[44]
$&$ (9,39^*)      
$&$      $&$       $&$       $&$        
$&$ 16      $&$        $&$  $\\      
\hline
$86      
$&${ 4}         $&${ 5}           $&${ 18}  $&${ 27}$&$[54]
$&$ (10,35^*)      
$&$      $&$       $&$       $&$        $&$          $&$  11    $&$ $\\
\hline
$88      
$&${ 4}         $&${ 6}           $&${ 7}   $&${ 17}
$&$[34]
$&$ (8,31^*)      
$&$      $&$       $&$  11   $&$        
$&$          $&$        $&$  $\\
\hline
$91      
$&${ 5}         $&${ 6}           $&${ 8}   $&${ 19}
$&$[38]
$&$ (7,35^*)      
$&$      $&$       $&$ 12  $&$        
$&$          $&$        $&$  $\\
\hline
$92      
$&${ 5}            $&${ 6}           $&${ 22}                $&${ 33}
$&$[66]
$&$ (9,39^*)       
$&$                $&$               $&$                      $&$        
$&$                $&$ 12            $&$                        $\\
\hline
$93   
$&${ 5}            $&${ 7}           $&${ 8}   $&${ 20}
$&$[40]
$&$ (8,28^*)       
$&$                $&$               $&$                       $&$        
$&$                $&$               $&$                         $\\
\hline
$94    
$&${ 7}            $&${ 8}           $&${ 10}                 $&${ 25}
$&$[50]
$&$ (6,39^*)      
$&$                $&$               $&$                      $&$        
$&$                $&$               $&$                        $\\
\hline
\end{tabular}
\end{table}

\normalsize

In this chain, tshown in Table~\ref{TabXLIII}, he dimension
${ d=2M+2N+2L}$, there is $M \leftrightarrow N \leftrightarrow L $
symmetry, the eldest vector ${ \vec {k}_{eld}=(1,1,1,3)}$, and the
invariant monomials are ${ (2,2,2,0)+(0,2,2,2)}$. We see 
in the Table the appearance of the following two-vector chains
\begin{eqnarray}
V:\,&&\,m\cdot (1,1,0,2)\,+\,n \cdot (0,0,1,1)\,
=\,(m,m,n,2m+n)\,\rightarrow                        \nonumber\\
&&M\,=\,N\,=\,m\,\leq\,[dim]\{(0,0,1,1)\}\,=\,2,   \nonumber\\
&&L\,=\,n\,\leq\,[dim]\{(0,0,1,1)\}\,=\,4.         \nonumber\\
\end{eqnarray}

\begin{eqnarray}
VI:\,&&\,m\cdot (1,0,1,2)\,+\,n \cdot (0,1,1,2)\,
=\,(m,n,m+n,2m+2n)\,\rightarrow                        \nonumber\\
&&M\,=\,m\,\leq\,[dim]\{(0,1,1,2)\}\,=\,4,             \nonumber\\
&&N\,=\,n\,\leq\,[dim]\{(1,0,1,2)\}\,=\,4,             \nonumber\\
&&L\,=\,m+n.                                           \nonumber\\
\end{eqnarray}

\begin{eqnarray}
IX:\,&&\,m\cdot (1,0,1,2)\,+\,n \cdot (0,2,1,3)\,
=\,(m,2n,m+n,2m+3n)\,\rightarrow                       \nonumber\\
&&M\,=\,m\,\leq\,[dim]\{(0,2,1,3)\}\,=\,6,             \nonumber\\
&&N\,=\,2n\,\leq\,2[dim]\{(1,0,1,2)\}\,=\,8,           \nonumber\\
&&L\,=\,m+n.                                           \nonumber\\
\end{eqnarray}

\begin{eqnarray}
XVI:\,&&\,m\cdot (1,0,2,3)\,+\,n \cdot (0,1,0,1)\,
=\,(m,n,2m,3m+n)\,\rightarrow                         \nonumber\\
&&M\,=\,m\,\leq\,[dim]\{(0,1,0,1)\}\,=\,2,            \nonumber\\
&&N\,=\,n\,\leq\,[dim]\{(1,0,2,3)\}\,=\,6,            \nonumber\\
&&L\,=\,2m.                                           \nonumber\\
\end{eqnarray}

\begin{eqnarray}
XVIII:\,&&\,m\cdot (1,0,1,2)\,+\,n \cdot (0,1,2,3)\,
=\,(m,n,m+2n,2m+3n)\,\rightarrow                      \nonumber\\
&&M\,=\,m\,\leq\,[dim]\{(0,1,2,3)\}\,=\,6,            \nonumber\\
&&N\,=\,n\,\leq\,[dim]\{(1,0,1,2)\}\,=\,4,            \nonumber\\
&&L\,=\,m+2n.                                         \nonumber\\
\end{eqnarray}

\begin{eqnarray}
XIX:\,&&\,m\cdot (1,0,2,3)\,+\,n \cdot (0,1,2,3)\,
=\,(m,n,2m+2n,3m+3n)\,\rightarrow                     \nonumber\\
&&M\,=\,m\,\leq\,[dim]\{(0,1,2,3)\}\,=\,6,            \nonumber\\
&&N\,=\,n\,\leq\,[dim]\{(1,0,2,3)\}\,=\,6,            \nonumber\\
&&L\,=\,2m+2n.                                        \nonumber\\
\end{eqnarray}

\begin{eqnarray}
XX:\,&&\,m\cdot (2,0,1,3)\,+\,n \cdot (0,2,1,3)\,
=\,(2m,2n,m+n,3m+3n)\,\rightarrow                     \nonumber\\
&&M\,=\,2m\,\leq\,2[dim]\{(0,1,2,3)\}\,=\,6,          \nonumber\\
&&N\,=\,2n\,\leq\,2[dim]\{(1,0,2,3)\}\,=\,6,          \nonumber\\
&&L\,=\,m+n.                                          \nonumber\\
\end{eqnarray}


\normalsize
\subsection{The Three-Vector Chain $IV_3$: $\vec{k}_4=(M,\,N,\,M+L,\,N+L)$}

\scriptsize    
\begin{table}[!ht]
\centering
\caption{\it The {8} 
$K3$ hypersurfaces in the $IV_3$ chain: $\vec {k} = (M,N,M+L,N+L)=
M \cdot (1,0,1,0) + N \cdot (0,1,0,1) + L \cdot (0,0,1,1)$.}
\label{tripleIV}
\vspace{.05in}
\begin{tabular}{|c||c|c|c|c||c||c||c|c|c|c|c|} 
\hline
${ \aleph} 
$&$  M     $&$  N       $&$   M+L      $&$ N+L   
$&$ { [d]} $&$ (\Delta, \Delta ^*)
$&$    VII $&$      X   $&$   XII  
$&$  XXI   $&$    XXII  $ \\ \hline\hline
$ -     
$&$     -  $&$     -    $&$  -         $&$  -   
$&$ -      $&$      { J/ (\Delta)} 
$&$ { 7 }  $&$ { 9 }    $&$  { 5 }      
$&${  7}   $&$ { 7 }    $\\ \hline 
$ -     
$&$     -  $&$     -    $&$  -         $&$  -   
$&$ -      $&$     { \pi(\Delta^*)}    
$&$ { 7 }  $&$ { 5 }    $&$  { 9 }      
$&${  7 }  $&$ { 7 }    $\\ \hline \hline
$-     
$&$     -  $&$     -    $&$    -       $&$    -  
$&$ -      $&$ { \pi(\Delta)}      
$&$ { -  } $&$ { - }    $&$  { 5'}      
$&${ 7' }  $&$ { - }    $\\ \hline 
$-     
$&$     -  $&$     -    $&$    -       $&$    -  
$&$ -      $&$  {  J(\Delta^*)}    
$&$ {  -}  $&$ { - }    $&$  { 9'}       
$&${  7'}  $&$ { - }    $\\ \hline \hline
$1            
$&$ { 1}   $&$ { 1}     $&$ { 1}       $&${ 1} 
$&${[ 4]}  $&$ (35, 5^*) 
$&$  { 1}  $&$ { 1}     $&$  { 1}       
$&$     -  $&$      -   $ \\ \hline
$4            
$&$ { 1}   $&$ { 1}     $&$ {  2}      $&${ 2 } 
$&${[ 6]}  $&$ (30, 6^*) 
$&$    -   $&$ { 2}     $&$      -  
$&$ { 1}   $&$ { 1}     $ \\ \hline
$10            
$&$ { 1}   $&$ { 2}     $&$ {  2}      $&${ 3 } 
$&${[ 8]}  $&$ (24, 8^*) 
$&$ { 2}   $&$  -       $&$ -        
$&$ -      $&$ -        $ \\ \hline
$13           
$&$ { 1}   $&$ { 2}     $&$ {  3}      $&$ { 4 } 
$&${[10]}  $&$ (23, 11^*) 
$&$  -     $&$  -       $&$ { 2}        
$&$ -      $&$ { 2}     $ \\ \hline
$16            
$&$ { 1}   $&$ { 2}     $&$ {  4}      $&$ { 5 } 
$&${[12]}  $&$ (24, 12^*) 
$&$  -     $&$  -       $&$ -      
$&$ { 2}   $&$ { 3}     $ \\ \hline
$44            
$&$ { 2}   $&$ { 3}     $&$ {  3}      $&$ { 4 } 
$&${[12]}  $&$ (15, 9^*) 
$&$ { 3}   $&$  -       $&$  -     
$&$ { 3}   $&$  -       $ \\ \hline
$45            
$&$ { 2}   $&$ { 3}     $&$ {  4}      $&$ { 5 } 
$&${[14]}  $&$ (13, 16^*) 
$&$  -     $&$  -       $&$  { 3}     
$&$  -     $&$ { 4}     $ \\ \hline
$66           
$&$ { 3}   $&$ { 4}     $&$ {  5}      $&$ { 6 } 
$&${[18]}  $&$ (10, 17^*) 
$&$  -     $&$       -  $&$  { 4}     
$&$ { 4}   $&$  -       $ \\ \hline
\end{tabular}
\end{table}
\normalsize

In this case (see Table \ref{tripleIV}), we have the dimension ${ d=2M+2N+2L}$, the eldest vector
$\vec {k}=(1,1,1,1)$, and the
invariant monomials are ${ (2,0,0,2)+(0,2,2,0)}$. This three-vector chain
includes the following vectors form the two-vector construction:

\begin{eqnarray}
VII:\,&&\,m\cdot (2,1,1,0)\,+\,n \cdot (0,1,1,2)\,
=\,(2m,m+n,m+n,2n)\,\rightarrow                       \nonumber\\
&&M\,=\,2m \, \leq \,4,                               \nonumber\\
&&N\,=\,m+n \,\leq \,4,                               \nonumber\\
&&L\,=\,n-m \,\geq \,0.                               \nonumber\\
\end{eqnarray}

\begin{eqnarray}
X:\,&&\,m\cdot (1,1,0,0)\,+\,n \cdot (0,0,1,1)\,
=\,(m,m,n,n)\,\rightarrow                             \nonumber\\
&&M\,=\,m \, \leq \,2,                                \nonumber\\
&&N\,=\,m \,\leq \,2,                                 \nonumber\\
&&L\,=\,n-m \,\geq \,0.                               \nonumber\\
\end{eqnarray}

\begin{eqnarray}
XII:\,&&\,m\cdot (3,2,1,0)\,+\,n \cdot (0,1,2,3)\,
=\,(3m,2m+n,m+2n,3n)\,\rightarrow                      \nonumber\\
&&M\,=\,3m,                                            \nonumber\\
&&N\,=\,2m+n,                                          \nonumber\\
&&L\,=\,2n-2m,                                         \nonumber\\
&& (m,n)=(1,2),\,(2,1)\,;\,(1,1),\,(1,4),\,(4,1),
\,(2,5),\,(5,2).                                       \nonumber\\
\end{eqnarray}

\begin{eqnarray}
XXI:\,&&\,m\cdot (1,2,3,0)\,+\,n \cdot (1,2,0,3)\,
=\,(m,2m,3m,3n)\,\rightarrow                           \nonumber\\
&&M\,=\,m,                                             \nonumber\\
&&N\,=\,2m,                                            \nonumber\\
&&L\,=\,m,                                             \nonumber\\
&& (m,n)=(1,1),(1,2),\,(2,1),\,,(1,5),\,(5,1),
\,(4,5),\,(5,4).                                       \nonumber\\
\end{eqnarray}

\begin{eqnarray}
XXII:\,&&\,m\cdot (1,0,2,1)\,+\,n \cdot (0,1,0,1)\,
=\,(m,n,2m,m+n)\,\rightarrow                          \nonumber\\
&&M\,=\,m\,\leq [dim]\{(1,0,2,1)\}\,=\,4,             \nonumber\\
&&N\,=\,n\,\leq [dim]\{(1,0,2,1)\}\,=\,4,             \nonumber\\
&&L\,=\,m.                                            \nonumber\\
\end{eqnarray}


%
%
%
%

\section{The Dual $K3$ Algebra from Four-Dimensional Extended Vectors}

As discussed in the Introduction, the enumeration of $K3$
reflexive polyhedra obtained at level zero
from pairs of projective vectors
(Section 5) and triples (Section 6) is not quite complete. The one
remaining example, corresponding to ${\vec k}_4 = (7,8,9,12)$,
can be found using the intersection-projection and duality
properties outlined in Section 3, as we now discuss.
This method can be used to build projective-vector chains
using the rich projective structure of $K3$ vectors.
For example, one can construct a chain with, as youngest vector,
${\vec k}_4 = (7,8,10,25)$, which is dual to the eldest vector
${\vec k}_4 = (1,1,1,3)$ contained in the triple chain $III_3$. Similarly,
one can
consider other cases, e.g., building a chain with
youngest vector ${\vec k}_4 = (5,6,8,11)$, contained in the triple chain
$II_3$.

\subsection{The Dual ${ \vec{\pi}}$ Projective-Vector Structure
of $K3$ Hypersurfaces} 

We obtained in section 6, as an interesting application of the ${K3}$
algebra, all the $1\,+\,(4 \times 2)\,+22$ invariant monomials of the 
{22} double-intersection $K3$ chains via the triple intersections of
${K3}$ extended vectors.  These invariant monomials correspond to
particular directions relative to the reflexive polyhedra, which can be
used to
find the projection structures of the vectors. In particular, they can be
used to find all the
projective vectors which have no planar-intersection structure at all.
Because of duality, their polyhedra have sufficient invariant
directions
that the projections on the corresponding perpendicular planes
give reflexive planar polyhedra. Examples include youngest 
vectors which are dual to eldest vectors as well as other
relations in the corresponding chain, e.g., as we shall see, the
remaining $K3$ vector (7,8,9,12) is dual to (1,1,1,1),  (7,8,10,25)
is dual to (1,1,1,3), etc..

To understand this more deeply, we consider triple 
chains built using a special subalgebra of the four-dimensional 
extended vectors:
${ \vec {k}_3^{ex(i)}}$ ${ =  (0,0,0,1)}$,
${  (0,0,1,1)}$, ${ (0,1,1,1)}$,
${  (0,1,1,2)}$ and ${ (0,1,2,3)}$, with all possible permutations.
We consider triples  ${ \vec {k}_3^{ex(i,j,l)}}$
of these vectors with the property that each pair
${ (i,j)},  { (j,l)}, { (l,i)}$ gives a reflexive planar polyhedron:

\begin{eqnarray}
[\vec {k}_3^{ex(i)}]\,\bigcap \,
[\vec {k}_3^{ex(j)}]\, =\, [\vec {k}_3].
\end{eqnarray} 
We note that the triple intersections of these triples of extended vectors
always define an invariant direction, ${ \vec {\pi} }$. In some cases,
the triple intersection contains just two monomial vectors,
and ${ \vec {\pi} }$ is simply defined by their difference:
\begin{eqnarray}
[\vec {k}_3^{ex(i)}]\,\bigcap \,
[\vec {k}_3^{ex(j)}]\,\bigcap \,
[\vec {k}_3^{ex(r)}]\, \Rightarrow {\vec {\pi}_N} =
\{ \vec {\mu}\, -\,\vec {\mu}_{0} \},
\end{eqnarray}
where ${ \vec {\mu}_{0}} = (1,1,1,1)$ is the basic
monomial  ${ z^{\vec {\mu}_{0}+\vec {1}}=x \cdot y \cdot z \cdot u }$.
These cases are listed in Table \ref{TabInv}.

\begin{table}[!ht]
\centering
\caption{\it Triples of $\vec {k_3}^{ex}$ vectors giving 
invariant directions ${  \vec {\pi}_N} = {\vec \mu}_N - {\vec \mu}_0$
defined by pairs of monomials.
Also indicated are the sizes of the corresponding polyhedra $\Delta$
and the two-vector chains to which they belong.}
 \label{TabInv}
\scriptsize
\vspace{.05in}
\begin{tabular}{|c||c|c|c||c|c|c||c|c|c|} \hline
${ \vec {\pi}_N^{(\alpha)}}   
$&$ {\vec k_3}^{ex(i)}$&$ {\vec k_3}^{ex(j)}     $&$ {\vec k_3}^{ex(p)} 
$&${\Delta}_{J_{ij}}  $&$ {\Delta}_{J_{jp}}      $&$ {\Delta}_{J_{pi}}
$&$ inv. monom        $&$\vec {\mu}_{N} 
$ \\ \hline\hline 
$\vec {\pi}_1^{(1)}   
$&$ (0,0,1,1)         $&$ (1,2,0,1) $&$ (1,2,1,0)
$&$  7_{XXII}         $&$ 9_{VI}    $&$ 7_{XXII}   
$&$x^3\cdot  z\cdot u $&$ (3,0,1,1) 
$ \\ 
$\vec {\pi}_1^{(2)}   
$&$ (0,0,1,1)         $&$ (1,2,0,3) $&$ (1,2,3,0)
$&$  7_{XVI}          $&$ 7_{XXI}   $&$ 7_{XVI}   
$&$x^3\cdot  z\cdot u $&$ (3,0,1,1) 
$ \\ \hline\hline
$\vec {\pi}_2^{(1)}   
$&$ (0,1,1,1)         $&$ (1,0,1,1) $&$ (1,1,0,1)
$&$  10_{II}          $&$ 10_{II}   $&$ 10_{II}   
$&$ u^3               $&$ (0,0,0,3)             
$ \\
$\vec {\pi}_2^{(2)}   
$&$ (0,1,1,1)         $&$ (1,0,1,1) $&$ (1,3,0,2)
$&$  10_{II}          $&$  4_{III}  $&$  7_{XVII} 
$&$ u^3               $&$ (0,0,0,3) 
$ \\
$\vec {\pi}_2^{(3)}   
$&$ (0,1,1,1)         $&$ (1,0,3,2) $&$ (1,3,0,2)
$&$   7_{XVII}        $&$  7_{XXI}  $&$  7_{XVII} 
$&$ u^3               $&$ (0,0,0,3) 
$ \\
$\vec {\pi}_2^{(4)}   
$&$ (0,1,1,1)         $&$ (3,0,1,2) $&$ (3,1,0,2)
$&$    4_{III}        $&$  7_{XIX}  $&$  4_{III}  
$&$ u^3               $&$ (0,0,0,3) 
$ \\ \hline\hline
$\vec {\pi}_3^{(1)}   
$&$ (1,1,1,0)         $&$ (0,1,2,1) $&$ (2,1,0,1)
$&$   8_{XIII}        $&$  9_{VII}  $&$  8_{XIII}  
$&$ y^3 \cdot u       $&$ (0,3,0,1) 
$ \\
$\vec {\pi}_3^{(2)}   
$&$ (1,1,1,0)         $&$ (0,1,2,1) $&$ (2,1,0,3)
$&$   8_{XIII}        $&$  6_{XIV}  $&$  4_{III} 
$&$ y^3 \cdot u       $&$ (0,3,0,1) 
$ \\
$\vec {\pi}_3^{(3)}   
$&$ (1,1,1,0)         $&$ (0,1,2,3) $&$ (2,1,0,3)
$&$   4_{III}         $&$  7_{XX}   $&$ 4_{III} 
$&$ y^3 \cdot u       $&$ (0,3,0,1) 
$ \\
$\vec {\pi}_3^{(4)}   
$&$ (0,1,2,1)         $&$ (1,2,3,0) $&$ (2,1,0,3)
$&$   7_{XVIII}       $&$  5_{XII}  $&$ 6_{XIV} 
$&$ y^3 \cdot u       $&$ (0,3,0,1) 
$ \\
\hline \hline
$\vec {\pi}_4^{(1)}   
$&$ (0,1,1,2)         $&$ (1,0,1,2) $&$ (1,2,1,0)
$&$   9_{VI}          $&$  9_{VII}  $&$ 5_{VIII} 
$&$ y^4               $&$ (0,0,4,0) 
$ \\
$\vec {\pi}_{4}^{(2)} 
$&$ (0,1,1,2)         $&$ (1,2,1,0) $&$ (2,0,1,1)
$&$   5_{VIII}        $&$ 5_{VIII}  $&$ 5_{VIII} 
$&$ y^4               $&$ (0,0,4,0) 
$ \\ 
\hline \hline
$\vec {\pi}_{5}^{(1)} 
$&$(1,0,1,2,)         $&$ (1,2,0,3) $&$ (1,2,3,0)
$&$   5_{IX}          $&$  7_{XXI}  $&$ 6_{XIV} 
$&$ x^4 \cdot y       $&$ (4,1,0,0) 
$ \\ 
\hline
\end{tabular}
\end{table}

\normalsize

These pairs of invariant monomials
correspond to directions $\vec {\pi_i}  \,=\,\vec {\mu}_i\,-\,\vec
{\mu}_0$ in the exponent/monomial hyperspace given by
the following vectors ${ \vec {\mu}_{N} }: {N=1,2,3,4,5}$:

\begin{eqnarray}
\vec {\mu}_1\,&=&\,(3,0,1,1), \nonumber\\
\vec {\mu}_2\,&=&\,(0,0,0,3), \nonumber\\
\vec {\mu}_3\,&=&\,(0,3,0,1), \nonumber\\ 
\vec {\mu}_4\,&=&\,(0,0,4,0), \nonumber\\ 
\vec {\mu}_5\,&=&\,(4,1,0,0). \nonumber\\ 
\end{eqnarray}
as can be seen in Table~\ref{TabInv}.

In the other cases, the triple intersections contain three points
which form a degenerate linear polyhedron, which also
defines a unique direction ${ \vec {\pi}}$ determined by three points,
one of which $({ \vec {\mu}_{0} })$ corresponds to the origin: 
\begin{eqnarray}
[\vec {k}_3^{ex(i)}]\,\bigcap \,
[\vec {k}_3^{ex(j)}]\,\bigcap \,
[\vec {k}_3^{ex(r)}]\, \Rightarrow 
\,\vec{\pi}_N\,=\,\{\vec {\mu}_{+}\,-\,\vec {\mu}_0 \}
\,=\,\{ \vec {\mu}_{0}\,-\,\vec{\mu}_{-} \},
\end{eqnarray}
as seen in Table \ref{Tab3Inv}.

\begin{table}[!ht]
\centering
\caption{\it Triples of $\vec {k_3}^{ex}$ vectors defining
directions $\vec {\pi}_N$ determined by three invariant monomials.}
\label{Tab3Inv}
\scriptsize
\vspace{.05in}
\begin{tabular}{|c||c|c|c||c|c|c||c|c|} \hline
${ \vec {\pi}_N^{(\alpha)}}   
$&$ {\vec k_3}^{ex(i)}  $&$ {\vec k_3}^{ex(j)} $&$ {\vec k_3}^{ex(p)} 
$&${\Delta}_{J_{ij}}    $&$ {\Delta}_{J_{jp}}  $&$ {\Delta}_{J_{pi}}
$&$\vec {\mu}_{+}       $&$\vec {\mu}_{-}            
$ \\ \hline\hline 
${ \vec {\pi}_6^{(1)}} 
$&$ (0,0,1,1)           $&$ (1,1,0,1)          $&$ (1,1,1,0)
$&$  9_{XI}             $&$ 10_{II}            $&$ 9_{XI}   
$&$ (0,2,1,1)           $&$ (2,0,1,1)            
$ \\ 
${ \vec {\pi}_6^{(2)}}
$&$ (0,0,1,1)           $&$ (1,1,0,2)          $&$ (1,1,2,0)
$&$  9_{V}              $&$ 9_{VII}            $&$ 9_{V}   
$&$ (0,2,1,1)           $&$ (2,0,1,1)            
$ \\ \hline
${ \vec {\pi}_7^{(1)}} 
$&$ (0,1,1,1)           $&$ (1,0,1,1)          $&$ (1,1,0,2)
$&$  10_{II}            $&$ 8_{XIII}           $&$   8_{XIII}  
$&$ (0,0,1,2)           $&$ (2,2,1,0)
$ \\ 
${ \vec {\pi}_7^{(2)}} 
$&$ (0,1,1,1)           $&$ (1,0,2,1)          $&$ (1,1,0,2)
$&$    8_{XIII}         $&$ 5_{VIII}           $&$   8_{XIII}  
$&$ (0,0,1,2)           $&$ (2,2,1,0)
$ \\ 
${ \vec {\pi}_7^{(2)}} 
$&$ (0,1,1,1)           $&$ (1,0,2,1)          $&$ (1,2,0,3)
$&$    8_{XIII}         $&$ 6_{XIV}            $&$   7_{XVII}  
$&$ (0,0,1,2)           $&$ (2,2,1,0)
$ \\ 
${ \vec {\pi}_7^{(2)}}
$&$ (0,1,2,1)           $&$ (1,0,2,1)          $&$ (1,1,0,2)
$&$  9_{VI}             $&$ 5_{VIII}           $&$  5_{VIII}  
$&$ (0,0,1,2)           $&$ (2,2,1,0)
$ \\ \hline
${ \vec {\pi}_8^{(1)}} 
$&$ (0,1,1,2)           $&$ (1,0,1,2)          $&$ (1,1,0,2)
$&$  9_{VI}             $&$ 9_{VI}             $&$  9_{VI}
$&$ (0,0,0,2)           $&$ (2,2,2,0)
$ \\ 
${ \vec {\pi}_8^{(2)}} 
$&$ (0,1,1,2)           $&$ (1,0,1,2)          $&$ (1,2,0,3)
$&$  9_{VI}             $&$ 5_{IX}             $&$  7_{XVIII}  
$&$ (0,0,0,2)           $&$ (2,2,2,0)
$ \\ 
${ \vec {\pi}_8^{(3)}} 
$&$ (0,1,1,2)           $&$ (1,0,2,3)          $&$ (1,2,0,3)
$&$  7_{XVIII}          $&$ 7_{XX}             $&$  7_{XVIII}  
$&$ (0,0,0,2)           $&$ (2,2,2,0)
$ \\ 
${ \vec {\pi}_8^{(4)}}
$&$ (0,1,1,2)           $&$ (2,0,1,3)          $&$ (2,1,0,3)
$&$  5_{IX}             $&$ 7_{XIX}            $&$  5_{IX}  
$&$ (0,0,0,2)           $&$ (2,2,2,0)
$ \\ \hline
\end{tabular}
\end{table}

\normalsize

It is easy to see that five of the invariant monomials from
Table~\ref{TabInv} produce a 
reflexive three-dimensional polyhedron. For example, from
${\vec \mu_2}$, ${\vec \mu_3}$, ${\vec \mu_4}$ and  ${\vec \mu_5}$
one obtains the following 
exceptional vector whose associated polyhedron
has no intersection substructure:

\begin{eqnarray}
\vec {\mu}_{\alpha} \cdot \vec {k}_4\,&=&\,
d\,=\, k_1\,+\,k_2\,+\,k_3\,+\,k_4,\,\,\,\,
\alpha=0,2,3,4,5  \nonumber\\
\vec {k}_4\,&=&\,(7,8,9,12)[d=36],
\end{eqnarray}
where we used the constraint
\begin{eqnarray}
\vec {\mu}_0\,=\,\frac{1}{4} \cdot (\vec {\mu}_2\,+\,
\vec {\mu}_3\,+\,\vec {\mu}_4\,+\,\vec {\mu}_5\,).
\end{eqnarray}
Thus duality enables us to identify the missing 95th
$K3$ vector, which was not generated previously in our
systematic study of the two- and three-vector chains. We
recall that they contain totals of 90 and 91 vectors,
respectively, of which only 94 were distinct.

Similarly, using these invariant monomials,
one can find the rest of the exceptional ${ \vec{k}_4}$ vectors,
${ (3,5,6,7)}$, ${ (3,6,7,8)}$, ${ (5,6,7,9)}$  which
were not included in the triple chains, together with
${ (3,4,5,6)}$. They have
intersection polyhedra that are not linear.
These other exceptional ${\vec k}_4$ vectors are defined as follows:
\begin{eqnarray}
\vec {\mu}_{\alpha} \cdot \vec {k}_4\,&=&\,d\,\,\,\,\,\,
\alpha=0,1,2,3,3'  \nonumber\\
\vec {k}_4\,&=&\,(3,5,6,7)[d=21],
\end{eqnarray}
where again the following constraint has been used:
\begin{eqnarray}
&&\vec {\mu}_0\,=\,\frac{1}{4} \cdot (\vec {\mu}_1\,+\,
\vec {\mu}_2\,+\,\vec {\mu}_3\,+\,\vec {\mu}_3'\,)\,=\nonumber\\
&&(3,1,0,1)\,+\,(0,0,0,3)\,+(1,0,3,0)\,+(0,3,1,0);\nonumber\
\end{eqnarray}
and
\begin{eqnarray}
\vec {\mu}_{\alpha} \cdot \vec {k}_4\,&=&\,d\,\,\,\,\,\,
\alpha=0,1,2,3,4  \nonumber\\
\vec {k}_4\,&=&\,(3,6,7,8)[d=24],
\end{eqnarray}
with the constraint:
\begin{eqnarray}
&&\vec {\mu}_0\,=\,\frac{1}{4} \cdot (\vec {\mu}_1\,+\,
\vec {\mu}_2\,+\,\vec {\mu}_3\,+\,\vec {\mu}_4\,)\,=\nonumber\\
&&(3,0,1,1)\,+\,(0,0,0,3)\,+(1,0,3,0)\,+(0,4,0,0).\nonumber\
\end{eqnarray}
We also find

\begin{eqnarray}
\vec {\mu}_{\alpha} \cdot \vec {k}_4\,&=&\,d\,\,\,\,\,\,
\alpha=2,3,3',5  \nonumber\\
\vec {k}_4\,&=&\,(5,6,7,9)[d=27],
\end{eqnarray}
where the following constraint also has been used:
\begin{eqnarray}
&&\vec {\mu}_0\,=\,\frac{1}{4} \cdot (\vec {\mu}_2\,+\,
\vec {\mu}_3\,+\,\vec {\mu}_3'\,+\,\vec {\mu}_5\,)\,=\nonumber\\
&&(0,0,0,3)\,+\,(0,3,0,1)\,+(0,1,3,0)\,+(4,0,1,0).\nonumber\
\end{eqnarray}
and
\begin{eqnarray}
\vec {\mu}_{\alpha} \cdot \vec {k}_4\,&=&\,d\,\,\,\,\,\,
\alpha=1,2,3,3'  \nonumber\\
\vec {k}_4\,&=&\,(3,4,5,6)[d=18],
\end{eqnarray}
where the following constraint also has been used:
\begin{eqnarray}
&&\vec {\mu}_0\,=\,\frac{1}{4} \cdot (\vec {\mu}_1\,+\,
\vec {\mu}_2\,+\,\vec {\mu}_3\,+\,\vec {\mu}_3'\,)\,=\nonumber\\
&&(3,1,1,0)\,+\,(0,0,0,3)\,+(1,0,3,0)\,+(0,3,0,1).\nonumber\
\end{eqnarray}

\subsection{Projective Chains of $K3$ Spaces Constructed from
$\vec{\pi}_N$ Vectors}

Using the invariant directions found in the previous Subsection,
one  can  construct new triple chains:
\begin{eqnarray}
  p \cdot [\vec {k}_4]_{\vec {\pi}_N} =
  m \cdot \vec {k}^{ex(i)}            \,+\, 
  n \cdot \vec {k}^{ex(j)}            \,+\,
  r \cdot \vec {k}^{ex(l)}
\end{eqnarray}
each corresponding to a direction ${\vec {\pi}}$ determined by 
an intersection of invariant monomial pairs. Each good projective vector
in such a chain, determined by an invariant direction,
contains the monomial/projective direction in
its polyhedron. With respect to this direction, the polyhedron  is   
projected onto a `good' planar reflexive polyhedron. 
If the projective vector appears in several
different chains, its polyhedron will have 
`good' projections corresponding to each of these chains. 
This property can be used to make a classification by their
projections of the projective vectors and their
reflexive polyhedra. One finds that {78} projective ${K3}$ vectors
out of {95} have such aprojective property. 
Taking into account the rest of the
vectors which already were known from
double-intersection ${J=\Pi}$-symmetric chains,
one can recover all {95} projective ${K3}$ vectors.

The distribution of the 3-dimensional set of positive-integer
numbers ${m,n,r}$ depends on the dimension of the three 
extended vectors
$d^{(i)}=\sum_{\alpha} \{\vec{k}_3^{ex(i)}\}_{\alpha}$,  $i=1,2,3$,
participating in the construction of the chain, can have some
`blank spots', corresponding to `false vectors' which do not correspond to
any reflexive polyhedron.
The origin of this phenomenon is connected with the structure of 
Calabi-Yau algebra, i.e., some of the projective vectors have
different
expansions (double-, triple-,...) in terms of the extended vectors. So,
for example, 
if a vector is {\it forbidden} in two-vector expansions, it 
should also be forbidden in triple, etc., expansions, which is
what we call a false vector.
The self-consistency of the algebra entails the absences of 
some combinations of integer numbers  ${m,n,r}$, even though all
of them are below their maximum values.
We already have met and discussed this phenomenon in the
classification of triple-vector chains.

As seen in Table~\ref{TabInv},
one can give examples of  triple intersections
giving just one good vector which has three different projections
with $\Pi=4$: 
\begin{eqnarray} 
    {[\vec {k}_4]}_{{\vec {\pi}}_1^{(2)}} \bigcap 
    {[\vec {k}_4]}_{{\vec {\pi}}_2^{(2)}} \bigcap  
    {[\vec {k}_4]}_{{\vec {\pi}}_3^{(4)}}    
   & \Rightarrow  &  (3,5,6,7)[21]                         \nonumber\\
    {[\vec {k}_4]}_{{\vec {\pi}}_1^{(2)}  } \bigcap 
    {[\vec {k}_4}]_{{\vec {\pi}}_2^{(2)}  } \bigcap  
    {[\vec {k}_4]}_{{\vec {\pi}}_4^{(2)}  }    
   & \Rightarrow  &  (3,6,7,8)[24]                         \nonumber\\
    {[\vec {k}_4]}_{{\vec {\pi}}_2^{(2)}  } \bigcap 
    {[\vec {k}_4]}_{{\vec {\pi}}_3^{(3)}  } \bigcap  
    {[\vec {k}_4]}_{{\vec {\pi}}_3^{(1)}  }    
   & \Rightarrow  &  (5,6,7,9)[27].                        \nonumber\\
\end{eqnarray}
Moreover, the exceptional vector, which has four different
projections with $\Pi=4$, is given by the intersection of
four such chains, i.e.: 
\begin{eqnarray} 
 {[\vec {k}_4]}_{{\vec {\pi}}_2^{(1)}  } \bigcap
   {[\vec {k}_4]}_{{\vec {\pi}}_3^{(2)}  } \bigcap  
    {[\vec {k}_4]}_{{\vec {\pi}}_4^{(2)}  } \bigcap  
    {[\vec {k}_4]}_{{\vec {\pi}}_5^{(1)}  }    
   & \Rightarrow  &  (7,8,9,12)[36].                      
\end{eqnarray}

\noindent
To understand this in more detail, we consider one chain
with projection ${\Pi=4}$,
which is determined by the  invariant direction
$\vec{\pi}_2^{(1)}$. 
The vectors of this chain are represented as linear combinations
with positive-integer coefficients, ${M,N,L,}$
of the following three projective vectors, taken from the third
line in Table~\ref{TabInv}:

\begin{eqnarray}
 \vec {k}_4(\vec{\pi}_2^{(1)}) & =&
 M \cdot (0,1,1,1) + N \cdot (1,0,1,1) + L \cdot (1,1,0,1)\nonumber\\
&=&(\,N\,+\,L,\,M\,+\,L,\,M\,+\,N,\,
M\,+\,N\,+\,L\,)\nonumber\\
\end{eqnarray}
The basis is constructed out of the  exceptional invariant
monomials determining the $\vec{\pi}$ directions. Projecting on 
the perpendicular plane gives us planar reflexive polyhedra,
so the  third basis vector  
\begin{equation}
\vec {e}_3=(-1,-1,-1,2)\,\,\,\Rightarrow \,\,\, (0,0,0,3).
\end{equation}
is  common to all the chains discussed in this Subsection.

\normalsize
\begin{table}[!ht]
\centering
\caption{\it Extended vectors ${\vec{k}_4}$ in the chain 
$\vec{\pi}_2^{(1)}$ with $ \Pi=4 $
with: $ {Q \cdot \vec {k}}$ $= (N+L,M+L,M+N,M+N+L)=
M \cdot (0,1,1,1) + N \cdot (1,0,1,1) + L \cdot (1,1,0,1)$ and
${d=3M+3N+3L}$, whose youngest vector is
${\vec {k}_{young}=(7,8,9,12)}$.}
\label{TabPI1}
\scriptsize
\vspace{.05in}
\begin{tabular}{|c||c|c|c|c|c||c||c|c|} 
\hline
${\aleph}       $&$ \vec{k}_4       $&$ {[det]}             
$&${M}          $&${N}          $&${L}           
$&$(\Delta,\Delta ^*)
$&$ \Pi - J^*       $&$ {Chain}                  $\\ 
\hline   \hline
${95}            $&$(7,8,9,12)       $&$ [36]            
$&${5}          $&$ {4}         $&${3}          
$&$  (5,35)             
$&$ {4 -10}     $&$ -                      $\\
${89 }           $&$(5,6,7,9)        $&$ [27]  
$&${4}          $&$ {3}         $&${2}        
$&$  (6,30)          
 $&$ {4-10}      $&$ {III}    $\\
${80 }           $&$(3,6,7,8)        $&$ [24]   
$&${5}          $&$ {2}         $&${1}         
$&$  (9,21)            
$&$ {4- 10}     $&$ {III}     $\\
${76 }           $&$(3,5,6,7)        $&$ [21]
$&${4}          $&$ {2}         $&${1}          
$&$  (9,21)            
$&$ {4 - 10}     $&$ {III}     $\\
${66 }           $&$(3,4,5,6)        $&$ [18]
$&${3}          $&$ {2}         $&${1}           
$&$  (10,17)            
$&$ {4- 10}     $&${III}      $\\
${65 }           $&$ (3,3,4,5)       $&$ [15]
$&${2}          $&$ {2}         $&${1}           
$&$ (12,12)           
$&$ {4- 10}     $&$ {III}     $\\
${44}           $&$(2,3,3,4)        $&$ {[24]}
$&${2}          $&$ {1}         $&${1}           
$&$ (7,26)             
$&$ {4- 10}     $&$ {III}     $\\ 
\hline
\end{tabular}
\end{table}

\normalsize

Looking at the distribution of allowed integers ${M,N,L}$,
we see `blank spots' such as ${M=N=L=1}$, corresponding to
the `false vector' $(2,2,2,3)$, 
which is forbidden by the double-vector classification: it
would require $m = 2$ in the chain
$(2,2,2,3)\,=\,m\,(1,1,1,0)\,+ \,n\,(0,0,0,1)$,
but actually $m_{max}= 1$ for this chain.
Also, all the polyhedra corresponding to  these projective vectors 
have the other invariant directions
 $\vec{\pi}_3^{(2)}\rightarrow (1,0,3,0)$
with $\Pi=4$
and should  produce the
following  triple-vector expansion chain: 

\begin{eqnarray}
 \vec {k}_4(\vec{\pi}_3^{(2)}) & =&
 M \cdot (0,1,1,1) + N \cdot (1,0,1,2) + L \cdot (3,2,1,0)\nonumber\\
&=&(\,N\,+\,3L,\,M\,+\,2\,L,\,M\,+\,N\,+\,L,\,
M\,+\,2\,N\,)\nonumber\\
\end{eqnarray}
 Projecting on 
the perpendicular plane to the vector
\begin{equation}
\vec {e}_3=(0,-1,2,-1)\,\,\,\Rightarrow \,\,\, (1,0,3,0). 
\end{equation}
gives us planar reflexive polyhedron with {4} points.
This chain is a little longer and contains other projective vectors.
Similarly, one can find using the other projective 
directions, $\vec{\pi}_4^{(\alpha)}$ and $\vec{\pi}_5^{(1)}$,
two new triple expansion chains. Together these four 
invariant directions, $\vec{\pi}_i^{(\alpha)}$, (i=2,3,4,5),
with the constructions of the corresponding triple projective chains 
contain {40} projective vectors  (see Table~\ref{list95}).


\begin{table}[!ht]
\centering
\caption{\it The $K3$ hypersurfaces in chain  
{III} with intersection {J=4}:
 $\vec {k}(III)$ $= (3n,m,m+n,m+2n)$ $=  m \cdot (0,1,1,1)$ $+ 
n \cdot (3,0,1,2)$: ${d=3m + 6n}$ with level ${l = m+n}$,
${m_{max}=6, n_{max}=2}$.}
\label{TabIII}
\scriptsize
\vspace{.05in}
\begin{tabular} {|c|c|c|c|c|} \hline
${\aleph} $&${{\vec k}_i[dim]} $&$ 
\Delta(J= {4})
$&$ \Delta^*(\Pi= {10}) 
 $&$({\Pi}, {J^*}) $\\ \hline\hline
${12} $&${(3,1,2,3)[ 9]} 
$&$ {23= 4_L+ {4}_J+15_R}
$&${8^*=3_L^*+ {4}_J^*+1_R^*}  
$&$ (10,4^*)$  \\
${44} $&$ {(3,2,3,4)[12]} 
$&$ {15= 7_L+ {4}_J+ 4_R}
$&${9^*=4_L^*+{3}_C^*+2_R^*} 
$&$( 9,5^*);(7,7^*)$ \\
${25} $&${(6,1,3,5)[15]} 
$&$ {21=16_L+ {4}_J+ 1_R} 
$&${15^*=7_L^*+ { 7}_J^*+1_R^*} 
$&$( 7,7^*)$ \\
${65} $&$ {(3,3,4,5)[15]} 
$&$ {12= 4_L+ {4}_J+ 4_R}  
$&${12^*=1_L^*+ {10}_J^*+1_R^*}
$&$(4,10)$ \\
${66} $&${(3,4,5,6)[18]} 
$&$ {10= 2_L+ {4}_J+ 4_R}
$&${17^*=8_L^*+3_C^*+6_R^*}  $&$(7,7)$ \\
${76} $&$ {(3,5,6,7)[21]} 
$&$ {9= 1_L+ {4}_J+ 4_R}
$&${21^*=11_L^*+3_C^*+7_R^*} $&$(4,10)$ \\
${80} $&$ {(3,6,7,8)[24]} 
$&$ {9= 1_L+ {4}_J+ 4_R}
$&${21^*=12_L^*+3_C^*+6_R^*} 
$&$(4,10)$ \\
${89} $&${(6,5,7,9)[27]} 
$&$ {6= 1_L+ {4}_J+ 1_R}
$&${30^*=17_L^*+3_C^*+10_R^*}$&$(4,10)$ \\
\hline
\end{tabular}
\end{table}

\normalsize

One can compare the  projection set, $\vec{\pi}_2^{(1)}$  
and  ${\Pi=4}$,
with the double-vector-intersection  chain with ${J=4}$.
It is interesting to note that six vectors from the projective chain
shown in Table~\ref{TabPI1} also appear in
the ${III}$-intersection chain with 
${J=4}$ shown in Table~\ref{TabIII}. Conversely, the chain shown in this 
latter Table has just two vectors: $(3,1,2,3), (1,3,5,6)$ that
are not contained in Table~\ref{TabPI1}.
The intersection structure of the ${III}$ chain shown in
Table~\ref{TabIII} is obtained from the following two vectors:
\begin{eqnarray}
 {\vec {k}_4(III)} & =&
{m \cdot (0,1,1,1) + n \cdot (3,0,1,2) } \nonumber\\
&=&{(\,n,\,m,\,\,m\,+\,n,\,m\,+\,n\,)}\nonumber\\
&&{1\,\leq\, m\,\leq \,6,\,\,\,\,\,\,\,1\,\leq \,n\,\leq\,2.}\nonumber\\ 
\end{eqnarray}
The corresponding four invariant monomials are:
\begin{eqnarray}
\vec{\mu}_0^1\,&=&\,(0,0,0,3) \Rightarrow 
                             u^3  \nonumber\\
\vec{\mu}_0^2\,&=&\,(1,0,3,0) \Rightarrow 
x                \cdot z^3        \nonumber\\
\vec{\mu}_0^3\,&=&\,(2,3,0,0) \Rightarrow 
x^2    \cdot y^3                  \nonumber\\
\vec{\mu}_0^4\,&=&\,(1,1,1,1) \Rightarrow 
x      \cdot y   \cdot z  \cdot u. 
\end{eqnarray}
and the corresponding basis can be chosen in the form: 
\begin{eqnarray} 
\vec{e}_1\,&=&\, (0,-m-n,m,0)  \nonumber\\
\vec{e}_2\,&=&\, (0,-1,2,-1)   \nonumber\\
\vec{e}_3\,&=&\, (-1,-1,-1,2)  \nonumber\\
\end{eqnarray}
The canonical expression for the determinant of this lattice is
\begin{eqnarray}
{det} (\vec {e}_1, \vec {e}_2, \vec {e}_3,
 \vec {e}_0)\,=\,{3\cdot m\, +\, 6\cdot n }\,=\, {d},
\end{eqnarray}
where $\vec {e}_0 \equiv (1,1,1,1)$.

\begin{table}[!ht]
\centering
\caption{\it The $K3$ hypersurfaces in 
chain {II} with intersection {J=10}:
$\vec {k}(II) =$ ${(n,m,m+n,m+n)}$ $= m \cdot (0,1,1,1) +$ 
$n \cdot (1,0,1,1)$ with ${d=3m + 3n }$, ${q=1}$,
$m_{max}=3, n_{max}=3$.}
\label{TabII}
\scriptsize
\vspace{.05in}
\begin{tabular}{|c|c|c|c|c|} \hline
${\aleph} $&${{\vec k}_i[dim]}$&$ 
\Delta(J= {10})  
$&$ \Delta^*(\Pi= 4) 
$&$({\Pi},{J^*} )$\\ \hline \hline
${4}  $&${(1,1,2,2)[ 6]}
$&$ {30=10_L+ {10}_J+10_R} 
$&${6^*=1_L^*+ {4}_J^*+1_R^*}  
$&$(10,4^*)$ \\
${12}  $&${(1,2,3,3)[ 9]}
$&$ {23=10_L+ {10}_J+ 3_R} 
$&${8^*=3_L^*+ {4}_J^*+1_R^*}  
$&$(10,4^*)$ \\
${21}  $&${(1,3,4,4)[12]}
$&$ {21=10_L+ {10}_J+ 1_R} 
$&${9^*=4_L^*+ {4}_J^*+1_R^*}  
$&$(10,4^*)$ \\
${48}  $&${(2,3,5,5)[15]}
$&$ {14= 3_L+ {10}_J+ 1_R} 
$&${11^*=4_L^*+ {4}_J^*+3_R^*}  
$&$(10,4^*)$ \\
\hline
\end{tabular}
\end{table}
\normalsize

\subsection{Example of a $J, \Pi = 10$ Double-Intersection Chain}

To see another aspect of mirror symmetry and duality,
consider the ${II}$ chain with  intersection 
 ${J(\Delta)=\Pi(\Delta)=10}$
and ${J(\Delta^*)=\Pi(\Delta^*)=4}$ shown in Table~\ref{TabII}. 
The decomposition of this  chain is in terms of the following
two vectors:
\begin{eqnarray}
 {\vec {k}_4} & =&
{m \cdot (0,1,1,1) + n \cdot (1,0,1,1) } \nonumber\\
&=&{(\,n,\,m,\,\,m\,+\,n,\,m\,+\,n\,)}\nonumber\\
&&{1\,\leq\,m\,\leq\,3,\,\,\,\,\,\,\,1\,\leq\,n\,\leq\,3.}
\nonumber\\
\end{eqnarray}
The basis of the lattice in which the polyhedral intersection with
the set of positive-integer points corresponds to Table~\ref{TabII} is
the 
following:
\begin{eqnarray} 
\vec{e}_1\,&=&\, (-m,n,0,0)\nonumber\\
\vec{e}_2\,&=&\, (-1,-1,1,0)\nonumber\\
\vec{e}_3\,&=&\, (-1,-1,0,1)
\end{eqnarray}
and the corresponding determinant is
\begin{eqnarray}
{det} (\vec {e}_1, \vec {e}_2, \vec {e}_3,
\vec {e}_0)\,=\, {3 \cdot m\, +\, 3\cdot n} \,=\, {d},
\end{eqnarray}
where $\vec {e}_0 =(1,1,1,1)$ again.
The ten corresponding invariant monomials are:
\begin{eqnarray}
\vec{\mu}_0^1   \,&=&\,(3,3,0,0) \Rightarrow  x^3   \cdot y^3                          \nonumber\\       
\vec{\mu}_0^2   \,&=&\,(2,2,1,0) \Rightarrow x^2   \cdot y^2  \cdot z                \nonumber\\   
\vec{\mu}_0^3   \,&=&\,(1,1,2,0) \Rightarrow  x     \cdot y^   \cdot z^2              \nonumber\\   
\vec{\mu}_0^4   \,&=&\,(0,0,3,0) \Rightarrow                         z^3              \nonumber\\   
\vec{\mu}_0^5   \,&=&\,(2,2,0,1) \Rightarrow  x^2   \cdot y^2  \cdot u,  
\nonumber\\
\vec{\mu}_0^6   \,&=&\,(1,1,1,1) \Rightarrow  x     \cdot y    \cdot z \cdot u,
 \nonumber\\  
\vec{\mu}_0^7   \,&=&\,(0,0,2,1) \Rightarrow                   z^2    \cdot u,
  \nonumber\\  
\vec{\mu}_0^8   \,&=&\,(1,1,0,2) \Rightarrow  x     \cdot y   \cdot u^2,
\nonumber\\
\vec{\mu}_0^9   \,&=&\,(0,0,1,2) \Rightarrow           z      \cdot u^2,
\nonumber\\ 
\vec{\mu}_0^{10}\,&=&\,(0,0,0,3) \Rightarrow          u^3.                    
\end{eqnarray}
For the vector ${\vec {k}_4=(1,1,2,2)}$, one can consider the basis  
\begin{eqnarray} 
\vec{e}_1\,&=&\, (-3,3,0,0)\nonumber\\
\vec{e}_2\,&=&\, (-1,-1,1,0)\nonumber\\
\vec{e}_3\,&=&\, (-1,-1,0,1)
\end{eqnarray}
with determinant {18}, in which the dual pair of polyhedra:

\begin{eqnarray}
1_L  \,+\,  {10}_{J}  \,+\,1_R  \,=\,12, \nonumber\\
4_L^*\,+\,  { 4}_{J}^*\,+\,4_R^*\,=\,12^*.
\end{eqnarray}
both contain 12 points and 12 mirror points, respectively.

\normalsize

\subsection{Example of a Chain with $\Pi = 5$ and Eldest Vector
$\vec{k}_4=(7,8,10,25)$}

Now we present in Table~\ref{TabPI3}
a projective chain with $\Pi = 5$, constructed from the 
invariant direction $\vec{\pi}_8^{(1)}$ with the  invariant monomials
$(0,0,0,2)+ (2,2,2,0)$.
The {14} projective vectors of this chain are represented as 
linear combinations with positive-integer coefficients, 
${M,N,L,Q}$: ${Q=2,1}$ of the following three vectors:
 Projecting on 
the perpendicular plane gives us planar reflexive polyhedra,
so the  third basis vector  
\begin{equation}
\vec {e}_3=(-1,-1,-1,1)\,\,\,\Rightarrow \,\,\, (0,0,0,2).
\end{equation}
is  common to all the chains discussed in this Subsection.

\begin{table}[!ht]
\centering
\caption{\it The $K3$ hypersurfaces
in the ${\vec{\pi}_8^{(1)}}$ chain with projection $\Pi=5$
related to the  {IX  } with ${J=5}$:
 $ {Q \cdot \vec {k}}$ $= (N+L,M+L,M+N,2M+2N+2L)$ $=
M \cdot (0,1,1,2)$ $+ N \cdot (1,0,1,2)$ $+ L \cdot (1,1,0,2)$
with ${d=4M+4N+4L}$, ${\vec {k}_{eld}=(1,1,1,3)}$,
${\vec {k}_{young}=(7,8,10,25)}$, ${Q=2 \, or \,1}$.}
\label{TabPI3}
\scriptsize
\vspace{.05in}
\begin{tabular}{|c||c|c|c|c|c|c||c||c||c|} 
\hline
${\aleph}       $&$ \vec{k}_4       $&$ {[det]}             
$&${M}          $&${N}          $&${L}    $&${Q}           
$&$(\Delta,\Delta ^*)
$&$ \Pi - J^*       $&${chain}       $\\ 
\hline   \hline
${94}            $&$(7,8,10,25)      $&$ [50]            
$&${11}         $&$ {9}         $&${5}    $&$  {2}    
$&$  (6,39^*)
$&$ {5 -9}      $&$ -                $\\
${93 }           $&$ (8,7,5,20)      $&$ [40]  
$&${2}          $&$ {3}         $&${5}    $&$  {1}        
$&$  (8,28^*)            
$&$ {5-9}       $&$  -              $\\
${91 }           $&$ (5,6,8,19)      $&$ [38]  
$&${9}          $&$ {7}         $&${3}    $&$  {2}        
$&$  (7,35^*)            
$&$ {5-9}       $&$  IX              $\\
${88}            $&$ (7,6,4,17)      $&$ [34]  
$&${9}          $&$ {5}         $&${3}    $&$  {2}        
$&$  (8,31^*)            
$&$ {5-9}       $&$ IX              $\\
${84 }           $&$ (4,5,7,16)      $&$ [32]  
$&${4}          $&$ {3}         $&${1}    $&$  {1}        
$&$  (9,27^*)            
$&$ {5-9}       $&$  IX              $\\
${82 }           $&$ (4,5,6,15)      $&$ [30]  
$&${7}          $&$ {5}         $&${3}    $&$  {2}        
$&$  (10,20^*)            
$&$ {5-9}       $&$ IX               $\\
${69 }           $&$ (3,4,5,12)      $&$ [48]  
$&${3}          $&$ {2}         $&${1}    $&$  {1}        
$&$  (7,35^*)            
$&$ {5-9}       $&$ IX               $\\
${64 }           $&$ (2,6,7,15)      $&$ [30]  
$&${6}          $&$ {3}         $&${1}    $&$  {2}        
$&$ (13,23^*)             
$&$ {5-9}       $&$ IX               $\\
${60 }           $&$ (2,5,6,13)      $&$ [26]  
$&${9}          $&$ {3}         $&${1}    $&$  {2}        
$&$ (13,23^*)             
$&$ {5-9}       $&$ IX               $\\
${58 }           $&$ (2,4,5,11)      $&$ [22]  
$&${7}          $&$ {3}         $&${1}    $&$  {2}        
$&$ (14,19^*)             
$&$ {5-9}       $&$ IX               $\\
${47 }           $&$ (2,3,4,9)      $&$ [36]  
$&${5}          $&$ {3}         $&${1}    $&$  {2}        
$&$ (9,27^*)             
$&$ {5-9}       $&$ IX               $\\
${43 }          $&$ (2,2,3,7)      $&$ [14]  
$&${3}          $&$ {3}         $&${1}    $&$  {2}        
$&$ (19,11^*)           
$&$ {5-9}       $&$ IX               $\\
${11 }           $&$ (1,2,2,5)      $&$ [20]  
$&${3}          $&$ {1}         $&${1}    $&$  {2}        
$&$ (15,15^*)             
$&$ {5-9}       $&$ IX               $\\
${3 }           $&$ (1,1,1,3)      $&$ [12]  
$&${1}          $&$ {1}         $&${1}    $&$  {2}        
$&$ (39,6^*)            
$&$ {5-9}       $&$ IX               $\\
\hline
\end{tabular}
\end{table}
\normalsize
The vectors of this chain are represented as linear combinations
with positive-integer coefficients, ${M,N,L,Q}$: ${Q=2,1}$
of the following three vectors:

\begin{eqnarray}
 {Q \cdot \vec {k}_4(\vec{\pi}_8^{(1)})} & =&
 {M \cdot (0,1,1,1) + N \cdot (1,0,1,2) + L \cdot (1,1,0,2)}\nonumber\\
&=&{(\,N\,+\,L,\,M\,+\,L,\,M\,+\,N,\,
2 \cdot M\,+\,2 \cdot N\,+\,2 \cdot L\,)}\nonumber\\
\end{eqnarray}
where the third basis vector,
\begin{equation}
\vec {e}_3=(-1,-1,-1,1)\,\,\,\Rightarrow \,\,\, (0,0,0,2).
\end{equation}
is  common to all the chain.

There can be constructed additional three chains, 
$\vec{\pi}_8^{(2,3,4)}$, with the same invariant direction,
(0,0,0,2)-(2,2,2,0),  and the same youngest vector, but with the
different triple intersections and therefore with the different
projective chains. Together one can find inside all of 
four projective  chains,
$\vec{\pi}_8^{(\alpha)}, \, \alpha=1,2,3,4$, 
a total of {33} projective vectors
(see Table~\ref{list95}).

It is interesting to note that the chain $\vec{\pi}_8^{(1)}$  has {11}
${\vec{k}_4}$ vectors   with  ${\Pi=5}$
in common with the ${IX_J}$ chain
where ${J=5}$, whose
structure
is obtained from the following two vectors:
\begin{eqnarray}
 {\vec {k}_4(IX)} & =&
{m \cdot (0,1,1,2) + n \cdot (2,1,0,3) } \nonumber\\
&=&{(\,2n,\,m\,+\,n,\,m,\,2m\,+\,3n\,)}\nonumber\\
&&{1\,\leq\, m\,\leq \,6,\,\,\,\,\,\,\,1\,\leq \,n\,\leq\,4.}\nonumber\\ 
\end{eqnarray}
The  {chain} $IX$ of ${\vec {k}_4}$ projective vectors
with the structure ${5_{J=\triangle} \leftrightarrow 9_{\Pi=\triangle}}$
is presented in Table~\ref{TabIX}.
\begin{table}[!ht]
\centering
\caption{\it  The $K3$ hypersurfaces in the chain
$IX$:  $\vec {k} = (2n,m+n,m,2m+3n)= m \cdot (0,1,1,2) + 
n \cdot (2,1,0,3)$ with ${d= 4m + 6n }$, ${m_{max}=6, n_{max}=3}$ and
${\vec {k}_{eld}=(1,2,2,5)[10]}$.}
\label{TabIX}
\scriptsize
\vspace{.05in}
\begin{tabular}{|c|c|c|c|c|c|} \hline
${ N} $&${\vec k}[dim]  
$&${\Delta}(J=\underline {5})       
$&${\Delta}^{*} (\Pi= {9}) 
$&$({\Delta}_{J}, {\Delta^*}_{\Pi}) $\\ 
\hline\hline
${11}$&${(2,2,1,5)[10]}        $&$ {28 =7_L + {5}_J+ 16_R}
$&$ 8^*= 3_L^*+ {4}_C^*+1_R^*  $&$(10')$ \\ \hline
${43}$&${(2,3,2,7)[14]}        $&$ {19 = 7_L +{5}_J+ 7_R}
$&${11^* = 1_L +{9}_J+ 1_R} 
$&$(5_{\Pi},9_{J})$ \\ 
${24}$&${(4,3,1,8)[16]}        $&$ {24=3_L +{5}_J+ 16_R}
$&${12^* =1_L +{5}_J+ 6_R}
$&$(5_{\Pi},9_{J})$ \\
${33}$&${(6,4,1,11)[22]}       $&$ {22=1_L +{5}_J+ 16_R}
$&${20^* =1_L +{5}_J+ 14_R}
$&$(7_{\Pi}\in 9_{\Pi}) $ \\
${47}$&${(2,4,3,9)[18]}        $&$ {16=7_L +{5}_J+ 4_R}
$&${14^*  =6_L +{7}_J+ 1_R}
$&$(7_{\Pi}\in 9_{\Pi})$ \\
${58}$&${(2,5,4,11)[22]}       $&$ {14=7_L +{5}_J+ 2_R}
$&${19^* =9_L +{9}_J+ 11_R}
$&$(5_{\Pi},9_{J})$ \\
${60}$&${(6,5,2,13)[26]}       $&$ {13=1_L +{5}_J+ 7_R}
$&${23^*=1_L +{9}_J+ 13_R} 
$&$(5_{\Pi},9_{J})$ \\
${69}$&${(4,5,3,12)[24]}       $&${12=3_L +{5}_J+ 4_R}
$&${18^*  =6_L +{7}_J+ 5_R}
$&$(7_{\Pi}\in 9_{\Pi})$ \\
${64}$&${(2,7,6,15)[30]}       $&$ {13=7_L +{5}_J+ 1_R}
$&${23^*=13_L +{9}_J+ 1_R}
$&$(5_{\Pi},9_{9})$ \\
${84}$&${(4,7,5,16)[32]}       $&$ {9=3_L +{5}_J+ 1_R}
$&${27^*=13_L +{9}_J+ 5_R} 
$&$(5_{\Pi},9_{J})$ \\
${88}$&${(6,7,4,17)[34]}       $&$ {8=1_L +{5}_J+ 2_R}
$&${31^* =9_L +{9}_J+ 13_R}
$&$(5_{\Pi},9_{J})$ \\
${91}$&${(6,8,5,19)[38]}       $&${ 7 =9_L +{5}_J+ 1_R}
$&${35^* =16_L +{7}_J+ 12_R} 
$&$(5_{\Pi},9_{J})$ \\
\hline
\end{tabular}
\end{table}
\normalsize
The lattice determinant and the basis are given
by the 
following expressions:
\begin{eqnarray} 
\vec{e}_1\,&=&\, (0,-m,m+n,0)\nonumber\\
\vec{e}_2\,&=&\, (-1,2,-2,0)\nonumber\\
\vec{e}_3\,&=&\, (-1,-1,-1,1),
\end{eqnarray}
and
\begin{eqnarray}
{det} (\vec {e}_1, \vec {e}_2, \vec {e}_3,
 \vec {e}_0)\,=\, {4\cdot m\, +\, 6\cdot n} \,=\, {d},
\end{eqnarray}
where $\vec {e}_0 =(1,1,1,1)$.

The possible values of ${m}$ and ${n}$ for this chain are
also  determined by the 
dimensions of the extended vectors,
$d(\vec {k}^{ex(i)})=6$ and $d(\vec {k}^{ex(j)})=4$,
with the additional constraint $n_{max}=3< dim (0,1,1,2)$
(see Table~\ref{TabIX}):
\begin{eqnarray}
p\cdot \vec{k}_4(IX)\,&=&\,m\cdot (0,1,1,2)\,
+\,n\cdot (2,0,1,3)\nonumber\\
p\,=\,1\,& \rightarrow &\,1\,\leq\,m\,\leq\,6;,\,\,1\,\leq\,n\,\leq \,3
\end{eqnarray} 
The {5} invariant monomials for this chain are the following:
\begin{eqnarray}
\vec{\mu}_0^1\,=\,(1,4,0,0) & \Rightarrow & 
x     \cdot y^4                             \nonumber\\
\vec{\mu}_0^2\,=\,(2,2,2,0) & \Rightarrow & 
x^2   \cdot y^2     \cdot z^2               \nonumber\\
\vec{\mu}_0^3\,=\,(3,0,4,0) & \Rightarrow & 
x^3                 \cdot z^4               \nonumber\\
\vec{\mu}_0^4\,=\,(1,1,1,1) & \Rightarrow & 
x     \cdot y       \cdot z     \cdot u     \nonumber\\  
\vec{\mu}_0^5\,=\,(0,0,0,2) & \Rightarrow & 
                                      u^2. 
\end{eqnarray}

\subsection{Example of a $J =\Pi=9$ Chain }

To see another aspect of mirror symmetry and duality,
we now consider the chain $VI$ with intersection 
 ${J(\Delta)=\Pi(\Delta)=9}$
and ${J(\Delta^*)=\Pi(\Delta^*)=5}$ shown in Table~\ref{TabII},
which is constructed from the extended vectors
${\vec {k}^{i}=(0,1,1,2)}$ and ${\vec {k}^{j}=(1,0,1,2)}$.
In this case, duality gives very simple connections
between the numbers of integer points in the dual polyhedron pair,
as seen in Table~\ref{TabVI}.

\begin{table}[!ht]
\centering
\caption{\it The $K3$ hypersurfaces in the
chain $VI$: $\vec {k}(VI) = (n,m,m+n,2m+2n)= m \cdot (0,1,1,2) + 
n \cdot (1,0,1,2)$.}
\label{TabVI}
\scriptsize
\vspace{.05in}
\begin{tabular}{|c|c|c|c|c|}\hline
$ {N}                              $&$  \vec{k}_4
$&$ {\Delta(J= {9})}                $&$ {\Delta^*(\Pi= {5})} 
$&${\Delta}_{\Pi}, {\Delta^*}_{J}   
$\\ \hline \hline
$ {6}                             $&$ (1,1,2,4)[8]
$&$ 35=13_L+ {9}_{J,\Pi}+13_R
$&$ 7^*=1_L^*+  {5}_{\Pi,J}^*+1_R^* 
$&$(9_{\Pi,J},5_{J,\Pi})$ \\ \hline
$ {15}  $&$ (2,1,3,6)[12] 
$&$  27=5_L+ {9}_{J,\Pi}+13_R
$&$  9^*=1_L^*+ {5}_{\Pi,J}^*+4_R^* 
$&$(9_{\Pi,J},5_{J,\Pi})$ \\
$ {24} $&$ (3,1,4, 8)[16] 
$&$  24=2_L+ {9}_{J,\Pi}+13_R
$&$  12^*=1_L^*+ {5}_{\Pi,J}^*+6_R
$&$(9_{\Pi,J},5_{J,\Pi})$ \\
$ {31} $&$ (4,1,5,10)[20] 
$&$  23=1_L+ {9}_{J,\Pi}+13_R
$&$  13^*=1_L^*+ {5}_{\Pi,J}^*+7_R 
$&$(9_{\Pi,J},5_{J,\Pi})$ \\
${ 24}  $&$ (3,2,5,10)[20] 
$&$  16=2_L+ {9}_{J,\Pi}+5_R
$&$  14^*=3_L^*+ {5}_{\Pi,J}^*+6_R^*  
$&$(9_{\Pi,J},5_{J,\Pi})$ \\
$ {71} $&$ (3,4,7,14)[28] 
$&$  12=2_L+ {9}_{J,\Pi}+1_R
$&$ 18^*=7_L^*+ {5}_{\Pi,J}^*+6_R^* 
$&$(9_{\Pi,J},5_{J,\Pi})$ \\
\hline\hline
\end{tabular}
\end{table}
\normalsize

The canonical basis for chain $VI$ is:
\begin{eqnarray} 
\vec{e}_1\,&=&\, (-m,n,0,0)\nonumber\\
\vec{e}_2\,&=&\, (-1,-1,1,0)\nonumber\\
\vec{e}_3\,&=&\, (-1,-1,-1,1)
\end{eqnarray}
with the following restriction on the determinant
\begin{eqnarray}
 det (\vec  {e}_1, \vec  {e}_2, \vec  {e}_3,
 \vec  {e}_0)\,=\,{ 4\cdot m\, +\, 4\cdot n}\,=\,  d,
\end{eqnarray}
where $\vec  {e}_0 =(1,1,1,1)$.

The possible values of  ${m}$ and  ${n}$ for this chain are
determined by the 
dimensions of the extended vectors, without any unexpected puzzles:
\begin{eqnarray}
p \cdot \vec{k}_4(VI)\,&=&\,m\cdot ({0,1,}1,2)\,
+\,n\cdot ( {1,0,}1,2)\nonumber\\
p\,=\,1\, & \rightarrow & \,1\,\leq\,m\,\leq\,4;\,\,\,\,1\,\leq\,n\,\leq\,4.
\end{eqnarray} 
and the following:
\begin{eqnarray}
\vec{\mu}_0^1\,=\,(4,4,0,0) & \Rightarrow & 
x^4   \cdot y^4                            \nonumber\\
\vec{\mu}_0^2\,=\,(3,3,1,0)& \Rightarrow & 
x^3   \cdot y^3    \cdot z                 \nonumber\\
\vec{\mu}_0^3\,=\,(2,2,2,0)& \Rightarrow & 
x^2   \cdot y^2    \cdot z^2               \nonumber\\
\vec{\mu}_0^4\,=\,(1,1,3,0)& \Rightarrow & 
x     \cdot y      \cdot z^3               \nonumber\\
\vec{\mu}_0^5\,=\,(0,0,4,0)& \Rightarrow & 
                         z^4               \nonumber\\ 
\vec{\mu}_0^6\,=\,(2,2,0,1)& \Rightarrow & 
x^2   \cdot y^2                \cdot u     \nonumber\\
\vec{\mu}_0^7\,=\,(1,1,1,1)& \Rightarrow & 
x     \cdot y      \cdot z     \cdot u     \nonumber\\
\vec{\mu}_0^8\,=\,(0,0,2,1)& \Rightarrow & 
                        z^2   \cdot u      \nonumber\\            
\vec{\mu}_0^9\,=\,(0,0,0,2)& \Rightarrow & 
                                    u^2. 
\end{eqnarray}
are the  {9} invariant monomials  ${ \Psi_{inv}}$
for this chain.

Analogously, one can consider the projective chain
$\vec{\pi}_7^{(\alpha)}$ ($\Pi=5$) with youngest vector (5,6,8,11), and
compare it with
the double-intersection chain $VIII$, constructed from the extended
vectors
 $\vec {k}(VIII) =  m \cdot (0,1,1,2) + 
n \cdot (1,1,2,0)$; $( d=4m + 4n$. $ m_{max}=3,n_{max}=4)$,
$ \vec {k}_{eld}=(1,2,3,2)[8]$.
Among the {95} ${K3}$ projective vectors, {26}
have such an invariant-direction structure, and therefore can be found in
corresponding projective chains (see Table \ref{list95}).

\section{$K3$ Hypersurfaces and  
{ Cartan-Lie Algebra} Graphs}

We discuss in this Section more details of the emergence of
Cartan-Lie algebra graphs in our construction of CY spaces.

\subsection{{Cartan-Lie} Algebra Graphs and the 
Classification of Chains of Projective Vectors}

\begin{figure}[th!]
   \begin{center}
   \mbox{
   \epsfig{figure=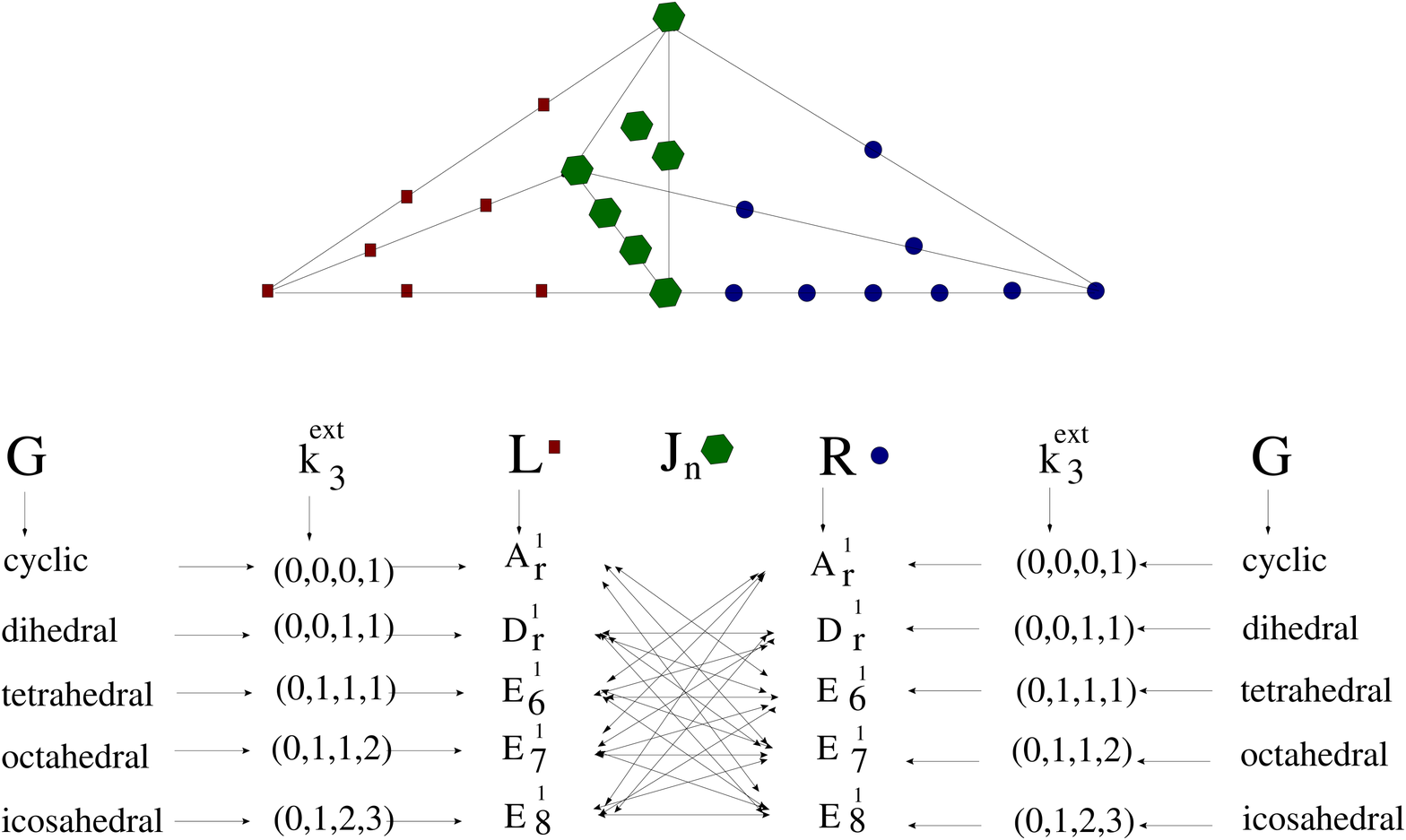,height=16cm,width=16cm}}
   \end{center}
   \caption{\it Illustration of the Cartan-Lie algebra diagram
classification of the 
${ 22 (=[13+1]+8^*)}$ chains of $K3$ polyhedra shown in Table~\ref{TabXL}.
Here ${G}$ denotes the cyclic, dihedral, tetrahedron, 
octahedron-cube, and icosahedron-dodecahedron subgroups of $SU(2)$,
${L/R}$ denote left/right integer points and ${ CLA_J}$ diagrams,
$ {J_n}$ the type of intersection, and the
${ \vec {k}_3}$ are all the possible planar vectors.
We find ${ {r_{max}}_L= 17}$ or  ${ {r_{max}}_R= 17}$ for 
the ${ A_r^{1}}$ series,  and ${{ r_{max}}_L= 16}$ or 
${ { r_{max}}_R= 16}$ for the ${ D_r^{1}}$ series.
In the example shown here, one can see the polyhedron with
the projective vector: ${ \vec{k}_4=(1,1,3,4)[9]}$.}
\label{cdjad}
\end{figure}

As we commented already in the Introduction and in Section 2, the
structure of the projective ${ \vec{k}_4}$ vectors in 
{ 22} chains leads to interesting relations with the five
classical regular dual polyhedron pairs in three-dimensional space:
the one-dimensional point, two-dimensional line segment
and three-dimensional tetrahedron, octahedron-cube and 
icosahedron-dodecahedron. There are also interesting
correspondences with
the Cartan-Lie algebra ${CLA}$ graphs for the
five types of groups in the ${ADE_{6,7,8}}$ series:
see Figure~\ref{cdjad}. The ${ CLA_{J,\Pi}}$ graphs, which can be 
seen in the polyhedra of the corresponding ${ \vec {k}_4}$ 
projective vectors, follow completely the structure of
the five possible extended vectors:
\begin{eqnarray}
\vec {k}_C^{ext}=(0,0,0,1)  & \leftrightarrow& A_r ;\nonumber\\
\vec {k}_D^{ext}=(0,0,1,1)  & \leftrightarrow& D_r ;\nonumber\\
\vec {k}_T^{ext}=(0,1,1,1)  & \leftrightarrow& E_6 ;\nonumber\\
\vec {k}_O^{ext}=(0,1,1,2)  & \leftrightarrow& E_7 ;\nonumber\\
\vec {k}_I^{ext}=(0,1,2,3)  & \leftrightarrow& E_8 .
\end{eqnarray}
We give in Table~\ref{TabXL} the {ADE} structures  and the
${ CD_J}$ diagrams  of all the eldest $K3$ projective 
vectors from the {22} double chains. An illustration is given in
Figure~\ref{cdjad}, and the rest of this Section discusses
the examples of chains $XV$ to $XIX$, illustrating the
power of our systematic approach.

 
\begin{table}[!ht]
\centering
\caption{\it The eldest vectors $(m = n = 1)$ for all the {22} chains
of $K3$ hypersurfaces,
and the corresponding Cartan-Lie algebra diagrams $CLA_J$.}
\label{TabXL}
\scriptsize
\vspace{.05in}
\begin{tabular}{|c|c|c|c||c||c|}
\hline
${ N} $&${ {\vec k}_i(eldest)} $&$ Structure$&$
max \,\Delta( J)$&${ CLA_J} 
$&$min \, \Delta^*( \Pi)$\\
\hline\hline
$I$&$(1,1,1,1)[4]
$&$ (0,1,1,1)_{e6}+(1,0,0,0)_{a}
$&$35={ 10_{e6}}+{10}_{J={\triangle}}+{ 15_a}
$&${  E_{6}^{(1)}\leftrightarrow  A_{12}^{(1)}}
$&$5^*=1_{e6}^*+3_C^*+1_a^*$ \\
$II $&$(1,1,2,2)[6]
$&$(0,1,1,1)_{e6}+(1,0,1,1)_{e6}
$&$ 30={ 10_{e6}}+ {10}_{J=\Pi={\triangle}}
+ { 10_{e6}}
$&${ E_6^{(1)} \leftrightarrow E_6^{(1)}}
$&$6^*={1}_{e6}^*+{4}_{\Pi=J={\triangle}}^*+1_{e6}^*$ \\
$III $&$(3,1,2,3)[9] $&$(0,1,1,1)_{e6}+(3,0,1,2)_{e8}
$&$23={ 4_{e6}}+ {4}_{J={\triangle}}+ { 15_{e8}}
$&$ { {G_2}^{(1)}\leftrightarrow E_8^{(1)}}
$&$8^*=3_{e6}^*+ {4}_{J={\triangle}}^*+1_{e8}^*$ \\
\hline\hline
$IV  $&$(1,1,1,2)[5] $&$(0,1,1,2)_{e7}+(1,0,0,0)_{a}
$&$34={ 13_{e7}}+ {9}_{J={\triangle}}+ { 12_{a}}
$&$   {  E_7^{(1)}\leftrightarrow A_{9}^{(1)}}
$&$6^*=1_{e6}^*+ {4}_C^*+1_a^*$ \\
$V $&$(1,1,1,3)[6] $&$(0,1,1,2)_{e7}+(1,0,0,1)_{d}
$&$39={ 13_{e7}}+{9}_{J={\triangle}}+{ 17_{d}}
$&${ E_7^{(1)}\leftrightarrow D^{(1)}_{10}}
$&$6^*=1_{e7}^*+4_C^*+1_{d}$ \\
$VI $&$(1,1,2,4)[8]  $&$(0,1,1,2)_{e7}+(1,0,1,2)_{e7}
$&$ 35={ 13_{e7}}+ {9}_{J=\Pi={\triangle}}+{ 13_{e7}}
$&${ E_7^{(1)}\leftrightarrow  E_7^{(1)}}
$&$  7^*=1_{e7}^*+  {5}_{\Pi={\triangle}}^*+1_{e7}^*$ \\
$VII $&$(1,1,1,1)[4] $&$(0,1,1,2)_{e7}+(2,1,1,0)_{e7}
$&$ 35={ 13_{e7}}+ {9}_{J={\triangle}}+{ 13_{e7}}
$&${ E_7^{(1)}\leftrightarrow E_7^{(1)}}
$&$ 5^*=1_L^*+3_C^*+1_R^*$ \\
$VIII  $&$ (1,2,3,2)[8] $&$(0,1,1,2)_{e7}+(1,1,2,0)_{e7} 
$&$ 24={ 12_{e7}} + {5}_{J={\triangle}}+{ 7_{e7}}
$&${ E_7^{(1)}\leftrightarrow F_4^{(1)}}
$&$  8^*=3_{e7}^*+4_C^*+1_{e7}^*$\\
$IX  $&$ (2,2,1,5)[10] $&$(0,1,1,2)_{e7}+(2,1,0,3)_{e8} 
$&$ 28={ 7_{e7}}+  {5}_{J={\triangle}} +{ 16_{e8}}
$&$ {  F_4^{(1)}\leftrightarrow   E_8^{(1)}}
$&$  8^*=3_{e7}^*+4_C^*+1_{e8}^*$\\
\hline\hline 
$X   $&$(1,1,1,1)[4]  $&$(0,0,1,1)_{d}+(1,1,0,0)_{d} 
$&$ 35={ 13_d}+ {9}_{J={\Box}}+{ 13_{d}}
$&${ D_8^{(1)}\leftrightarrow D_8^{(1)}}
$&$  5_{\Pi={\diamond}}^*=1_d^*+3_C^*+1_d^*$\\
$XI    $&$(1,1,1,2)[5] $&$(0,0,1,1)_d + (1,1,0,1)_{e6} 
$&$ 34={ 15_d}+ {9}_{J={\Box}}+{ 10_{e6}}
$&$ { D_8^{(1)}\leftrightarrow E_6^{(1)} }
$&$  6^*=1_d^*+4_C^*+1_{e6}^*$\\
$XII  $&$(1,1,1,1)[4] $&$(0,1,2,3)_{e8}+(3,2,1,0)_{e8} 
$&$ 13={ 4_{e8}}+ {5}_{J=\Pi={\Box}}+{ 4_{e8}}
$&$ { G_2^{(1)}\leftrightarrow G_2^{(1)} }
$&$ 11^*=1_{e8}^*+9_{\Pi=J={\Box}}^*+1_{e8}^*$\\
$XIII  $&$(1,1,2,3)[7] $&$(0,1,1,1)_{e6}+(1,0,1,2)_{e7} 
$&$ 31={ 10_{e6}}+ {8}_{J=\Pi={\Box}}+{ 13_{e7}}
$&${ E_6^{(1)}\leftrightarrow E_7^{(1)}}
$&$  8^*=1_{e6}^*+6_{\Pi^,J}^*+1_{e7}^*$\\
$XIV  $&$(1,1,1,2)[5]  $&$(0,1,1,2)_{e6}+(2,1,3,0)_{e8} 
$&$ 18={ 7_{e6}}+ {6}_{J=\Pi={\Box}}+{ 5_{e8}}
$&$ { F_4^{(1)}\leftrightarrow G_2^{(1)} }
$&$ 10^*=1_{e6}^*+8_{\Pi=J={\Box}}^*+1_{e8}^*$\\
$XXII $&$ (1,2,1,2)[6]  $&$(0,1,1,2)_{e7}+(1,1,0,0)_{d} 
$&$ 30={ 13_{e7}}+ {7}_{J=\Pi={\Box}}+{ 10_{d}}
$&${ E_7^{(1)}\leftrightarrow  {D_7} }
$&$ 6^*=1_{e7}^*+4_{J}^*+1_{d}^*$\\
\hline\hline
$XV   $&$ (1,1,2,3)[7] $&$(0,1,2,3)_{e8}+ (1,0,0,0)_{a}
$&$ 31={ 16_{e8}}+ {7}_{J={\triangle}}+{ 8_a}
$&$ { E_8^{(1)}\leftrightarrow A_6^{(1)} }
$&$  8^*=1_{e8}^*+6_{C}^*+1_a^*$\\
$XVI  $&$ (1,1,2,4)[8] $&$(0,1,2,3)_{e8}+(1,0,0,1)_{d} 
$&$ 35={ 16_{e8}}+ {7}_{J={\triangle}}+{ 12_{d}}
$&$ { E_8^{(1)}\leftrightarrow D_8 }
$&$  7^*=1_{e8}^*+5_{C}^*+1_{e8}^*$\\
$XVII     $&$(1,1,3,4)[9]  $&$ (0,1,2,3)_{e8}+(1,0,1,1)_{e6} 
$&$ 33={ 16_{e8}}+ {7}_{J=\Pi={\triangle}}+{ 10_{e6}}
$&${ E_8^{(1)}\leftrightarrow  E_6^{(1)} }
$&$  9^*=1_{e8}^*+7_{\Pi=J={\triangle}}^*+1_{e6}^*$\\
$XVIII   $&$(1,1,3,5)[10]  $&$ (0,1,2,3)_{e8}+(1,0,1,2)_{e7} 
$&$ 36={ 16_{e8}}+ {7}_{J=\Pi={\triangle}}+{ 13_{e7}}
$&$ { E_8^{(1)}\leftrightarrow  E_7^{(1)} }
$&$ 9^*=1_{e8}^*+7_{\Pi=J={\triangle}}^*+1_{e7}^*$\\
$XIX   $&$ (1,1,4,6)[12]  $&$ (0,1,2,3)_{e8}+(1,0,2,3)_{e8} 
$&$ 39={ 16_{e8}}+ {7}_{J=\Pi={\triangle}}+{ 16_{e8}}
$&$ { E_8^{(1)}\leftrightarrow  E_8^{(1)} }
$&$ 9^*=1_{e8}^*+7_{\Pi=J={\triangle}}^*+1_{e8}^*$\\
$XX      $&$ (1,1,1,3)[6]    $&$(0,1,2,3)_{e8}+(2,1,0,3)_{e8} 
$&$ 21={ 7_{e8}}+ {7}_{J=\Pi={\triangle}}+{ 7_{e8}}
$&$ { {F_4^{(1)}}\leftrightarrow  {F_4^{(1)}} }
$&$ 9^*=1_{e8}^*+7_{\Pi=J={\triangle}}^*+1_{e8}^*$\\
$XXI     $&$ (3,2,4,3)[12] $&$(0,1,2,3)_{e8}+(3,1,2,0)_{e8} 
$&$ 15={ 4_{e8}}+ {7}_{J=\Pi={\triangle}}+{ 4_{e8}}
$&$ { {G_2^{(1)}} \leftrightarrow  {G_2^{(1)}} }
$&$ 9^*=1_{e8}^*+7_{\Pi=J={\triangle}}^*+1_{e8}^*$\\
\hline
\end{tabular}
\end{table}

\newpage 

\subsection{The  ${K3}$ Chain $XV$ with Graphs in the
${E_8^{(1)}-A_r^{(1)}}$ Series}

Here we give the list of ${\vec {k}_4}$ vectors which can be constructed
from the Weierstrass vectors ${\vec {k}_3 \equiv (1,2,3)}$ and  
${\vec {k}_1=(1)}$, shown as chain $XV$ in Table~\ref{TabXV}.
The number of ${\vec {k}_4}$ vectors in this  chain  is determined by 
the positive-integer numbers: ${m = 1, n  \leq 6}$,
 according to the dimensions of the corresponding component
${\vec {k}^{i}}$.
  
\begin{table}[!ht]
\centering
\caption{\it The $K3$ hypersurfaces in the
chain $XV$: $\vec {k} = (m,2 \cdot m,3 \cdot m,n)= 
m \cdot (1,2,3,0) + n \cdot (0,0,0,1)$:
${d=6m+n}$,  ${m_{max}=1, n_{max}=6}$,
${\vec {k}_{eld}=(1,2,3,1)[7]}$.}
\label{TabXV}
\scriptsize
\vspace{.05in}
\begin{tabular}{|c|c|c|c|c|c|}
\hline
${\aleph} $&$ m,n $&$ {\vec k}[d]  $&$ 
\Delta(J=7) $&${Group} $&$\Delta^*(\Pi=7) $\\
\hline\hline
${5} $&$ 1,1 $&${(1,2,3,1)[7]}   $&$31=8_L+7_{J}+16_R
$&$ {A_6^{(1)}}_L                             $&$8^*=1_L^*+6_C^*+1_R^* 
$ \\ \hline
${10} $&$ 1,2 $&${(1,2,3,2)[8]}  $&$24=10_L+7_{J}+7_R
$&${A_7^{(1)}}_L                              $&$8^*=3_L^*+4_C^*+1_R^* 
$ \\
${12}  $&$ 1,3 $&${(1,2,3,3)[9]}  $&$23=12_L+7_{J}+4_R
$&$ {A_8^{(1)}}_L                             $&$8^*=4_L^*+3_C^*+1_R^*  
$ \\
${13} $&$ 1,4 $&${(1,2,3,4)[10]} $&$23=14_L+7_{J}+2_R
$&${A_9^{(1)}}_L                              $&$11^*=3_L^*+3_C^*+1_R^* 
$ \\
${14} $&$ 1,5 $&${(1,2,3,5)[11]} $&$24=16_L+7_{J}+1_R
$&${A_{10}^{(1)}}_L                           $&$13^*=9_L^*+3_C^*+1_R^*  
$ \\
${15} $&$ 1,6 $&${(1,2,3,6)[12]} $&$27=19_L+7_{J}+1_R
$&${A_{11}^{(1)}}_L                           $&$9^*=5_L^*+3_C^*+1_R^* 
$ \\
\hline
\end{tabular}
\end{table}
\normalsize

\begin{figure}[h]
   \begin{center}
   \mbox{
   \epsfig{figure=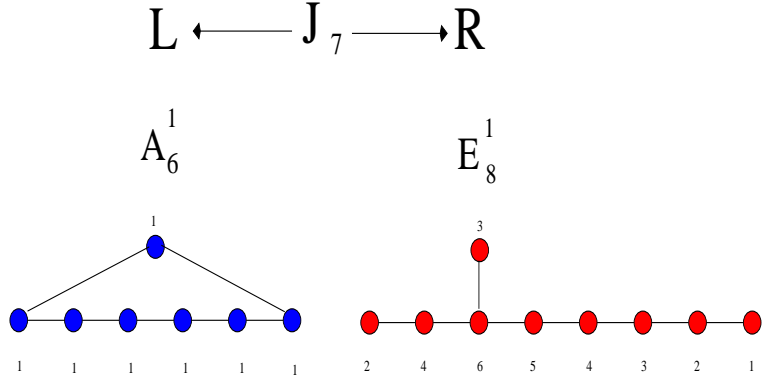,height=5cm,width=10cm}}
   \end{center}
   \caption{\it The $ {A_6^{(1)}}_L-{E_8^{(1)}}_R$   
graph from the eldest $(1,2,3,1)[7]$ polyhedron in chain $XV$:
${31=8_L+7_J+16_R}$.}
\label{XVe8xa6}
\end{figure}

The basis for this chain, 
see Figure~\ref{XVe8xa6}, can be written 
in the the following form:
\begin{eqnarray}
\vec{e}_1\,&=&\,(-n,0,0,m)\nonumber\\
\vec{e}_2\,&=&\, (-2,1,0.0)\nonumber\\
\vec{e}_3\,&=&\,(-3,0,1,0)
\end{eqnarray}
The determinant of this canonical basis coincides, of course, with the 
dimensions of the ${\vec {k}_4}$ vectors: 

\begin{eqnarray}
{det} (\vec {e}_1, \vec {e}_2, \vec {e}_3,
 \vec {e}_0)\,=\, {6\cdot m\, +\, 1\cdot n}\,=\,{d},
\end{eqnarray}
where $\vec {e}_0 =(1,1,1,1)$.
The decomposition of this  chain is again determined by the 
dimension of the extended vectors 
${d(\vec {k}^{ex(i)})=k_1^{ex(i)}+k_2^{ex(i)}+ 
k_3^{ex(i)}+ k_4^{ex(i)}}$, as seen in Table~\ref{TabXV}:
\begin{eqnarray}
\vec{k}_4(XV)\,=\,m\cdot 
( {1},2,3, {0})\,
&+&\,n\cdot ( {0},0,0, {1})\nonumber\\
m\,=\,1\,\,&\,\,\,&\,\,1\,\leq\,n\,\leq\,6.
\end{eqnarray} 
The seven invariant monomials corresponding to this chain are:
\begin{eqnarray}
\vec{\mu}_0^1\,=\, (6,0,0,1) &\Rightarrow & 
x^6                           \cdot u         \nonumber\\
\vec{\mu}_0^2\,=\, (4,1,0,1) &\Rightarrow & 
x^4  \cdot y                  \cdot u         \nonumber\\
\vec{\mu}_0^3\,=\, (2,2,0,1) &\Rightarrow & 
x^2  \cdot y^2                \cdot u         \nonumber\\
\vec{\mu}_0^4\,=\, (0,3,0,1) &\Rightarrow & 
           y^3                \cdot u         \nonumber\\
\vec{\mu}_0^5\,=\, (3,0,1,1) &\Rightarrow & 
 x^3    \cdot z                \cdot u         \nonumber\\
\vec{\mu}_0^6\,=\, (1,1,1,1) &\Rightarrow & 
x    \cdot y    \cdot z       \cdot u          \nonumber\\
\vec{\mu}_0^7\,=\, (0,0,2,1) &\Rightarrow & 
                      z^2     \cdot u.
\end{eqnarray}
Considering the dual pairs 
for these vectors, one can see that the singularities of the eldest
vector 
${\vec {k}_4=(1,2,3,1)}$ correspond to some  
graphs of the ${{A_6^{(1)}}_L-{E_8^{(1)}}_R}$ series,
as seen in Figure~\ref{XVe8xa6}. 
For instance, if one looks at the integer points in the edges 
of the polyhedron on the left (right) side of the intersection by the 
hyperplane ${\vec{k}^{i}= (0,1,2,3)}$, one sees 
graphs with $ {A_{6}^{(1)}}_L$ and ${E_8^{(1)}}_R$ Lie algebras. 
Going to the last 
minimal ${\vec {k}=(1,2,3,6)}$ of this chain, we find that the right 
graph degenerates and left points reproduce
${A_{11}^{(1)}}$
with the maximum possible rank in this chain. Thus, the {six} 
${\vec {k}}$ vectors in this chain produce the following 
graphs 
in the $A$ series: ${A_{6}^{(1)}, A_{7}^{(1)}, A_{8}^{(1)},
 A_{9}^{(1)}, A_{10}^{(1)}, A_{11}^{(1)}  }$.   

\subsection{The $K3$ Chain $XVI$ with Graphs
in the ${E_8^{(1)}-D_r}$ Series}

\begin{table}[!ht]
\centering
\caption{\it The $K3$ hypersurfaces in the
chain $XVI$: $\vec {k} = (n,m,2m,3m+n)= m \cdot (0,1,2,3) + 
n \cdot (1,0,0,1)$: $({d=6m + 2n})$, ${m_{max}=2,n_{max}=6}$,
${\vec {k}_{eld}=(1,1,2,4)[8]}$.}
\label{TabXVI}
\scriptsize
\vspace{.05in}
\begin{tabular} {|c|c|c|c|c|c|} \hline
${\aleph} $&$ {\vec k}[d]
$&$ \Delta(J=7)             $&${Group}      
$&$\Delta^*(\Pi=7)          $&$(\Pi(\Delta), J(\Delta^*))$\\ \hline\hline
${6} $&$ {(1,1,2,4)[8] } 
$&$ 35=16_L+7_J+12_R        $&$ D_{8R} 
$&$ 7^*=1_L^*+5_C^*+1_R^*   $&$(9,5^*) $\\ \hline
${11} $&$ {(2,1,2,5)[10]} 
$&$ 28=7_L+7_J+14_R         $&$D_{9R}
$&$ 8^*=1_L^*+4_C^*+3_R^*   $&$ (10,4^*)  $\\
${15} $&$ {(3,1,2,6)[12]}
$&$ 27=4_J+7_J+16_R         $&$D_{10R} 
$&$ 9^*=1_L^*+4_C^*+4_R^*   $&$ (9 ,5^*)  $\\
${17 }$&$ {(4,1,2,7)[14]}
$&$ 27=2_L+7_J+18_R         $&$D_{11R}
$&$12^*=1_L^*+4_C^*+7_R^*   $&$ (7 , 7^*)  $\\
${ 19} $&$ {(5,1,2,8)[16]}
$&$ 28=1_L+7_J+14_R         $&$D_{12R}
$&$14^*=1_L^*+4_C^*+9_R^*   $&$ (7 , 7^*) $\\
${20} $&$ {(6,1,2,9)[18]} 
$&$ 30=1_L+7_J+22_R         $&$D_{13R}
$&$12^*=1_L^*+4_C^*+7_R^*   $&$ (7 , 7^*)  $\\
${ 47} $&$ {(3,2,4,9)[18]} 
$&$16=4_L+7_J+5_R           $&$---
$&$14^*=3_L^*+5_C^*+6_R^*   $&$(7 , 7^*) $\\
${ 58} $&$ {(5,2,4,11)[22]} 
$&$ 14=1_L+7_J+6_R          $&$--- 
$&$19^*=3_L^*+4_C^*+12_R^*  $&$ (5 , 9^*)  $\\
\hline
\end{tabular}
\end{table}
\normalsize

\begin{figure}[h]
   \begin{center}
   \mbox{
   \epsfig{figure=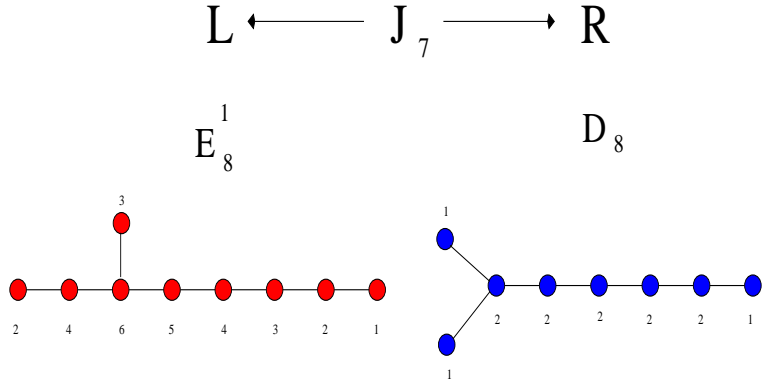,height=5cm,width=10cm}}
   \end{center}
   \caption{\it The $ {E_8^{(1)}}_L-D_{8R}$  
graph from the eldest $(1,1,2,4))[8]$ polyhedron
in the chain $XVI$:
${33 = 16_L+7_J+10_R}$.} 
\label{XVIe8xd8}
\end{figure}

The basis for the chain shown in Table~\ref{TabXVI} is
\begin{eqnarray} 
\vec{e}_1\,&=&\, (-m,n,0,0)\nonumber\\
\vec{e}_2\,&=&\, (0,-2,1,0)\nonumber\\
\vec{e}_3\,&=&\, (-1,-1,-1,1),
\end{eqnarray}
with
\begin{eqnarray}
{det} (\vec {e}_1, \vec {e}_2, \vec { e}_3,
 \vec {e}_0)\,=\, {6\cdot m\, +\, 2\cdot n}
 \,=\,{ d},
\end{eqnarray}
where $\vec {e}_0 =(1,1,1,1)$ again.
The decomposition of this  chain is completely determined by the 
dimensions of the vectors shown in Table~\ref{TabXVI}:

\begin{eqnarray}
p \cdot \vec{k}_4(XVI)\,&=&\,m\cdot ( {0,1,}2,3)\,
+\,n\cdot ( {1,0,}0,1)\nonumber\\
p\,=\,1^{*}\, &\rightarrow &
\,1\,\leq\,m\,\leq\,2;\,\,1\,\leq\,n\,\leq\,6\nonumber\\
p\,=\,2\, &\rightarrow &\, m\,=\,n\,=2.
\end{eqnarray} 
The seven invariant monomials corresponding to this chain 
are the following:

\begin{eqnarray}
\vec{\mu}_0^1\,=\,(2,6,0,0) & \Rightarrow & 
x^2 \cdot y^6                     \nonumber\\
\vec{\mu}_0^2\,=\,(2,4,1,0) & \Rightarrow & 
x^2 \cdot y^4 \cdot z             \nonumber\\
\vec{\mu}_0^3\,=\,(2,2,2,0) & \Rightarrow & 
x^2 \cdot y^2 \cdot z^2           \nonumber\\
\vec{\mu}_0^4\,=\,(2,0,3,0) & \Rightarrow & 
x^2 \cdot z^3                     \nonumber\\
\vec{\mu}_0^5\,=\,(1,3,0,1) & \Rightarrow & 
x \cdot y^3 \cdot u               \nonumber\\
\vec{\mu}_0^6\,=\,(1,1,1,1) & \Rightarrow & 
x \cdot y  \cdot z  \cdot u       \nonumber\\
\vec{\mu}_0^7\,=\,(0,0,0,2) & \Rightarrow & 
u^2.
\end{eqnarray}
The example of the $ {E_8^{(1)}}_L-D_{8R}$
graph associated with the eldest $(1,1,2,4))[8]$ polyhedron in
Table~\ref{TabXVI}
is shown in Figure~\ref{XVIe8xd8}.

\subsection{The ${J=\Pi}$ Symmetric Chain $XVII$
with Exceptional Graph $E_6 \times E_8$}

\begin{figure}[h]
   \begin{center}
   \mbox{
   \epsfig{figure=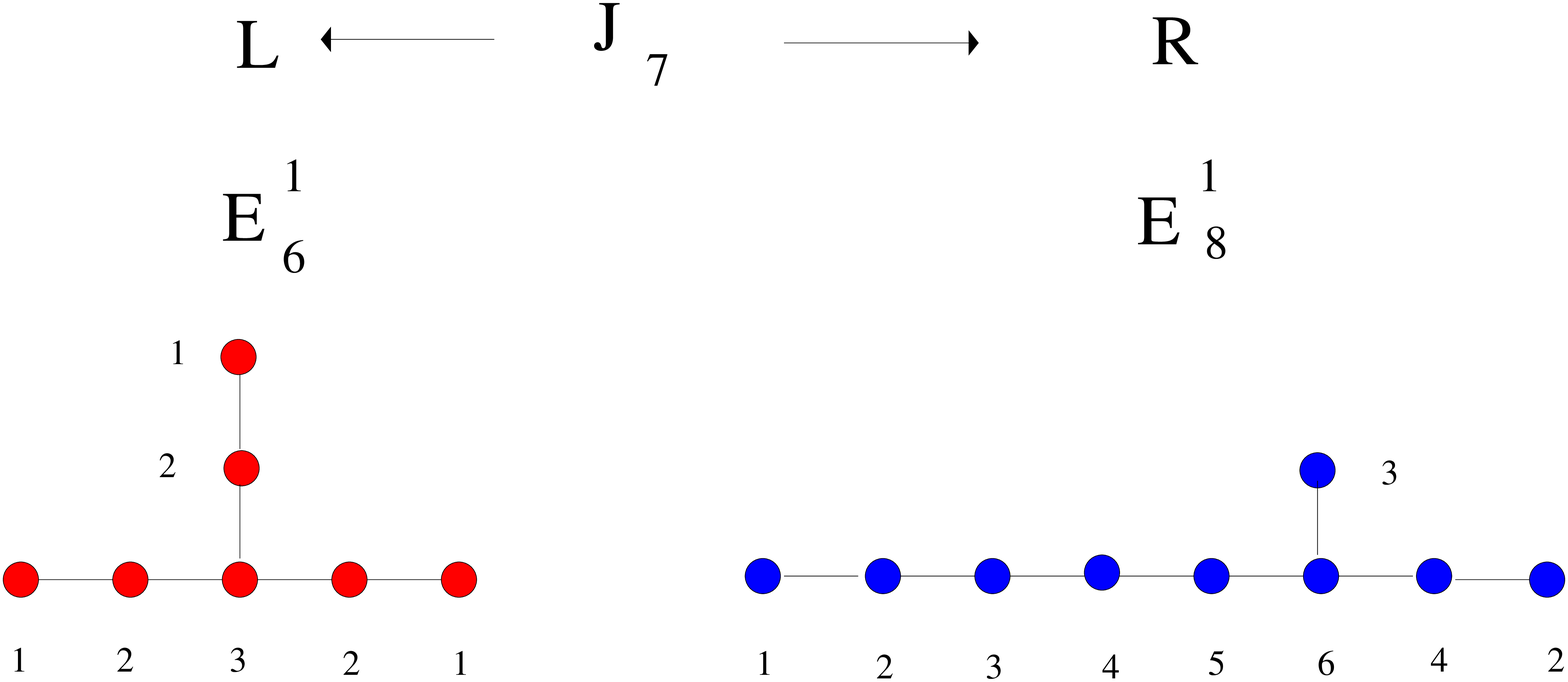,height=5cm,width=10cm}}
   \end{center}
   \caption{\it The $ {E_6^{(1)}}_L-{E_{8}^{(1)}}_R$ 
graph associated with the eldest $(1,1,3,4))[9]$ polyhedron in chain XVII:
${33=10_L+7_J+16_R}$.}
\label{XVIIe6xe81}
\end{figure}

We show in Table~\ref{TabXVII} the projective vectors
constructed from ${\vec{k}_3^{ex}= (0,1,1,1)}$ and
 $\vec{k}_3^{ex}= (1,0,2,3)$. In this case, the number of points
in the maximal polyhedron with ${m=n=1}$ can easily be calculated:
${33 = (10)_L +(7)_{int} + (16)_R}$.
The `right' ${ 15_R +1_R}$ points form the 
graph for the affine ${E_8^{(1)}}$ Lie algebra, as shown in
Figure~\ref{XVIIe6xe81}:

\begin{eqnarray}
{6}\,=\,1\,+\,1\,+\,1\,+\,1\,+1\,+1\, & \Rightarrow &\,
\{(P_{x_0})_1\,+\,(P_{x_1})_2\,+\,(P_{x_2})_3\,+\,(P_{x_3})_4\,
+\,(P_{x_4})_5\,+\,(P_{x_5})_6 \}                \nonumber\\
{3}\,=\,3\,                           & \Rightarrow &\, 
\{(P_{x_6,x_6^{'},x_6^{''}})_3 \}                \nonumber\\
{6}\,=\,4\,+\,2\,                     & \Rightarrow &\,
\{(P_{x_7,x_7^{'},x_7^{''},x_7^{'''}})_4\,+\,(P_{x_8,x_8^{'}})_2 \}.
\end{eqnarray} 
The `left' points in this polyhedron,
${9_L+1_L}$, correspond to the ${ E_6^{(1)}}$
affine series with the Coxeter numbers:

\begin{eqnarray}
{ 3}\,=\,1\,+\,1\,+\,1\,                   & \Rightarrow &\,
\{(P_{x_1})_1\,+\,(P_{x_2})_2\,+\,(P_{x_3})_3 \}     \nonumber\\
{ 3}\,=\,2\,+\,1\,                         & \Rightarrow &\,
\{(P_{x_4,x_4^{'}})_2\,+\,(P_{x_0})_1 \}     \nonumber\\
{ 3}\,=\,2\,+\,1\,                         & \Rightarrow &\,
\{(P_{x_5,x_5^{'}})_2\,+\,(P_{x_6})_1 \}.   
\end{eqnarray} 
For
${ m_{max}=d(\vec{k}(1,2,3))=6}$ and ${ n_{min}=1}$, the 
corresponding polyhedron 
contains {18} points: ${ 18 = (10)_L + (7)_{int} + (1)_R }$.
Conversely, for ${ m_{min}=1}$ and ${n_{max} = 3 = dim(\vec {k}(1,1,1))}$,
the self-dual vector ${ \vec {k}=(3,1,7,10)}$  has {24} integer points:
${ 24 = (1)_L + (7)_{int} + (16)_R}$. Finally, the  polyhedron
with ${m=5}$ and ${n=3}$ contains the minimal possible number
of integer points, namely ${ 9 = (1)_L + (7)_{int} + (1)_R}$.
This minimal vector ${(3,5,11,14)[33]}$ is the dual conjugate of the 
vector ${ \vec{k}=(1,1,4,6)[12]}$.
 
\begin{table}[!ht]
\centering
\caption{\it The $K3$ hypersurfaces in the
chain $XVII$: $\vec {k} = (n,m,m+2n,m+3n)= m \cdot (0,1,1,1) + 
n \cdot (1,0,2,3)$: ${ d=3m + 6n}$, ${ max (m,n)=(6,3)}$.}
\label{TabXVII}
\scriptsize
\vspace{.05in}
\begin{tabular}{|c|c|c|c|} \hline
${ \aleph} $&$ {\vec k}[d] 
$&$ \Delta    
$&$ \Delta^*                            $\\ \hline\hline
${ 7} $&$ { (1,1,3,4)[9]} 
$&$   33=10_L  +7_{J=\Pi}  +16_R
$&$  9^*=1_L^* +7_{\Pi=J}^*+1_R^*         $ \\ \hline
${ 16} $&$ { (1,2,4,5)[12]} 
$&$   24=10_L  +7_{J=\Pi}  +7_R
$&$ 12^*=4_L^* +7_{\Pi=J}^*+1_R^*            $ \\
${ 25}  $&$ {(1,3,5,6)[15]} 
$&$   21=10_L  +7_{J=\Pi}  +4_R       
$&$ 15^*=7_L^* +7_{\Pi=J}^*+1_R^*        $ \\
${ 32 }$&$ {(1,4,6,7)[18]}
$&$   19=10_L  +7_{J=\Pi}  +2_R
$&$ 20^*=12_L^*+7_{\Pi=J}^*+1_R^*        $ \\
${ 36} $&$ {(1,5,7,8)[21]} 
$&$   18=10_L  +7_{J=\Pi}  +1_R
$&$ 24^*=16_L^*+7_{\Pi=J}^*+1_R^*           $ \\
${ 39} $&$ {(1,6,8,9)[24]} 
$&$   18=10_L  +7_{J=\Pi}  +1_R
$&$ 24^*=16_L^*+7_{\Pi=J}^*+1_R^*       $ \\
${ 18} $&$ { (2,1,5,7)[15]} 
$&$   26=3_L   +7_{J=\Pi}  +16_R
$&$ 17^*=1_L^* +7_{\Pi=J}^*+9_R^*      $ \\
${ 27} $&$ {(3,1,7,10)[21]} 
$&$   24=1_L   +7_{J=\Pi}  +16_R
$&$ 24^*=1_L^* +7_{\Pi=J}^*+16_R^*       $ \\
${ 52} $&$ {(2,3,7,9)[21]} 
$&$   14=3_L   +7_{J=\Pi}  + 4_R
$&$ 23^*=7_L^* +7_{\Pi=J}^*+9_R^*         $ \\ 
${ 54} $&$ {(3,2,8,11)[24]}
$&$   15=1_L   +7_{J=\Pi}  + 7_R
$&$ 27^*=4_L^* +7_{\Pi=J}^*+16_R^*         $ \\
${ 61} $&$ {(2,5,9,11)[27]} 
$&$   11=1_L   +7_{J=\Pi}  + 3_R
$&$ 32^*=9_L^* +7_{\Pi=J}^*+16_R^*          $ \\
${ 72} $&$ {(3,4,10,13)[30]}
$&$   10=1_L   +7_{J=\Pi}  + 2_R
$&$ 35^*=12_L^*+7_{\Pi=J}^*+16_R^*         $ \\
${ 77} $&$ {(3,5,11,14)[33]} 
$&$    9=1_L   +7_{J=\Pi}  + 1_R
$&$ 39^*=16_L^*+7_{\Pi=J}^*+16_R^*         $ \\
\hline
\end{tabular}
\end{table}
\normalsize

The canonical basis of the chain shown in Table~\ref{TabXVII} is:
\begin{eqnarray} 
\vec{e}_1\,&=&\, (-m,n,0,0)\nonumber\\
\vec{e}_2\,&=&\, (-2,-1,1,0)\nonumber\\
\vec{e}_3\,&=&\, (-1,0,-1,1),
\end{eqnarray}
with
\begin{eqnarray}
{det} (\vec { e}_1, \vec { e}_2, \vec { e}_3,
 \vec {e}_1)\,=\, { 3\cdot m\, +\, 6\cdot n} \,=\, { d},
\end{eqnarray}
where $\vec {e_1} =(1,1,1,1)$.
The possible values of ${m}$ and ${n}$ for this chain 
are determined in the standard way from
the dimensions of the extended vectors,
${d(\vec {k}^{ex(j)})=6}$ and ${d(\vec {k}^{ex(i)})=3}$, as seen in
Table~\ref{TabXVII}:

\begin{eqnarray}
p \cdot \vec{k}_4(XVII)\,&=&\,m\cdot ({0,1,}1,1)\,
+\,n\cdot ( {1,0,}2,3)\nonumber\\
p\,=\,1^{*}\,&\rightarrow & 
1\,\leq\,m \, \leq \,6;\,;\,\,1\,\leq \,n\,\leq\,3; \nonumber\\
p\,=\,2 \,&\rightarrow & \, m\,=\,n\,=\,2;\nonumber\\ 
p\,=\,3 \,&\rightarrow & \, m\,=\,n\,=\,3. 
\end{eqnarray} 

\begin{table}[!ht]
\centering
\caption{\it The group singularities 
of the  dual pairs of elliptic polyhedra in chain $XVII$.}
\label{TabXVIIb}
\scriptsize
\vspace{.05in}
\begin{tabular}{|c|c|c|c|c|c|c|} \hline
$P^3( \vec{k})$&$H(\Delta)$&$H(\Delta^*) 
$&$ G_{L}(\Delta)$&$G_{R}(\Delta)$&$ G_L(\Delta^*)$&
$G_{R}(\Delta^*)$\\ \hline\hline
$(1,1,3,4) $&$m_1+m_2+m_3=0 $&$m_1^*=0
$&$ E_6 $&$   E_8   $&$   SU(1)      $&$  SU(1) $\\
$(1,2,4,5) $&$m_1+m_2+m_3=0 $&$m_1^*=0
$&$ E_6 $&$   F_4   $&$   G_2       $&$  SU(1) $\\
$(1,3,5,6) $&$m_1+m_2+m_3=0 $&$m_1^*=0
$&$ E_6 $&$   G_2   $&$   F_4       $&$  SU(1) $\\
$(1,4,6,7) $&$m_1+m_2+m_3=0 $&$m_1^*=0
$&$ E_6 $&$   SU(2)  $&$   E_7       $&$  SU(1) $\\
$(1,5,7,8) $&$m_1+m_2+m_3=0 $&$m_1^*=0
$&$ E_6 $&$   SU(1)  $&$   E_8       $&$  SU(1) $\\
$(1,6,8,9) $&$m_1+m_2+m_3=0 $&$m_1^*=0
$&$ E_6 $&$   SU(1)  $&$   E_8       $&$  SU(1) $\\
\hline
\end{tabular}
\normalsize
\end{table}

The seven invariant monomials corresponding to this chain 
are the following:
\begin{eqnarray}
\vec{\mu}_0^1\,=\,(6,3,0,0,) &\Rightarrow & 
x^6 \cdot y^3                     \nonumber\\
\vec{\mu}_0^2\,=\,(4,2,1,0,) &\Rightarrow & 
x^4 \cdot y^2 \cdot z             \nonumber\\
\vec{\mu}_0^3\,=\,(2,1,2,0,) &\Rightarrow & 
x^2 \cdot y   \cdot z             \nonumber\\
\vec{\mu}_0^4\,=\,(0,0,3,0,) &\Rightarrow & 
                    z^3           \nonumber\\
\vec{\mu}_0^5\,=\,(3,2,0,1,) &\Rightarrow & 
x^3 \cdot y^2         \cdot u     \nonumber\\
\vec{\mu}_0^6\,=\,(1,1,1,1,) &\Rightarrow & 
x   \cdot y   \cdot z \cdot u     \nonumber\\
\vec{\mu}_0^7\,=\,(0,1,0,2,) &\Rightarrow & 
          y           \cdot u^2 
\end{eqnarray}
and the corresponding $ {E_6^{(1)}}_L-{E_{8}^{(1)}}_R$
graph associated with the eldest $(1,1,3,4))[9]$ polyhedron in chain
$XVII$
is shown in Table~\ref{TabXVIIb} and Figure~\ref{XVIIe6xe81}.

\subsection{The ${ J=\Pi}$ Symmetric Chain $XVIII$ with 
Exceptional Graph $E_7 \times E_8$}

\begin{figure}[h]
   \begin{center}
   \mbox{
   \epsfig{figure=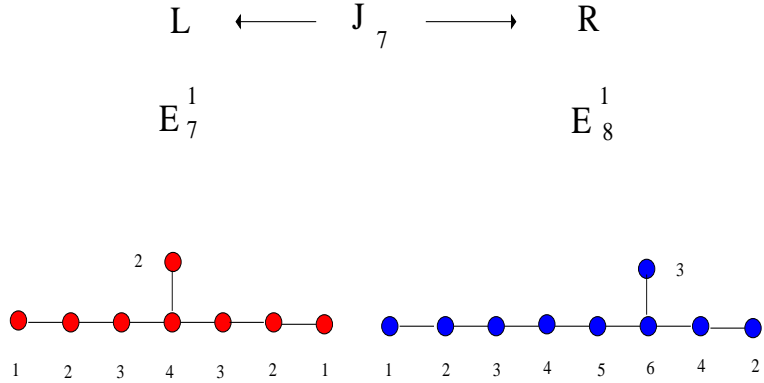,height=5cm,width=10cm}}
   \end{center}
   \caption{\it The $ {E_7^{(1)}}_L-{E_{8}^{(1)}}_R$ graphs
associated with the $(1,1,3,5))[10]$ polyhedron in chain $XVIII$:
${36=13_L+7_J+16_R}$.} 
\label{XVIIIe7xe8}
\end{figure}

This chain can be built from the vectors
${\vec{k}_4^{ex^{i}}=(0,1,1,2)}$ and 
${\vec{k}_4^{ex^{j}}=(1,0,2,3)}$,
with positive integers ${ m \leq 6}$ and $ { n \leq 4}$.
The maximal (${m=n=1}$) polyhedron in this chain is again completely 
determined by the dimensions {4} and {6} of the projective vectors  
${\vec{k}_4^{ex^{i}}}$ and ${ \vec{k}_4^{ex^{j}}}$,
respectively: 
\begin{equation}
36 = (13)_L + (7)_{J=\Pi} + (16)_R.
\end{equation}
The `right' ${ 15_R + 1_R}$ and `left' ${ 12_L+1_L}$
points produce the
graphs for the affine ${ E_8^{(1)}}$ and ${ E_7^{(1)}}$ Lie algebras,
respectively, as seen in Figure~\ref{XVIIIe7xe8}.
The vector ${ \vec {k}=(3,4,9,14)[28]}$ is self-dual with
${ E_8^{(1)}}$  graphs for the dual polyhedron pair.
The `minimal' vector
${\vec{k}}$ gives the following set of integer
lattice points in the polyhedron:
\begin{equation}
 (1)_L + (7)_{int} + (1)_R = 9.
\end{equation}

\begin{table}[!ht]
\centering
\caption{\it The $K3$ hypersurfaces in the
chain $XVIII$: $\vec {k} = (n, m,m+2n,2m+3n)= m \cdot (0,1,1,2) + 
n \cdot (1,0,2,3)$: ${ d=4m + 6n }$, ${ m_{max}=6, n_{max}=4}$.}
\label{TabXVIII}
\scriptsize
\vspace{.05in}
\begin{tabular}{|c|c|c|c|} \hline
${\aleph} $&$ {\vec k}[d] 
$&$ \Delta    
$&$ \Delta^*                           $\\ \hline\hline
${ 8} $&$ { (1,1,3,5)[10]} 
$&$   36=13_L  +7_{J=\Pi}  +16_R
$&$  9^*=1_L^* +7_{\Pi=J}^*+1_R^*         $ \\\hline
${ 17} $&$ { (1,2,4,7)[14]} 
$&$   27=13_L  +7_{J=\Pi}  +7_R
$&$ 12^*=4_L^* +7_{\Pi=J}^*+1_R^*          $ \\
${ 26}  $&$ {(1,3,5,9)[18]} 
$&$   24=13_L  +7_{J=\Pi}  +4_R       
$&$ 15^*=7_L^* +7_{\Pi=J}^*+1_R^*         $ \\
${ 33 }$&$ {(1,4,6,11)[22]}
$&$   22=13_L  +7_{J=\Pi}  +2_R
$&$ 20^*=12_L^*+7_{\Pi=J}^*+1_R^*          $ \\
${ 37} $&$ {(1,5,7,13)[26]} 
$&$   21=13_L  +7_{J=\Pi}  +1_R
$&$ 24^*=16_L^*+7_{\Pi=J}^*+1_R^*          $ \\
${ 40} $&$ {(1,6,8,15)[30]} 
$&$   21=13_L  +7_{J=\Pi}  +1_R
$&$ 24^*=16_L^*+7_{\Pi=J}^*+1_R^*         $ \\
${ 19} $&$ { (2,1,5,8)[16]} 
$&$   28=5_L   +7_{J=\Pi}  +16_R
$&$ 14^*=1_L^* +7_{\Pi=J}^*+6_R^*         $ \\
${ 28} $&$ {(3,1,7,11)[22]} 
$&$   25=2_L   +7_{J=\Pi}  +16_R
$&$ 20^*=1_L^* +7_{\Pi=J}^*+12_R^*        $ \\
${ 34} $&$ {(4,1,9,14)[28]} 
$&$   24=1_L   +7_{J=\Pi}  + 16_R
$&$ 24^*=1_L^* +7_{\Pi=J}^*+16_R^*          $ \\ 
${ 55} $&$ {(3,2,8,13)[26]}
$&$   16=2_L   +7_{J=\Pi}  + 7_R
$&$ 23^*=4_L^* +7_{\Pi=J}^*+12_R^*       $ \\
${ 53} $&$ {(2,3,7,12)[24]} 
$&$   16=5_L   +7_{J=\Pi}  + 4_R
$&$ 20^*=7_L^* +7_{\Pi=J}^*+6_R^*         $ \\
${ 74} $&$ {(4,3,11,18)[36]}
$&$   12=1_L   +7_{J=\Pi}  + 4_R
$&$ 30^*=7_L^*+7_{\Pi=J}^*+16_R^*         $ \\
${ 73} $&$ {(3,4,10,17)[34]} 
$&$    11=2_L   +7_{J=\Pi}  + 2_R
$&$ 31^*=12_L^*+7_{\Pi=J}^*+12_R^*       $ \\
${ 62} $&$ {(2,5,9,16)[32]} 
$&$    13=5_L   +7_{J=\Pi}  + 1_R
$&$ 29^*=16_L^*+7_{\Pi=J}^*+6_R^*         $ \\
${ 78} $&$ {(3,5,11,19)[38]} 
$&$    10=2_L   +7_{J=\Pi}  + 1_R
$&$ 35^*=16_L^*+7_{\Pi=J}^*+12_R^*          $ \\
${ 85} $&$ {(4,5,13,22)[44]} 
$&$    9=1_L   +7_{J=\Pi}  + 1_R
$&$ 39^*=16_L^*+7_{\Pi=J}^*+16_R^*         $ \\
\hline
\end{tabular}
\end{table}
\normalsize
\noindent
The canonical basis 
for the chain shown in Table~\ref{TabXVIII} is:
\begin{eqnarray} 
\vec{e}_1\,&=&\, (-m,n,0,0)\nonumber\\
\vec{e}_2\,&=&\, (-2,-1,1,0)\nonumber\\
\vec{e}_3\,&=&\, (-1,-1,-1,1),
\end{eqnarray}
with
\begin{eqnarray}
{det} (\vec { e}_1, \vec { e}_2, \vec { e}_3,
 \vec { 1})\,=\,{ 4\cdot m\, +\, 6\cdot n} \, =\, { d},
\end{eqnarray}
The possible values of ${m}$ and ${ n}$ for this chain 
fill up 
the dimensions of the extended vectors
${d(\vec {k}^{ex(j)})=6}$ and ${d(\vec {k}^{ex(i)})=4}$, as seen in
Table~\ref{TabXVIII}:

\begin{eqnarray}
p \cdot \vec{k}_4(XVIII)\,&=&\,m\cdot ({0,1,}1,2)\,
+\,n\cdot ( {1,0,}2,3)\nonumber\\
p\,=\,1^{*}\,&\rightarrow & 
1\,\leq\,m \, \leq \,6;\,\,\,1\,\leq \,n\,\leq\,4; 
\nonumber\\
p\,=\,2\,    &\rightarrow & \, m\,=\,n\,=\,2; \nonumber\\
p\,=\,3\,    &\rightarrow & \, m\,=\,n\,=\,3; \nonumber\\
p\,=\,4\,    &\rightarrow & \, m\,=\,n\,=\,4.
\end{eqnarray} 

\begin{table}[!ht]
\centering
\caption{\it The group singularities
of the dual pairs of elliptic polyhedra in chain XVIII.}
\label{TabXVIIIb}
\scriptsize
\vspace{.05in}
\begin{tabular}{|c|c|c|c|c|c|c|} \hline
$P^3( \vec{k})$&$H(\Delta)$&$H(\Delta^*) 
$&$ G_{L}(\Delta)$&$G_{R}(\Delta)$&$ G_L(\Delta^*)$&
$G_{R}(\Delta^*)$\\ \hline\hline
$(1,1,3,5) $&$m_1+m_2+2 m_3=0 $&$m_1^*=0
$&$ E_7 $&$   E_8      $&$   SU(1)      $&$  SU(1) $\\
$(1,2,4,7) $&$m_1+m_2+2 m_3=0 $&$m_1^*=0
$&$ E_7 $&$   F_4      $&$   G_2       $&$  SU(1) $\\
$(1,3,5,9) $&$m_1+m_2+2 m_3=0 $&$m_1^*=0
$&$ E_7 $&$   G_2      $&$   F_4       $&$  SU(1) $\\
$(1,4,6,11)$&$m_1+m_2+2 m_3=0 $&$m_1^*=0
$&$ E_7 $&$   SU(2)     $&$   E_7       $&$  SU(1) $\\
$(1,5,7,13)$&$m_1+m_2+ 2m_3=0 $&$m_1^*=0
$&$ E_7 $&$   SU(1)     $&$   E_8       $&$  SU(1) $\\
$(1,6,8,15)$&$m_1+m_2+2 m_3=0 $&$m_1^*=0
$&$ E_7 $&$   SU(1)     $&$   E_8       $&$  SU(1) $\\
\hline
\end{tabular}
\normalsize
\end{table}
The seven invariant monomials corresponding to this chain 
are the following:
\begin{eqnarray}
\vec{\mu}_0^1\,=\,(6,4,0,0,) &\Rightarrow & 
x^6 \cdot y^4                   \nonumber\\
\vec{\mu}_0^2\,=\,(4,3,1,0,) &\Rightarrow & 
x^4 \cdot y^3 \cdot z           \nonumber\\
\vec{\mu}_0^3\,=\,(2,2,2,0,) &\Rightarrow & 
x^2 \cdot y^2 \cdot z^2           \nonumber\\
\vec{\mu}_0^4\,=\,(0,1,3,0,) &\Rightarrow & 
          y   \cdot z^3         \nonumber\\
\vec{\mu}_0^5\,=\,(3,2,0,1,) &\Rightarrow & 
x^3 \cdot y^2          \cdot     u   \nonumber\\
\vec{\mu}_0^6\,=\,(1,1,1,1,) &\Rightarrow & 
x   \cdot y   \cdot z \cdot u   \nonumber\\
\vec{\mu}_0^7\,=\,(0,1,0,2,) &\Rightarrow & 
                    y  \cdot u^2 
\end{eqnarray}
The $ {E_7^{(1)}}_L-{E_{8}^{(1)}}_R$ graph
associated with the eldest $(1,1,3,5))[10]$ polyhedron in chain $XVIII$
can be seen in Table~\ref{TabXVIIIb} and Figure~\ref{XVIIIe7xe8}.

\subsection{Chain $XIX$ with $ {(7_{J},7_{\Pi})}$
Weierstrass Triangle Fibrations}
 
\begin{figure}[h]
   \begin{center}
     \mbox{
     \epsfig{figure=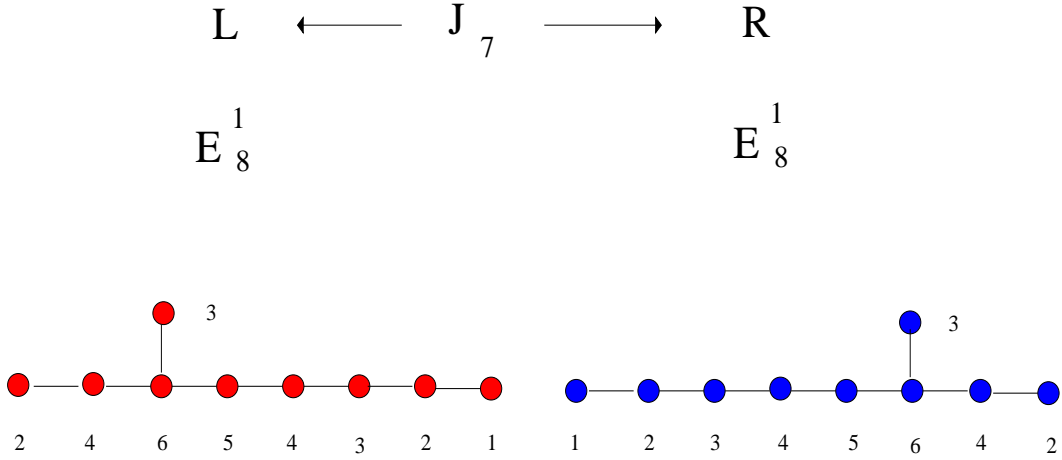,height=6cm,width=14cm}}
   \end{center}
   \caption{\it The ${E^{(1)}_8}_L - {E^{(1)}_8}_R$ graph obtained from
the eldest $(1,1,4,6)[12]$ polyhedron in chain $XIX$:
${39=16_L+7_J+16_R}$.}
\label{XIXe8xe8}
\end{figure}

We now consider the chain $XIX$ of ${\vec {k}_4}$ projective 
vectors with ${{E_8}_L}$ and ${ {E_8}_R}$  graphs.
This chain starts from the  ${ m=n=1}$ polyhedron, which is
left-right symmetric with 
respect to the intersection ${ P^2(1,2,3)}$. This polyhedron 
${P^3(1,1,4,6)}$ contains
${39 = 16_L + (7)_{J=\Pi} + 16_R }$ integer points:
see Table~\ref{TabXIX} and Figure~\ref{XIXe8xe8}. 
The minimal vector ${\vec {k}=(5,6,22,33)[66]}$ is the dual conjugate of
the eldest vector ${\vec {k}=(1,1,4,6)[12]}$, the
vector  ${\vec {k}=(1,6,14,21)[42]}$ is self-dual,
and its dual pair of ${K3}$ polyhedra yield the self-dual 
${E_8^{(1)}}$ graph.

\scriptsize  
\begin{table}[!ht]
\centering
\caption{\it The $K3$ hypersurfaces in the ${ {J=\Pi}}$ symmetric
chain $XIX$ with $\vec {k} = (n,m,2m+2n,3m+3n)
= m \cdot (0,1,2,3) + n \cdot (1,0,2,3)$:
${ d=6m+6n})$, ${ m_{max}=6, n_{max} =6}$,
${ \vec{k}_{eld}=(1,1,4,6)[12]}$.}
\label{TabXIX}
\vspace{.05in}
\begin{tabular}{|c|c|c|c|} 
\hline
$   { \aleph}                         $&${ \vec {k}_4} 
$&$ \Delta(J=\Pi=7)    
$&$ \Delta^* (\Pi=J=7)                           $ \\ 
\hline\hline
${   9}                         $&${ (1,1,4,6)[12]} 
$&$ 39=16_L+7_{J=\Pi}+16_R
$&$ 9^*=1^*_L+7_{\Pi=J}^*+1_R^*      $ \\ \hline
$   { 20}                        $&${ (1,2,6,9)[18]} 
$&$ 30=16_L+7_{J=\Pi}+7_R
$&$ 12^*=4_L^* +7_{\Pi=J}^*+1_R^*     $ \\
$   { 29}                        $&${ (1,3,8,12)[24]} 
$&$ 27=16_L+7_{J=\Pi}+4_R
$&$ 15^*=7_L^* +7_{\Pi=J}^*+1_R^*      $ \\
$   { 35}                        $&${ (1,4,10,15)[30]} 
$&$ 25=16_L+7_{J=\Pi}+2_R
$&$ 20^*=12_L^* +7_{\Pi=J}^*+1_R^*       $ \\  
$   { 38}                        $&${ (1,5,12,18)[36]} 
$&$ 24=16_L+7_{J=\Pi}+1_R
$&$ 24^*=16_L^*+7_{\Pi=J}^*+1_R^*       $ \\
$   { 41}                        $&${ (1,6,14,21)[42]} 
$&$ 24=16_L+7_{J=\Pi}+1_R
$&$ 24^*=16_L^*+7_{\Pi=J}^*+1_R^*       $ \\
$   { 56}                        $&${ (2,3,10,15)[30]} 
$&$ 18=7_L+7_{J=\Pi}+4_R
$&$ 18^*=7_L^*+7_{\Pi=J}^*+4_R          $ \\
$   { 75}                        $&${ (3,4,14,21)[42]} 
$&$ 13=4_L+7_{J=\Pi}+2_R            
$&$ 26^*=12_L^*+7_{\Pi=J}^*+7_R      $ \\
$   { 63}                        $&${ (2,5,14,21)[42]} 
$&$ 15=7_L+7_{J=\Pi}+1_R
$&$ 27^*=16_L^*+7_{\Pi=J}^*+4_R^*    $ \\    
$   { 79}                        $&${ (3,5,16,24)[48]} 
$&$ 12=4_L+7_{J=\Pi}+1_R
$&$ 30^*=16_L^*+7_{\Pi=J}^*+7_R^*    $ \\
$   { 86}                        $&${ (4,5,18,27)[54]} 
$&$ 10=2_L+7_{J=\Pi}+1_R
$&$ 35^*=16_L^*+7_{\Pi=J}^*+12_R^*     $ \\
$   { 92}                        $&${ (5,6,22,33)[66]} 
$&$ 9=1_L+7_{J=\Pi}+1_R
$&$39^*=16_L^*+7_{\Pi=J}^*+16_R^*     $ \\
\hline
\end{tabular}
\end{table}
\normalsize
The basis of the chain shown in Table~\ref{TabXIX} is the following:
\begin{eqnarray} 
\vec{e}_1\,&=&\, (-m,n,0,0)\nonumber\\
\vec{e}_2\,&=&\, (-2,-2,1,0)\nonumber\\
\vec{e}_3\,&=&\, (-1,-1,-1,1),
\end{eqnarray}
with
\begin{eqnarray}
{det} (\vec { e}_1, \vec { e}_2, \vec { e}_3,
 \vec { e}_0)\,=\,{  6\cdot m\, +\, 6\cdot n} \,=\, { d},
\end{eqnarray}
where $\vec { e}_0 =(1,1,1,1)$.
The possible values of ${m}$ and ${n}$ for this chain 
are completely determined by
the dimensions of the vectors
${d(\vec {k}^{ex(j)})=6}$ and ${d(\vec {k}^{ex(i)})=6}$
(see Table~\ref{TabXIX}):

\begin{eqnarray}
p \cdot \vec{k}_4(XIX)\,&=&\,m\cdot ({0,1,}2,3)\,
+\,n\cdot ( {1,0,}2,3)\nonumber\\
p\,=\,1^{*}\,&\rightarrow & 
1\,\leq\,m \, \leq \,6;\,\,\,1\,\leq \,n\,\leq\,6; 
\nonumber\\
p\,=\,2\,    &\rightarrow & \, m\,=\,n\,=\,2; \nonumber\\
p\,=\,3\,    &\rightarrow & \, m\,=\,n\,=\,3; \nonumber\\
p\,=\,4\,    &\rightarrow & \, m\,=\,n\,=\,4; \nonumber\\
p\,=\,6\,    &\rightarrow & \, m\,=\,n\,=\,6.
\end{eqnarray} 
The seven invariant monomials corresponding to this chain  
are the following:
\begin{eqnarray}
\vec{\mu}_0^1\,=\,(6,6,0,0) & \Rightarrow &  
x^6  \cdot y^6                                    \nonumber\\
\vec{\mu}_0^2\,=\,(4,4,1,0) & \Rightarrow & 
x^4  \cdot y^4 \cdot z                           \nonumber\\
\vec{\mu}_0^3\,=\,(2,2,2,0) & \Rightarrow & 
x^2  \cdot y^2 \cdot z^2                         \nonumber\\
\vec{\mu}_0^4\,=\,(0,1,3,0) & \Rightarrow & 
           y   \cdot z^3                         \nonumber\\
\vec{\mu}_0^5\,=\,(3,3,0,1) & \Rightarrow & 
x^3  \cdot y^3                \cdot u            \nonumber\\
\vec{\mu}_0^6\,=\,(1,1,1,1) & \Rightarrow & 
x    \cdot y   \cdot z        \cdot u            \nonumber\\
\vec{\mu}_0^7\,=\,(0,0,0,2) & \Rightarrow &  
                                     u^2.
\end{eqnarray}
Using these invariant monomials and basis the {CY} equations
for all the ${\vec{k}(l=m+n)}$ projective vectors of this chain can be 
written in the following form:

\begin{eqnarray}
F(\vec {z})_{m,n}                                         \,=\, 
\sum_{j=1}^{j=7}  
\vec {z}^{\vec {\mu}_0^j}                                 \{ 
\sum_{p=1}^{p={\Pi}_{jL}} 
a_{\vec {\mu}_0^j}^{pL}                                   \cdot 
\vec {z}^{n_{pL}                                          \cdot 
(-\vec {e}_1)} \,+\, 
\sum_{p=1}^{p={\Pi}_{jR}} 
a_{\vec {\mu}_0^j}^{pR}                                   \cdot 
\vec {z}^{-n_{pR}                                         \cdot 
(-\vec {e}_1)}                                            \},
\end{eqnarray}
where the basis vector ${\vec {e}_1= (m,-n,0,0)}$.
The ${E^{(1)}_8}_L - {E^{(1)}_8}_R$ graph obtained from
the eldest $(1,1,4,6)[12]$ polyhedron in chain $XIX$ is shown in
Table~\ref{ellipXIX} and Figure~\ref{XIXe8xe8}.

\begin{table}[!ht]
\centering
\caption{\it The group singularities of the  
dual pairs of elliptic polyhedra in chain $XIX$.}  
\label{ellipXIX}
\scriptsize
\vspace{.05in}
\begin{tabular}{|c|c|c|c|c|c|c|} \hline
$  P^3( \vec{k})   
$&$ H(\Delta)            $&$  H(\Delta^*)  
$&$ G_{L}(\Delta)        $&$  G_{R}(\Delta)  
$&$ G_L(\Delta^*)        $&$  G_{R}(\Delta^*)            $\\ \hline\hline
$  (1,1,4,6) 
$&$ m_1+2 m_2+3 m_3=0    $&$  m_1^*=0
$&$ E_8                  $&$  E_8   
$&$ SU(1)                $&$  SU(1)                      $\\
$ (1,2,6,9) 
$&$ m_1+2 m_2+3 m_3=0    $&$  m_1^*=0
$&$ E_8                  $&$  F_4   
$&$ G_2                 $&$  SU(1)                      $\\
$  (1,3,8,12) 
$&$ m_1+2 m_2+3 m_3=0    $&$  m_1^*=0
$&$ E_8                  $&$  G_2   
$&$ F_4                 $&$  SU(1)                      $\\
$(1,4,10,15) 
$&$ m_1+2 m_2+3 m_3=0    $&$  m_1^*=0
$&$ E_8                  $&$  SU(2)  
$&$ E_7                  $&$  SU(1)                      $\\
$ (1,5,12,18) 
$&$ m_1+2 m_2+3 m_3=0    $&$  m_1^*=0
$&$ E_8                  $&$  SU(1)  
$&$ E_8                  $&$  SU(1)                      $\\
$ (1,6,14,21) 
$&$ m_1+2 m_2+3 m_3=0    $&$  m_1^*=0
$&$ E_8                  $&$  SU(1)  
$&$ E_8                  $&$  SU(1)                      $\\
\hline
\end{tabular}
\normalsize
\end{table}

%
%


%
%
%
%

\section{Perspectives on the Further Classification of $CY_3$ and $K3$
Spaces}

Although a fuller study of $CY_3$ spaces lies outside the scope of
this paper, a preliminary study is of interest here, for
the following reason.
In addition to the 95 $K3$ spaces (Table~\ref{list95}) related to the 
zeroes of single 
polynomials discussed in previous Sections, others can be found by
`higher-level' 
constructions as the intersections of the loci of zeroes of 
quasihomogeneous polynomials, which are obtainable from
$CY_3$ spaces, as we now discuss.

When going on to consider the general construction of ${\vec{k}_5}$ 
projective vectors in ${CP^4}$ that
describe $CY_3$ hypersurfaces, we start from
the {95} simple extensions of these $K3$ vectors
as well as 5 multiple extensions of lower-dimensional vectors,
together with all their possible permutations.
Corresponding to the previous five
forms of extended vectors, one finds the following sets and 
permutations:
quadruply-extended basic vectors with the cyclic ${C_5}$
group of permutations:
\begin{equation}
\vec {k}_1^{ex}= (0,0,0,0,1):\,\,\,\,\,|C_5|=5;
\end{equation}
triply-extended composite vectors with the dihedral ${D_5}$
group of permutations:
\begin{equation}
\vec {k}_2^{ex}= (0,0,0,1,1):\,\,\,\,\,|D_5|=10;
\end{equation}
the following doubly-extended composite vectors  with the
${D_5'}, {A_5'}$ and  ${A_5}$
groups of permutations:
\begin{eqnarray}
\vec {k}_3^{ex}\, &=&\, (0,0,1,1,1):\,\,\,\,\, |D_5'|=10, \\
\vec {k}_3^{ex}\, &=&\, (0,0,1,1,2):\,\,\,\,\, |A_5'|=30, \\ 
\vec {k}_3^{ex}\, &=&\, (0,0,1,2,3);\,\,\,\,\, |A_5 |=60.
\end{eqnarray}
we recall that the alternating group of
permutations ${A_5}$ can be identified with
the icosahedral symmetry group ${I}$. All the other 
extended ${\vec{k}_5}$ vectors can be obtained similarly from {95}
${K3}$ vectors, utilising the symmetric group ${S_5}$ or some subgroups.
The full set of extended ${\vec{k}_5}$ vectors is displayed in
Table~\ref{Tab100}. As noted in its caption, the total number of
extended vectors is 10~270.

\begin{table}[!ht]
\centering
\caption{ \it The {100} distinct types of five-dimensional 
`extended' projective vectors used to construct $CY_3$ spaces, listed
together with the
orders of their permutation groups. Including these
permutations, the total number of extended vectors is {10~270}.}
\label{Tab100}
\scriptsize
\vspace{.05in}
\begin{tabular}{|c|c|c||c|c|c|} 
\hline
$   { \aleph} $&$ { {\vec k_{5ex}}^{(i)}} $&${ G(perm)}
$&$ { \aleph} $&$ { {\vec k_{5ex}}^{(i)}} $&${ G(perm)}$\\ 
\hline\hline
$   { i}      $&$  {(0, 0, 0, 0, 1)}          $&$ 5  
$&$ { 46}     $&$  {(0, 2, 3, 4, 7)}          $&$ 120    
$\\ \hline
$   { ii}      $&$  {(0, 0, 0, 1, 1)}          $&$ 10               
$&$ { 47}     $&$  {(0 ,2 ,3 ,4 ,9)}          $&$ 120 
$\\ \hline
$   { iii}      $&$  { (0, 0, 1, 1, 1)}         $&$ 10               
$&$ { 48}     $&$  { (0 ,2 ,3 ,5, 5)}         $&$ 60      
$\\ \hline
$   { iv}      $&$  {(0, 0, 1, 1 ,2)}          $&$ 30               
$&$ { 49}     $&$  {(0 ,2 ,3 ,5 ,7)}          $&$ 120     
$\\ \hline
$   { v}      $&$  {(0 ,0 ,1 ,2 ,3)}          $&$ 60               
$&$ { 50}     $&$  {(0 ,2 ,3 ,5 ,8)}          $&$ 120     
$\\ \hline
$   { 1}      $&$  {(0 ,1 ,1 ,1 ,1)}          $&$ 5                
$&$ { 51}     $&$  {(0 ,2 ,3 ,5 ,10)}         $&$ 120     
$\\ \hline
$   { 2}      $&$  {(0 ,1 ,1 ,1 ,2)}          $&$ 20               
$&$ { 52}     $&$  {(0 ,2 ,3 ,7 ,9)}          $&$ 120     
$\\ \hline
$   { 3}      $&$  {(0, 1, 1, 1, 3)}          $&$ 20               
$&$ { 53}     $&$  {(0 ,2 ,3 ,7 ,12)}         $&$ 120     
$\\ \hline
$   { 4}      $&$  {(0 ,1 ,1 ,2 ,2)}          $&$ 30               
$&$ { 54}     $&$  {(0 ,2 ,3 ,8 ,11)}         $&$ 120     
$\\ \hline
$   { 5}     $&$  {(0 ,1 ,1 ,2 ,3)}           $&$    60                  
$&$ { 55}     $&$  {(0 ,2 ,3 ,4 ,7)}           $&$    120        
$\\ \hline
$   { 6}     $&$  {(0 ,1 ,1 ,2 ,4)}           $&$    60                  
$&$ { 56}     $&$ { (0 ,2 ,3 ,10 ,15)}         $&$    120        
$\\ \hline
$   { 7}     $&$  {(0 ,1 ,1 ,3 ,4)}           $&$    60                  
$&$ { 57}     $&$  {(0 ,2 ,4 ,5 ,9)}           $&$    120        
$\\ \hline
$   { 8}     $&$  {(0 ,1 ,1 ,3 ,5)}           $&$    60                  
$&$ { 58}     $&$  {(0 ,2 ,4 ,5 ,11)}          $&$    120        
$\\ \hline
$   { 9}     $&$  {(0, 1, 1 ,4 ,6)}           $&$    60                 
$&$ { 59}     $&$  {(0 ,2 ,5 ,6 ,7)}           $&$    120        
$\\ \hline
$   { 10}     $&$  {(0 ,1 ,2 ,2 ,3)}           $&$    60                  
$&$ { 60}     $&$  {(0 ,2 ,5 ,6 ,13)}          $&$    120        
$\\ \hline
$   { 11}     $&$ { (0 ,1 ,2 ,2 ,5)}           $&$    60                  
$&$ { 61}     $&$  {(0 ,2 ,5 ,9 ,11) }         $&$    120        
$\\ \hline
$   { 12}     $&$  {(0 ,1 ,2 ,3 ,3)}           $&$    60                 
$&$ { 62}     $&$  {(0 ,2 ,5 ,9 ,16)}          $&$    120        
$\\ \hline
$   { 13}     $&$  {(0 ,1 ,2, 3, 4)}           $&$    120                 
$&$ { 63}     $&$  {(0 ,2 ,5 ,14 ,21)}         $&$    120        
$\\ \hline
$   { 14}     $&$  {(0 ,1 ,2 ,3 ,5)}           $&$    120                 
$&$ { 64}     $&$  {(0, 2 ,6 ,7 ,15)}          $&$    120        
$\\ \hline
$   { 15}     $&$  {(0 ,1 ,2 ,3 ,6)}           $&$    120                 
$&$ { 65}     $&$  {(0 ,3 ,3 ,4 ,5)}           $&$    60         
$\\ \hline
$   { 16}     $&$ { (0, 1 ,2 ,4 ,5)}           $&$    120                 
$&$ { 66}     $&$  {(0 ,3 ,4 ,5 ,6)}           $&$    120        
$\\ \hline
$   { 17}     $&$  {(0 ,1 ,2 ,4, 7)}           $&$    120                 
$&$ { 67}     $&$  {(0, 3, 4, 5, 7)}           $&$    120        
$\\ \hline
$   { 18}     $&$  {(0 ,1 ,2 ,5 ,7)}           $&$    120                 
$&$ { 68}     $&$  {(0 ,3 ,4 ,5 ,8) }          $&$    120        
$\\ \hline
$   { 19}     $&$  {(0 ,1 ,2 ,5 ,8)}           $&$    120                
$&$ { 69}     $&$  {(0 ,3 ,4 ,5 ,12) }         $&$    120        
$\\ \hline
$   { 20}     $&$  {(0 ,1 ,2 ,6 ,9) }          $&$    120                
$&$ { 70}     $&$  {(0 ,3 ,4 ,7 ,10)}          $&$    120        
$\\ \hline
$   { 21}     $&$  {(0 ,1 ,3 ,4, 4)}           $&$    60                  
$&$ { 71}     $&$  {(0 ,3 ,4 ,7, 14)}          $&$    120        
$\\ \hline
$   { 22}     $&$  {(0 ,1 ,3 ,4 ,5)}           $&$    120                 
$&$ { 72}     $&$  {(0 ,3 ,4 ,10 ,13) }        $&$    120        
$\\ \hline
$   { 23}     $&$  {(0 ,1 ,3 ,4 ,7)}           $&$    120                 
$&$ { 73}     $&$  {(0 ,3 ,4 ,10 ,17)}         $&$    120        
$\\ \hline
$   { 24}     $&$ { (0 ,1 ,3 ,4 ,8) }          $&$    120                 
$&$ { 74}     $&$  {(0 ,3, 4 ,11 ,18)}         $&$    120        
$\\ \hline
$   { 25}     $&$  {(0, 1, 3, 5, 6)}           $&$    120                 
$&$ { 75}     $&$  {(0 ,3 ,4 ,14 ,21)}         $&$    120        
$\\ \hline
$   { 26}     $&$  {(0 ,1, 3, 5, 9)}           $&$    120                 
$&$ { 76}     $&$  {(0 ,3 ,5 ,6 ,7)}           $&$    120       
$\\ \hline
$   { 27}     $&$  {(0 ,1, 3, 7 ,10)}          $&$    120                 
$&$ { 77}     $&$  {(0 ,3 ,5 ,11 ,14) }         $&$    120       

$\\ \hline
$   { 28}     $&$ { (0, 1 ,3 ,7 ,11)}             $&$    120              
$&$ { 78}     $&$  {(0 ,3 ,5 ,11 ,19)}           $&$    120      
$\\ \hline
$   { 29}     $&$ { (0 ,1 ,3 ,8 ,12) }         $&$    120                 
$&$ { 79}     $&$ { (0 ,3 ,5 ,16 ,24)  }        $&$    120      
$\\ \hline
$   { 30}     $&$  {(0 ,1 ,4 ,5 ,6)  }          $&$    120                
$&$ { 80}     $&$ { (0 ,3 ,6 ,7 ,8) }           $&$    120       
$\\ \hline
$   { 31}     $&$  {(0 ,1 ,4 ,5 ,10) }          $&$    120               
$&$ { 81}     $&$  {(0 ,4 ,5 ,6 ,9) }           $&$    120       
$\\ \hline
$   { 32}     $&$  {(0 ,1 ,4 ,6 ,7) }            $&$    120              
$&$ { 82}     $&$  {(0 ,4 ,5 ,6 ,15) }           $&$    120       
$\\ \hline
$   { 33}     $&$  {(0 ,1 ,4 ,6 ,11)  }          $&$    120                
$&$ { 83}     $&$  {(0 ,4 ,5 ,7 ,9)  }           $&$    120       
$\\ \hline
$   { 34}     $&$  {(0 ,1 ,4 ,9 ,14)  }          $&$    120                
$&$ { 84}     $&$ { (0 ,4 ,5 ,7 ,16)  }            $&$    120       
$\\ \hline
$   { 35}     $&$  {(0 ,1 ,4 ,10, 15) }        $&$    120                 
$&$ { 85}     $&$  {(0 ,4 ,5 ,13 ,22)}         $&$    120         
$\\ \hline
$   { 36}     $&$  {(0 ,1 ,5 ,7 ,8) }          $&$    120                  
$&$ { 86}     $&$  {(0 ,4 ,5 ,18 ,27)}         $&$    120         
$\\ \hline
$   { 37}     $&$ { (0 ,1 ,5 ,7 ,13)}          $&$    120                  
$&$ { 87}     $&$  {(0 ,4 ,6 ,7 ,11) }         $&$    120         
$\\ \hline
$   { 38}     $&$  {(0 ,1 ,5 ,12, 18)}         $&$    120                  
$&$ { 88}     $&$  {(0 ,4 ,6 ,7 ,17) }         $&$    120         
$\\ \hline
$   { 39}     $&$  {(0 ,1 ,6 ,8, 9)  }         $&$    120                  
$&$ { 89}     $&$  {(0 ,5 ,6 ,7 ,9)}           $&$    120         
$\\ \hline
$   { 40}     $&$  {(0 ,1 ,6, 8, 15)}          $&$    120                  
$&$ { 90}     $&$  {(0 ,5, 6 ,8 ,11)}          $&$    120         
$\\ \hline
$   { 41}     $&$ { (0 ,1, 6, 14, 21)}         $&$    120                 
$&$ { 91}     $&$ { (0 ,5 ,6 ,8 ,19)}          $&$    120         
$\\ \hline
$   { 42}     $&$  {(0 ,2 ,2 ,3 ,5)}           $&$    60                  
$&$ { 92}     $&$  {(0 ,5 ,6 ,22 ,33)}         $&$    120         
$\\ \hline
$   { 43}     $&$  {(0 ,2 ,2, 3, 7) }          $&$    60                  
$&$ { 93}     $&$  {(0 ,5 ,7 ,8 ,20)}          $&$    120         
$\\ \hline
$   { 44}     $&$ { (0 ,2 ,3 ,3 ,4)}           $&$    60                   
$&$ { 94}     $&$  {(0 ,7 ,8 ,10 ,25)}         $&$    120         
$\\ \hline
$   { 45}     $&$  {(0, 2, 3, 4, 5 )}          $&$    120                  
$&$ { 95}    $&$  {(0 ,7 ,8 ,9 ,12)}          $&$    120         
$\\ \hline
\end{tabular}
\end{table}
\normalsize

As an illustration how our method may be used to classify $CY_3$
manifolds, we now describe briefly how to obtain the
complete list of ${\vec {k}_5}$ vectors with ${K3}$
intersections, which we find to be distributed in {4242} chains.
To build the chains for ${CY_3}$ which have a double-vector structure, 
each of which is parametrized by a pair of positive integers,
should find the `good' pairs of `extended' vectors (i.e., those whose
intersection gives a reflexive ${K3}$ hypersurface), which involves
checking
all the ${10~270 \times 10~271 / 2 = 52~731~315}$ possible pairs of
vectors from
Table~\ref{Tab100}. It was just such a search by computer that led to
the {4242} double chains mentioned above,
together with their eldest vectors. For more complete information about
these chains, see~\cite{FrancoFTP}.

These chains give many $CY_3$
projective vectors, but not all. The complete list
also includes the `good' triples 
which have elliptic fibres. This requires looking for good triples
among the following five types of five-dimensional 
extended vectors:
\begin{eqnarray}
&1.&(0,0,0,0,1)\,\Rightarrow\, 5,\nonumber\\ 
&2.&(0,0,0,1,1)\,\Rightarrow\, 10,\nonumber\\ 
&3.&(0,0,1,1,1)\,\Rightarrow\, 10,\nonumber\\ 
&4.&(0,0,1,1,2)\,\Rightarrow\, 30,\nonumber\\ 
&5.&(0,0,1,2,3)\,\Rightarrow\, 60,
\label{noperms}
\end{eqnarray}
where the number after the arrow on each line of (\ref{noperms})
corresponds to the number of permutations in each case.
We have found 259 such good triples, 
together with their eldest vectors, corresponding to
{259} elliptic chains. The union of the ${ K3}$ and elliptic
projective vectors still does not yield the full dual set of ${
\vec{k_5}}$ projective vectors. We must also construct another
set of chains using quadruples from among the following
multiply-extended vectors:
\begin{eqnarray}
&1.&(0,0,0,0,1)\,\Rightarrow\, 5,\nonumber\\ 
&2.&(0,0,0,1,1)\,\Rightarrow\, 10,
\end{eqnarray}
The number of $CY_3$ chains found in this way is just six.

In addition to these $4242$ double, $259$ triple and
$6$ quadruple $CY_3$ chains
(to be compared with the $22$ double and $4$ triple 
$K3$ chains found previously), one must find all the vectors whose
intersection contains only one central interior point
(to be compared with the exceptional $K3$ vector $(7,8,9,12)$). We
have found just two such examples in the case of
$CY_3$, namely $(41,48,51,52,64)$ and $(51,60,64,65,80)$,
again using the intersection-projection duality technique.
The eldest vectors for all the
$CY_3$ projective vector chains we have found can be obtained
from~\cite{FrancoFTP}.

In the cases of dimension higher than 
three, the concept of intersection-projection duality is richer,
and leads to one important and by now well-known
consequence~\cite{Batyrev,Greene1}, namely the
isomorphism between different homology groups
for dual pairs of $CY_d$ manifolds $M, M^*$, 
and specifically the following relation:
\begin{equation}
H^{p,q}(M) \sim H^{d-p,q}(M^*)   
\end{equation}
for $0 \leq p,q \leq d$. We leave a more complete discussion
of duality of $CY_3$ spaces to future work, limiting our
discussion here of their ramifications for the classification of $K3$

Our construction based on the {10~270} extended vectors obtained from the
$100 (= 95 + 5)$ types of projective vectors in lower dimensions
$n = 1,2,3,4$ shown in Table~\ref{Tab100} yielded all the
$4242~(259, ...)$ eldest vectors representing
$CY_3$ spaces with $K3$ (elliptic, ...) fibers. 
However, this method of construction simultaneously 
provides a new {\it higher-level} list of $K3$ spaces defined by planar
polyhedra. To explain this, let us first assign
to all $K3$ spaces defined by $n$-dimensional projective vectors {\it
level zero}, and denote them by $\Pi_0$. Then,
{\it level one} $K3$ spaces consist of all the `good' 
intersections\footnote{In the sense that they give $n$-dimensional reflexive
polyhedra.} of 
two $(n+1)$-dimensional extended vectors, denoted by $\Pi_1$.
This yields a list of reflexive polyhedra that is more complete than the
previous list of polyhedra obtained from
$n$-dimensional projective vectors,
i.e., $ \Pi_0\subseteq \Pi_1$. Continuing, one may define
the set of all `good' intersections of {\it level two},
$\Pi_2$, by considering
the intersections of three $(n+2)$-dimensional extended vectors,
and similarly for the higher levels $3,4, ...$:
\begin{eqnarray}
{\Pi}_0 \subseteq {\Pi}_1 \subseteq {\Pi}_2 
\subseteq ... \subseteq {\Pi}_{last} 
\end{eqnarray}
until this  process gives us no new reflexive polyhedra.
Since the number of distinct reflexive polyhedra in 
any dimension is finite, e.g., the maximal number of
integer points for planar polyherdra is {10}, for $K3$ polyhedra it is
{39}, etc.,
there exists a maximum last level, after which one cannot find any new 
types of polyhedra. 

Following this approach in the simple case of $CY_1$ spaces, we 
recall that we found three planar polyhedra (triangles)
at level zero, determined by the three projective vectors $(1,1,1)$,
$(1,1,2)$ and $(1,2,3)$. At level one, constructing the {22}
chains of $K3$ projective vectors via the {22} `good' intersections
of the five types of four-dimensional extended vectors, we now
find {7} new planar polyhedra
in  {9} of the {22} two-vector ${K3}$
chains, differing from the previous three triangles
by the numbers of vertices $(V, V^*)$ and/or by the numbers of integer
points $(N, N^*)$ and/or by the areas of these planar polyhedra,
as shown in Table~\ref{Tab7plane}. To look for further new polyhedra
at level {2}, one should consider the five following types of
vectors: $(1), (1,1), (1,1,1), (1,1,2)$ and $(1,2,3)$, extended to
five dimensions. Taking into account all the 50 possible permutations, 
and looking for the `good' triple
intersections, we find among the {259} `good' planar reflexive
polyhedra mentioned above just three distinct
new polyhedra, which are exhibited in Table~\ref{Tab3plane}. 

 \begin{table}[!ht]
\centering
\caption{ \it The 7 distinct new planar polyhedra, representing new
$CY_1$ spaces, that are found as double intersections involving 9 of the 22
two-vector $K3$
chains. Two realizations each are given for 2 of the new polyhedra.}
\label{Tab7plane}
\scriptsize
\vspace{.05in}
 \begin{tabular}{|c||c|c||c|c|}
 \hline
 $ \aleph $&$ {\vec k_{4ex}}^{(i)} $&${\vec k_{4ex}}^{(i)} $&$ N,N^*$&$V,V^*$\\
 \hline
 \hline
 $1$&$(0,0,1,1)$&$(1,1,0,0)$&$9,5^*$&$4,4^*$\\
 $ $&$(0,0,1,1)$&$(1,1,0,1)$&$9,5^*$&$4,4^*$\\
 \hline
 $2$&$(0  ,0  ,1  ,1)$&$  (    1,   2,   0,   1)$&$  7  , 7^* $&$       4  ,4^*$\\
 \hline
 $3$&$ ( 0,   1,   1,   1)$&$(      1 ,  0,   1 ,  2)$&$   8 ,  6^*$&$       4 ,  4^*$\\
 \hline
 $4$&$  (0 ,  1,   1,   1)$&$(      3 ,  0,   1,   2)$&$  4 , 10^*$&$       3 ,  3^*$\\
 \hline
 $5$&$  (0 ,  1,   1,   2)$&$(      1 ,  1,   2,   0)$&$   5, 9^*$&$       3,   3^*$\\
 $ $&$ ( 0 ,  1,   1,   2)$&$(      2 ,  0,   1,   3)$&$   5, 9^*$&$       3 ,  3^*$\\
 \hline
 $6$&$ ( 0,   1,   1,   2)$&$(      2 ,  1,   3,   0)$&$   6 , 8^*$&$       4 , 4^*$\\
 \hline
 $7$&$  (0,   1,   2,   3)$&$(      3,   2,   1,   0)$&$     5 ,  9^*$&$       4 ,  4^*$\\
 \hline
 \end{tabular}
 \end{table}
\normalsize

\scriptsize
 \begin{table}[!ht]
\centering
\caption{ \it The 3 distinct new planar polyhedra, representing new
$CY_1$ spaces, that are obtainable as triple
intersections of five-dimensional extended projective vectors, the sum of
which gives an eldest $CY_3$ projective vector. Three realizations 
each are given for 2 of the new polyhedra.}
\label{Tab3plane}
\vspace{.05in}
\begin{tabular}{|c||c|c|c||c|c|}
\hline
 $ \aleph $&$ {\vec k_{5ex}}^{(i)} $&${\vec k_{5ex}}^{(i)} 
$&${\vec k_{5ex}}^{(i)} $&$ N,N^*$&$V,V^*$\\
 \hline\hline
 $1$&$  (0,   0,   0,   1,   1)$&$(  0,   1,   1,   0,   0)$&$( 1,   0,   1,   0,   1)$&$  
     8 ,  6^*$&$       5 ,  5^*$\\
 \hline 
 $2$&$  (0,   0,   0,   1,   1)$&$( 0 , 1,   1,   0,   1)$&$(1,   0,   1,   1,   0)$&$      
 7 ,  7^*$&$       5 ,  5^*$\\
 $ $&$  (0,   0,   0,   1,   1)$&$(  0 ,  1,   1,   0,   1)$&$(    1,   1,   2,   0,   0) $&$
       7 ,  7^*$&$       5 ,  5^*$\\
 $ $&$  (0,   0,   1,   1,   1)$&$(  1,   1,   0,   0,   1)$&$( 0,   1,   0,   1,   2)$&$
       7 ,  7^*$&$       5 ,  5^*$\\
 \hline 
 $3$&$  (0 ,  0,   0,   1,   1)$&$(   1 ,  1,   1,   0,   0)$&$(  0,   1,   2,   0,   1) $&$
       6 ,  8^*$&$       5  , 5^*$\\
 $ $&$  (0,   0,   0,   1,   1)$&$(     0,   1,   2,   0,   1)$&$(     2,   1,   0,   1,   0) $&$
       6 ,  8^*$&$       5 ,  5^*$\\
 $ $&$ ( 0 ,  0 ,  1 ,  1 ,  1)$&$(     0,   1,   0,   1,   2)$&$(     1,   0,   2,   1,   0) $&$
       6 ,  8^* $&$       5 ,  5^*$\\
 \hline
 \end{tabular}
 \end{table}
\normalsize

\begin{figure}[th!]
   \begin{center}
   \mbox{
   \epsfig{figure=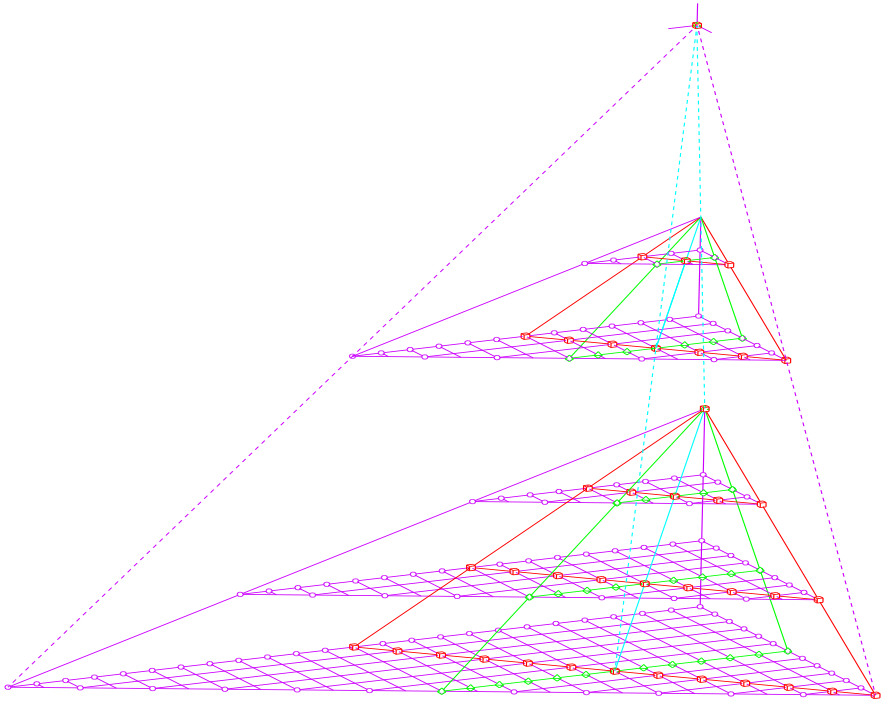,height=14 cm,width=14cm}}
   \end{center}
   \caption{\it The 4-dimensional pentahedroid corresponding to the
$CY_3$ space specified by the
eldest vector $\vec{k}_5=(1,1,2,8,12)[24]$ in the two-vector chain
$m(0,1,1,4,6)+n(1,0,1,4,6)$.
The number of integer points in this pentahedroid is
$N(S)=335$, and the volume $S=72$.
$SL(4,Z)$ transformations produce an infinite
number of polyhedroids, conserving the volume.}
\label{vec112812}
\end{figure}

Extending this procedure, we found among the 4242 chains of $CY_3$
spaces with `good'
intersections 730 new $K3$ polyhedra at level one, many with
multiple realizations as in Tables~\ref{Tab7plane} and~\ref{Tab3plane}. As
an example
how such new $K3$ spaces emerge, consider the following two-vector
$CY_3$ chain:  $m(0,1,1,4,6)+n(1,0,1,4,6)$.
The maximum values of ${m}$ and ${n}$
are determined  by the dimensions of these extended vectors, namely
$d = 12$. This chain contains 46 different $\vec{k}_5$
projective vectors. The four-dimensional pentahedroid 
corresponding to the eldest vector in this chain is shown in
Figure~\ref{vec112812}. 
As can be seen there, in addition to its 5 vertices, the
pentahedroid has 10 one-dimensional edges, 10 two-dimensional 
triangular faces, and 5 three-dimensional tetrahedral facets. This
pentahedroid contains two realizations of the tetrahedron
corresponding to ${\vec k}_4 = (1,1,4,6)$, whose intersection
contains an elliptic fibre corresponding to ${\vec k}_3 = (1,2,3)$.

\begin{figure}[th!]
   \begin{center}   
   \mbox{
   \epsfig{figure=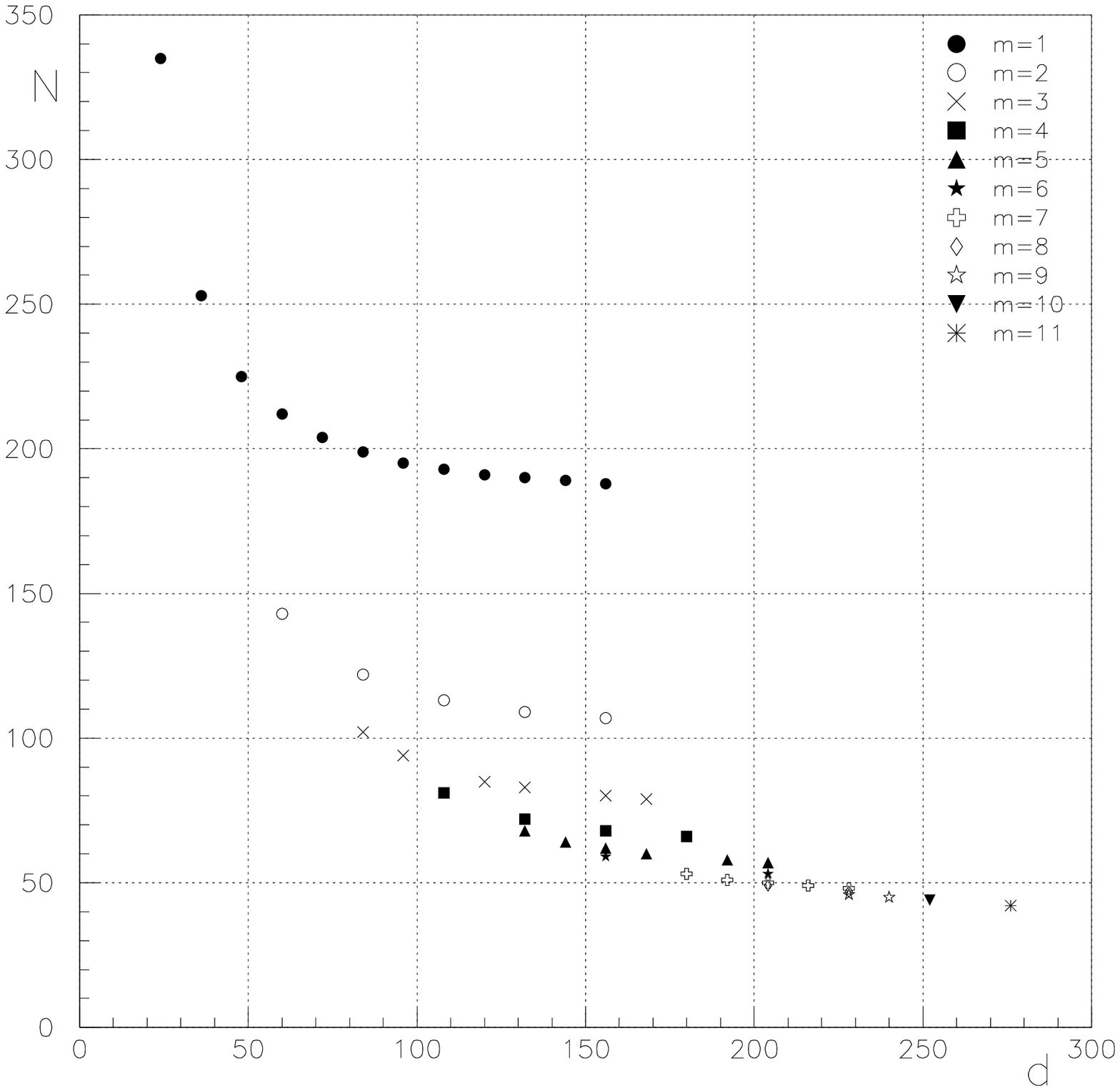,height=14 cm,width=14cm}}
   \end{center}
   \caption{\it The number of points $N$ found in
different members of the chain $m(0,1,1,4,6)+n(1,0,1,4,6)$,
plotted as a function of $d = k_1 + k_2 + k_3 + k_4 + k_5$ for different
values of $m$.}
\label{dchain}
\end{figure}

\begin{figure}[th!]
   \begin{center}   
   \mbox{
   \epsfig{figure=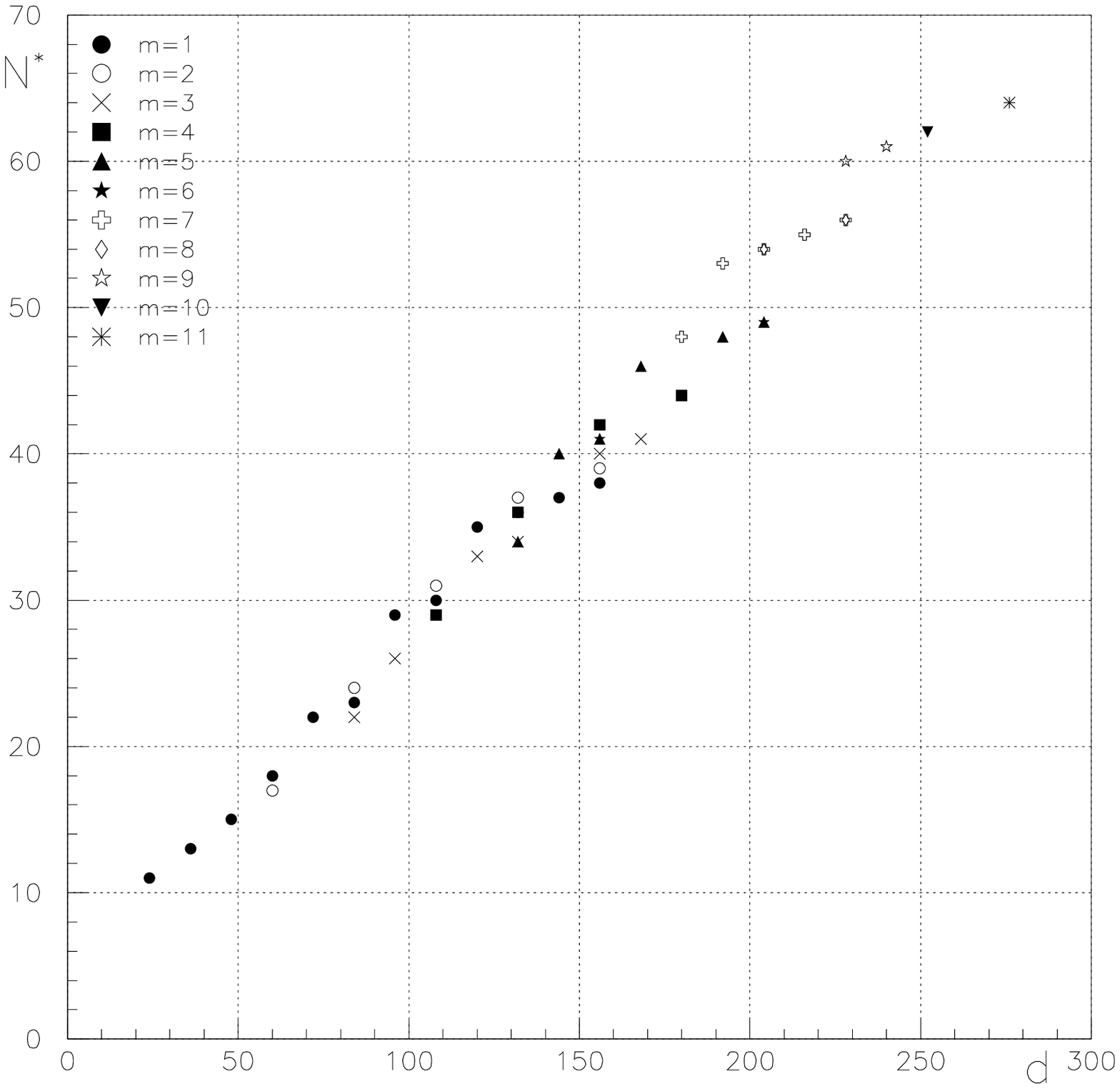,height=14 cm,width=14cm}}
   \end{center}
   \caption{\it The number of points $N^*$ found in
the polyhedra dual to the previous
$m(0,1,1,4,6)+n(1,0,1,4,6)$ chain,
plotted as a function of $d = k_1 + k_2 + k_3 + k_4 + k_5$ for different
values of $m$.}
\label{dstarchain}
\end{figure}

A snapshot of the complete $m(0,1,1,4,6)+n(1,0,1,4,6)$ chain
is shown in Figure~\ref{dchain}, where the number of points $N$ in
each member of the chain is plotted as a function of 
$d = k_1 + k_2 + k_3 + k_4 + k_5$ for each
of the allowed values of $m$. We note a systematic tendency for
$N$ to {\it decrease} as $d$ increases. (The structure of
the chain is, of course, symmetric under the interchange: $n
\leftrightarrow m$). The corresponding plot for the dual
polyhedra is shown in Figure~\ref{dstarchain}: here we see that
the number of points $N^*$ {\it increases} as $d$ increases.

To get another impression of the rich new structures emerging
at levels one and above, we consider  
a `tetrahedron subalgebra' of our $K3$ algebra, i.e., we
consider only those projective vectors corresponding to point- and
segment-polyhedra, triangles and tetrahedra.
With this restriction, we start from only 32 $K3$ projective vectors, 
corresponding to four-vertex tetrahedra and five of our previous
extended
vectors. In this way, the number of reflexive polyhedra at
level one is reduced to just 632, consisting of
460 tetrahedra and 172 reflexive polyhedra with numbers of
vertices between 5 and 10.
In this list of 632 polyhedra, there are actually only 146
distinct new types of polyhedra, as shown in Table~\ref{Tab146}. 
More information about them can be obtained from~\cite{FrancoFTP}: we
leave their more detailed study to later work.

 \begin{table}[!ht]
\centering
\caption{ \it The 146 distinct new polyhedra, representing new
$K3$ spaces, that are obtainable as double
intersections of projective vectors in the `tetrahedron
subalgebra' containing only point- and segment-polyhedra,
triangles and tetrahedra. Many of these have several different
realizations as double intersections: more details can be found
in~\cite{FrancoFTP}.}
\label{Tab146}
 \scriptsize
\vspace{.05in}
\begin{tabular}{|c||c|c|c|||c||c|c|c|||c||c|c|c|}
 \hline
$\aleph$&$N,N^*$&$V,V^*$&$Pic,Pic^*$&$\aleph$&$N,N^*$&$V,V^*$&$Pic,Pic^*$&$\aleph$&$N,N^*$&$V,V^*$&$Pic,Pic^*$\\
 \hline \hline
 $ 1$&$ 31, 6^*$&$ 6, 5^*$&$ 2, 18^*$&$ 51$&$ 14, 19^*$&$ 7, 6^*$&$ 13,
 8^*$&$ 101$&$ 22, 20^*$&$ 5, 5^*$&$ 10, 11^*$\\
 \hline
 $ 2$&$ 28, 9^*$&$ 7, 6^*$&$ 4, 16^*$&$ 52$&$ 26, 8^*$&$ 6, 5^*$&$ 4,
 17^*$&$ 102$&$ 20, 16^*$&$ 5, 5^*$&$ 10, 13^*$\\
 \hline
 $ 3$&$ 29, 7^*$&$ 6, 5^*$&$ 4, 17^*$&$ 53$&$ 25, 14^*$&$ 6, 6^*$&$ 7,
 13^*$&$ 103$&$ 24, 18^*$&$ 5, 5^*$&$ 9, 12^*$\\
 \hline
 $ 4$&$ 22, 8^*$&$ 6, 5^*$&$ 7, 16^*$&$ 54$&$ 15, 21^*$&$ 5, 5^*$&$ 12,
 10^*$&$ 104$&$ 15, 21^*$&$ 4, 4^*$&$ 14, 10^*$\\
 \hline
 $ 5$&$ 31, 9^*$&$ 6, 5^*$&$ 3, 17^*$&$ 55$&$ 22, 16^*$&$ 6, 6^*$&$ 9,
 11^*$&$ 105$&$ 21, 15^*$&$ 4, 4^*$&$ 10, 14^*$\\
 \hline
 $ 6$&$ 21, 12^*$&$ 7, 6^*$&$ 8, 13^*$&$ 56$&$ 12, 18^*$&$ 6, 7^*$&$ 13
 , 10^*$&$ 106$&$ 10, 26^*$&$ 5, 6^*$&$ 15, 7^*$\\
 \hline
 $ 7$&$ 17, 20^*$&$ 7, 7^*$&$ 11, 9^*$&$ 57$&$ 17, 13^*$&$ 6, 6^*$&$ 10
 , 13^*$&$ 107$&$ 10, 32^*$&$ 6, 6^*$&$ 16, 4^*$\\
 \hline
 $ 8$&$ 22, 14^*$&$ 6, 6^*$&$ 8, 13^*$&$ 58$&$ 24, 12^*$&$ 5, 5^*$&$ 7,
 14^*$&$ 108$&$ 19, 14^*$&$ 5, 5^*$&$ 11, 13^*$\\
 \hline
 $ 9$&$ 24, 12^*$&$ 6, 5^*$&$ 7, 14^*$&$ 59$&$ 15, 15^*$&$ 4, 4^*$&$ 14
 , 12^*$&$ 109$&$ 16, 26^*$&$ 5, 5^*$&$ 13, 8^*$\\
 \hline
 $ 10$&$ 20, 12^*$&$ 6, 6^*$&$ 10, 13^*$&$ 60$&$ 20, 11^*$&$ 7, 6^*$&$
 9, 14^*$&$ 110$&$ 12, 27^*$&$ 5, 5^*$&$ 15, 7^*$\\
 \hline
 $ 11$&$ 20, 20^*$&$ 6, 6^*$&$ 10, 10^*$&$ 61$&$ 10, 20^*$&$ 5, 6^*$&$
 16, 9^*$&$ 111$&$ 15, 15^*$&$ 5, 5^*$&$ 12, 13^*$\\
 \hline
 $ 12$&$ 13, 14^*$&$ 6, 6^*$&$ 13, 11^*$&$ 62$&$ 11, 14^*$&$ 6, 6^*$&$
 14, 12^*$&$ 112$&$ 10, 23^*$&$ 6, 6^*$&$ 15, 7^*$\\
 \hline
 $ 13$&$ 26, 8^*$&$ 6, 5^*$&$ 5, 17^*$&$ 63$&$ 24, 18^*$&$ 5, 5^*$&$ 8,
 12^*$&$ 113$&$ 6, 34^*$&$ 5, 6^*$&$ 18, 2^*$\\
 \hline
 $ 14$&$ 26, 7^*$&$ 6, 5^*$&$ 5, 17^*$&$ 64$&$ 16, 17^*$&$ 6, 6^*$&$ 11
 , 11^*$&$ 114$&$ 25, 11^*$&$ 5, 5^*$&$ 8, 15^*$\\
 \hline
 $ 15$&$ 18, 8^*$&$ 6, 5^*$&$ 9, 16^*$&$ 65$&$ 8, 26^*$&$ 5, 6^*$&$ 17,
 5^*$&$ 115$&$ 15, 15^*$&$ 4, 4^*$&$ 13, 13^*$\\
 \hline
 $ 16$&$ 24, 10^*$&$ 6, 6^*$&$ 6, 15^*$&$ 66$&$ 14, 11^*$&$ 7, 6^*$&$
 12, 14^*$&$ 116$&$ 14, 16^*$&$ 5, 5^*$&$ 12, 13^*$\\
 \hline
 $ 17$&$ 11, 11^*$&$ 4, 4^*$&$ 15, 15^*$&$ 67$&$ 8, 26^*$&$ 6, 7^*$&$
 17, 3^*$&$ 117$&$ 9, 27^*$&$ 5, 5^*$&$ 16, 6^*$\\
 \hline
 $ 18$&$ 21, 17^*$&$ 7, 7^*$&$ 9, 11^*$&$ 68$&$ 21, 19^*$&$ 6, 6^*$&$
 10, 10^*$&$ 118$&$ 10, 26^*$&$ 6, 6^*$&$ 16, 6^*$\\
 \hline
 $ 19$&$ 14, 15^*$&$ 6, 6^*$&$ 12, 11^*$&$ 69$&$ 12, 12^*$&$ 4, 4^*$&$
 14, 14^*$&$ 119$&$ 22, 14^*$&$ 5, 5^*$&$ 9, 14^*$\\
 \hline
 $ 20$&$ 23, 11^*$&$ 5, 5^*$&$ 7, 15^*$&$ 70$&$ 10, 17^*$&$ 5, 6^*$&$
 15, 11^*$&$ 120$&$ 7, 31^*$&$ 5, 6^*$&$ 17, 3^*$\\
 \hline
 $ 21$&$ 10, 20^*$&$ 7, 7^*$&$ 15, 7^*$&$ 71$&$ 9, 15^*$&$ 4, 4^*$&$ 16
 , 12^*$&$ 121$&$ 15, 15^*$&$ 5, 5^*$&$ 13, 12^*$\\
 \hline
 $ 22$&$ 7, 23^*$&$ 5, 6^*$&$ 17, 5^*$&$ 72$&$ 8, 23^*$&$ 5, 6^*$&$ 16,
 8^*$&$ 122$&$ 15, 15^*$&$ 4, 4^*$&$ 12, 14^*$\\
 \hline
 $ 23$&$ 10, 14^*$&$ 5, 6^*$&$ 15, 12^*$&$ 73$&$ 24, 12^*$&$ 6, 6^*$&$
 8, 14^*$&$ 123$&$ 19, 11^*$&$ 5, 5^*$&$ 10, 14^*$\\
 \hline
 $ 24$&$ 12, 12^*$&$ 6, 6^*$&$ 13, 13^*$&$ 74$&$ 19, 11^*$&$ 4, 4^*$&$
 11, 14^*$&$ 124$&$ 12, 18^*$&$ 6, 6^*$&$ 14, 10^*$\\
 \hline
 $ 25$&$ 6, 30^*$&$ 4, 4^*$&$ 18, 4^*$&$ 75$&$ 11, 19^*$&$ 4, 4^*$&$ 17
 , 10^*$&$ 125$&$ 11, 17^*$&$ 5, 5^*$&$ 14, 11^*$\\
 \hline
 $ 26$&$ 25, 11^*$&$ 6, 6^*$&$ 6, 14^*$&$ 76$&$ 19, 11^*$&$ 4, 4^*$&$
 10, 17^*$&$ 126$&$ 20, 14^*$&$ 6, 6^*$&$ 7, 14^*$\\
 \hline
 $ 27$&$ 12, 12^*$&$ 4, 4^*$&$ 16, 14^*$&$ 77$&$ 8, 24^*$&$ 5, 6^*$&$
 16, 7^*$&$ 127$&$ 14, 16^*$&$ 5, 5^*$&$ 13, 12^*$\\
 \hline
 $ 28$&$ 21, 9^*$&$ 4, 4^*$&$ 9, 17^*$&$ 78$&$ 31, 11^*$&$ 5, 5^*$&$ 5,
 16^*$&$ 128$&$ 19, 17^*$&$ 5, 5^*$&$ 11, 12^*$\\
 \hline
 $ 29$&$ 15, 15^*$&$ 5, 6^*$&$ 11, 12^*$&$ 79$&$ 20, 22^*$&$ 5, 5^*$&$
 11, 10^*$&$ 129$&$ 12, 24^*$&$ 5, 5^*$&$ 15, 8^*$\\
 \hline
 $ 30$&$ 12, 12^*$&$ 4, 4^*$&$ 14, 16^*$&$ 80$&$ 26, 10^*$&$ 6, 5^*$&$
 3, 17^*$&$ 130$&$ 12, 20^*$&$ 6, 6^*$&$ 13, 10^*$\\
 \hline
 $ 31$&$ 31, 8^*$&$ 5, 5^*$&$ 4, 17^*$&$ 81$&$ 26, 10^*$&$ 5, 5^*$&$ 7,
 16^*$&$ 131$&$ 12, 24^*$&$ 5, 5^*$&$ 14, 9^*$\\
 \hline
 $ 32$&$ 17, 11^*$&$ 6, 5^*$&$ 9, 16^*$&$ 82$&$ 19, 11^*$&$ 4, 4^*$&$
 10, 16^*$&$ 132$&$ 7, 26^*$&$ 5, 6^*$&$ 17, 5^*$\\
 \hline
 $ 33$&$ 20, 10^*$&$ 5, 5^*$&$ 9, 16^*$&$ 83$&$ 16, 14^*$&$ 5, 5^*$&$
 12, 14^*$&$ 133$&$ 11, 28^*$&$ 7, 7^*$&$ 15, 5^*$\\
 \hline
 $ 34$&$ 18, 12^*$&$ 5, 5^*$&$ 11, 14^*$&$ 84$&$ 14, 16^*$&$ 6, 6^*$&$
 12, 12^*$&$ 134$&$ 9, 33^*$&$ 5, 5^*$&$ 16, 4^*$\\
 \hline
 $ 35$&$ 15, 12^*$&$ 4, 4^*$&$ 13, 13^*$&$ 85$&$ 23, 13^*$&$ 5, 5^*$&$
 9, 14^*$&$ 135$&$ 14, 28^*$&$ 5, 5^*$&$ 14, 7^*$\\
 \hline
 $ 36$&$ 9, 21^*$&$ 4, 4^*$&$ 17, 9^*$&$ 86$&$ 23, 10^*$&$ 5, 5^*$&$ 8,
 15^*$&$ 136$&$ 10, 29^*$&$ 6, 6^*$&$ 15, 5^*$\\
 \hline
 $ 37$&$ 25, 17^*$&$ 6, 6^*$&$ 8, 12^*$&$ 87$&$ 14, 16^*$&$ 6, 5^*$&$
 14, 11^*$&$ 137$&$ 11, 25^*$&$ 5, 5^*$&$ 15, 8^*$\\
 \hline
 $ 38$&$ 15, 21^*$&$ 5, 5^*$&$ 13, 10^*$&$ 88$&$ 12, 18^*$&$ 6, 6^*$&$
 15, 10^*$&$ 138$&$ 17, 26^*$&$ 6, 6^*$&$ 12, 8^*$\\
 \hline
 $ 39$&$ 17, 10^*$&$ 6, 5^*$&$ 11, 15^*$&$ 89$&$ 29, 13^*$&$ 5, 5^*$&$
 6, 15^*$&$ 139$&$ 15, 18^*$&$ 5, 5^*$&$ 13, 11^*$\\
 \hline
 $ 40$&$ 10, 23^*$&$ 6, 6^*$&$ 16, 7^*$&$ 90$&$ 17, 19^*$&$ 5, 5^*$&$
 12, 11^*$&$ 140$&$ 11, 19^*$&$ 5, 5^*$&$ 16, 10^*$\\
 \hline
 $ 41$&$ 13, 28^*$&$ 7, 7^*$&$ 14, 6^*$&$ 91$&$ 11, 19^*$&$ 4, 4^*$&$
 16, 10^*$&$ 141$&$ 20, 25^*$&$ 5, 5^*$&$ 11, 9^*$\\
 \hline
 $ 42$&$ 24, 21^*$&$ 5, 5^*$&$ 9, 11^*$&$ 92$&$ 14, 16^*$&$ 6, 6^*$&$
 13, 11^*$&$ 142$&$ 10, 26^*$&$ 5, 5^*$&$ 16, 7^*$\\
 \hline
 $ 43$&$ 9, 24^*$&$ 5, 5^*$&$ 17, 7^*$&$ 93$&$ 10, 24^*$&$ 6, 6^*$&$ 15
 , 6^*$&$ 143$&$ 11, 25^*$&$ 6, 6^*$&$ 15, 7^*$\\
 \hline
 $ 44$&$ 12, 30^*$&$ 6, 6^*$&$ 15, 5^*$&$ 94$&$ 8, 34^*$&$ 5, 6^*$&$ 17
 , 3^*$&$ 144$&$ 9, 33^*$&$ 5, 5^*$&$ 17, 4^*$\\
 \hline
 $ 45$&$ 21, 9^*$&$ 5, 5^*$&$ 8, 16^*$&$ 95$&$ 14, 16^*$&$ 5, 5^*$&$ 14
 , 12^*$&$ 145$&$ 11, 13^*$&$ 5, 5^*$&$ 14, 13^*$\\
 \hline
 $ 46$&$ 16, 11^*$&$ 6, 5^*$&$ 11, 13^*$&$ 96$&$ 16, 15^*$&$ 7, 6^*$&$
 12, 12^*$&$ 146$&$ 9, 36^*$&$ 5, 5^*$&$ 17, 3^*$\\
 \hline
 $ 47$&$ 11, 16^*$&$ 7, 7^*$&$ 13, 9^*$&$ 97$&$ 11, 31^*$&$ 5, 5^*$&$
 16, 5^*$&$$&$ $&$ $&$ $\\
 \hline
 $ 48$&$ 26, 10^*$&$ 6, 5^*$&$ 6, 15^*$&$ 98$&$ 9, 30^*$&$ 6, 7^*$&$ 16
 , 4^*$&$$&$ $&$ $&$ $\\
 \hline
 $ 49$&$ 18, 12^*$&$ 6, 5^*$&$ 11, 13^*$&$ 99$&$ 14, 10^*$&$ 6, 5^*$&$
 12, 15^*$&$$&$ $&$ $&$ $\\
 \hline
 $ 50$&$ 12, 22^*$&$ 6, 7^*$&$ 14, 8^*$&$ 100$&$ 9, 28^*$&$ 6, 7^*$&$
 16, 5^*$&$$&$ $&$ $&$ $\\
 \hline
 \hline
 \end{tabular}
 \end{table}
\normalsize

The method described here has a very simple geometrical
interpretation. According to the chain structure, each ${ CY_3}$  
can have a complex internal structure, and correspondingly its vector can
be extended as a sum of two $K3$, three elliptic, four two-component
or five single-component extended vectors. 
Another nice feature of this chain structure is that it gives us
complete information about the integer lattice which determines all the
CY equations. Moreover, it also gives us the possibility of summarizing
the singularity structure of ${ CY_3}$ spaces. 
As we discussed in Section 8,
the ${ K3}$ polyhedron structure gives us a systematic
way of classifying the corresponding
Cartan-Lie algebra graphs.
It will be interesting to make a full corresponding analysis for ${ CY_3}$
hypersurfaces, taking duality into account.
This method could also provide the full classification of
Betti-Hodge topological numbers for ${CY_3}$ manifolds.
Moreover, this
algebraic method enables us to `walk' between different dimensions,
e.g., to classify ${CY_4}$, ${..._5}$,... manifolds (Figure \ref{gen1}). 
The greatest limitations may be our abilities to analyze this
algebra and/or the available computer resources.

A fuller analysis of our structural classification of 
the ${\vec {k}_5}$ vectors for $CY_3$ manifolds will be given in 
later work. An important aspect of this procedure is that we can study the
structures
of the positive-integer lattices which correspond to the ${\vec
{k}}$ vectors, introducing the corresponding modular (for
two-dimensional sublattices) and 
hypermodular  (for 3-, 4- or higher-dimensional lattices)
transformations. These yield duality groups that are
more general than the well-known $S, T$ and $U$ dualities,
including them as subgroups. 
Moreover, the study of the geometric properties of the 
one-dimensional complex torus, two-dimensional ${K3}$
hypersurfaces
and Calabi-Yau manifolds with dimensions $d = 3, 4, ...$
gives insight into the
possible rank and dimensions of the Lie algebras which may be important
for the understanding of the nature of the symmetries
used in high-energy physics.

\begin{center}

{\bf Acknowledgements}\\

{~~}\\

\end{center}

G.V. thanks H. Dahmen, G. Harigel, L. Fellin, V. Petrov, A. Zichichi 
and the CERN Theory Division for support. The work of D.V.N. was
supported in part by DOE grant no. DE-FG-0395ER40917.


\end{document}